\documentclass[floats,floatfix,amssymb,prd,twocolumn,nofootinbib,superscriptaddress]{revtex4-1}
\usepackage{amssymb,amsmath,verbatim,mathtools,needspace,enumitem,etoolbox,physics,microtype,ulem}
\usepackage[utf8]{inputenc}
\usepackage{graphicx}

\usepackage{xspace}
\usepackage{subcaption}
\usepackage{csvsimple}
\usepackage{pifont}
\usepackage{appendix}
\usepackage{multirow}
\usepackage[dvipsnames, usenames]{xcolor}
\usepackage{pgfplotstable}
\usepackage[unicode, colorlinks=true, linkcolor=linkcolor, citecolor=linkcolor, filecolor=linkcolor,urlcolor=linkcolor, pdfusetitle]{hyperref}
\usepackage{aas_macros}
\usepackage{multirow}
\usepackage{subcaption}
\usepackage{cleveref}
\usepackage{natbib,tabularx}
\usepackage{booktabs,subcaption}%

\usepackage{gensymb} %
\usepackage{amsbsy} %

\definecolor{linkcolor}{rgb}{0.0,0.3,0.5}
\definecolor{urlcolor}{rgb}{0.27,0.55,0.}
\definecolor{funcolor}{rgb}{0.65, 0.16, 0.16}

\newcolumntype{b}{>{\hsize=0.6\hsize\centering\arraybackslash}X}
\newcolumntype{X}{>{\hsize=0.3\hsize\centering\arraybackslash}\centering X}
\newcolumntype{Y}{>{\hsize=1.1\hsize\centering\arraybackslash}\centering X}

\newcommand{\chieff}{\ensuremath{\chi_{\mathrm{eff}}}\xspace}

\newcommand{\msun}{\ensuremath{M_{\odot}}\xspace}
\newcommand{\mchirp}{\ensuremath{\mathcal{M}}\xspace}
\newcommand{\mcs}{\ensuremath{\mathcal{M}^\mathrm{source}}\xspace}
\newcommand{\mcd}{\ensuremath{\mathcal{M}^\mathrm{det}}\xspace}
\newcommand{\mBHs}{\ensuremath{m_1^\mathrm{source}}\xspace}
\newcommand{\mNSs}{\ensuremath{m_2^\mathrm{source}}\xspace}
\newcommand{\mBHd}{\ensuremath{m_1^\mathrm{det}}\xspace}
\newcommand{\mNSd}{\ensuremath{m_2^\mathrm{det}}\xspace}
\newcommand{\mts}{\ensuremath{M_\mathrm{total}^\mathrm{source}}\xspace}

\newcommand{\tidal}{\ensuremath{\Lambda_2}\xspace}

\newcommand{\NRSurT}{\texttt{NRHybSur3dq8Tidal}\xspace}
\newcommand{\SEOBNSBH}{\texttt{SEOBNSBH}\xspace}
\newcommand{\IMRp}{\texttt{IMRp}\xspace}
\newcommand{\IMRNSBH}{\texttt{IMRNSBH}\xspace}
\newcommand{\IMRpT}{\texttt{IMRpT}\xspace}
\newcommand{\SEOB}{\texttt{SEOB}\xspace}
\newcommand{\SEOBT}{\texttt{SEOBT}\xspace}
\newcommand{\LEAplus}{\texttt{LEA+}\xspace}
\newcommand{\LEApluslong}{Lackey\_Tidal\_2013\_SEOBNRv2\_ROM\xspace}
\newcommand{\XHM}{\texttt{IMRXHM}\xspace}

\newcommand{\BSN}{\ensuremath{\mathrm{ln}\mathcal{B}}\xspace}

\newcommand{\dg}{\ensuremath{^\circ}\xspace}
\newcommand{\dl}{\ensuremath{D_L}\xspace}
\newcommand{\rns}{\ensuremath{r_\mathrm{NS}}\xspace}
\newcommand{\inc}{\ensuremath{\theta_{JN}}\xspace}
\newcommand{\si}{\ensuremath{\sim}\xspace}
\newcommand{\LNS}{\ensuremath{\Lambda_\mathrm{NS}}\xspace}
\newcommand{\sBH}{\ensuremath{|s_1|}\xspace}
\newcommand{\sBHz}{\ensuremath{s_{1,z}}\xspace}
\newcommand{\sNS}{\ensuremath{|s_2|}\xspace}
\newcommand{\sNSz}{\ensuremath{s_{2,z}}\xspace}

\newcommand{\beqn}{\begin{eqnarray}}
\newcommand{\enqn}{\end{eqnarray}}
\newcommand{\beq}{\begin{equation}}
\newcommand{\eeq}{\end{equation}}

\newcommand{\LIGOlabMIT}{\affiliation{LIGO Laboratory, Massachusetts Institute of Technology, 185 Albany St, Cambridge, MA 02139, USA}}
\newcommand{\MKI}{\affiliation{Department of Physics and Kavli Institute for Astrophysics and Space Research, Massachusetts Institute of Technology, 77 Massachusetts Ave, Cambridge, MA 02139, USA}}
\newcommand{\UNH}{\affiliation{Department of Physics \& Astronomy, University of New Hampshire, 9 Library Way, Durham NH 03824, USA}}
\newcommand\CIT{\affiliation{TAPIR 350-17, California Institute of
Technology, 1200 E California Boulevard, Pasadena, CA 91125, USA}}

\newcommand\CornellPhyAndAstro{\affiliation{Department of Physics, and Cornell
Center for Astrophysics and Planetary Science, Cornell University, Ithaca, New
York 14853, USA}}

\begin{document}
\title{Statistical and systematic uncertainties in extracting the source properties of neutron star - black hole binaries with gravitational waves}

\author{Yiwen Huang} 
\email{ywh@mit.edu}
\LIGOlabMIT \MKI

\author{Carl-Johan Haster} \LIGOlabMIT \MKI
\email{haster@ligo.mit.edu}

\author{Salvatore Vitale} \LIGOlabMIT \MKI
\email{salvatore.vitale@ligo.org}

\author{Vijay Varma}
\thanks{Klarman fellow}
\CIT \CornellPhyAndAstro

\author{Francois Foucart} \UNH

\author{Sylvia Biscoveanu} \LIGOlabMIT \MKI

\date{\today}

\begin{abstract}
Gravitational waves emitted by neutron star black hole mergers encode key properties of neutron stars -- such as their size, maximum mass and spins -- and black holes.
However, the presence of matter and the high mass ratio makes generating long and accurate waveforms from these systems hard to do with numerical relativity, and not much is known about systematic uncertainties due to waveform modeling.
We simulate gravitational waves from neutron star black hole mergers by hybridizing numerical relativity waveforms produced with the SpEC code with a recent numerical relativity surrogate \NRSurT. 
These signals are analyzed using a range of available waveform families, and statistical and systematic errors are reported.
We find that at a network signal-to-noise ratio (SNR) of 30, statistical uncertainties are usually larger than systematic offsets, while at an SNR of 70 the two become comparable. 
The individual black hole and neutron star masses, as well as the mass ratios, are typically measured very precisely, though not always accurately at high SNR. 
At a SNR of 30 the neutron star tidal deformability can only be bound from above, while for louder sources it can be measured and constrained away from zero. 
All neutron stars in our simulations are non-spinning, but in no case we can constrain the neutron star spin to be smaller than \si0.4 (90\% credible interval).
Waveform families whose late inspiral has been tuned specifically for neutron star black hole signals typically yield the most accurate characterization of the source parameters. Their measurements are in tension with those obtained using waveform families tuned against binary neutron stars, even for mass ratios that could be relevant for both binary neutron stars and neutron star black holes mergers.

\end{abstract}

\maketitle

\section{Introduction}
\label{Sec:Intro}

With the detection of the binary neutron star (BNS) merger GW170817, and the associated counterparts across all of the electromagnetic (EM) spectrum (AT2017gfo/GRB170817A)~\cite{TheLIGOScientific:2017qsa,Abbott:2018wiz,Coulter:2017wya,GBM:2017lvd,Soares-Santos:2017lru,Hallinan:2017woc,Monitor:2017mdv}, the era of multi-messenger astrophysics %
based on photons and gravitational waves (GWs) has begun, providing new tools to explore the universe.
The multi-messenger observation of GW170817 yielded constraints on the the neutron star equation of state (EoS)~\cite{Abbott:2018wiz} as well as on the heavy metal production in neutron star mergers~\cite{Monitor:2017mdv, GBM:2017lvd, Metzger:2016pju, Fernandez:2015use, Metzger:2011bv, Metzger:2010sy}, demonstrated that at least a fraction of the short gamma-ray bursts are produced by BNSs~\cite{Monitor:2017mdv}, and enabled the first-ever measurement of the Hubble constant based on standard sirens~\cite{Schutz:1986gp,Holz:2005df,Abbott:2017xzu}.
Even when an EM counterpart is not found, as was the case for the second BNS detection, GW190425~\cite{Abbott:2020uma}, GWs alone provide precious information, for example, about the masses and spins of neutron stars in binaries.
All of the component objects in GW170817 and GW190425 were consistent with having small or no spin, and the total mass of GW190425 was found to be significantly higher than that of any known galactic binary pulsar~\cite{Abbott:2020uma}. 

Heterogeneous binary systems made of one neutron star and one black hole (NSBH) have yet to be discovered. While they are usually expected to exist, to date no uncontroversial observational evidence has been found, and a 90\% upper limit on their merger rate has been set by advanced LIGO~\cite{TheLIGOScientific:2014jea} and Virgo~\cite{TheVirgo:2014hva} in the first and the second observing runs to be 610 Gpc$^{-3}$ y$^{-1}$~\cite{LIGOScientific:2018mvr}. 
The recently announced source GW190814~\cite{Abbott:2020khf}, might have been the first NSBH detected with GWs, but the unusually high mass of the lighter object in the binary, a 2.6~\msun compact object, leaves intact the possibility that GW190814 is in fact a BBH.
When detected, NSBHs will come with features that makes them potentially unique laboratories for physics, astrophysics and cosmology.
The large mass ratio\footnote{Note that two conventions exist for the mass ratio. 
The LIGO-Virgo collaboration usually defines the mass ratio in the range $[0,1]$. 
We will follow the opposite convention (primarily used in the numerical relativity community) and define $q=m_1/m_2$, with $m_1\geq m_2$.\label{Foot.MassRatio}} will enhance the effect of eventual spin precession~\cite{Apostolatos:1994mx}, making it easier to measure the black hole spin with good precision~\cite{Vitale:2014mka,Vitale:2016avz,Fairhurst:2019srr,Fairhurst:2019vut}.
Similarly, the impact of higher multipoles is larger for systems with large mass ratios, paving the way to tests of the multipolar structure of general relativity~\cite{Thorne:1980ru, LIGOScientific:2020stg}. 
Tidal effects might also be present, if the black hole is light enough, or with significant spin~\cite{Foucart:2010eq, Pannarale:2011pk, Kyutoku:2011vz, Foucart:2012nc, Foucart:2012vn, Kawaguchi:2015bwa, Foucart:2016vxd, Foucart:2018rjc, Pannarale:2015jka, Ascenzi:2018mwp} (otherwise the neutron star will cross the event horizon before it can be significantly disrupted).
Furthermore, the potential presence of significant spin-induced orbital precession would break the degeneracy between luminosity distance and orbital inclination, which could make NSBHs significant contributors to the measurement of the Hubble constants with standard sirens~\cite{Chen:2018omi}.
Finally, a precise localization of the host galaxy and the study of the light-curves associated with the EM emission, which is expected as long as the neutron star is tidally disrupted~\cite{Lee:2007js,Nakar:2007yr,Clark:2014jpa,Pannarale:2014rea,Ascenzi:2018mwp}, will give precious information about the environment in which NSBHs form.

Key to the interpretation of GW detections and signal analysis is the development of accurate and computationally efficient GW waveforms, which are used to measure the parameters of the signal by matching the model waveform\footnote{In what follows we will use ``waveform model'' and ``waveform approximant'' as synonyms, as they are both commonly used in GW literature.} against the GW data.
GW models are usually calibrated against waveforms obtained directly with numerical relativity (NR) codes, which solve Einstein's equations on a computer~\cite{Pretorius:2005gq,Campanelli:2005dd,Baker:2005vv,Boyle:2019kee}.
Unfortunately, the very features that make NSBH exciting systems to study also make their resulting GW emission challenging to simulate numerically. 
Currently, there are 7 high-resolution NSBH waveforms~\cite{Foucart:2013psa,Foucart:2014nda,Hinderer:2016eia,Chakravarti:2018uyi,Foucart:2018lhe} publicly available in the Simulating eXtreme Spacetimes (SXS) GW database\footnote{As opposed to 2018 binary black hole (BBH) waveforms in the latest SXS catalogue~\cite{Foucart:2012vn,Boyle:2019kee}.}~\cite{SXSDatabase}, and 134 lower-resolution NSBH waveforms generated using the SACRA code, which have been used to calibrate various waveform approximants but are not publicly available~\cite{Lackey:2013axa,Yamamoto:2008js}.

Furthermore, one usually does not directly use NR waveforms to measure the parameters of detected compact binary coalescences (CBCs) (but see~\cite{Pankow:2015cra,Lange:2018pyp}), due to their high individual computational cost and sparsity across the parameter space.
Instead, surrogate, phenomenological (IMRPhenom) or effective-one-body (EOB) GW models are produced, which are calibrated against NR simulations. 
To make these waveform models fast enough to be calculated millions of times, as required by stochastic samplers, and in some cases due to limitations in the very NR simulations that the models are calibrated against, only some of the relevant physical features are included (e.g. spin precession but not tidal deformability). 
Due to the lack of a large NR database, and the fact that all of the physics that is relevant to describe a CBC can induce \emph{measurable} effects in NSBHs\footnote{For example, higher order modes are formally present in all CBC signals, but are suppressed for systems close to equal mass. This is the reason why waveform models that do not model them perform well with most of the binary black holes detected to date. The same will not necessarily be true for NSBHs.}, these systems are potentially very prone to systematic errors due to waveform modeling. 

In this paper we create hybrids from recent NSBH NR simulations to produce full inpiral-merger-ringdown waveforms that are then added to the data stream of a three-detector gravitational-wave network made of the two advanced LIGO~\cite{Harry:2010zz} and the advanced Virgo~\cite{TheVirgo:2014hva} detectors. 
We simulate signals at various mass ratios, signal-to-noise ratios (SNRs) and orbital orientation, and measure their parameters with stochastic samplers, using a suite of phenomenological and effective-one-body models. 
Our work significantly extends what done by Ref.~\cite{Chakravarti:2018uyi}, which explored NSBH waveform systematics by only looking at waveform overlaps, instead of performing a full Markov Chain Monte Carlo (MCMC) measurement of all of the binary parameters. 
Ref.~\cite{Kumar:2016zlj} looked at parameter estimation for NSBH sources, but only used a single waveform model, \LEApluslong~\cite{Lackey:2011vz,Taracchini:2013rva,Lackey:2013axa}, and did not measure extrinsic parameters or source-frame masses.

The rest of the paper is organized as follows: in Sec.~\ref{Sec.Method} we describe the generation of simulated signals and the source characterization algorithm; we report our main findings in Sec.~\ref{Sec.Results} and then conclude in Sec.~\ref{Sec.Conclusions}.
Given the large volume of data produced, we have to make drastic cuts in what can be reported in the main body of the paper without disrupting the flow: the appendices provide extensive summary tables for all of the analyses we performed and additional details about their setup.

\section{Method}
\label{Sec.Method}

Gravitational waves emitted by a binary of compact objects in a quasi-circular orbit can be described by 15 parameters, including masses, spin vectors, sky position, luminosity distance and orbital orientation. 
Each neutron star in the binary adds one extra parameter of tidal deformability\footnote{In this work, we will assume that black holes are not tidally deformed, and that all finite-size effects can be accurately modeled as functions of the tidal deformability of neutron stars.}.

As mentioned above, our goal is to verify if current waveform approximants can be used to accurately constrain the unknown parameters of NSBH systems. 
If not, we wish to check which parameters are more susceptible to biases, and at which SNR these biases become significant compared to the statistical uncertainties. 
In this section, we describe the generation and construction of inspiral-merger-ringdown NSBH waveforms used in this work and the data analysis approaches to measure their parameters.

\subsection{Simulated signals}\label{Sec.NR}

One can decompose the GW strain into a sum of spin-weighted spherical harmonics~\cite{Thorne:1980ru} as:

\begin{equation}\label{Eq.HM}
h(t,\iota,\psi_0)= \sum^{\infty}_{l=2}\sum^{l}_{m=-l}h_{lm}(t)^{-2}Y_{lm}(\iota,\psi_0)
\end{equation}
where $\iota$ is the angle between the line of sight and the orbital angular momentum, ${}^{-2}Y_{lm}$ are spin -2 weighted spherical harmonics, and $\psi_0$ is the initial binary phase~\cite{Maggiore}. 

In general, the $(l=2,|m|=2)$ mode is dominant~\cite{Thorne:1980ru,Blanchet:2013haa,Ruiz:2007yx,Blanchet:2008je, Kidder:2007rt,Healy:2013jza} and higher order modes (HOMs) are suppressed. 
This is particularly true for low-mass systems with mass ratio close to unity~\cite{Kidder:2007rt, Pan:2011gk, Varma:2016dnf, Capano:2013raa, Varma:2014jxa, Bustillo:2015qty, Graff:2015bba, Shaik:2019dym}.
The impact of HOMs is also reduced in systems which are observed close to ``face-on'' i.e. with the orbital angular momentum aligned with the line-of-sight, since in this case the angular structure of the spin-weighted spherical harmonics suppresses the magnitude of the higher order terms relative to the dominant $(l=2,|m|=2)$ mode.
This is consistent with the fact that GW190412~\cite{LIGOScientific:2020stg} and GW190814~\cite{Abbott:2020khf}, the first two CBC detections with visible imprints of HOMs, are also the GW events with the most asymmetric mass ratio reported to date, at around 4 and 9, respectively.

\begin{table*}[t!]
\centering
\begin{tabularx}{0.90\textwidth}{b|b|bbbb|b}
\toprule\toprule
  NR Waveforms  & $M_{BH}$/$M_\odot$ & $M_{NS}$/$M_\odot$ & NS equation of state & $\Lambda_{NS}$ & $r_{NS}$/km & Modes $(l,|m|)$\\ [0.5ex]
 \midrule
  q6 ~\cite{Foucart:2013psa} & 8.4 & 1.4 & $\Gamma2$($\kappa=92.12$)\footnote{$P=\kappa \rho^2+\rho T$} & 526 & 13.3 & (2,2) (3,3) (2,1) (4,4) (5,5)\\ [0.5ex]
   \midrule
 q3~\cite{Foucart:2018lhe} & 4.05 & 1.35 & H1\footnote{Piecewise polytrope equation of state defined in~\cite{Read:2008iy}.}%
 & 624 & 12.3 & (2,2) (3,3) (2,1) (4,4) \\ [0.5ex]
   \midrule
 q2~\cite{Foucart:2018lhe}& 2.8 & 1.4 & $\Gamma2$($\kappa=101.45$)  & 791 &14.4& (2,2) (3,3) \\ [0.5ex]
\bottomrule\bottomrule
\end{tabularx}
\caption{Numerical relativity waveforms used for the post-inspirial part of the simulated signals. Full waveforms are obtained by hybridizing with \NRSurT waveforms. See the body for mode details. Note all the neutron stars and black holes are non-spinning. }
\label{Table.WF}
\end{table*} 

To generate realistic GWs emitted from NSBHs we use NR simulations carried out by the SXS collaboration. We consider 3 different NR simulations~\cite{Foucart:2013psa,Foucart:2018lhe} produced by the SXS collaboration using the SpEC code~\cite{Foucart:2012vn,Duez:2008rb} at the highest available resolution. 
In Table~\ref{Table.WF}, we report their corresponding masses, spins, HOMs, as well as the tidal deformability of the neutron star, defined as: 
\begin{equation}\label{Eq.Lambda2}
\Lambda_{\mathrm{NS}}=\frac{2}{3}k_2\left(\frac{R_{\mathrm{NS}}}{M_{\mathrm{NS}}}\right)^5
\end{equation}
where $R_{\mathrm{NS}}$ and $M_{\mathrm{NS}}$ are the radius and mass of the neutron star, $k_2$ is its tidal Love number describing the susceptibility of its shape to changes in response to a tidal potential~\cite{Mora:2003wt}, and we assume G=c=1. 

The NR waveforms used in this work are for non-spinning neutron stars and black holes\footnote{We will report on spinning NSBHs in a follow-up study, as more NR simulations become available.}. 
This limitation does not make our analysis less relevant or urgent since most of the black holes and all of the neutron stars discovered with gravitational waves to date are consistent with having small or no spin~\cite{LIGOScientific:2018mvr,Roulet:2018jbe}.
In the rest of this paper we will often use the mass ratios of the systems, as reported in the first column of Table~\ref{Table.WF}, to refer to the individual NSBH simulations. 

The EoS for cold, supranuclear matter in these simulations is such that the resulting tidal deformability is toward the high end of the region still allowed by previous GW observations~\cite{Abbott:2018wiz,Abbott:2020uma}.
Specifically, the q6 and q2 NR waveforms use a simple $\Gamma$-law EoS, where the pressure $P$, density $\rho$, temperature $T$, and specific internal energy $\epsilon$ are related by $P=\kappa \rho^\Gamma+\rho T$, $\epsilon=P/\rho/(\Gamma-1)$. 
Both simulations use $\Gamma=2$, while $\kappa=92.12$ for q6 and $\kappa=101.45$ for q2\footnote{Simulations using a $\Gamma$-law EoS can in theory be rescaled to any mass, at constant mass ratio and neutron star compactness. 
However, we do have to choose a mass scale when injecting the waveform into detector data. 
In this work, we set the mass of the neutron star to $1.4M_\odot$.}. 
This results in the neutron star having a tidal deformability of $\Lambda=526$ for q6, and $\Lambda=791$ for q2. 
Finally, q3 uses a piece-wise polytropic equation of state (H1, defined in~\cite{Read:2008iy}). 
For the $1.35M_\odot$ neutron star considered in the simulation, this leads to a tidal deformability of $\Lambda=624$. 

We note that while q2 and q6 are relatively long waveforms by the standard of hydrodynamic simulations ($>20$ cycles), q3 is comparatively short (13.3 cycles). 
In all cases, these simulations contain only the last few cycles before the two compact objects merge, while we need to simulate the full GW signals starting at the low frequency cut-off of gravitational-wave detectors (i.e. 20~Hz) for the purpose of this study.

We use the hybridization scheme described in~\cite{Varma:2018mmi} to combine the NR simulations with models for the early-inspiral section of the waveform.
The late and post-inspiral phases from NR are smoothly attached to the early-inspiral section predicted by the \texttt{NRHybSur3dq8Tidal}\footnote{\texttt{NRHybSur3dq8Tidal} is constructed by adding post-Newtonian tidal effects to the underlying BBH model \texttt{NRHybSur3dq8}~\cite{Varma:2018mmi}. Therefore it only includes the inspiral part of the waveform.} model~\cite{Barkett:2019tus}.

We use the infrastructure described in~\cite{Schmidt:2017btt} to project the hybrid signals into the data streams of a network of 2 aLIGO detectors (Hanford and Livingston) and the Virgo detector.
To better isolate biases due to waveform systematics from offsets due to gaussian noise fluctuations, we work with a zero-noise realization of the data~\cite{Vallisneri:2007ev}, i.e. a data stream where the noise is zero at each time or frequency bin (whereas the noise power spectral density (PSD) itself is non-zero).

We probe the effect of the orbital orientation on the results by simulating every source at two different inclinations\footnote{Defined in this work as \inc, the angle between the line-of-sight vector and the total angular momentum.}, representative of a ``typical''~\cite{Schutz:2011tw} detection (30\dg, ``face-on'') and of a high-inclination system (70 degrees, close to ``edge-on''). 
As mentioned above, larger inclinations should make the effects of HOMs more visible and, conversely, increase the bias when these effects are not taken into account but where they would have a significant contribution to the overall signal~\cite{Varma:2014jxa,Bustillo:2015qty}. 
Finally, all of these systems are put at two distances to give a network SNR of 30 (comparable to the loudest CBC discovered to date) and 70.
The masses given in Table~\ref{Table.WF} are to be interpreted as detector-frame masses, with the astrophysically relevant source-frame masses being smaller by a factor of $(1+z)$, with $z$ being the redshift of the source.
Strictly speaking, the masses reported by NR simulations should be treated as defined in the source-frame, but 
 given the proximity of our sources (Appendix ~\ref{App.FullPe}) these differ by at most a few percent, affecting the mapping between masses and NS radius by an amount much smaller than either statistical or systematic uncertainties. 
In the rest of the paper we will only report on the measurement of the source-frame masses.
Finally, the sky location of all sources is fixed to the (arbitrary) value of $60$\dg for both right ascension and declination. 

\subsection{Source characterization and waveform models}

With the data stream from all detectors in hand, $\boldsymbol{d}$, we want to measure the unknown source parameters, $\boldsymbol {\theta}$.
We perform Bayesian inference~\cite{Veitch:2014wba, TheLIGOScientific:2016wfe}, and calculate the posterior probability density function (PDF) for all source parameters:

\begin{equation}
\label{Eq.BayesTheorem}
p({\boldsymbol {\theta}}|{\boldsymbol {d}},H) = \frac{p({\boldsymbol {\theta}}|H)p({\boldsymbol {d}}|{\boldsymbol {\theta}},H)}{p({\boldsymbol {d}}|H)}
\end{equation}
where $p({\boldsymbol {\theta}}|H)$ is the prior probability density of ${\boldsymbol {\theta}}$, under the model $H$, while the second term in the numerator is the likelihood:

\begin{equation}
\label{Eq.Likelihood}
p({\boldsymbol {d}}|{\boldsymbol {\theta}},H) \propto \exp(-\frac{1}{2} \langle {\boldsymbol d}- h({\boldsymbol\theta})| {\boldsymbol d} -h({\boldsymbol\theta})\rangle)
\end{equation}
where we have defined the noise weighted scalar product:

\begin{equation}
\label{Eq.Inner Product}
\langle a(\boldsymbol {\theta}, f)|b(\boldsymbol {\theta}, f) \rangle \equiv 4\int^{f_\mathrm{high}}_{f_\mathrm{low}}\frac{a(\boldsymbol {\theta},f) b(\boldsymbol {\theta},f)^*}{S_n(f)}df
\nonumber\end{equation}
with ${S_n(f)}$ being the PSD of the detector noise. 
In this work, we use the design sensitivity for aLIGO and Virgo~\cite{Aasi:2013wya,UpdatedaLIGOdesign-T1800044}. 

Finally, the normalization constant $p({\boldsymbol {d}}|H )$ is the evidence for the model $H$:

\begin{equation}
\label{Eq.Evidence}
Z = p({\boldsymbol {d}}| H) = \int \mathrm{d}{\boldsymbol {\theta}} p({\boldsymbol {d}}|{\boldsymbol {\theta}},H) p(\boldsymbol{\theta}|H) 
\end{equation}

The choice of the waveform approximants, i.e. the waveform models we use to characterize the hybrid waveforms described in the previous section, enters the analysis through the term $h(\boldsymbol{\theta})$ in the likelihood. 

At the time of writing, no waveform was available in the LIGO Algorithm library~\cite{lalsuite} that accounts for all of tidal effects, spin precession and higher order modes. We therefore use a range of approximants that have some but not all of these features. 
These are reported in Table~\ref{Table.Approx}, together with a list of the physical features that are included and, when relevant, their range of validity (which usually restricts the mass ratio and/or the black hole spin).
Details on the priors assumed in the analyses are given in~\Cref{App.Priors}.

\begin{table*}[t!]
\centering
\begin{tabularx}{0.99\textwidth}{c|bbbbc}
\toprule\toprule
  Waveform Approximant & Tides & Precession & HOMs $(l,|m|)$ & NSBH-amplitude correction & Validity Range \\ [0.5ex]
 \midrule
 SEOBNRv4\_ROM (\SEOB)~\cite{Bohe:2016gbl, Purrer:2014fza} & &  &  &  &  \\ 
  [0.5ex]
   \midrule
  SEOBNRv4\_NRTidal\_ROM (\SEOBT)~\cite{Bohe:2016gbl,Purrer:2014fza, Dietrich:2017aum, Dietrich:2018uni} & \checkmark &  &  &  & \\ 
  [0.5ex]
  \midrule
 SEOBNRv4\_ROM\_NRTidalv2\_NSBH & \multirow{2}{*}{\checkmark} &  &  & \multirow{2}{*}{\checkmark} & \multirow{2}{*}{BH spin=[-0.5,0.8]}  \\ 
    (\SEOBNSBH)~\cite{Bohe:2016gbl, Purrer:2014fza,SEOBNRNSBH} &&&&&\\ 
  [0.5ex]
   \midrule
    \LEApluslong  & \multirow{2}{*}{\checkmark} &  & & \multirow{2}{*}{\checkmark} & q=[2,5],  \\ 
    (\LEAplus)~\cite{Lackey:2011vz,Taracchini:2013rva,Lackey:2013axa,Kumar:2016zlj}& & & & & BH spin=[-0.5,0.5] \\
  [0.5ex]
   \midrule
IMRPhenomNSBH (\IMRNSBH)~\cite{PhenomNSBH} & \checkmark &  &  & \checkmark & BH spin=[-0.5,0.5] \\
 [0.5ex]
   \midrule
IMRPhenomPv2 (\IMRp)~\cite{Hannam:2013oca, Husa:2015iqa, Khan:2015jqa} &  & \checkmark &  & &  \\ 
  [0.5ex]
   \midrule
  IMRPhenomPv2\_NRTidal (\IMRpT)~\cite{Hannam:2013oca, Husa:2015iqa, Khan:2015jqa,Dietrich:2017aum, Dietrich:2018uni} & \checkmark & \checkmark &  &  &  \\ 
  [0.5ex]
   \midrule
 \multirow{2}{*}{IMRPhenomXHM (\XHM)~\cite{Garcia-Quiros:2020qlt,Garcia-Quiros:2020qpx}} &  &  & (2,1),(3,2), &  & \\ 
    \vspace{-1\baselineskip}
 &&&(3,3),(4,4)&&\\
  [0.5ex]

\bottomrule\bottomrule
\end{tabularx}
\caption{Waveform approximants used to characterize the source parameters of the simulated NSBH signals. We report their full name and a short label (in typewriter fonts) used in the body of the paper, whether they support tides, spin precession, HOMs (if yes, which modes), correction to their amplitude tuned for NSBH sources, and eventual restrictions in their parameter space.}
\label{Table.Approx}
\end{table*}

The waveform approximants used in this study are constructed following either the EOB formalism~\cite{Buonanno:1998gg,Pan:2011gk,Barausse:2011ys,Hinder:2013oqa,Hinderer:2013uwa, Bohe:2016gbl,Taracchini:2013rva}, or based on the phenomenological extension to analytical post-Newtonian waveforms (IMRPhenom)~\cite{Santamaria:2010yb,Husa:2015iqa, Khan:2015jqa, Pratten:2020fqn, Garcia-Quiros:2020qlt,Garcia-Quiros:2020qpx}.
Both approaches smoothly extend the inspiral waveforms with models of binary merger-ringdown, calibrated against a set of spin-aligned BBH NR waveforms.

SEOBNRv4\_ROM (henceforth:~\SEOB)~\cite{Bohe:2016gbl, Purrer:2014fza} describes BBH inspiral-merger-ringdown signals with general spins aligned to the orbital angular momentum. 
For computational efficiency, we evaluate the likelihood (c.f Eq.~\eqref{Eq.Likelihood}) using a reduced-order-quadrature rule (ROQ)~\cite{Smith:2016qas} version of SEOBNRv4 expressed in the frequency-domain.
SEOBNRv4\_ROM\_NRTidal (\SEOBT)~\cite{Bohe:2016gbl,Purrer:2014fza, Dietrich:2017aum, Dietrich:2018uni} builds on the baseline BBH model \SEOB by adding a correction of the waveform phase through a prescription of tidal effects found in BNS systems, calibrated against a set of BNS NR simulations~\cite{Dietrich:2017aum, Dietrich:2018uni}.
 \LEApluslong (\LEAplus)~\cite{Lackey:2011vz,Taracchini:2013rva,Lackey:2013axa,Kumar:2016zlj} adds both phase and amplitude corrections specific to NSBH systems, but is constructed from a reduced order model (ROM) of the older SEOBNRv2 BBH baseline waveform model~\cite{Taracchini:2013rva}. 
We also use a ROQ implementation for \LEAplus.
SEOBNRv4\_ROM\_NRTidalv2\_NSBH (\SEOBNSBH)~\cite{SEOBNRNSBH} is also based on \SEOB and dedicated to describing NSBH systems, but adds both a phase correction (through an updated formalism to the one included in \SEOBT~\cite{Dietrich:2019kaq}) as well as corrections to the waveform amplitude based on the model of~\cite{Pannarale:2015jka}.

From the IMRPhenom waveform family we use IMRPhenomNSBH (\IMRNSBH)~\cite{PhenomNSBH} constructed from the aligned-spin BBH baseline IMRPhenomC~\cite{Santamaria:2010yb} with updated phase~\cite{Dietrich:2019kaq} and NSBH-specific amplitude~\cite{Pannarale:2015jka} corrections similar to \SEOBNSBH.
As a comparison to other GW analyses, we also include a ROQ implementation of IMRPhenomPv2 (\IMRp)~\cite{Hannam:2013oca, Husa:2015iqa, Khan:2015jqa}, which has been used for the majority of CBC analyses in recent years.
This waveform is based on the newer aligned-spin IMRPhenomD~\cite{Husa:2015iqa,Khan:2015jqa} BBH baseline, extended to also capture spin-induced orbital precession through an effective precession formalism~\cite{Hannam:2013oca}. 
We also use IMRPhenomPv2\_NRTidal (\IMRpT)~\cite{Hannam:2013oca, Husa:2015iqa, Khan:2015jqa,Dietrich:2017aum, Dietrich:2018uni}, which further augments \IMRp by adding a phase correction based on the same BNS tidal description as used in \SEOBT\footnote{Note that the two BNS NRTidal models, \SEOBT and \IMRpT, do not include a description for the post-merger section (either a BH ringdown, or a NS remnant oscillation) of the waveform.}.
Finally, we use IMRPhenomXHM (\XHM)~\cite{Garcia-Quiros:2020qlt,Garcia-Quiros:2020qpx}, which is based off the recent IMRPhenomXAS model~\cite{Pratten:2020fqn}, and describes aligned-spin BBH waveforms including HOMs. 
Note that \XHM does not include any phase or amplitude corrections from the presence of a NS in the binary. 

Overall, the choice of waveform models used for this study is determined by a compromise between covering a large variety of families and physics, while keeping the computational cost reasonable. 
This second factor is the reason why we do not include other waveform approximants with HOMs, for example the time-domain \texttt{SEOBNRv4HM}~\cite{Cotesta:2018fcv,Ossokine:2020kjp}\footnote{We note that a frequency-domain ROM of the aligned-spin \texttt{SEOBNRv4HM}~\cite{Cotesta:2018fcv} model was made available as this study reached completion~\cite{Cotesta:2020qhw}.}.

We analyze all of the hybrid signals we generated with all of these waveform models (with some exception for the \LEAplus model, used for q3 only due to its limited range of validity on mass ratio). 
As mentioned above, none of the waveform families account for all of the relevant physical effects. 
Given that none of the hybrid waveform we are simulating has precessing spins, we do not expect large penalty for waveforms that do not model spins, while the lack of tides and HOMs might have a visible impact, depending on the mass ratio, inclination angle, and SNR of the source systems. 

\section{Results}
\label{Sec.Results}

In this section we summarize the main findings of our study, with sections dedicated to the most significant astrophysical parameters that can be inferred from GW observations. 
As mentioned above, each of the signals is analyzed at two different values of the inclination angle ($30$\dg and $70$\dg) and two values of network SNR ($30$ and $70$).
Unless otherwise stated, we quote the 90\% credible intervals (CI), either absolute or relative to the true value. 
Given the very large number of configurations, we will not report all uncertainties in the main body of the paper. 
The interested reader will find extensive tables in Appendix~\ref{App.FullPe}.

\subsection{Masses}
\label{Sec:Masses}

\subsubsection{Chirp mass}
\label{Sec:ChirpMass}

For low-mass CBCs, the best constrained parameter is the chirp mass \mchirp, defined as

\begin{equation}\label{Eq.mc}
\mathcal{M} = \frac{(m_1m_2)^{3/5}}{{(m_1+m_2)}^{1/5}}\;,
\end{equation}
where $m_1$ and $m_2$ are the component masses\footnote{Following the GW literature, we will use $m_1>m_2$ (cf. footnote~\ref{Foot.MassRatio}).}, as this parameter enters the phase of the GW signal already at the leading-order~\cite{Buonanno:2009zt}.

\mchirp also appears in the waveform amplitude, together with the luminosity distance \dl and the redshift $z$:

\begin{equation}\label{Eq.WFamplitude}
\mathcal{A}\sim {\left((1+z) \mchirp\right)}^{5/3}/{D_L}\;.
\end{equation}

While this would suggest that the two parameters are positively correlated~\cite{Roulet:2018jbe}, in practice for CBCs with NSs and/or stellar mass BHs, the phasing evolution determines the chirp mass so precisely that it can be treated as known in the amplitude of the signal (which is usually measured much less precisely).
Indeed, for NSBH systems like those reported here, there are $\mathcal{O}(1000)$ observable inspiral cycles, leading to a precise \mchirp measurement from the waveform phase alone.
For example, for the sources we analyze, the typical fractional uncertainty for \mchirp is $\lesssim 1\%$ whereas the luminosity distance has fractional uncertainties of $\sim 50\%$, due to its correlation with the orbital orientation, see~\Cref{Sec:ExtrinsicParams}.

\begin{figure}
\centering
\begin{subfigure}[b]{0.4\textwidth}
   \includegraphics[width=1\linewidth]{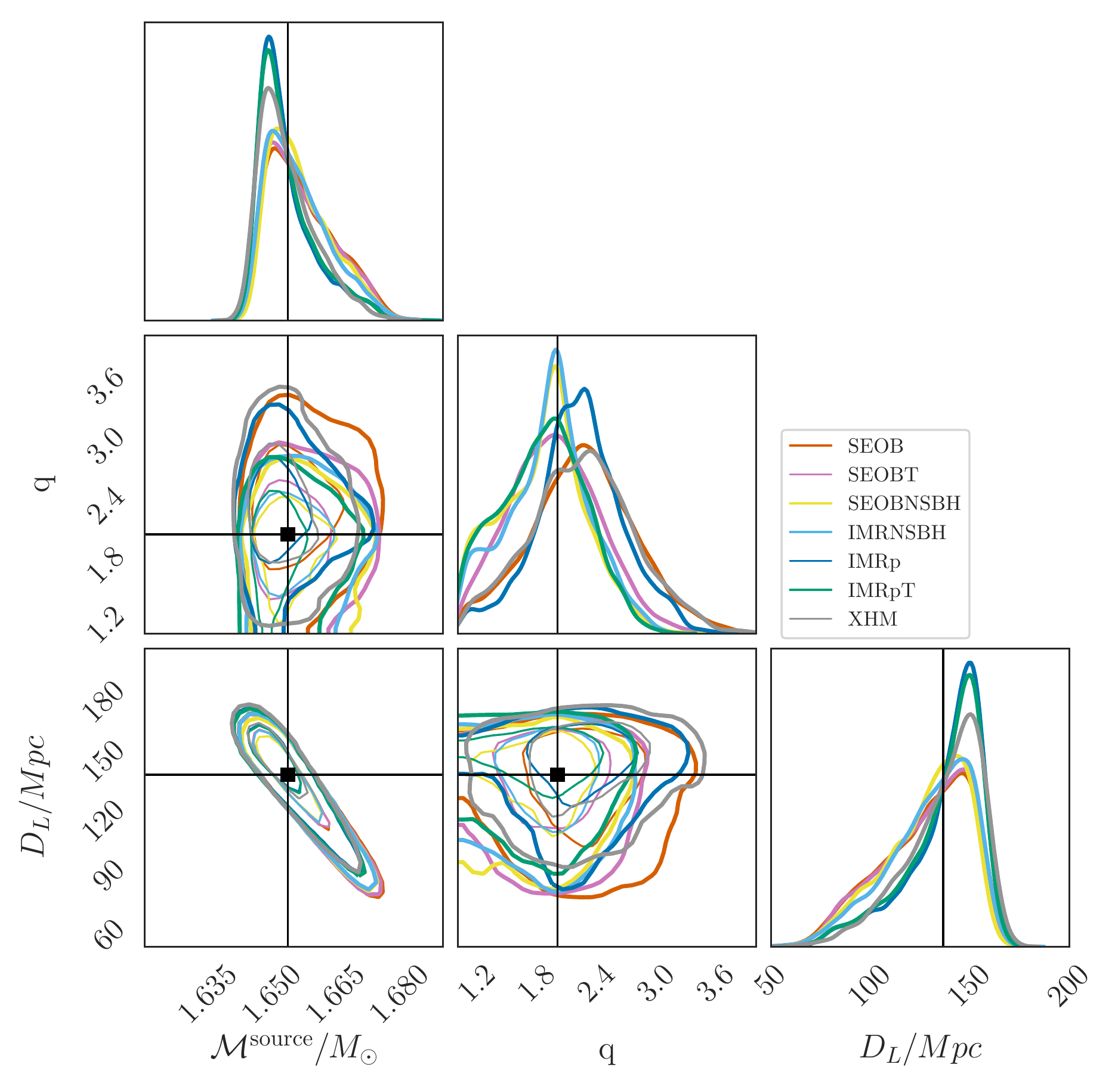}
   \vspace{-1.5\baselineskip}
   \caption{SNR 30.}
   \label{Fig.q2_mc_SNR30_inc30} 
\end{subfigure}
\begin{subfigure}[b]{0.4\textwidth}
   \includegraphics[width=1\linewidth]{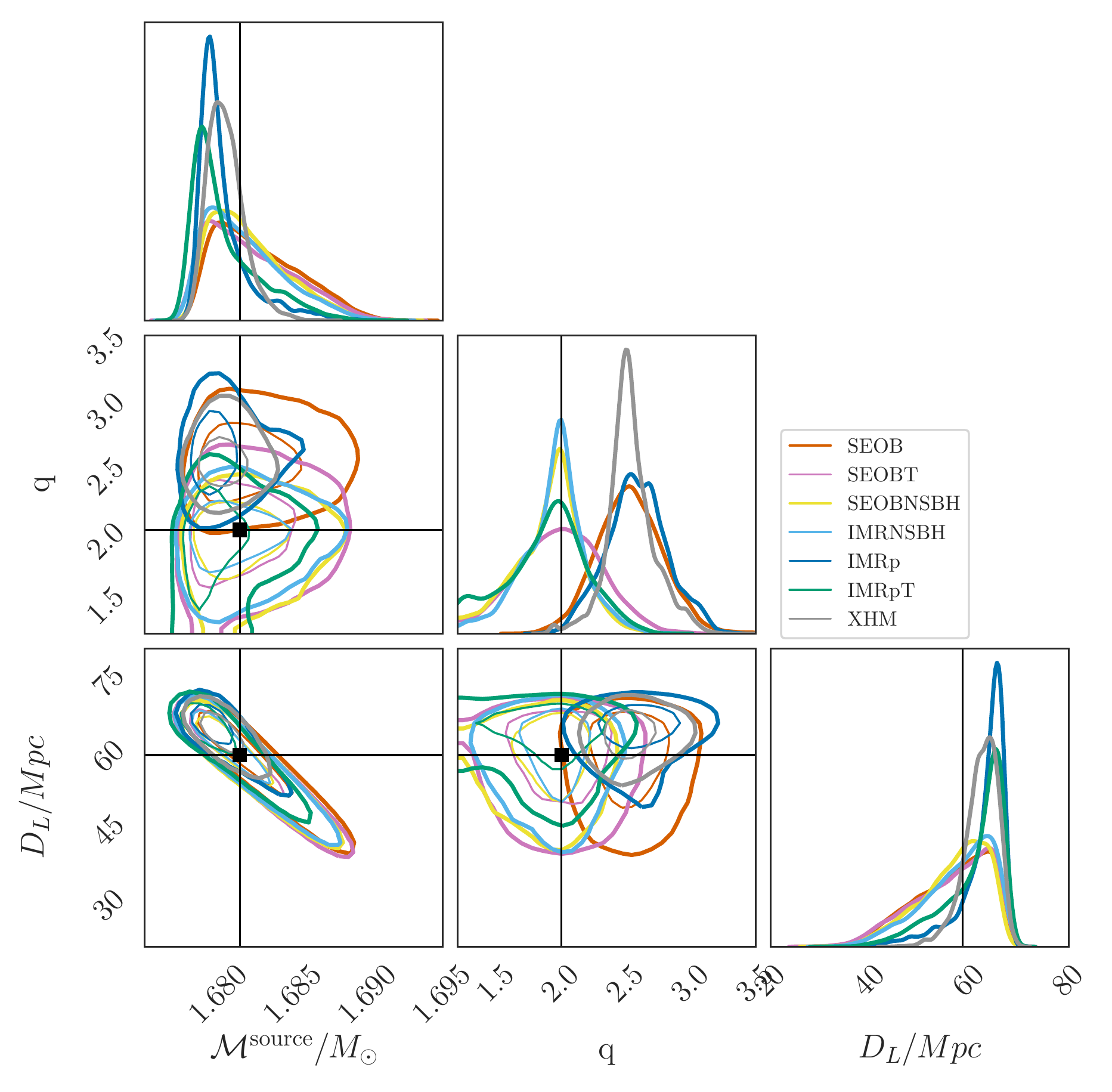}
   \vspace{-1.5\baselineskip}
   \caption{SNR 70.}
   \label{Fig.q2_mc_SNR70_inc30} 
\end{subfigure}
\caption{Corner plot of posterior distributions for chirp mass \mcs, mass ratio q, and luminosity distance \dl, recovered by different approximants for $q=2$, inclination $30^{\circ}$. The thin (thick) lines mark the 50\% (90\%) contour, same for all corner plots to follow.}
\label{Fig.mc_BHNSq2s0_inc30}
\end{figure}

\begin{figure}
\centering
\begin{subfigure}[b]{0.4\textwidth}
   \includegraphics[width=1\linewidth]{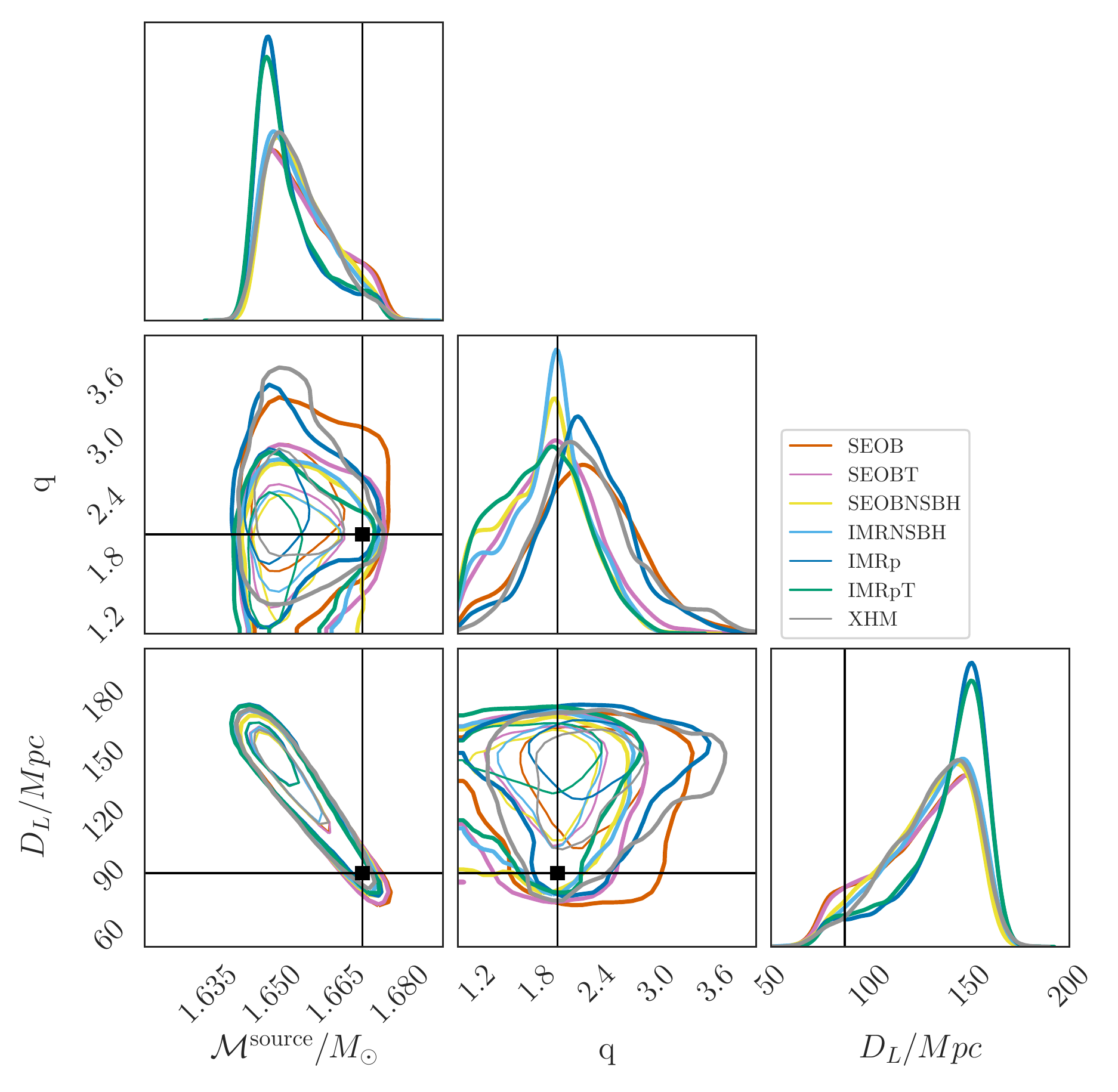}
   \vspace{-1.5\baselineskip}
   
   \caption{SNR 30.}
   \label{Fig.q2_mc_SNR30_inc70} 
\end{subfigure}
\begin{subfigure}[b]{0.4\textwidth}
   \includegraphics[width=1\linewidth]{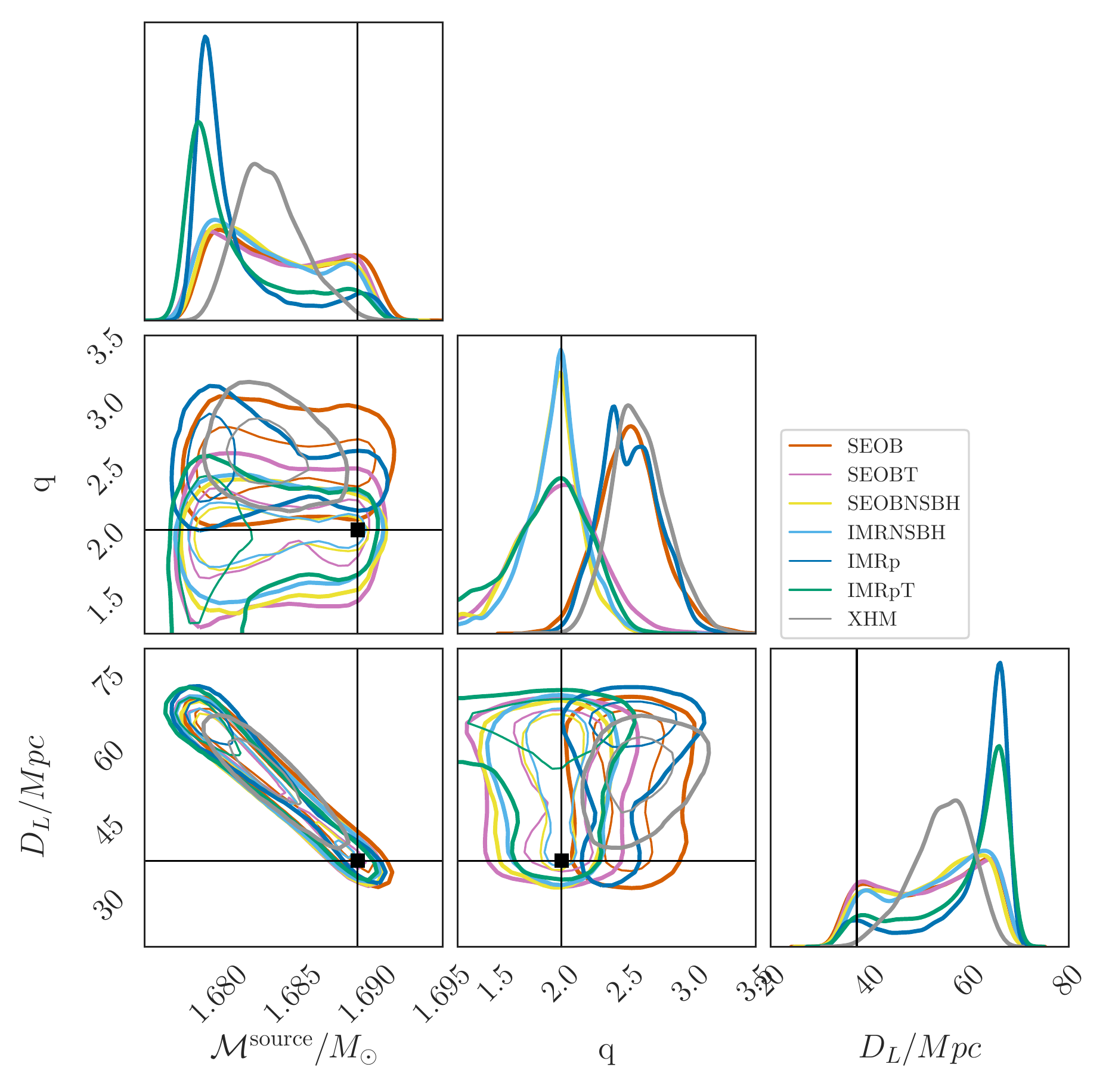}
   \vspace{-1.5\baselineskip}
   \caption{SNR 70.}
   \label{Fig.q2_mc_SNR70_inc70} 
\end{subfigure}
\caption{Corner plot of posterior distributions for chirp mass \mcs, mass ratio q, and luminosity distance \dl, recovered by different approximants for $q=2$, inclination $70^{\circ}$.}
\label{Fig.mc_BHNSq2s0_inc70}
\end{figure}

\begin{figure}
\centering
\begin{subfigure}[b]{0.4\textwidth}
   \includegraphics[width=1\linewidth]{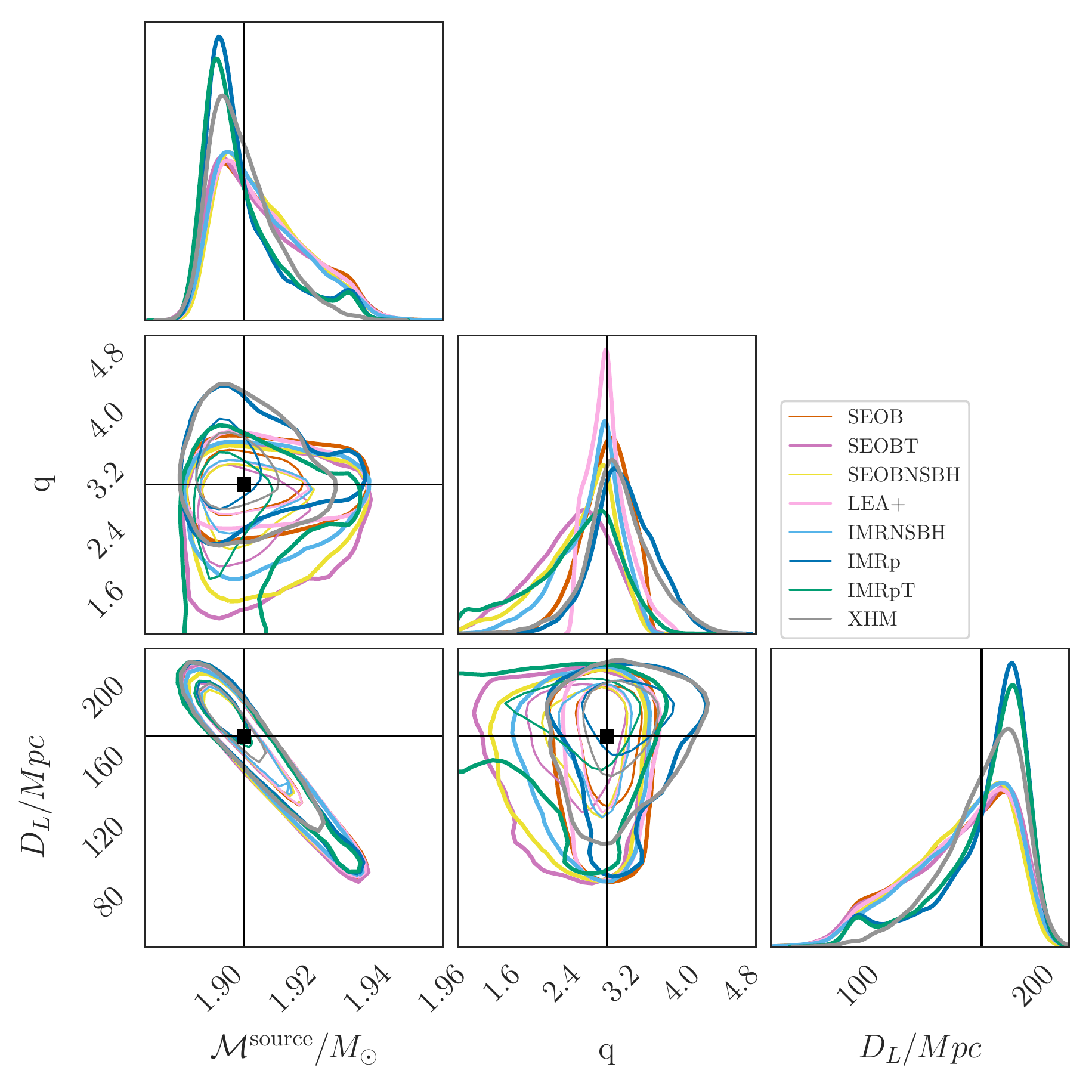}
   \vspace{-1.5\baselineskip}
   \caption{SNR 30.}
   \label{Fig.q3_mc_SNR30_inc30} 
\end{subfigure}
\begin{subfigure}[b]{0.4\textwidth}
   \includegraphics[width=1\linewidth]{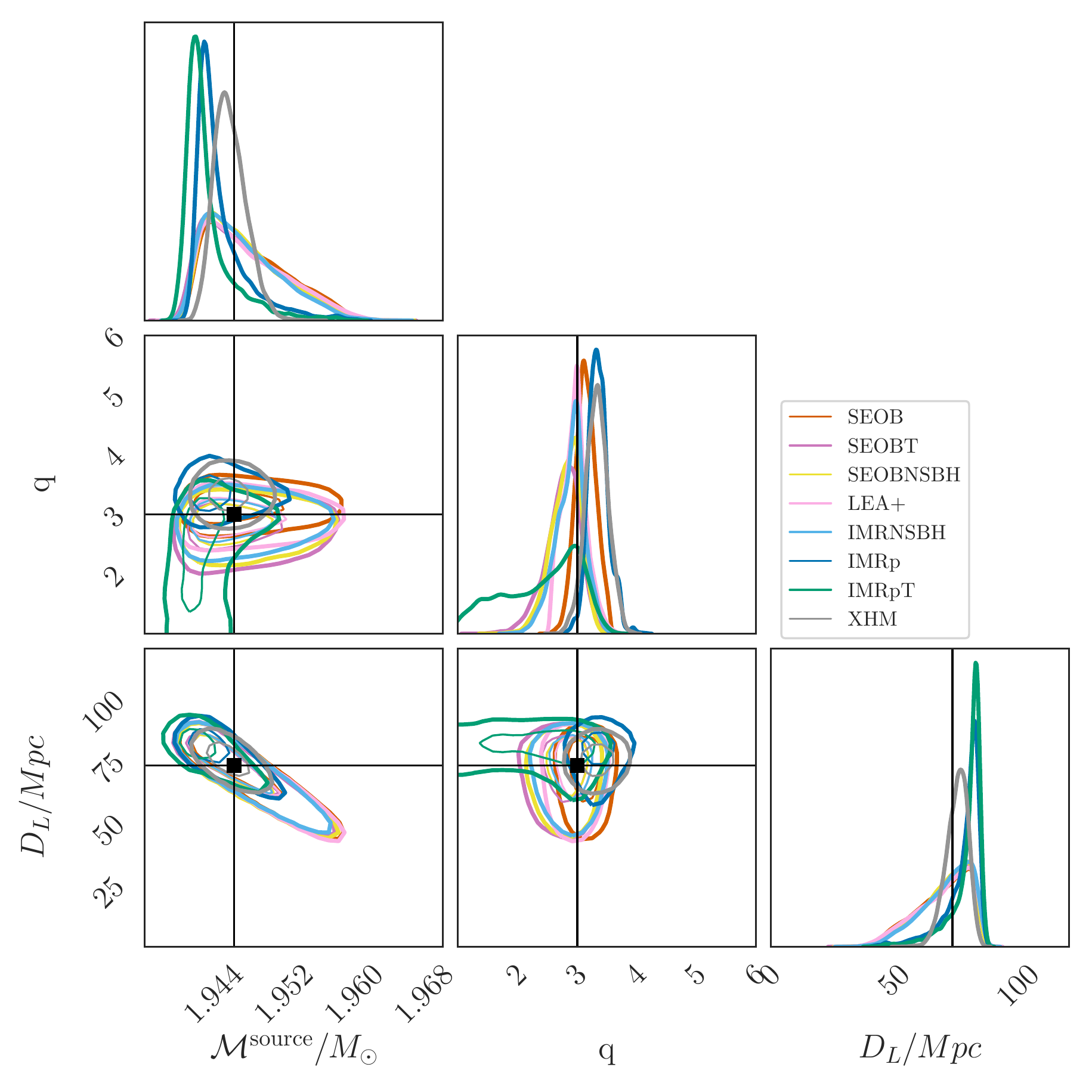}
   \vspace{-1.5\baselineskip}
   \caption{SNR 70.}
   \label{Fig.q3_mc_SNR70_inc30} 
\end{subfigure}
\caption{Corner plot of posterior distributions for chirp mass \mcs, mass ratio q, and luminosity distance \dl, recovered by different approximants for $q=3$, inclination $30^{\circ}$.}
\label{Fig.mc_BHNSq3s0_inc30}
\end{figure}

\begin{figure}
\centering
\begin{subfigure}[b]{0.4\textwidth}
   \includegraphics[width=1\linewidth]{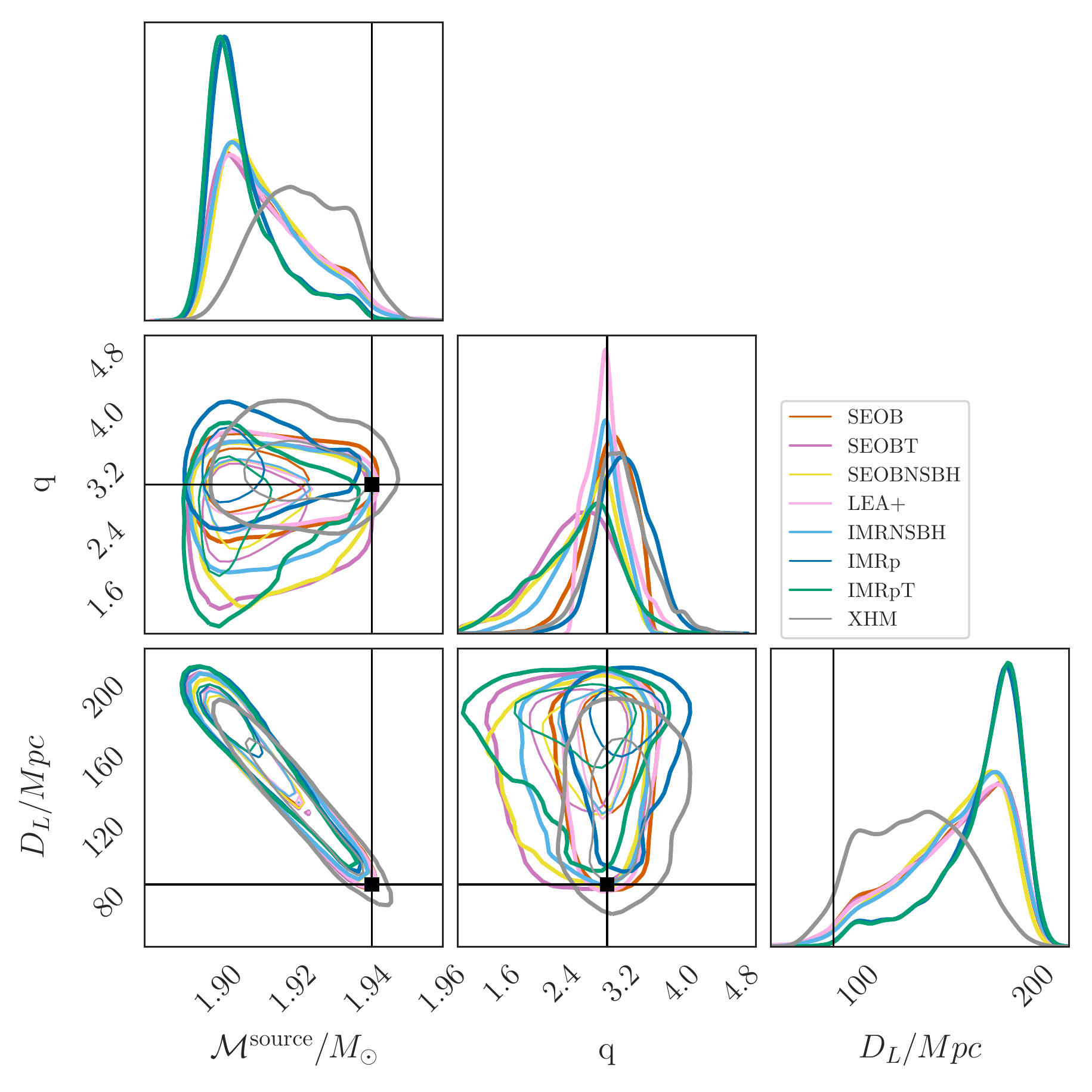}
   \vspace{-1.5\baselineskip}
   \caption{SNR 30.}
   \label{Fig.q3_mc_SNR30_inc70} 
\end{subfigure}
\begin{subfigure}[b]{0.4\textwidth}
   \includegraphics[width=1\linewidth]{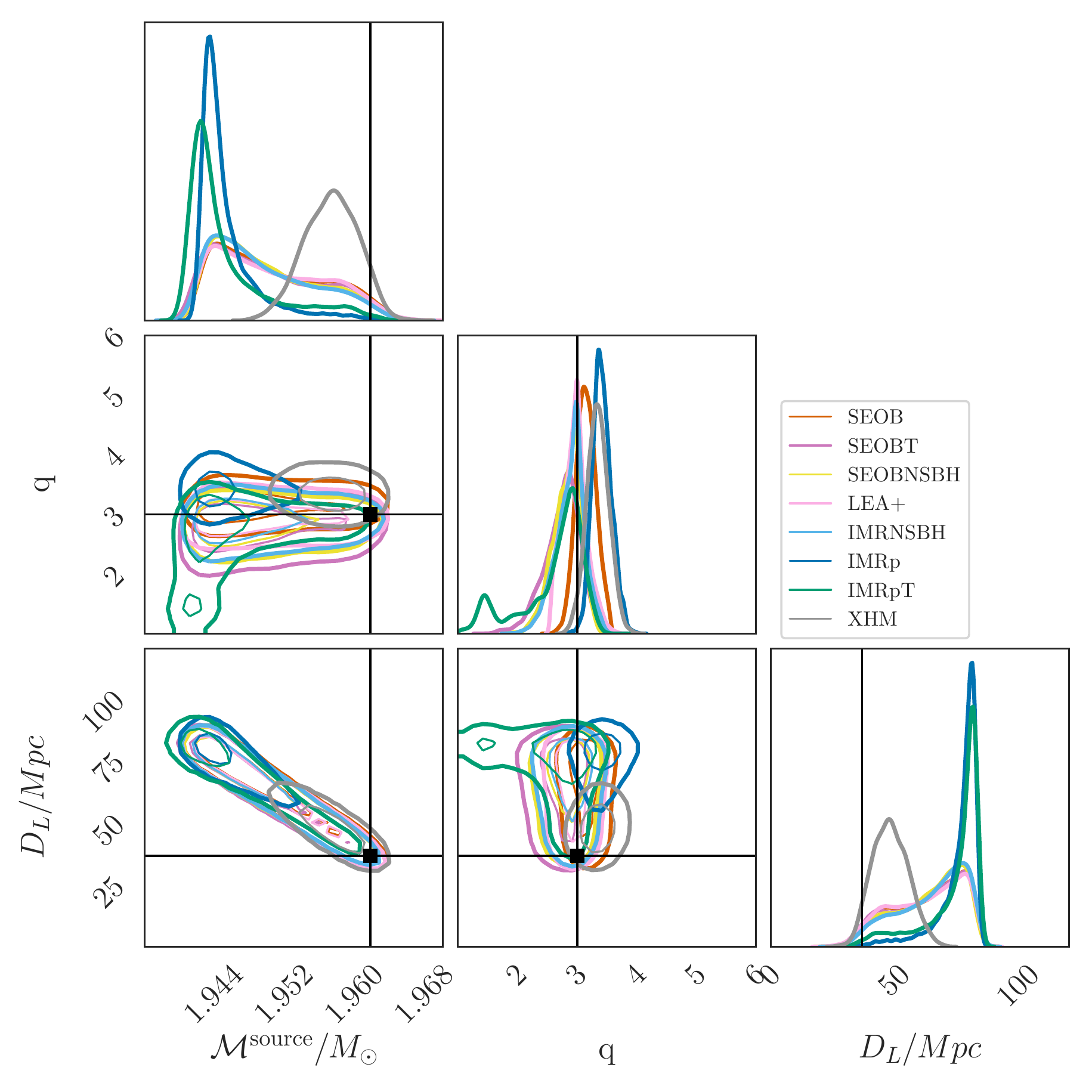}
   \vspace{-1.5\baselineskip}
   \caption{SNR 70.}
   \label{Fig.q3_mc_SNR70_inc70} 
\end{subfigure}
\caption{Corner plot of posterior distributions for chirp mass \mcs, mass ratio q, and luminosity distance \dl, recovered by different approximants for $q=3$, inclination $70^{\circ}$.}
\label{Fig.mc_BHNSq3s0_inc70}
\end{figure}

However, we do find a clear anti-correlation between \mcs and \dl, as seen in Fig.~\ref{Fig.mc_BHNSq2s0_inc30} -- \ref{Fig.mc_BHNSq6s0_inc70}, where \mcs is the rest frame (or source-frame) chirp mass of the NSBH binary: when one increases the other decreases. 
This behavior can be explained by the fact that what is measured from GW data is the detector-frame chirp mass, which is larger than the source-frame mass by a factor $(1+z)$. 
Thus, to convert detector-frame chirp mass to the astrophysically interesting source-frame chirp mass, one must use the measured luminosity distance (and assume a cosmology; we use the Planck 2015 cosmological parameters\footnote{We use the cosmology defined in the \texttt{TT+lowP+lensing+ext} column of Table 4 from~\cite{Ade:2015xua}. This corresponds to $\Omega_M = 0.3065$, $\Omega_\Lambda = 0.6935$, $w_0 = -1$ and $H_0 = 67.90~\mathrm{km} \mathrm{s}^{-1} \mathrm{Mpc}^{-1}$.}~\cite{Ade:2015xua}). 
For a given measured detector-frame mass, if the source were a bit farther away (higher $z$), the source-frame mass would have to be slightly smaller in order to yield the same detector-frame value. 
This is indeed what we find, and is worth stressing, as the uncertainty on the luminosity distance is often a significant factor in the statistical and systematic uncertainty for the source-frame chirp mass.

For the majority of the systems we analyze, especially for the inclination 30\dg binaries, the true \mcs value is recovered within the 90\% CI and little difference is seen between approximants. 
The situation is quite different for the systems with inclination equal to 70\dg: for those the source frame chirp mass is usually underestimated. 
In turn, this happens because the distance is overestimated (as explained in Sec.~\ref{Sec:ExtrinsicParams} below). 
This bias is reduced or even absent when using \XHM, at large inclinations and SNRs, since in that case HOMs become observable enough to help break the distance-inclination degeneracy, thus yielding unbiased chirp mass estimates (e.g. Fig.~\ref{Fig.mc_BHNSq6s0_inc70} and Sec.~\ref{Sec:ExtrinsicParams}).

On the other hand, there is only marginal difference in the recovery of \mcs between waveform models that do or do not include NS tidal effects. However, the same is not necessarily true for other parameters, as discussed below.

\subsubsection{Mass ratio}
\label{Sec:MassRatio}

As we will discuss further in Sec.~\ref{Sec:MatterEffects}, it is largely the binary mass ratio, together with the BH-spin and NS EoS, that determines if and to which extent the finite-size of the NS will leave an observable imprint in the detected GW signal~\cite{Foucart:2012nc,Pannarale:2015jka,Foucart:2018rjc}.
For the $q=2$ binaries, for which tidal effects are largest, the waveforms that do not model the tidal disruption of the NS recover a strongly biased posterior of $q$, with the true value of $q=2$ only barely included in the tail of the posterior for the SNR 70 sources, Fig.~\ref{Fig.q2_mc_SNR70_inc30} and~\ref{Fig.q2_mc_SNR70_inc70}.
For the $q=3$ binaries of Fig.~\ref{Fig.mc_BHNSq3s0_inc30} and~\ref{Fig.mc_BHNSq3s0_inc70}, the tidal effects present are less prominent and hence all waveform models show similar performance in recovering the true mass ratio, with the exception of non-tidal IMRPhenom-based waveforms, \IMRp and \XHM, which are only marginally consistent with the true value. 

\begin{figure}
\centering
\begin{subfigure}[b]{0.4\textwidth}
   \includegraphics[width=1\linewidth]{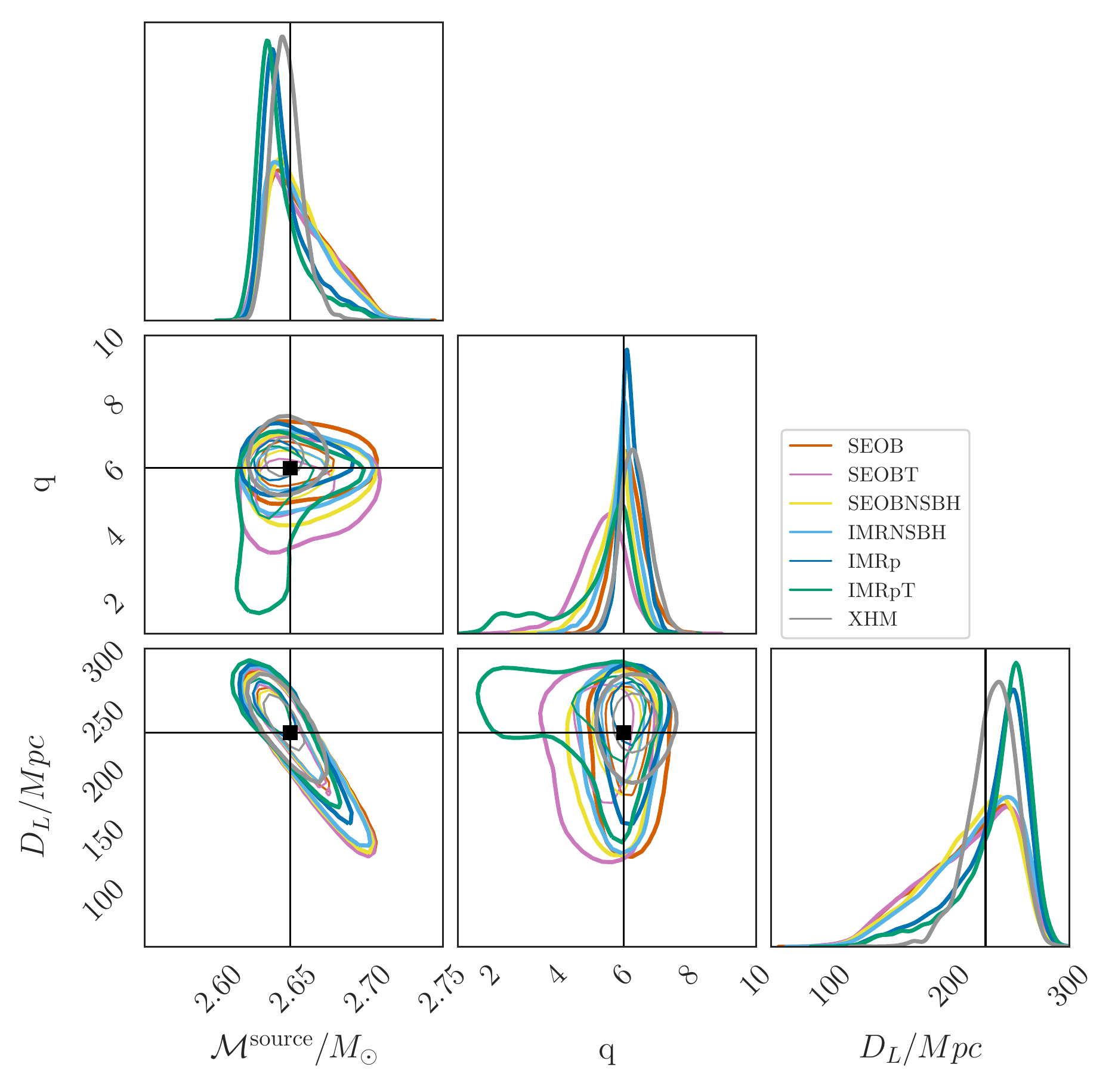}
   \vspace{-1.5\baselineskip}
   \caption{SNR 30.}
   \label{Fig.q6_mc_SNR30_inc30} 
\end{subfigure}
\begin{subfigure}[b]{0.4\textwidth}
   \includegraphics[width=1\linewidth]{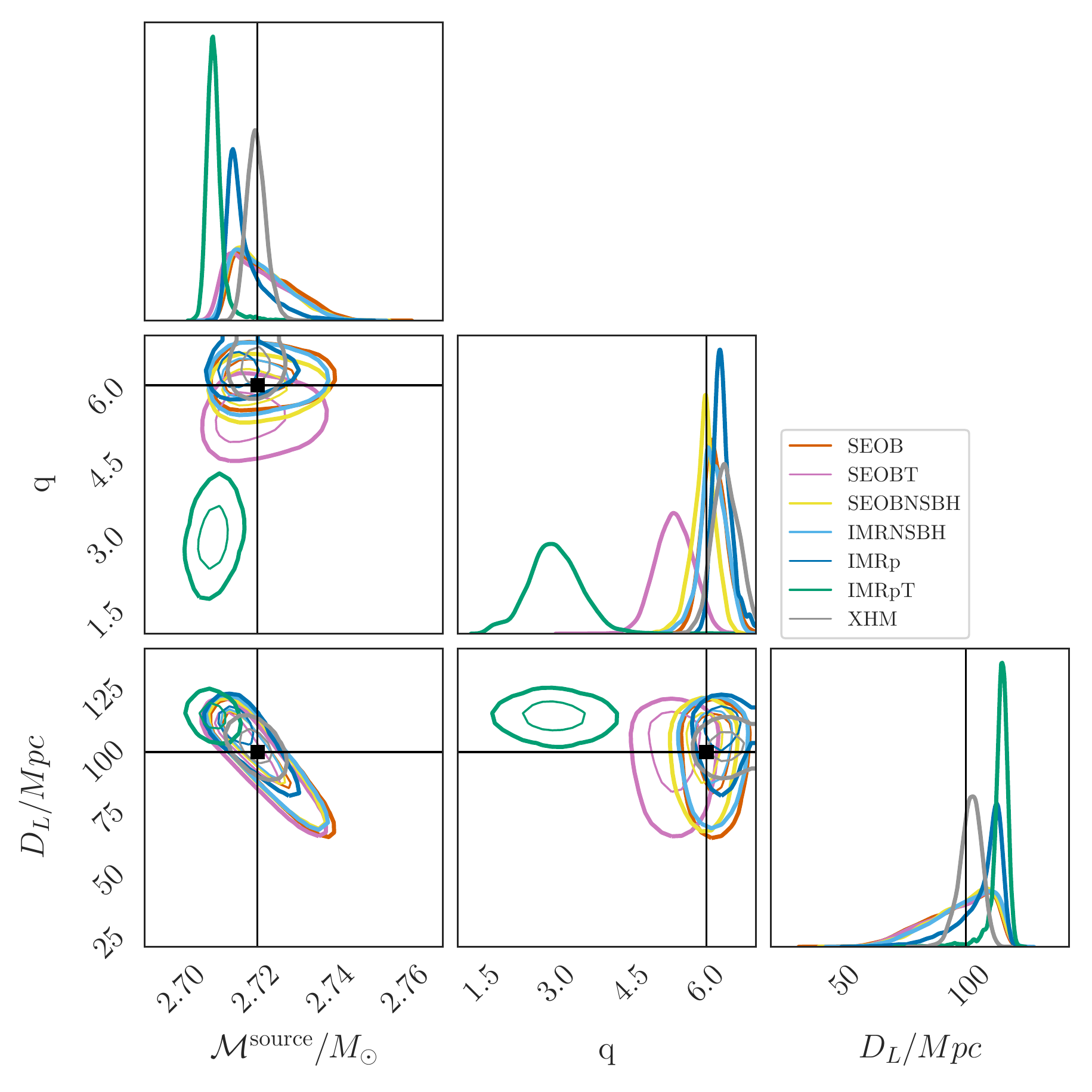}
   \vspace{-1.5\baselineskip}
   \caption{SNR 70.}
   \label{Fig.q6_mc_SNR70_inc30} 
\end{subfigure}
\caption{Corner plot of posterior distributions for chirp mass \mcs, mass ratio q, and luminosity distance \dl, recovered by different approximants for $q=6$, inclination $30^{\circ}$.}
\label{Fig.mc_BHNSq6s0_inc30}
\end{figure}

\begin{figure}
\centering
\begin{subfigure}[b]{0.4\textwidth}
   \includegraphics[width=1\linewidth]{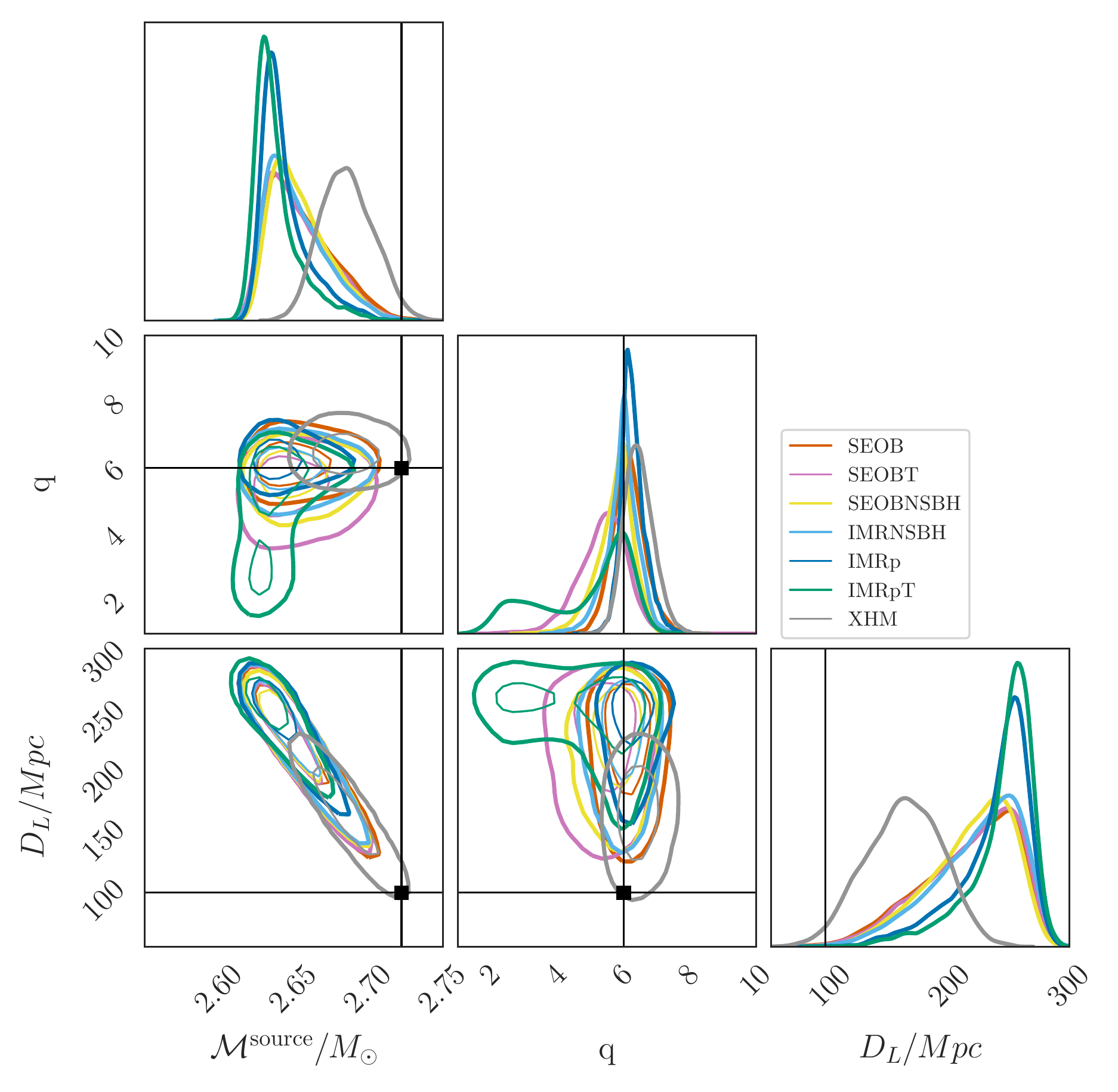}
   \vspace{-1.5\baselineskip}
   \caption{SNR 30.}
   \label{Fig.q6_mc_SNR30_inc70} 
\end{subfigure}
\begin{subfigure}[b]{0.4\textwidth}
   \includegraphics[width=1\linewidth]{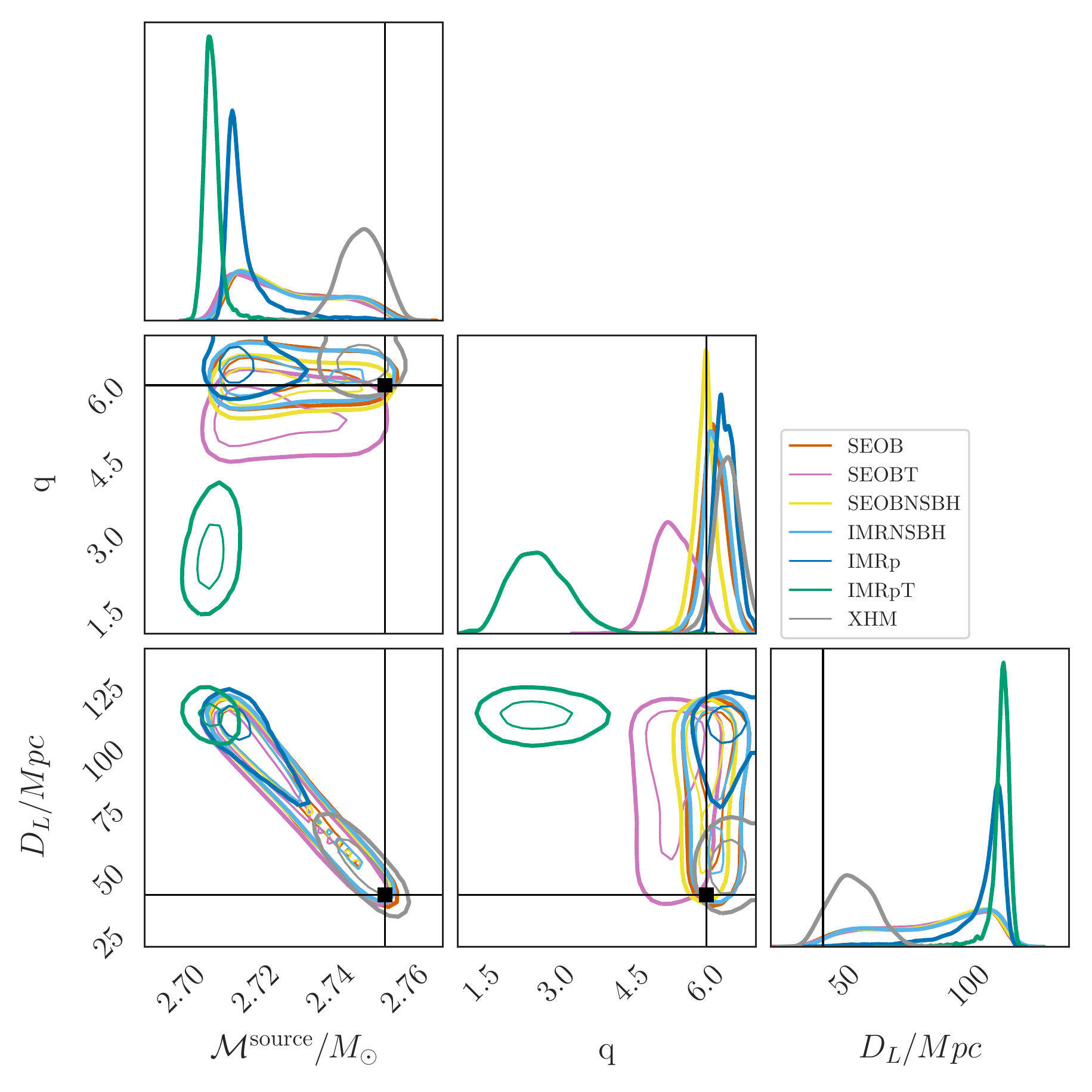}
   \vspace{-1.5\baselineskip}
   \caption{SNR 70.}
   \label{Fig.q6_mc_SNR70_inc70} 
\end{subfigure}
\caption{Corner plot of posterior distributions for chirp mass \mcs, mass ratio q, and luminosity distance \dl, recovered by different approximants for $q=6$, inclination $70^{\circ}$.}
\label{Fig.mc_BHNSq6s0_inc70}
\end{figure}

When $q=6$, Fig.~\ref{Fig.mc_BHNSq6s0_inc30} and~\ref{Fig.mc_BHNSq6s0_inc70}, the NS is not expected to disrupt before plunging into the BH, and hence tidal effects are unmeasurable (as shown in Fig.~\ref{Fig.q6_lambda2rNS_inc30} and~\ref{Fig.q6_lambda2rNS_inc70}). For these sources, it is the two waveform models that are explicitly calibrated against (near-equal-mass) BNS simulations, \SEOBT and \IMRpT, that produce biased mass-ratio posteriors, with \IMRpT being farther away from the true value. 
While at SNR 30, Fig.~\ref{Fig.q6_mc_SNR30_inc30} and ~\ref{Fig.q6_mc_SNR30_inc70}, a second peak at more equal mass ratios is already visible, the main peak is still present at the true value of $q=6$. 
It is only for the loud signals that the peak at the true value disappears resulting in a significant bias, especially for \IMRpT.
This further highlights the need for specialized NSBH waveforms, like \IMRNSBH and \SEOBNSBH, in the analysis of potential NSBH candidates in order to avoid significant astrophysical inference biases.

Overall, we find statistical uncertainties for $q$ of the order of $\si 0.8$ at SNR 70 for the NSBH-tuned models without much dependence on the true value of $q$.
While the absolute value of the 90\% CI stays roughly constant with $q$, the relative uncertainty is 3 times smaller for q6 than for q2.
For the SNR 30 signals, the uncertainties are naturally higher and %
fall into the range of 1.2-1.6 for all three systems.

\subsubsection{Neutron star and black hole masses}
\label{Sec:ComponentMasses}

One of the most attractive features of NSBH binaries is the potential of a precise measurement of the neutron star mass, including putting constraints on its maximum value, which is still under debate~\cite{Lawrence:2015oka,Alsing:2017bbc,Cromartie:2019kug,Bauswein:2020aag,Chatziioannou:2020msi}. 
Unfortunately, this is hard to achieve even at high SNRs with BNSs, due to their mass ratio being close to unity~\cite{Farr:2015lna}.

This is particularly true if one follows an agnostic approach, without assuming \emph{a priori} that a compact object lighter than 2 \msun is necessarily a NS, and allows for the object to assume spins larger than what a NS could nominally support~\cite{Hessels:2006ze,Tauris:2017omb}. 
In that case, a known spin-mass ratio degeneracy will significantly increase the uncertainty in both parameters~\cite{Ng:2018neg}. 
This was clearly shown with the first BNS source~\cite{TheLIGOScientific:2017qsa}, for which the upper value of the 90\% CI for the primary mass increases by \si40\% (\si18\%) when the spin magnitude limit is increased from $0.05$ to $0.89$ for spin-aligned (spin-precessing) waveforms. 
Similar differences have been reported for GW190425~\cite{Abbott:2020uma}.
For BNSs, the spin prior used will usually determine whether it is possible to set a significant upper bound on NS masses.

\begin{figure}
\centering
\begin{subfigure}[b]{0.4\textwidth}
   \includegraphics[width=1\linewidth]{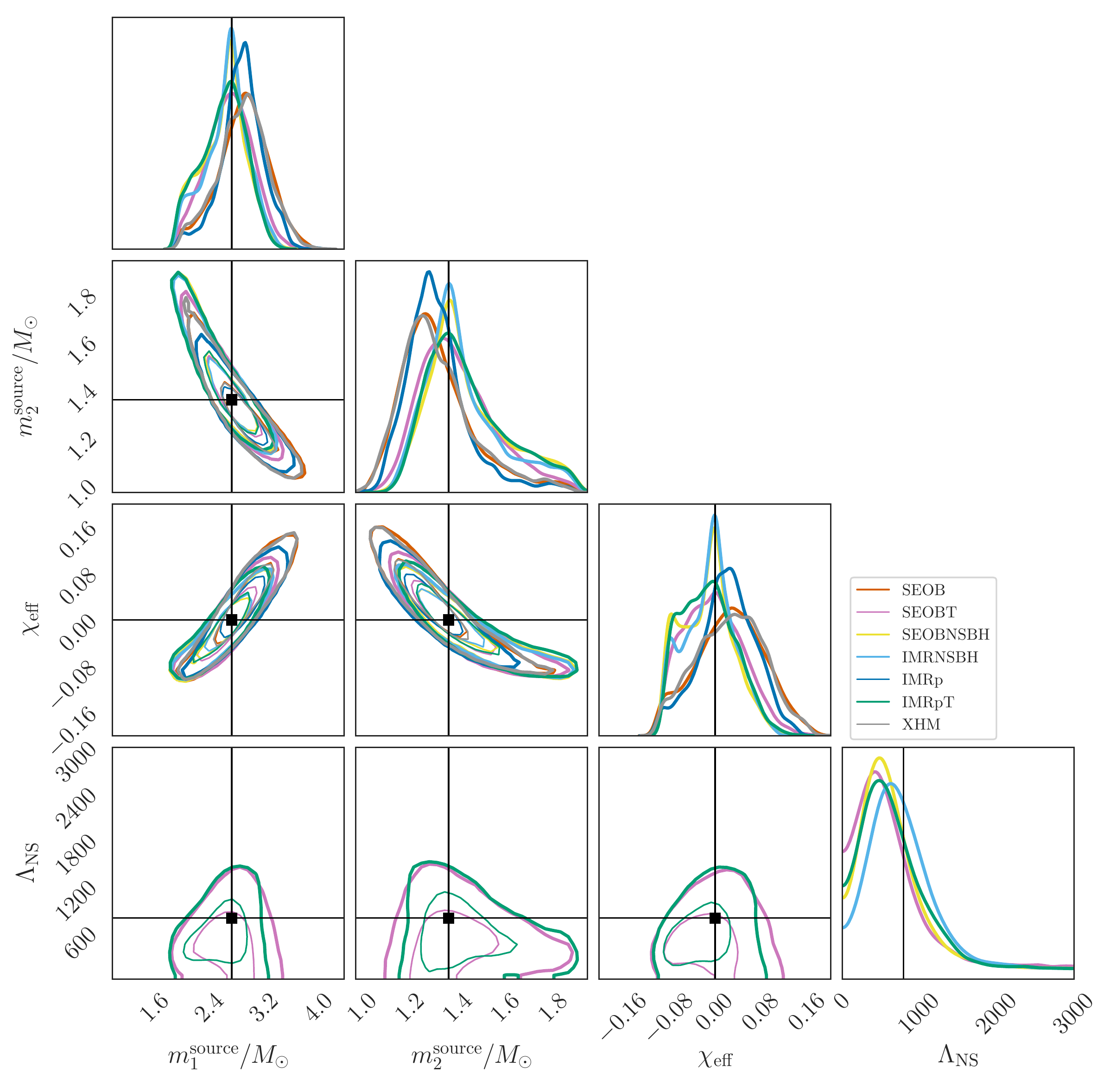}
   \vspace{-1.5\baselineskip}
   \caption{SNR 30.}
   \label{Fig.q2_compMass_SNR30_inc30} 
\end{subfigure}
\begin{subfigure}[b]{0.4\textwidth}
   \includegraphics[width=1\linewidth]{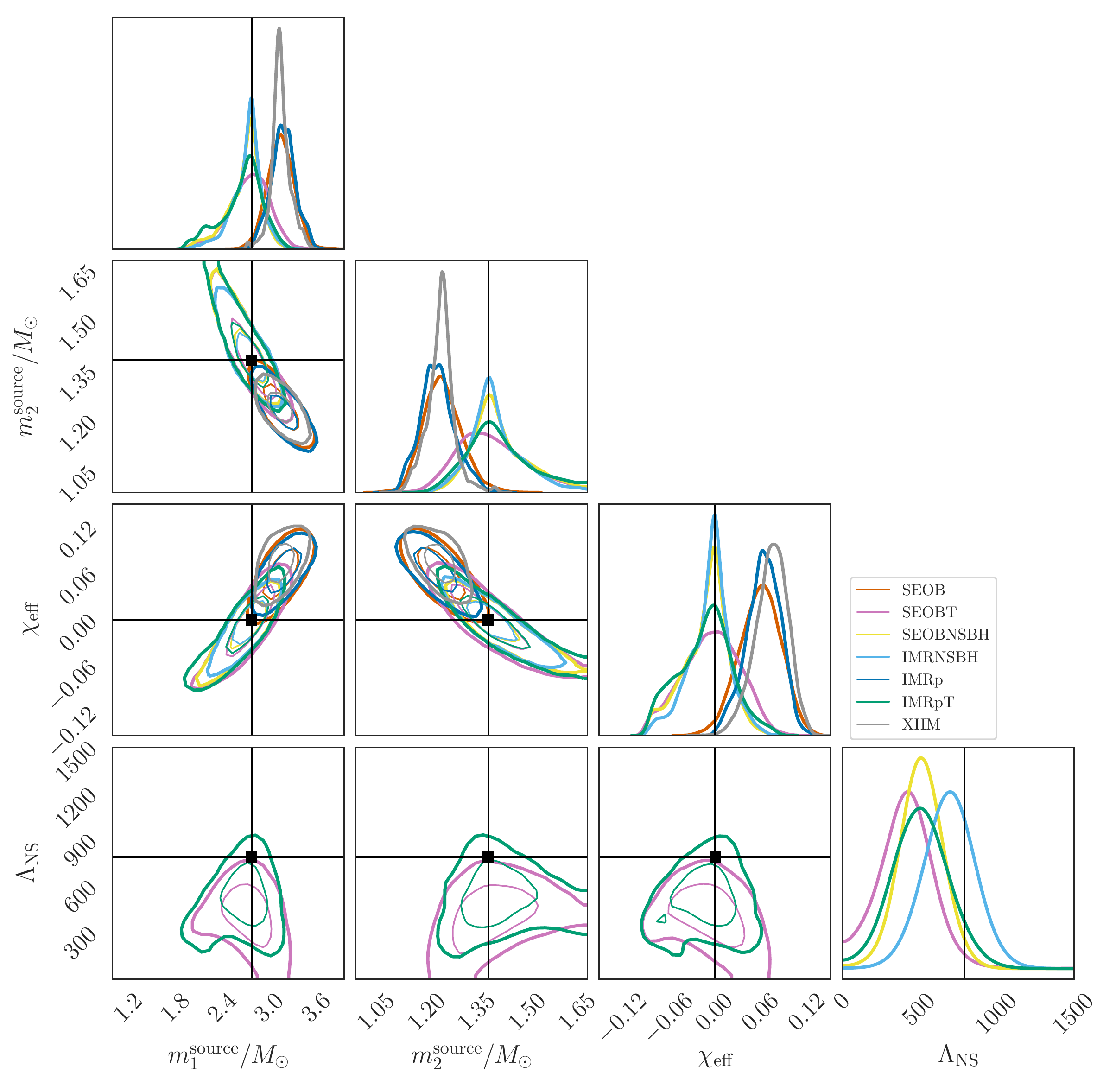}
   \vspace{-1.5\baselineskip}
   \caption{SNR 70.}
   \label{Fig.q2_compMass_SNR70_inc30} 
\end{subfigure}
\caption{Corner plot of posterior distributions for component masses \mBHs and \mNSs, the effective spin \chieff and the tidal deformability \LNS recovered by different approximants for $q=2$, inclination $30^{\circ}$.}
\label{Fig.compMass_BHNSq2s0_inc30}
\end{figure}

\begin{figure}
\centering
\begin{subfigure}[b]{0.4\textwidth}
   \includegraphics[width=1\linewidth]{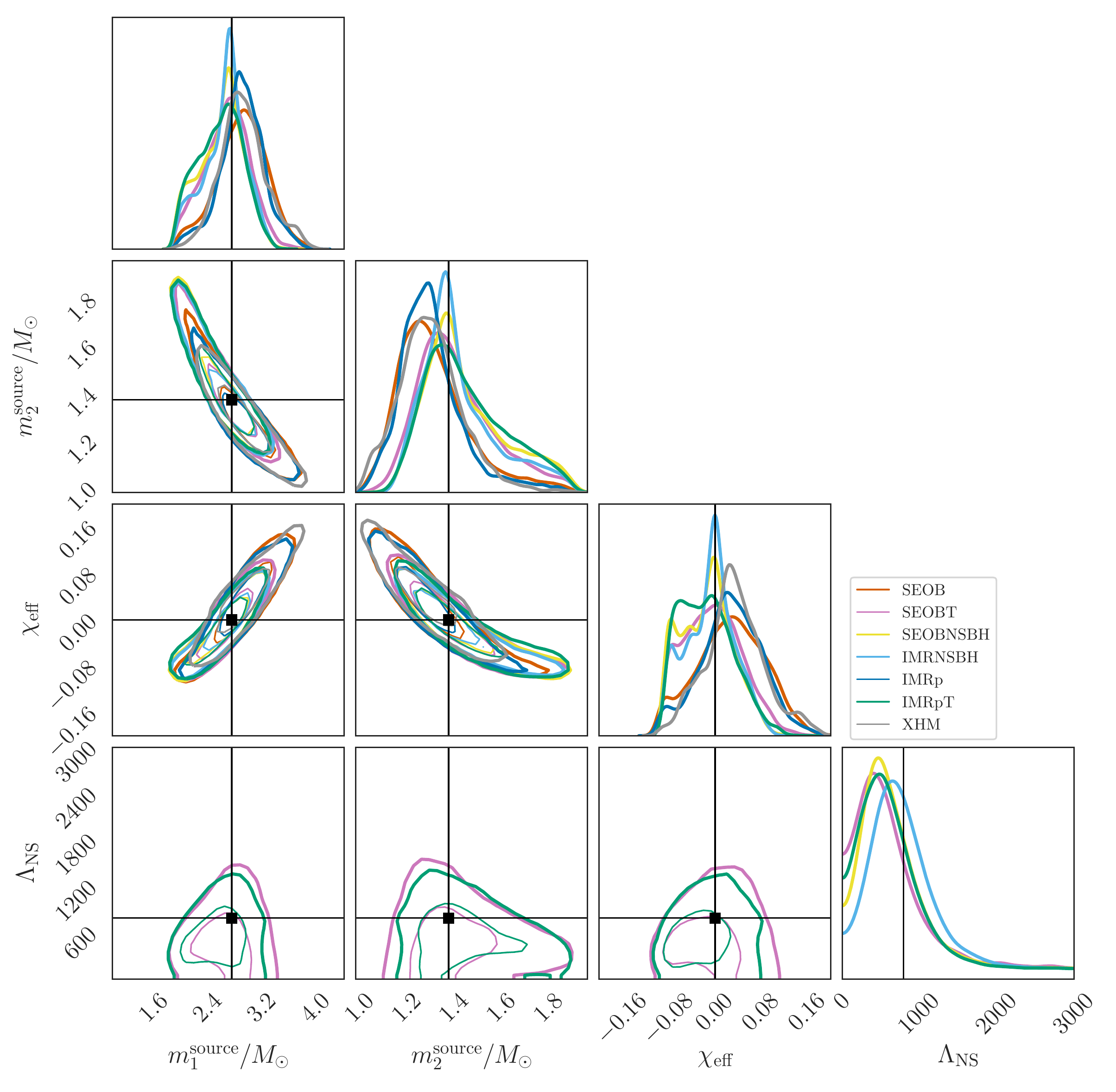}
   \vspace{-1.5\baselineskip}
   \caption{SNR 30.}
   \label{Fig.q2_compMass_SNR30_inc70} 
\end{subfigure}
\begin{subfigure}[b]{0.4\textwidth}
   \includegraphics[width=1\linewidth]{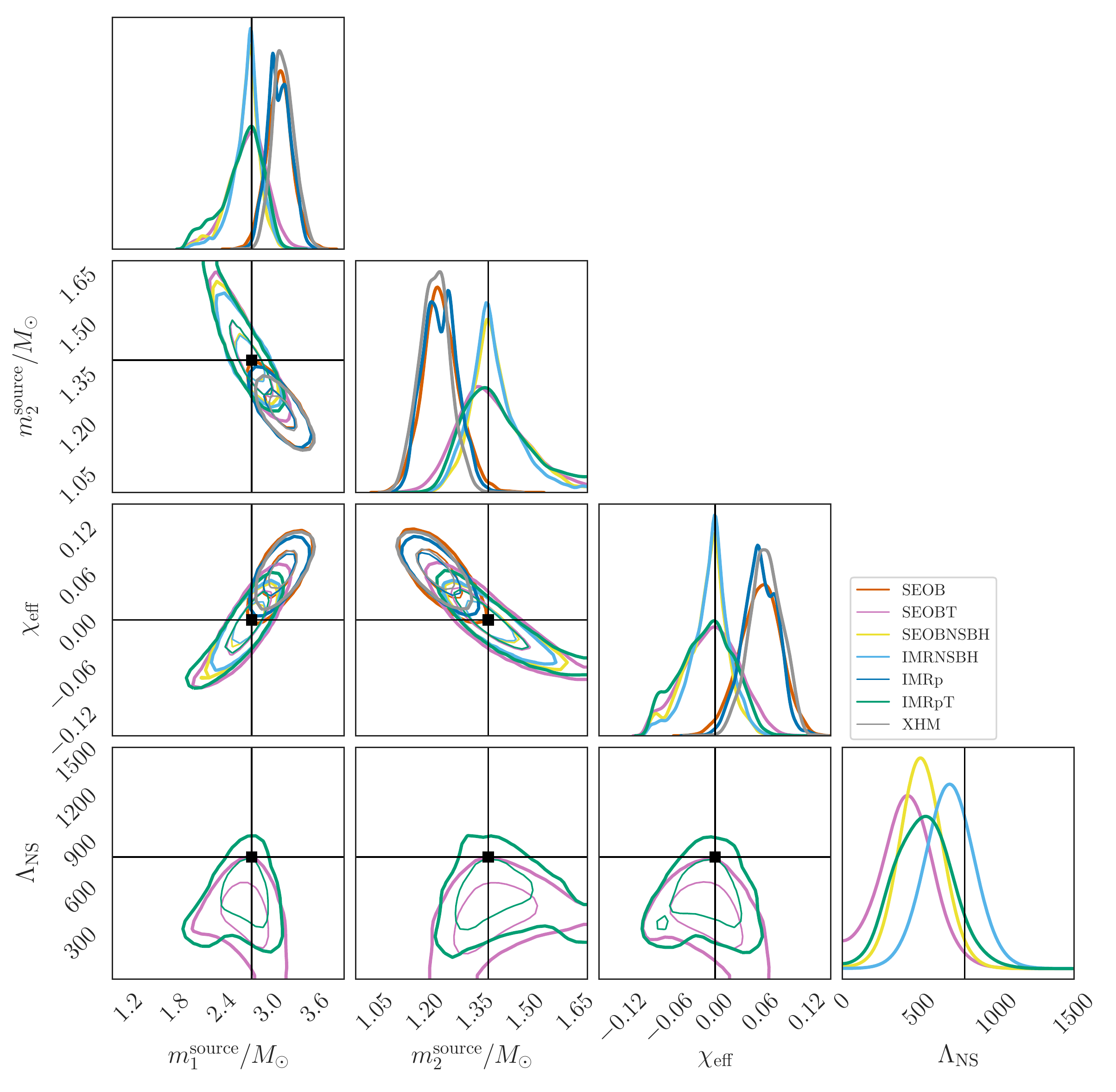}
   \vspace{-1.5\baselineskip}
   \caption{SNR 70.}
   \label{Fig.q2_compMass_SNR70_inc70} 
\end{subfigure}
\caption{Corner plot of posterior distributions for component masses \mBHs and \mNSs, the effective spin \chieff and the tidal deformability \LNS recovered by different approximants for $q=2$, inclination $70^{\circ}$.}
\label{Fig.compMass_BHNSq2s0_inc70}
\end{figure}

\begin{figure}
\centering
\begin{subfigure}[b]{0.4\textwidth}
   \includegraphics[width=1\linewidth]{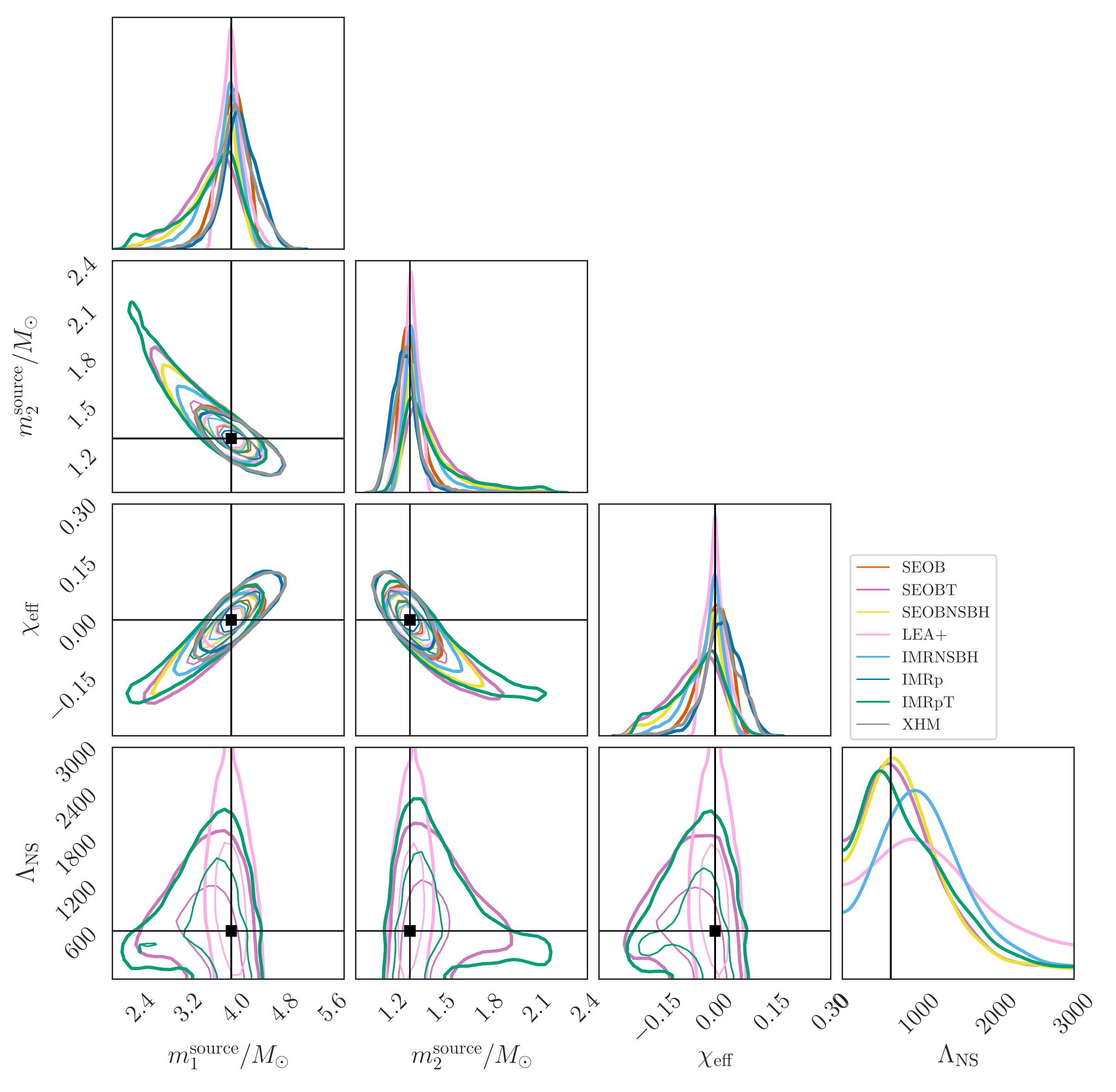}
   \vspace{-1.5\baselineskip}
   \caption{SNR 30.}
   \label{Fig.q3_compMass_SNR30_inc30} 
\end{subfigure}
\begin{subfigure}[b]{0.4\textwidth}
   \includegraphics[width=1\linewidth]{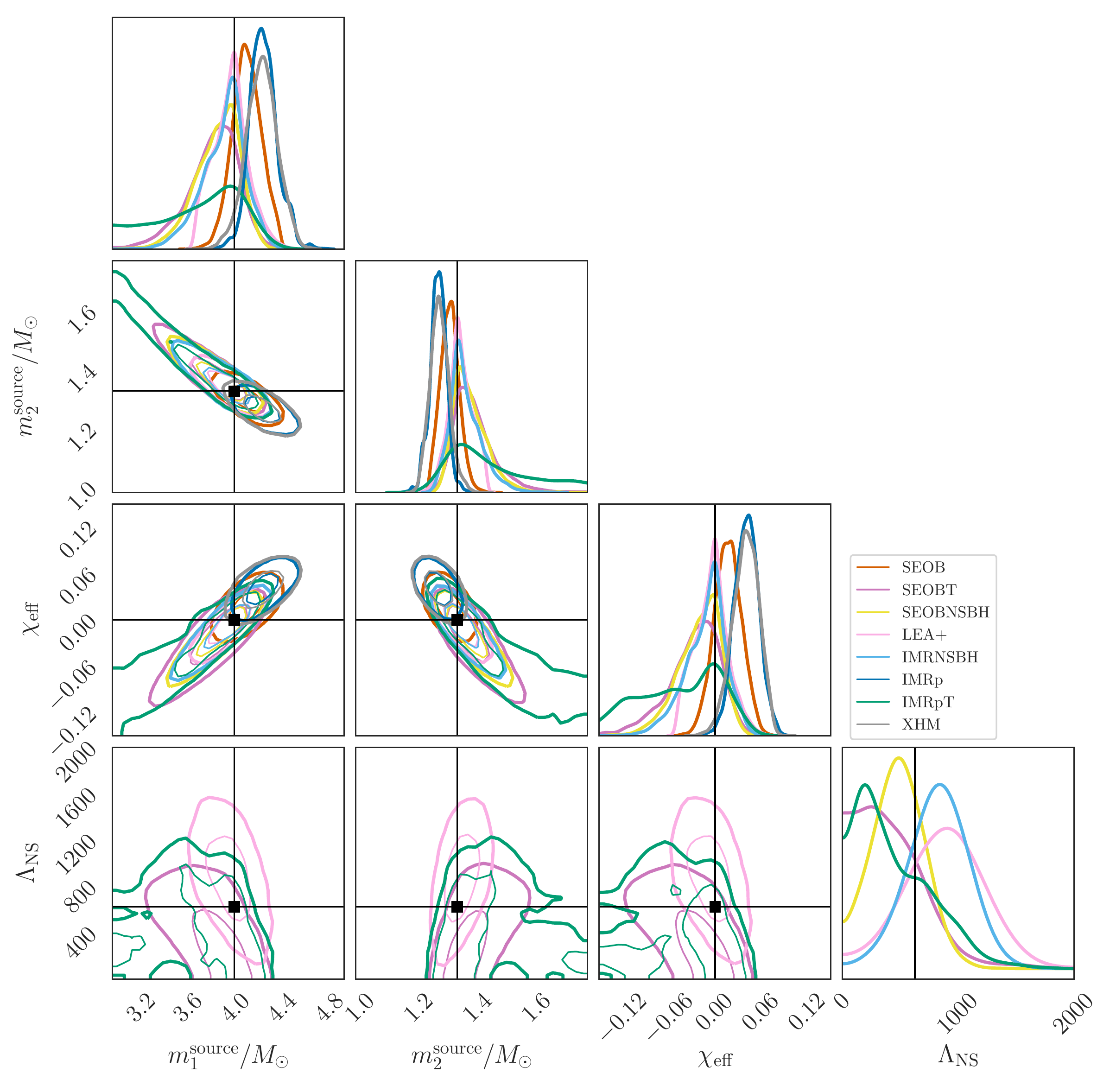}
   \vspace{-1.5\baselineskip}
   \caption{SNR 70.}
   \label{Fig.q3_compMass_SNR70_inc30} 
\end{subfigure}
\caption{Corner plot of posterior distributions for component masses \mBHs and \mNSs, the effective spin \chieff and the tidal deformability \LNS recovered by different approximants for $q=3$, inclination $30^{\circ}$.}
\label{Fig.compMass_BHNSq3s0_inc30}
\end{figure}

\begin{figure}
\centering
\begin{subfigure}[b]{0.4\textwidth}
   \includegraphics[width=1\linewidth]{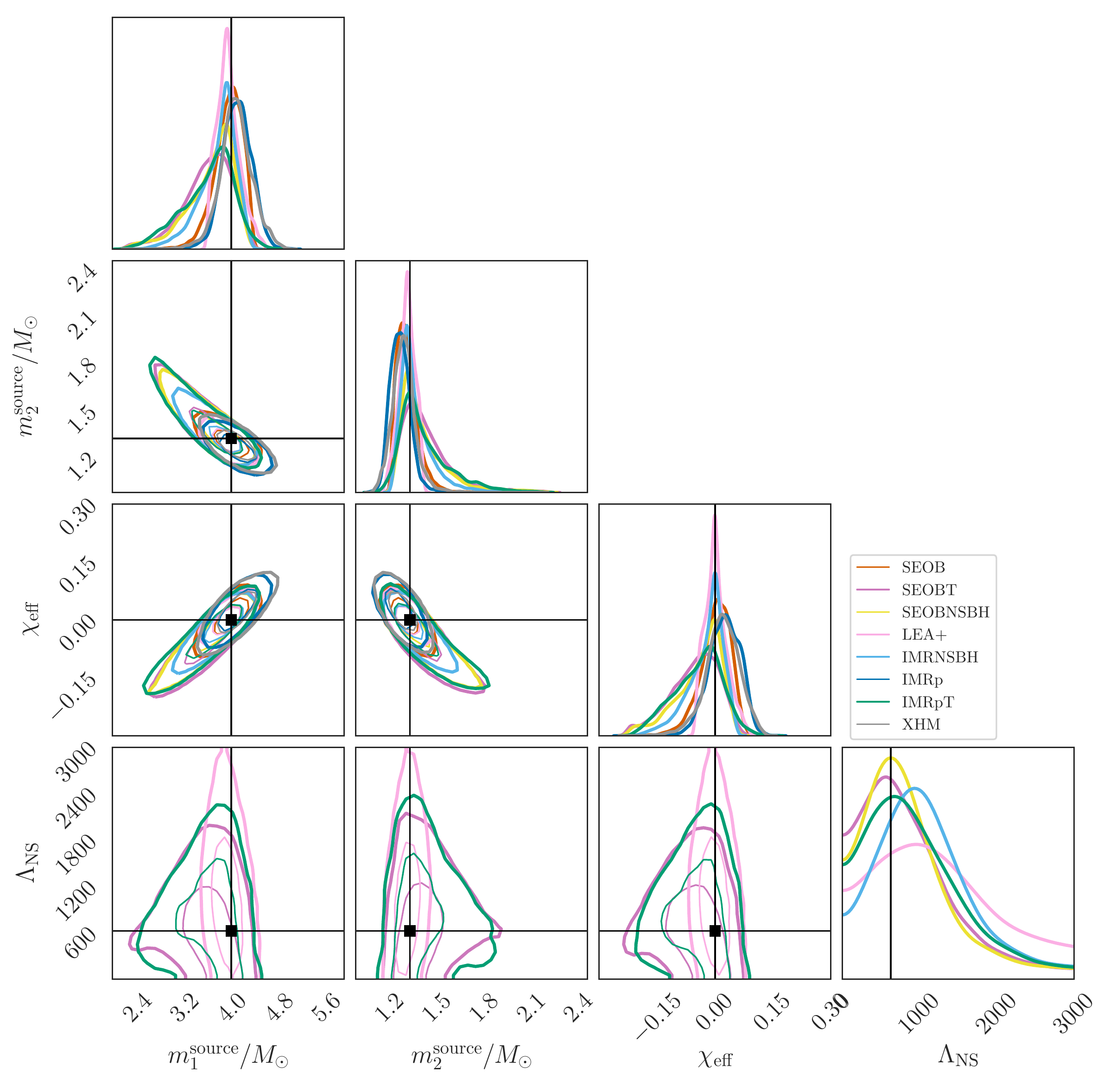}
   \vspace{-1.5\baselineskip}
   \caption{SNR 30.}
   \label{Fig.q3_compMass_SNR30_inc70} 
\end{subfigure}
\begin{subfigure}[b]{0.4\textwidth}
   \includegraphics[width=1\linewidth]{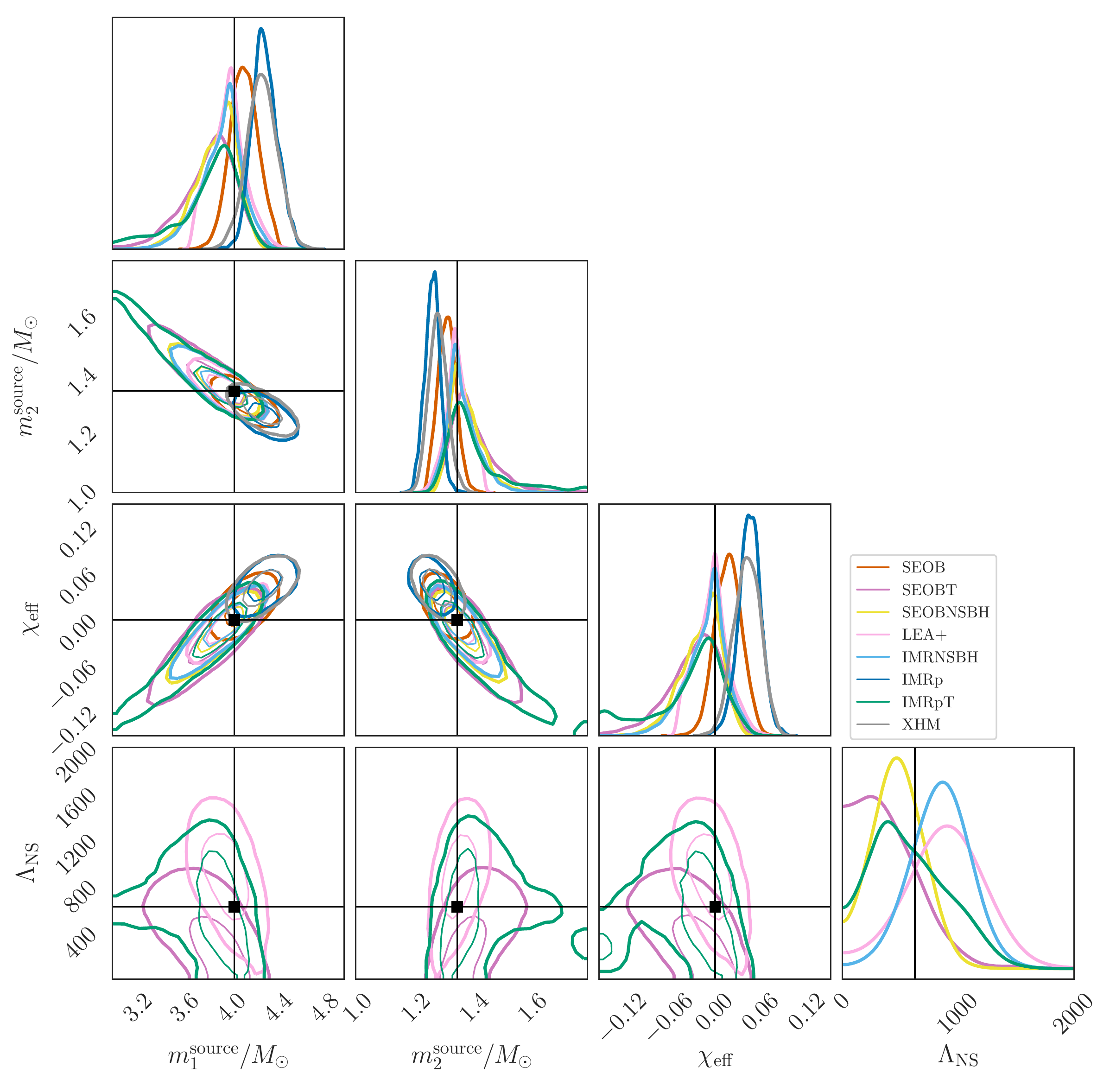}
   \vspace{-1.5\baselineskip}
   \caption{SNR 70.}
   \label{Fig.q3_compMass_SNR70_inc70} 
\end{subfigure}
\caption{Corner plot of posterior distributions for component masses \mBHs and \mNSs, the effective spin \chieff and the tidal deformability \LNS recovered by different approximants for $q=3$, inclination $70^{\circ}$.}
\label{Fig.compMass_BHNSq3s0_inc70}
\end{figure}

\begin{figure}
\centering
\begin{subfigure}[b]{0.4\textwidth}
   \includegraphics[width=1\linewidth]{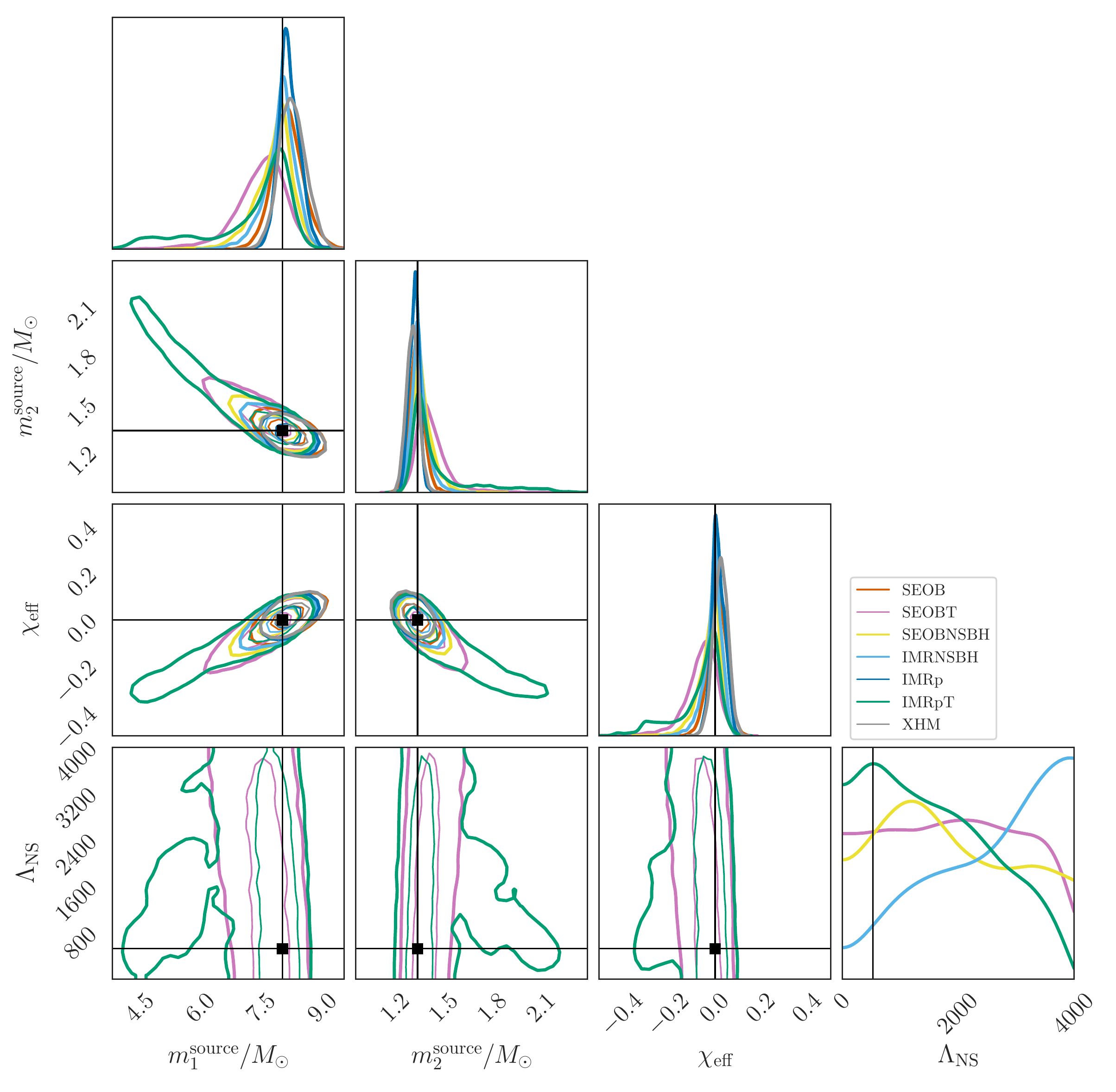}
   \vspace{-1.5\baselineskip}
   \caption{SNR 30.}
   \label{Fig.q6_compMass_SNR30_inc30} 
\end{subfigure}
\begin{subfigure}[b]{0.4\textwidth}
   \includegraphics[width=1\linewidth]{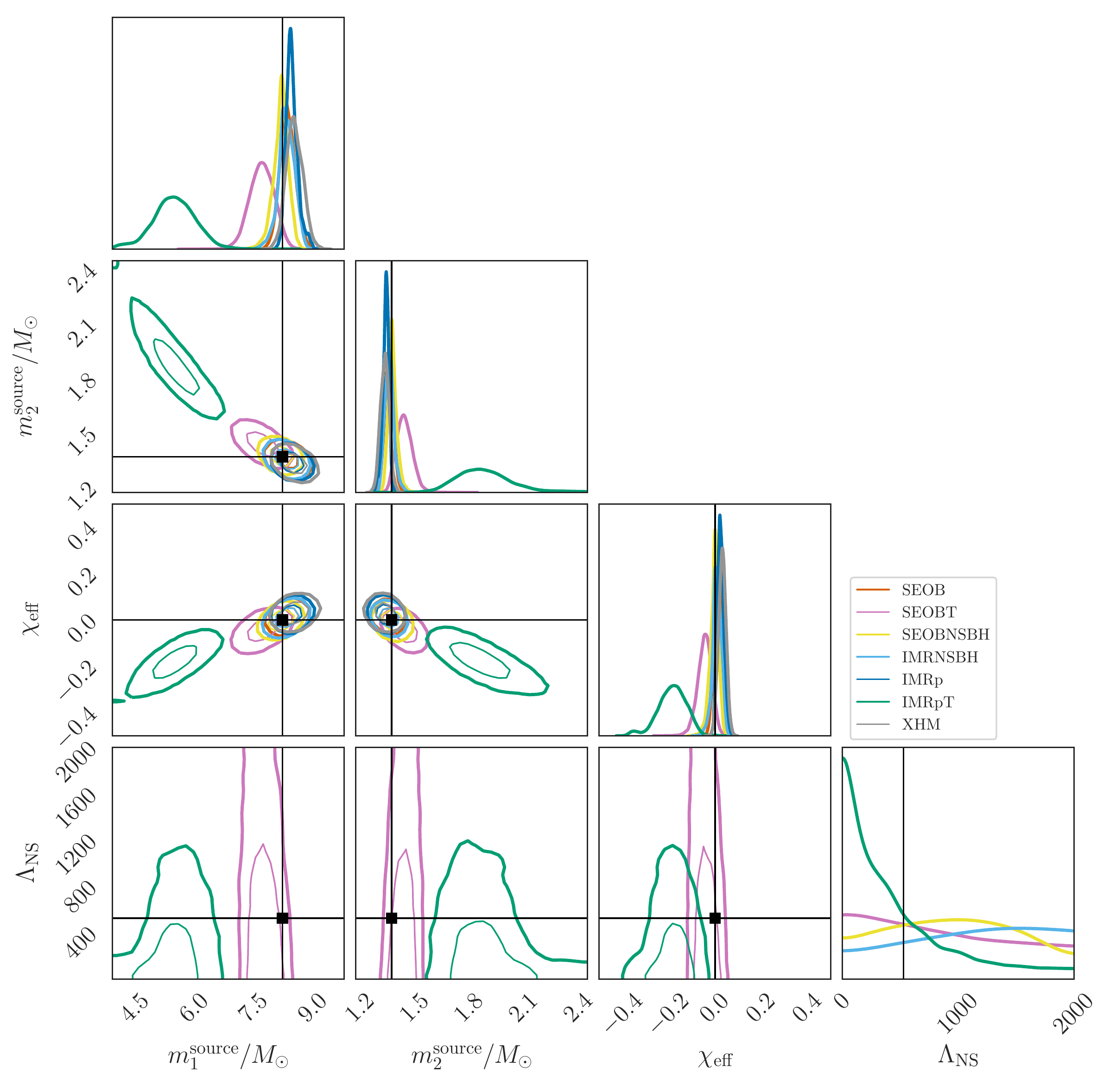}
   \vspace{-1.5\baselineskip}
   \caption{SNR 70.}
   \label{Fig.q6_compMass_SNR70_inc30} 
\end{subfigure}
\caption{Corner plot of posterior distributions for component masses \mBHs and \mNSs, the effective spin \chieff and the tidal deformability \LNS recovered by different approximants for $q=6$, inclination $30^{\circ}$.}
\label{Fig.compMass_BHNSq6s0_inc30}
\end{figure}

\begin{figure}
\centering
\begin{subfigure}[b]{0.4\textwidth}
   \includegraphics[width=1\linewidth]{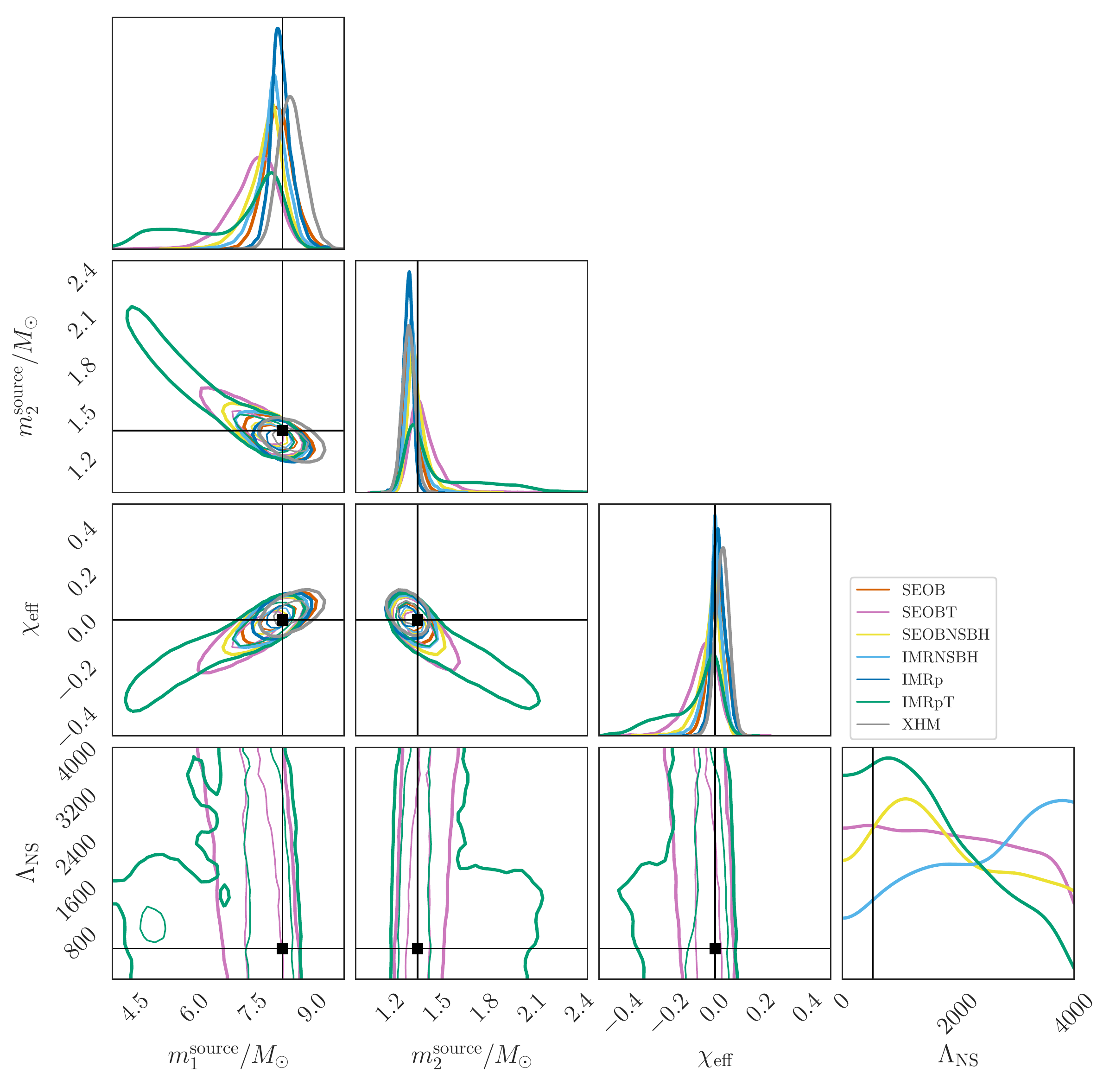}
   \vspace{-1.5\baselineskip}
   \caption{SNR 30.}
   \label{Fig.q6_compMass_SNR30_inc70} 
\end{subfigure}
\begin{subfigure}[b]{0.4\textwidth}
   \includegraphics[width=1\linewidth]{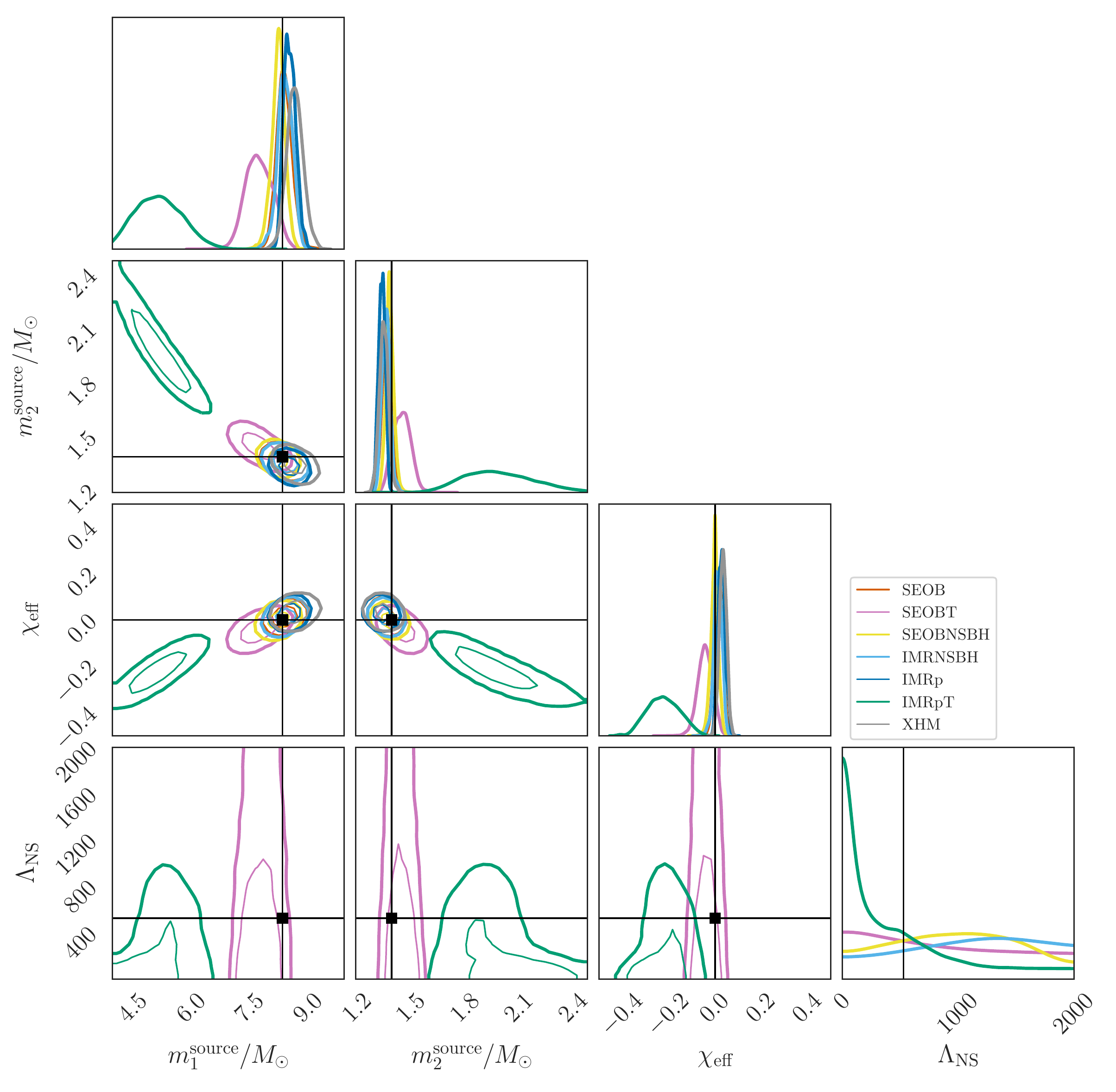}
   \vspace{-1.5\baselineskip}
   \caption{SNR 70.}
   \label{Fig.q6_compMass_SNR70_inc70} 
\end{subfigure}
\caption{Corner plot of posterior distributions for component masses \mBHs and \mNSs, the effective spin \chieff and the tidal deformability \LNS recovered by different approximants for $q=6$, inclination $70^{\circ}$.}
\label{Fig.compMass_BHNSq6s0_inc70}
\end{figure}

We want to verify if NS mass measurement obtained from NSBH sources are more precise, as one would expect from their larger mass ratios, and more accurate.
We find that for the SNR 30 binaries,~\Crefrange{Fig.q2_compMass_SNR30_inc30}{Fig.q6_compMass_SNR30_inc70}, all waveform models perform comparatively well in recovering the true binary masses.
The exception is the $q=3$ binaries, Fig.~\ref{Fig.q3_compMass_SNR30_inc30} and Fig.~\ref{Fig.q3_compMass_SNR30_inc70}: two of the models which allows for NS tidal effects and are dedicated to BNS systems (\SEOBT and \IMRpT) have wider tails towards more equal-mass binaries, and hence heavier NSs (Fig.~\ref{Fig.q3_mc_SNR30_inc30} and~\ref{Fig.q3_mc_SNR30_inc70}). 
This behavior is also seen for the $q=6$ binaries, Fig.~\ref{Fig.q6_compMass_SNR30_inc30} and~\ref{Fig.q6_compMass_SNR30_inc70}, with \IMRpT being especially offset.
On the other hand, note that the \IMRNSBH and \SEOBNSBH analyses are more constraining on the NS and BH masses compared to these ``BNS-tuned'' waveform models, similar to the discussion in Sec.~\ref{Sec:MassRatio}.

For the SNR 70 binaries, severe biases are visible, due to two different factors.
The true values are outside of the 90\% credible regions for the $q=2$ binaries, Fig.~\ref{Fig.q2_compMass_SNR70_inc30} and~\ref{Fig.q2_compMass_SNR70_inc70}, when using approximants that do not support NS matter effects. 
This is to be expected since tidal effects are most visible at small mass ratios, and in light of the fact that tides and mass ratios enter the GW phase in combination~\cite{DelPozzo:2013ala}.
This bias of $\sim 0.1 M_\odot$ for the recovered NS mass, though only a small fractional error, could be detrimental when propagated to the inference on the NS EoS, which is very sensitive to changes in NS mass.
It is also interesting to note that the HOMs included in the \XHM model do not affect the inferred masses for these binaries, and indeed recover the same biased masses as the other waveform models without NS matter effects.
A similar behavior is also seen for the $q=3$ binaries, Fig.~\ref{Fig.q3_compMass_SNR70_inc30} and~\ref{Fig.q3_compMass_SNR70_inc70}, where again the models without NS matter effects show larger biases in the NS and BH masses, though smaller than the those of the $q=2$ binaries.

For the $q=6$ binaries, Fig.~\ref{Fig.q6_compMass_SNR70_inc30} and~\ref{Fig.q6_compMass_SNR70_inc70}, we see even stronger biases, but the reason is now different. 
As the more unequal mass ratio reduces the observational impact of the tidal effects, the ``BBH-like'' models can describe the system quite well, and the models tuned to BNS-like (thus light and nearly equal-mass) tidal effects (\SEOBT and \IMRpT) greatly misestimate the NS and BH masses. This might be due to the conditioning applied to the end of \SEOBT and \IMRpT waveforms, which would be outside of the most sensitive part of the detector bandwidth for BNS-like systems, but might leave a detectable imprint for NSBH binaries with increased mass ratio and total masses as the binary merger now occurs at frequencies where the detectors are more sensitive.

Analyses of CBCs containing an object whose mass is reasonably consistent with being a neutron star can intuitively be expected to exhibit some form of tidal effects.
Waveform models that allow for such effects could therefore be believed to measure the source parameters better, as they nominally contain a more accurate description of all relevant physical effects.
Naively following these assumptions for the $q=6$ binaries would, as shown here, potentially lead to significant errors in the inferred astrophysics.
As an example, the \IMRpT analysis in Fig.~\ref{Fig.q6_compMass_SNR70_inc70} would, if taken in isolation, have a strong impact on the inferred maximum NS mass, a parameter which in turn affects the constraints on the NS EoS~\cite{Lawrence:2015oka, Alsing:2017bbc, Cromartie:2019kug, Bauswein:2020aag, Chatziioannou:2020msi, Miller:2019nzo}.
It is worth noticing that one can quantitatively assess the relative goodness of fit to the data of two models by computing Bayes factors. 
We find that the BNS-tuned tidal waveforms are strongly disfavored even when compared to non-tidal waveforms for $q=6$ and SNR 70, which could be used as a figure of merit to exclude them from parameter estimation for specific candidate events. 
Further details about Bayes factors are given below, in Sec.~\ref{Sec:TidalModelSelection}.

\subsection{Matter effects}
\label{Sec:MatterEffects}

Together with the mass, the radius is probably the most interesting astrophysical quantity one can infer from GW observations of neutron stars.
As seen above, in Eq.~\eqref{Eq.Lambda2}, this information is encoded in the tidal deformability of neutron stars, which directly enters the phase evolution of GW signals, though at high post-Netwonian orders~\cite{Hinderer:2007mb,Damour:2009vw, Binnington:2009bb, Vines:2011ud,Dietrich:2018uni}. 
In this section we will discuss the measurement of both radius and tidal deformability. 

\subsubsection{NS tidal deformability}
\label{Sec:LambdaNS}

While GWs carry information about the NS tidal deformability, whether these effects are in practice observable depends heavily on the binary parameters, for a fixed SNR.
Specifically, if the mass ratio is too large, the neutron star will cross the event horizon of the black hole before any significant tidal disruption can occur. The exact value of the mass ratio above which tidal effects are shut off also depends on the black hole spin (as this affects the position of the outer event horizon) and the neutron star compactness or, equivalently, its radius \rns, see App.~\ref{App.CalcRNS}~\cite{Foucart:2010eq, Pannarale:2011pk, Kyutoku:2011vz, Foucart:2012nc, Foucart:2012vn, Kawaguchi:2015bwa, Foucart:2016vxd,Foucart:2018rjc, Pannarale:2015jka, Ascenzi:2018mwp}.
Therefore, we do not expect to gain significant information about tides from the $q=6$ signals. 
We stress that, if one is agnostic and does not a priori exclude the existence of black holes with masses comparable to neutron stars, measuring the deformability of the secondary object as being non-zero would be the main way to \emph{prove} that it was not a BH (unless EM emission is detected, which would be an even stronger indication that a NS was involved in the merger). 

Indeed, at a mass ratio of $q=6$, NR simulations that nominally include the effects of a tidally disrupted NS are indistinguishable from ``pure BBH'' simulations, with the tidal signature on the generated waveform being comparable to, or smaller than, the numerical precision of current NR simulations. 
Thus, we do not expect to be able to constrain the tidal deformability for these types of high mass-ratio NSBH binaries. 

On the other hand, as discussed in Sec.~\ref{Sec.NR}, we should be able to constrain the tidal deformability better for binaries with less asymmetric mass ratio. In light of the above discussion, we expect the $q=2,3$ binaries to be the more favorable configurations in this analysis to measure \LNS: a low-mass black hole (within the putative mass gap between neutron stars and black holes~\cite{Farr:2010tu,Kreidberg:2012ud,Gupta:2019nwj}) with a relatively massive neutron star.  
We stress that black holes with masses in the gap have likely already been discovered:  the lighter component of GW190814 was a $\sim 2.6$~\msun compact object, making it either the heaviest NS or the lightest BH ever found~\cite{Abbott:2020khf}. Moreover, the total mass of GW190425 was $\sim 3.3$\msun~\cite{Abbott:2020uma,Kyutoku:2020xka,Foley:2020kus}: if the final product of the merger was a BH, which is likely, it would have masses in between the BHs of our q2 and q3 simulations\footnote{Whether such black hole would have a high probability of merging again, with a NS, is highly dependent on the environment where it formed.}.
The tidal deformability is indeed best constrained for the most equal-mass system in our study, the $q=2$ binaries from Fig.~\ref{Fig.q2_lambda2_SNR30_inc30} and~\ref{Fig.q2_lambda2_SNR30_inc70}. 
We note, however, that for these signals, \LNS is generally underestimated for all waveform models %
, while still containing the true value within the 90\% CI at SNR 30. 
For the SNR 70 sources, the statistical uncertainty shrinks, while the offsets remain comparable, %
with the exception of \IMRNSBH. 
Even for \SEOBNSBH, which is nominally tuned for this kind of source, we find that the true value of \LNS is outside of the 90\% CI. 
It is worth mentioning that \SEOBNSBH, \IMRpT and \SEOBT all roughly agree with each other, and underestimate \LNS by a similar amount. 

The situation is not too different for the $q=3$ binaries, for which we can additionally use the \LEAplus model, whose range of validity is limited to $q \in [2,5]$). 
At SNR 30, \Cref{Fig.q3_lambda2_SNR30_inc30,Fig.q3_lambda2_SNR30_inc70}, the peak of the \LNS posterior is close to the true value for all approximants, which also agree well with each other, with the exception of \LEAplus and \IMRNSBH whose posteriors are clearly separated and slightly overestimate the \LNS (while still containing the true value within their very large 90\% CIs).
A more complex picture emerges when SNR is 70, \Cref{Fig.q3_lambda2_SNR70_inc30,Fig.q3_lambda2_SNR70_inc70}. 
In this case, we observe that posteriors cluster around two values , one larger and one smaller than the true \LNS, with the true value roughly in between the two sets. 
\IMRpT yields the longest tail in the posteriors among the approximants, its 90\% CI is however only slightly wider than the others'.
It is worth stressing that the differences we see do not simply align with the underlying base model (IMRPhenom or EOB), as instead happens for, e.g., the luminosity distance, Sec.~\ref{Sec:ExtrinsicParams} below. 
We do not have a simple (or complicated) explanation for these features, which are due to the detailed way each approximant implements and calibrates tidal corrections, but we stress again that it also applies to the NSBH-tuned models we are using, and it makes a significant difference.
While based on \IMRNSBH (or \LEAplus, for q3) we would be able to place $\LNS=0$ at a very low confidence level, \SEOBNSBH finds a non-negligeable amount of posterior support there, and would not allow to rule out that the secondary is in fact a black hole, for which $\LNS=0$.

For the $q=6$ sources at SNR of 30, (Fig.~\ref{Fig.q6_lambda2rNS_inc30} and~\ref{Fig.q6_lambda2rNS_inc70}), the recovered posteriors on \LNS are not much different from the prior, explicitly showing that for non-spinning systems at such a large mass ratio, there simply is no information about the NS composition, since the NS plunges into the BH horizon before it is significantly deformed.
At SNR 70, this general behavior still persists but with a slightly more discernible fall off at high \LNS (not visible in the plots, due to the range we show in the horizontal axis, but conveyed by the 90\% CIs quoted in App.~\ref{App.FullPe}).
The clear exception is \IMRpT whose posterior has a visible peak at $\LNS=0$, and is significantly different from the prior. 
To a smaller extent, \SEOBT shows the same trend. 
However, as discussed in Sec.~\ref{Sec:Masses}, these approximants also recover significantly biased mass parameters, and have an unfavorable Bayes factors compared to other approximants at $q=6$.

\subsubsection{NS radius}
\label{Sec:rNS}

Another astrophysically important quantity, capable of constraining the NS EoS through observations with both gravitational and electromagnetic observations~\cite{Ozel:2006bv,Steiner:2010fz,Abbott:2018exr, Abbott:2020uma, Miller:2019cac, Bogdanov:2019ixe,Bogdanov:2019qjb}, is the radius of NSs, \rns.
Unlike \LNS, \rns is not directly encoded in the GW signal, but rather inferred from the measurements of the NS mass and \LNS using fitting formulae (see App.~\ref{App.CalcRNS}).
 
\begin{figure}
\centering
\begin{subfigure}[b]{0.4\textwidth}
   \includegraphics[width=1\linewidth]{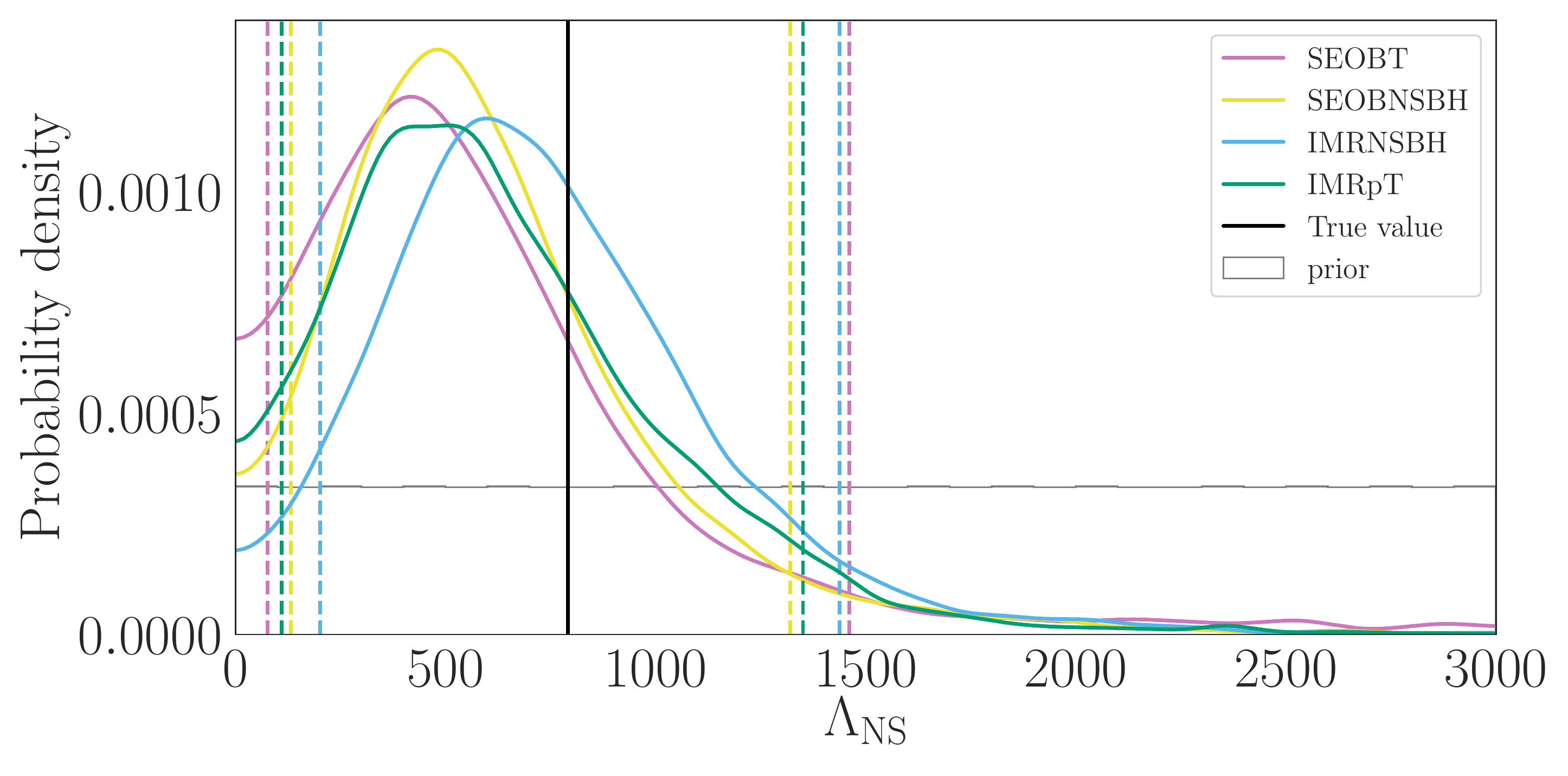}
   \vspace{-1.5\baselineskip}
   \caption{\LNS, SNR 30.}
   \label{Fig.q2_lambda2_SNR30_inc30} 
\end{subfigure}
\begin{subfigure}[b]{0.4\textwidth}
   \includegraphics[width=1\linewidth]{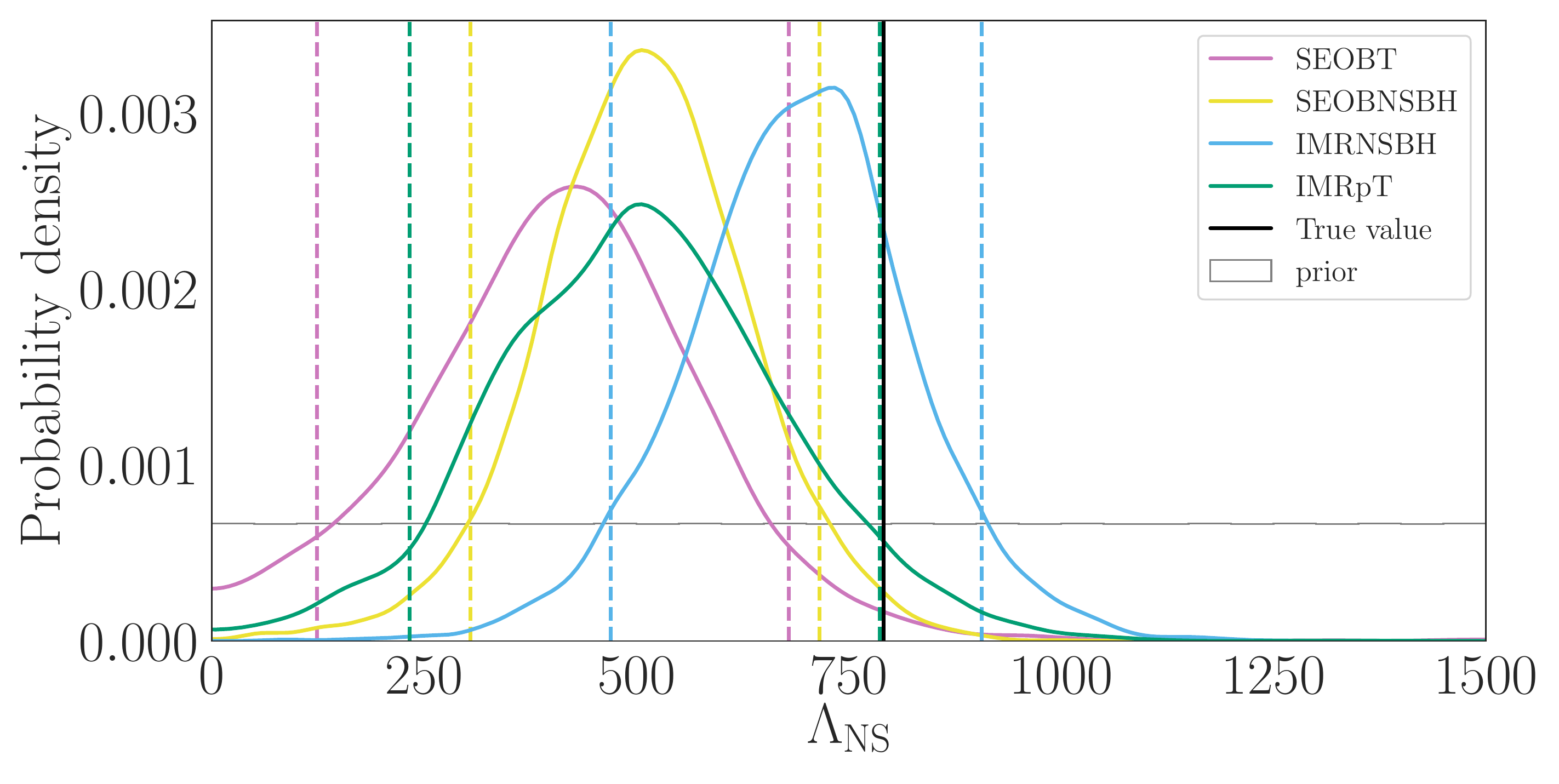}
   \vspace{-1.5\baselineskip}
   \caption{\LNS, SNR 70.}
   \label{Fig.q2_lambda2_SNR70_inc30} 
\end{subfigure}
\begin{subfigure}[b]{0.4\textwidth}
   \includegraphics[width=1\linewidth]{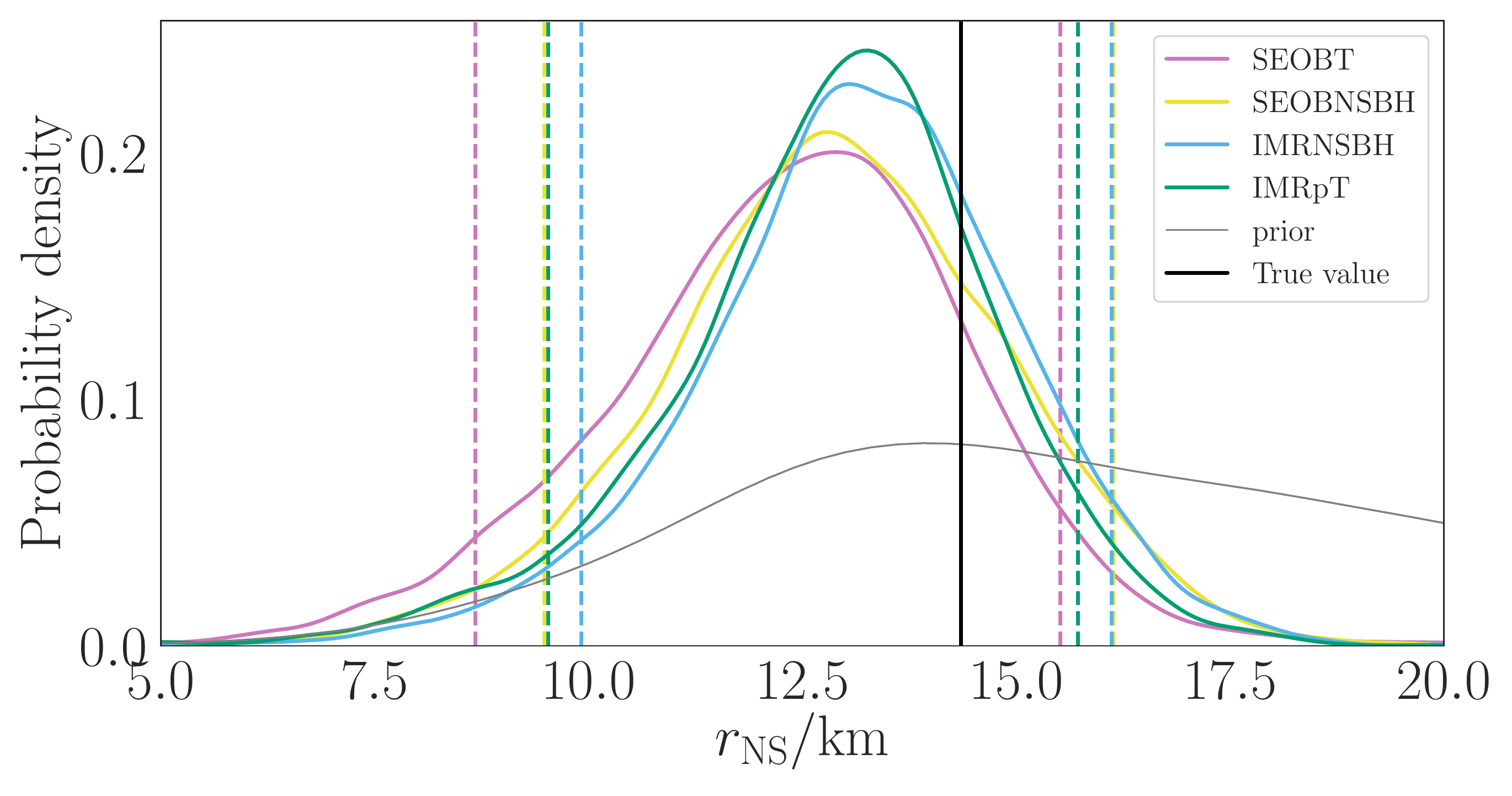}
   \vspace{-1.5\baselineskip}
   \caption{\rns, SNR 30.}
   \label{Fig.q2_rNS_SNR30_inc30} 
\end{subfigure}
\begin{subfigure}[b]{0.4\textwidth}
   \includegraphics[width=1\linewidth]{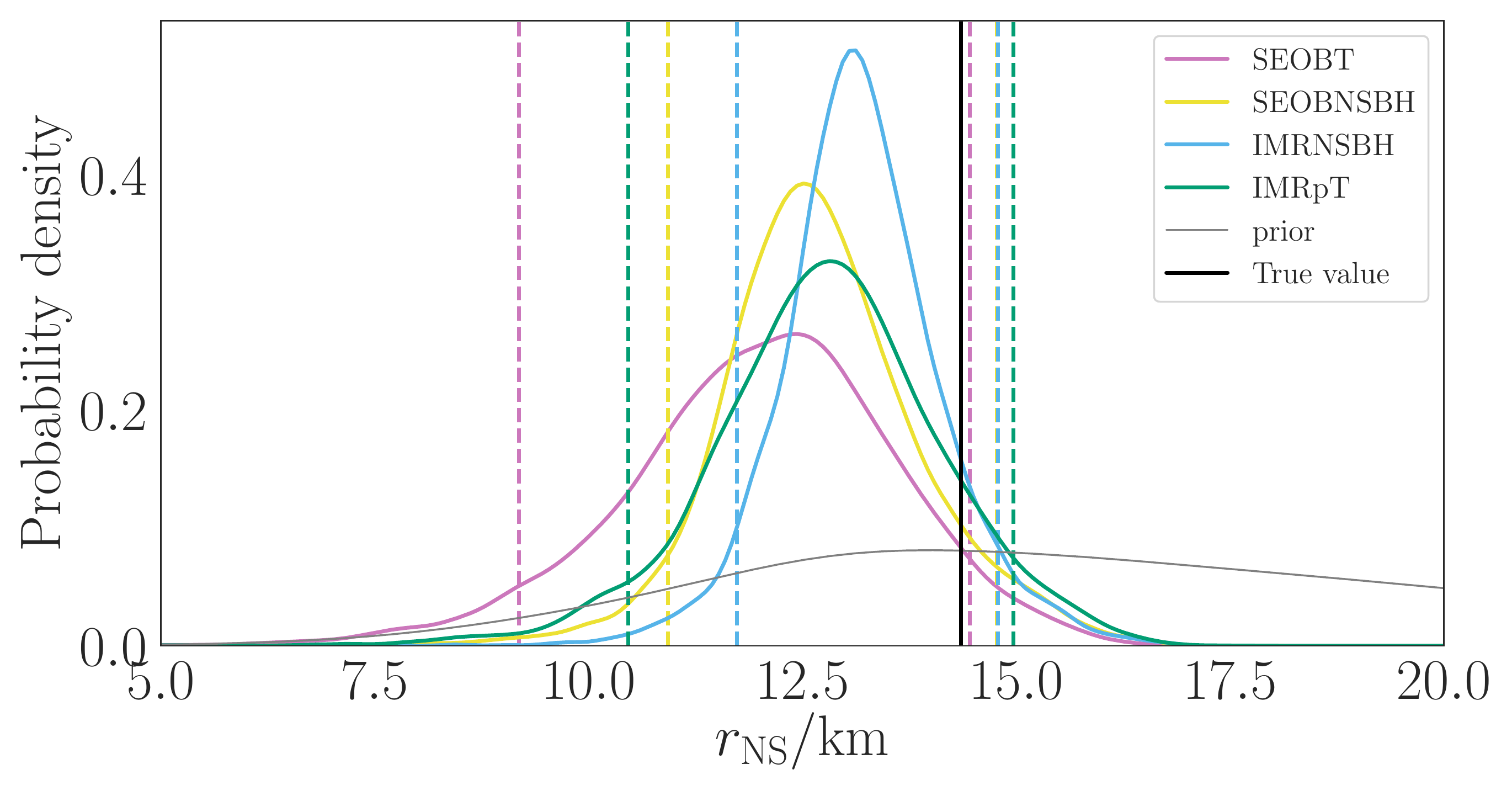}
   \vspace{-1.5\baselineskip}
   \caption{\rns, SNR 70.}
   \label{Fig.q2_rNS_SNR70_inc30} 
\end{subfigure}
\caption{Posterior distributions for \LNS and \rns recovered by different approximants for $q=2$, inclination $30^{\circ}$. 
The dashed lines mark the 90\% credible intervals, same for all 1D plots to follow.}
\label{Fig.q2_lambda2rNS_inc30}
\end{figure}

\begin{figure}
\centering
\begin{subfigure}[b]{0.4\textwidth}
   \includegraphics[width=1\linewidth]{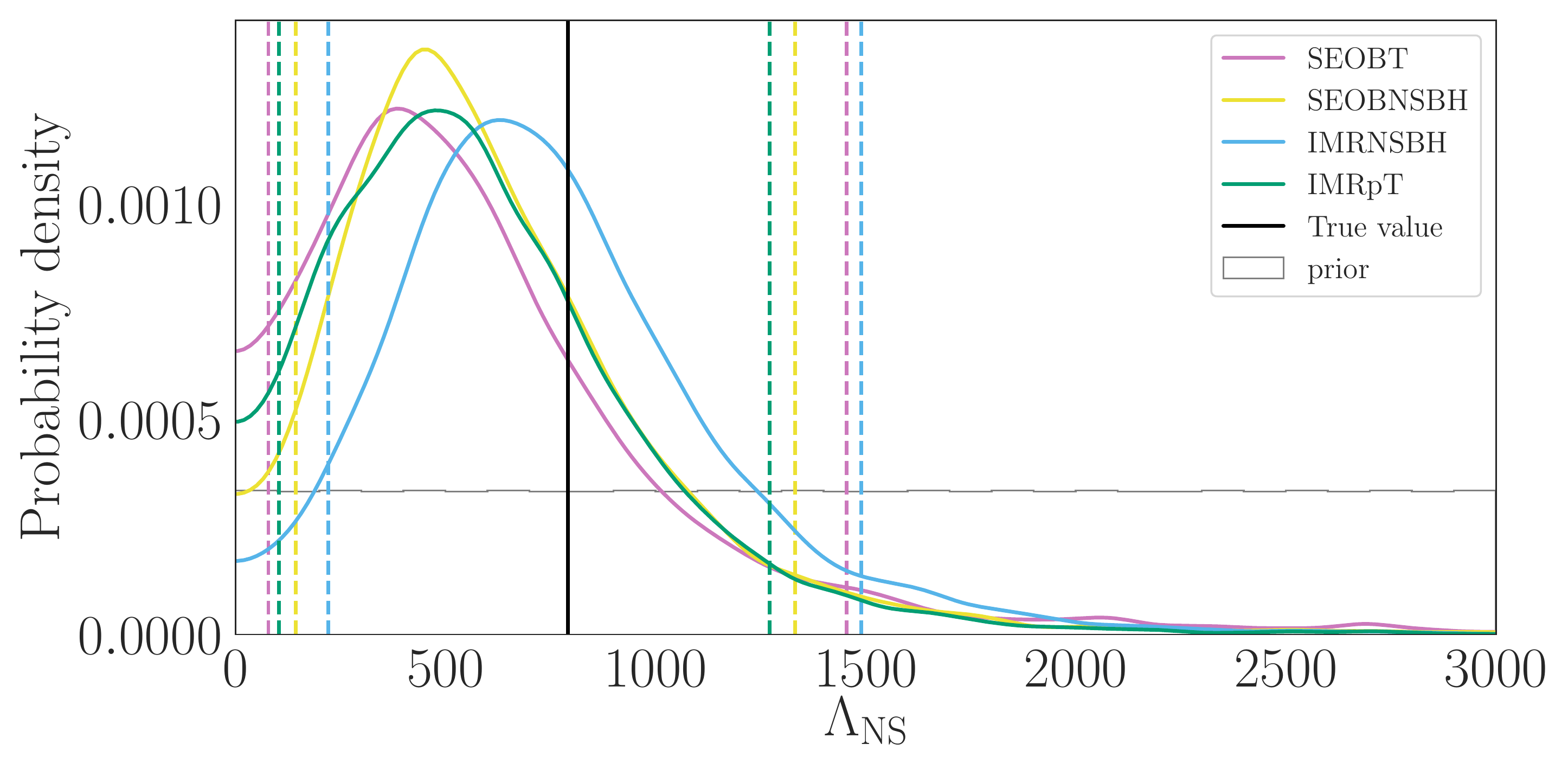}
   \vspace{-1.5\baselineskip}
   \caption{\LNS, SNR 30.}
   \label{Fig.q2_lambda2_SNR30_inc70} 
\end{subfigure}
\begin{subfigure}[b]{0.4\textwidth}
   \includegraphics[width=1\linewidth]{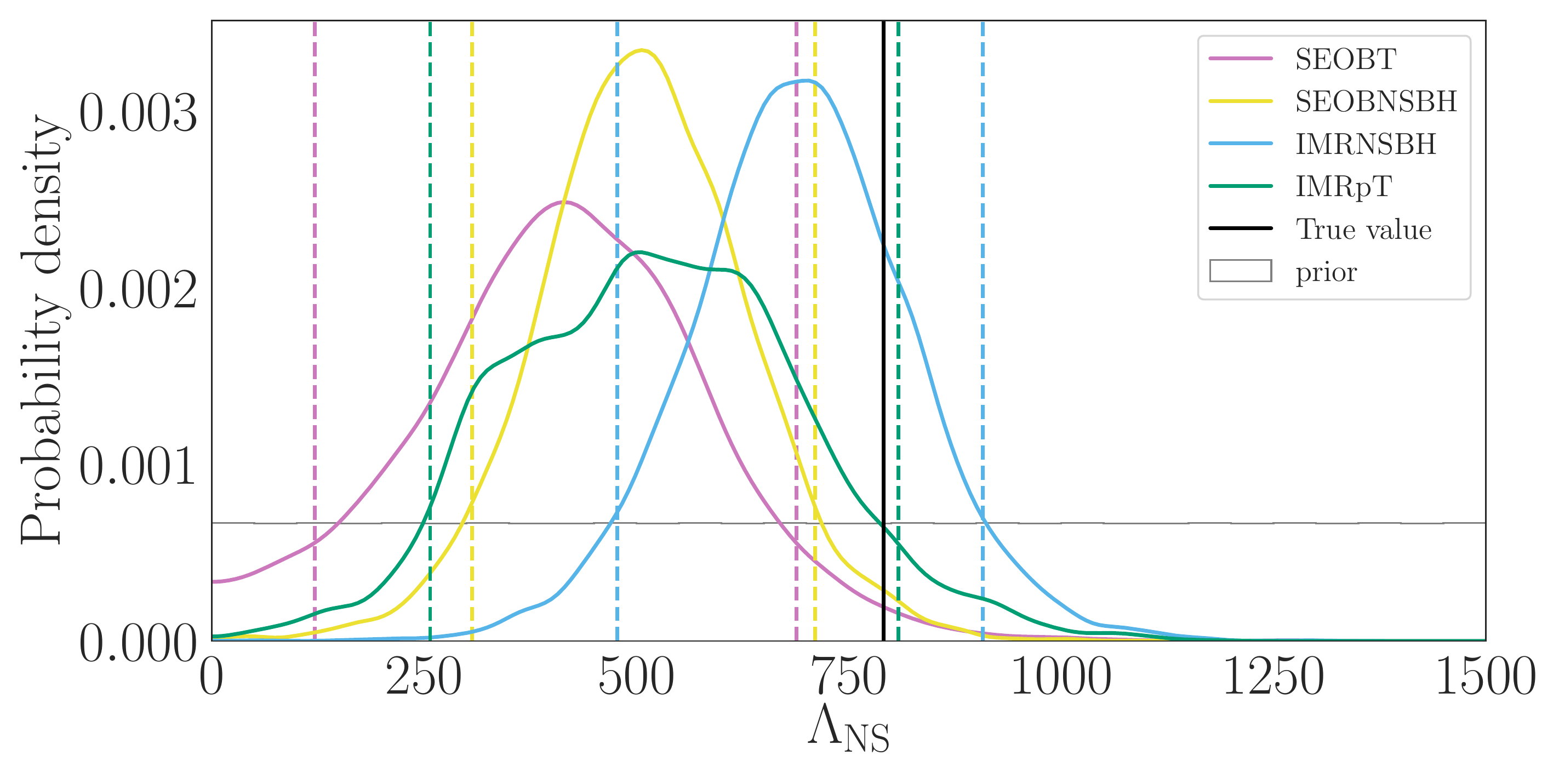}
   \vspace{-1.5\baselineskip}
   \caption{\LNS, SNR 70.}
   \label{Fig.q2_lambda2_SNR70_inc70} 
\end{subfigure}
\begin{subfigure}[b]{0.4\textwidth}
   \includegraphics[width=1\linewidth]{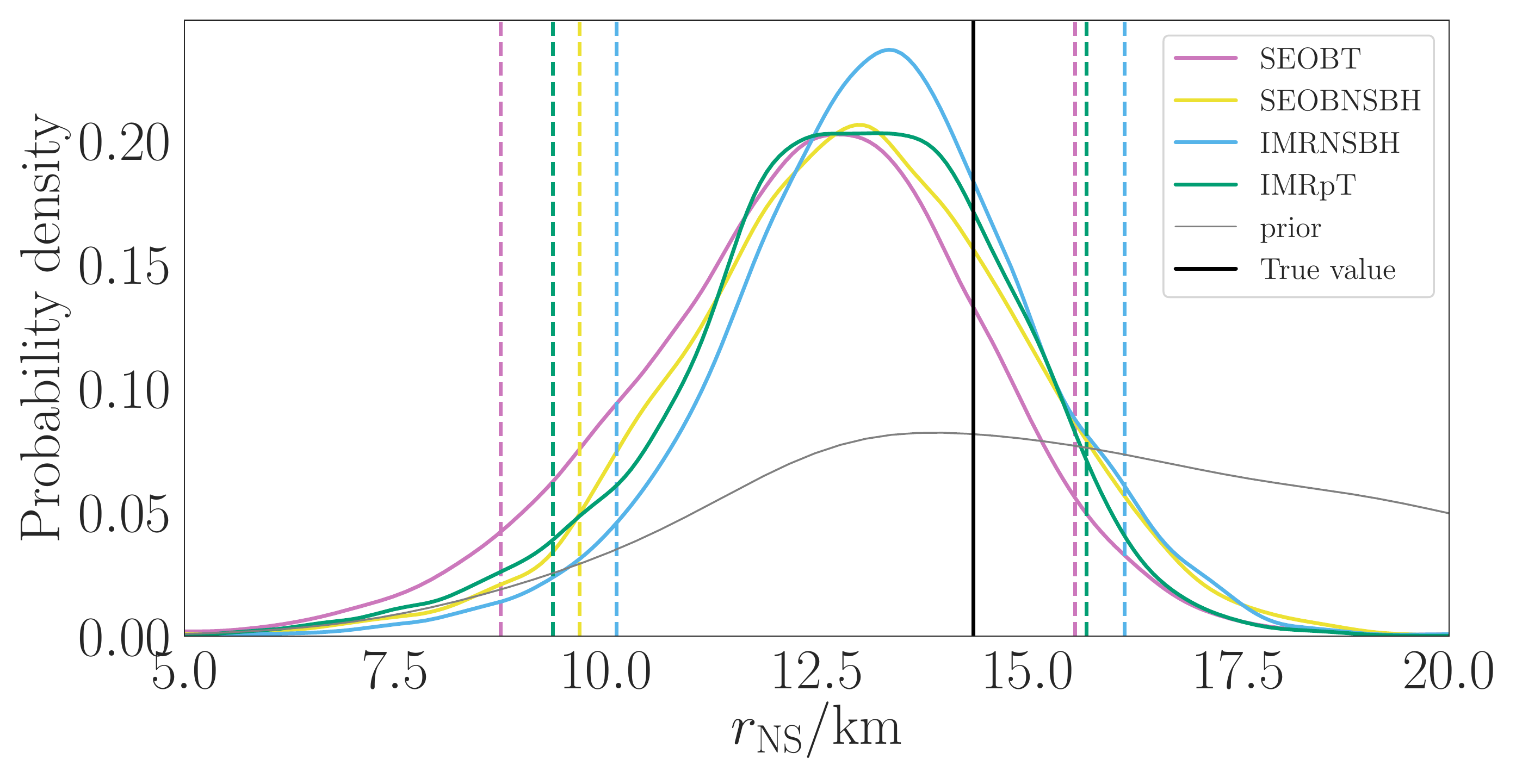}
   \vspace{-1.5\baselineskip}
   \caption{\rns, SNR 30.}
   \label{Fig.q2_rNS_SNR30_inc70} 
\end{subfigure}
\begin{subfigure}[b]{0.4\textwidth}
   \includegraphics[width=1\linewidth]{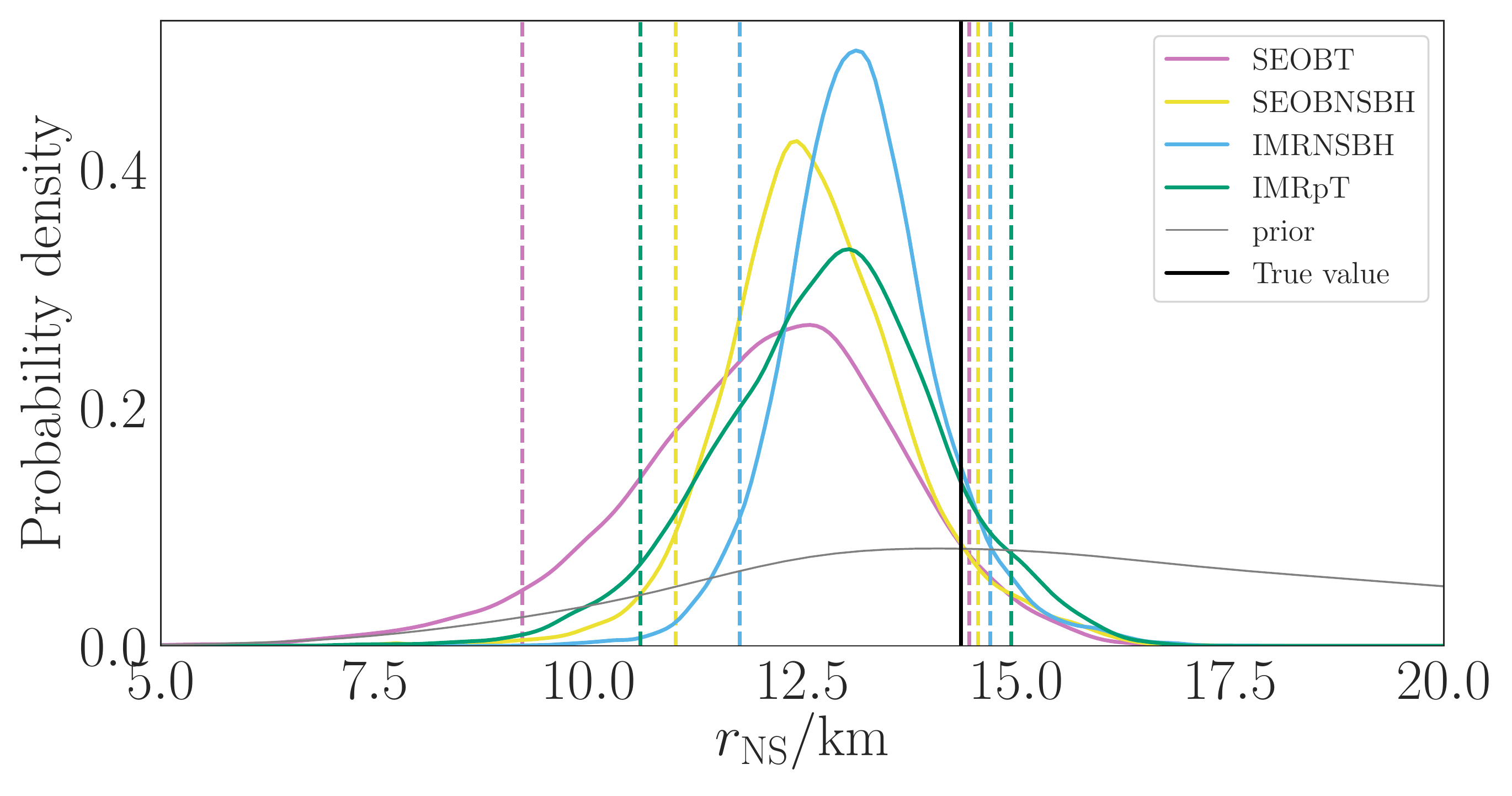}
   \vspace{-1.5\baselineskip}
   \caption{\rns, SNR 70.}
   \label{Fig.q2_rNS_SNR70_inc70} 
\end{subfigure}
\caption{Posterior distributions for \LNS and \rns recovered by different approximants for $q=2$, inclination $70^{\circ}$. }
\label{Fig.q2_lambda2rNS_inc70}
\end{figure}

We report these posteriors in panels (c) and (d) of Figs.~\ref{Fig.q2_lambda2rNS_inc30} -- ~\ref{Fig.q6_lambda2rNS_inc70}. Overall, the radius is not constrained with high precision at SNR 30, with typical widths of $\sim 7$~km (compare with $\lesssim 4$~km for the BNS GW170817~\cite{Abbott:2018exr}, which had a comparable SNR). The fact that the NSBH sources we study do not provide a radius measurement as precise as GW170817 is due to the dependency of the tidal terms on the mass ratio, and the fact that fewer waveform cycles are in band for CBCs with larger chirp masses.

For the $q=2$ binaries, the inferred \rns distributions show a smaller spread than the respective \LNS posteriors.
For SNR 70 especially (Fig.~\ref{Fig.q2_rNS_SNR70_inc30} and~\ref{Fig.q2_rNS_SNR70_inc70}), the inferred \rns is underestimated, though the true value is contained within the 90\% CI.

A similar behavior is seen for the $q=3$ binaries, again with a reduced spread compared to \LNS.
In the SNR 70 analyses (Fig.~\ref{Fig.q3_rNS_SNR70_inc30} and~\ref{Fig.q3_rNS_SNR70_inc70}), \SEOBNSBH and \SEOBT return accurate distributions for \rns, whereas \IMRNSBH and \LEAplus slightly overestimate \rns, while still including the true value at a high confidence level. 

For the SNR 70 sources, typical 90\% CIs are of $\sim 3-4$~km ($4-5$~km) for $q=2$ ($q=3$).

As with \LNS, the $q=6$ analyses recover very broad posteriors for \rns, and only exclude extremely large values of the radius ($\geq 20$~km), Figs.~\ref{Fig.q6_lambda2rNS_inc30} and ~\ref{Fig.q6_lambda2rNS_inc70}. It is worth stressing that most of this information does not come from \LNS, but rather from the measurement of the NS mass (cf. Sec.~\ref{Sec:ComponentMasses}).
Finally, while \IMRpT finds very biased posteriors for \LNS and NS mass at $q=6$, the two biases cancel out, giving a derived posterior on \rns not too different from what is obtained with other approximants.

\subsubsection{Model selection}
\label{Sec:TidalModelSelection}

As mentioned above, the relative goodness of fit of waveform models to the data in hand can be quantified by calculating the Bayes factors between them.
If one calls $H_1$ the model where the approximant $A_1$ is used to analyze the data, and $H_2$ the model where the approximant $A_2$ is used, the Bayes factor can be obtained as the ratio of the models' evidences, Eq~\eqref{Eq.Evidence}:
\begin{equation}
\label{Eq.BayesFactor}
 \mathcal{B}^{A_1}_{A_2}\equiv \frac{Z(H_1)}{Z(H_2)}
\end{equation}
with $\mathcal{B}^{A_1}_{A_2}>0$ if the model $H_1$, i.e. if the waveform model $A_1$, is preferred.

By comparing the ratio of evidences for competing models, one can quantify the relative belief that a given model represents the true signal in the data~\cite{Veitch:2009hd} in a way that also automatically penalizes models with more degrees of freedom, or larger priors.

In Table~\ref{Table.BSN_SNR30_inc30_comp} --~\ref{Table.BSN_SNR70_inc70_comp} we present the \emph{natural log} Bayes factor between some of the models used in this study. 

As one would expect, for the $q=2$ binaries, there is more support for models that include tidal effects. 
\SEOBNSBH and \IMRNSBH match the data equally well, and have large odds ratios relative to all other models, likely due to the fact that they have been tuned specifically for NSBH signals, and that they do not allow for spin in the neutron star, see~\Cref{Table.Prior_BHNSq3s0}. 
\emph{Because} the true NS spin is actually zero, they are not penalized by that limitation and their odds ratios are are boosted by the smaller prior volume. 
The same would not necessarily be true if the source contained a spinning NS. 
\SEOBT is significantly favored over \SEOB, while there is only a mild support for \IMRpT over \IMRp even at SNR 70. %

The results for the $q=3$ binaries are more nuanced, while still showing a mild preference for NSBH approximants. 
The \LEAplus model performs as well as the other two, newer, NSBH approximants, %
The comparisons between \SEOBT and \SEOB, as well as between \IMRpT and \IMRp, are rather inconclusive compared to the results for the $q=2$ binaries.

For the $q=6$ binaries, the NSBH approximants perform slightly better than non-tidal waveforms, and much better than the two tidal models that are tuned for BNS systems. 
It is interesting to observe that for $q=6$ \SEOB (\IMRp) does better than \SEOBT (\IMRpT). 
This is not due to the fact that tides are unmeasurable, and hence ``unnecessary'' in the model: as we have seen before, no significant constraints can be placed on \LNS (cf. Sec.~\ref{Sec:LambdaNS}) for these sources, and not much information is gained relative to the prior distribution.
In this case, no significant Occam penalty~\cite{mackay2003information} is assigned to the models with tides. 
Hence, the fact that the BNS-tuned tidal waveform models are disfavored over their related non-tidal models for the $q=6$ binaries must be attributed to them failing to properly describe the NSBH waveforms in that mass range.

\subsection{Spins}

There are multiple reasons why an accurate measurement of the spin of black holes in NSBH systems is important. 
First, the BH spin should be measured more precisely in NSBHs than in BBHs, since the potentially large mass ratio of NSBHs enhances the effect of spin precession and spin-orbit coupling, yielding a larger amount of phase and amplitude modulation than what would be present in an equal mass system with similar spins. 
NSBHs might very well be the systems that yield the most precise measurement of BH spins in the next few years. 
It is thus important that accuracy follows.
Second, spins are a good tracer of the formation channel of compact binaries~\cite{Vitale:2015tea, Farr:2017gtv, Farr:2017uvj, Zhu:2017znf, Miller:2020zox, Stevenson:2017dlk}.
A precise and accurate measurement of spins could be key to determine whether the formation pathways of BBH and NSBH systems are different. 
When the masses of the compact objects in a binary are comparable, GWs provide a good measurement of the effective spin, \chieff, the mass-weighted projection of the total spin along the orbital angular momentum~\cite{Damour:2001tu,Racine:2008qv,Santamaria:2010yb,Ajith:2009bn} (G=c=1)
\begin{equation}
\chi_{\mathrm{eff}}= \frac{\boldsymbol{S}_1/m_1+\boldsymbol{S}_2/m_2}{(m_1+m_2)}\cdot\frac{\boldsymbol{L}}{|\boldsymbol{L}|}
\end{equation}
where $\boldsymbol{S}_{1,2}$ are the individual spins and $\boldsymbol{L}$ the orbital angular momentum of the system; but not of the individual spins. 
However, the individual spins are poorly constrained~\cite{LIGOScientific:2018mvr,Purrer:2015nkh,Vitale:2016avz}, but see~\cite{Biscoveanu:2020are}.
As the mass ratio increases, the spin of the primary becomes the leading contribution to \chieff (this is even more true for an NSBH, as NSs are expected to have small spins), and one indeed finds that the spin of the primary becomes measurable~\cite{Vitale:2014mka,Haster:2015cnn,Mandel:2015spa}.

Because our simulations have non-spinning black holes and neutron stars, we will not be able to probe the quality of spin measurement for large spins. 
However, it is still very interesting to show if waveform systematics affect the measurement of small spins in NSBH because a) most of the BH found to date are consistent with having small or no spin~\cite{Farr:2017gtv, Farr:2017uvj, Miller:2020zox, LIGOScientific:2018jsj} and b) there is correlation between effective spin and mass ratio~\cite{Ng:2018neg}, as well as between mass ratio and tidal parameters~\cite{DelPozzo:2013ala}, hence different waveforms might produce visibly different posteriors. 
We expect biases to be more visible when the mass ratio is small enough that the NS can acquire significant tidal deformation and disruption before merging with the BH. 

Indeed, this is what our simulations show, in~\Cref{Fig.compMass_BHNSq2s0_inc30,Fig.compMass_BHNSq2s0_inc70} for $q=2$. 
While some differences in behavior between tidal and non-tidal approximants are already visible at SNR 30, it is only when the signals are very loud, SNR 70, that the tension becomes significant compared to the statistical uncertainties. 
For these loud simulations, Fig~\ref{Fig.q2_compMass_SNR70_inc30} and~\ref{Fig.q2_compMass_SNR70_inc70}, the models that include tides recover the true value of \chieff and $q$, with the 90\% CI of $\sim 0.1$ (with some small differences depending on inclination angle and the tidal waveform model). 
As discussed in Sec.~\ref{Sec:Masses}, biases in the models without tides are also visible for the mass ratio and hence the component masses. 
For all of these parameters, the true values are marginally included in, or excluded from, the 90\% CI. 
We do not observe significant differences between the NSBH-tuned waveforms and the other tidal waveforms.
It is also worth stressing that the \XHM waveform does not perform better, or even differently, than the other non-tidal waveforms, showing explicitly that even at this high SNR, the missing tidal terms have a dominant effect on the waveform systematics over the missing HOMs for small mass ratios.

As the mass ratio increases, the biases in the spin posteriors of the $q=2$ binaries become less and less apparent, as one would expect given that the effect of tides decreases with more unequal masses. 
However, for $q=3$ at SNR 70, Fig. \ref{Fig.q3_compMass_SNR70_inc30},~\ref{Fig.q3_compMass_SNR70_inc70}, we still see a bias for the non-tidal IMR models, whereas the EOB models are consistent with the true values of \chieff and masses. 
In general, we find that non-tidal IMR models tend to overestimate \chieff and, due to its correlation with $q$~\cite{Ng:2018neg}, to overestimate the mass ratio. 
Finally, for $q = 6$, Fig. \ref{Fig.compMass_BHNSq6s0_inc30} and~\ref{Fig.compMass_BHNSq6s0_inc70}, the true value of \chieff and masses are within the 90\% CIs even at SNR 70 for all non-tidal approximants. 
This suggests that for $q\simeq 5$, even higher SNRs would be needed for the measurement of \chieff and masses to be limited by waveform systematics.
On the other hand, we observe that \IMRpT, which has tidal effects, gives significantly biased results at this high mass ratio. 
While the statistical uncertainties are large enough at SNR 30 that the posterior is still consistent with the true values, the same is no longer true at SNR 70, and for both masses and \chieff the \IMRpT posteriors are in significant tension with the true values (even the component spin magnitude is heavily biased, as presented in Tab.~\ref{Table.Param_BHNSq6s0_SNR70}).
This can be explained with the fact that \SEOBT and \IMRpT are constructed with the goal of matching the late inspiral of BNSs, for which the mass ratios are close to 1. 
As the true mass ratio increases, we are using these two models further and further from their range of validity. 
In fact, while less pronounced than for \IMRpT, one can see that even \SEOBT starts diverging from the other approximants at $q=6$. 
This suggests that for mass ratios high enough, waveforms without tidal terms actually do better than waveforms with tidal terms tuned to only BNS mergers. 
This explanation for the biases is corroborated by the total lack of biases in \LEAplus (only used for the q3 analysis), \IMRNSBH and \SEOBNSBH, all of which have tidal terms that have phase and amplitude corrections tuned against NSBH systems.

For all of the simulations, we find that \LEAplus, \IMRNSBH and \SEOBNSBH yield the most precise estimates of \chieff. 
However, more than representing a true feature of these models, this is merely a consequence of their prior support as they do not allow spin in the NS and thus enable a narrower range of spin for the BH.
This reduces correlations in the GW phase, and hence yields a better measurement of the only spin parameter.

To summarize, we find that 90\% statistical uncertainties for \chieff are typically around $\sim 0.16$ for SNR=30 sources (with small variations depending on the mass ratio) and $\sim 0.08$ for the SNR 70 sources. 
In fact, the ratio of statistical uncertainties for any given source when measured at SNR 70 and at 30 is close to the ratio of SNRs, as one would expect for loud enough sources for which the Fisher matrix limit is valid~\cite{Vallisneri:2007ev,Zanolin:2009mk,Vitale:2010mr,Vitale:2011zx}.

Given that the mass ratio of these events is far from unity, one might hope to also measure the individual (BH) spins, and not only \chieff. 
In general, we find that \IMRp yields consistently larger uncertainties, followed by spin-aligned waveforms (\IMRpT,\SEOBT,\XHM,\SEOB) and by single-spin waveforms (\LEAplus, \IMRNSBH and \SEOBNSBH). 
As in the case of \chieff, these differences can be explained with the reduced parameter space covered by different models. 
\IMRp includes a prescription for effective spin-orbit precession, and covers a higher dimensionality than any other waveform in our set. 
Conversely, \LEAplus, \IMRNSBH and \SEOBNSBH only allow the black hole to be spinning (with a smaller maximum amplitude), and only along the orbital angular momentum (see Table~\ref{Table.Approx}), while setting the neutron star spin to be exactly 0.

If it worth stressing that for none of our configurations can we constrain the magnitude of the neutron star spin (for the waveform models that allow it to vary from 0) to below $\sim0.4$ at the 90\% confidence level. 
This suggests that even for NSBH sources, constraining the neutron star spin to values consistent with the range of spins of known pulsars will be challenging and require extremely loud sources. 
We do not expect this conclusion to depend significantly on the fact that our BHs did not have any spins, as similarly poor constraints on the NS spin in $10-1.4~\msun$ NSBH with precessing spins were reported by Ref.~\cite{Vitale:2014mka}, though they worked with inspiral-only waveforms that did not include tides or higher order modes.

\subsection{Extrinsic parameters}
\label{Sec:ExtrinsicParams}

In this section we focus on the measurement of the inclination angle and the luminosity distance, both of great importance for fully exploiting the scientific potential of NSBH sources.
At least some of the NSBHs are expected to produce EM radiation as they merge~\cite{Foucart:2010eq, Pannarale:2011pk, Kyutoku:2011vz, Foucart:2012nc, Foucart:2012vn, Kawaguchi:2015bwa, Foucart:2016vxd, Foucart:2018rjc, Pannarale:2015jka, Ascenzi:2018mwp,Lee:2007js,Nakar:2007yr,Clark:2014jpa,Pannarale:2014rea}, making an accurate measurement of their luminosity distance crucial for a successful EM follow-up program. 
Furthermore, the potentially small statistical uncertainty in their luminosity distance results allows NSBHs to be valuable standard sirens in a measurement of the Hubble constant~\cite{Chen:2018omi}.
Measurement of the orbital orientation could be used to distinguish between competing kilonova models~\cite{Metzger:2019zeh} and, more generally, to study their detailed emission angular pattern at all wavelengths.

We report the inclination/luminosity distance corner plots for the face-on (i.e. true inclination 30\dg) systems in Figs.~\ref{Fig.dist_inc_BHNSq2s0_inc30},~\ref{Fig.dist_inc_BHNSq3s0_inc30}, and~\ref{Fig.dist_inc_BHNSq6s0_inc30}. 
It is worth underlining a few common features (the full set of results can be found in tables in Appendix~\ref{App.FullPe}).
First, the only waveform model with HOMs among those we use, \XHM, yields both smaller statistical errors and smaller offsets relative to the true value. 
Smaller statistical errors are not unexpected, since the true signals \emph{do} have HOMs, which are known to help break the distance-inclination degeneracy~\cite{Bustillo:2016gid} hence reducing the statistical uncertainty. 
One might be surprised that systematic errors are smallest for \XHM even though it does not allow for tides, even when the mass ratios are small. 
This can be explained with the fact that HOMs affect the overall amplitude since they change the angular dependence on the orbital orientation, whilst tides only affect the late inspiral and mostly the phase of the waveform, thus not as directly related to distance and inclination.
This also explains why, while \XHM performs similarly to other IMR waveforms at small mass ratios for which HOMs are less important, it does significantly better at $q=6$.
For example, at SNR$=30$ and inclination 30\dg, the 90\% relative uncertainty for the luminosity distance is $45\%$ for all IMR models when $q=2$, but decreases to $32\%$ for \XHM only when $q=6$, while staying above $40\%$ for the other IMR approximants. 
Biases for the luminosity distance are usually constrained to lie within the $90\%$ credible intervals for the recovered posterior, with typical offsets of the order of \si$5-15\%$ of the statistical uncertainty for the runs with SNR 30. 
As the SNR increases, the statistical uncertainties shrink, making systematic offsets percentually more important, though usually still smaller than the corresponding statistical uncertainty. 
The only exception is the \IMRpT posterior for $q=6$, Fig.~\ref{Fig.q6_dist_inc_SNR70_inc30}, which is very narrow and only marginally consistent with the true value. 
As already discussed above, this approximant yields biases for most parameters at $q=6$, which is quite far from its intended region of validity.

The situation is quite different when the sources are simulated at an inclination angle of $70$\dg. 
We find that most waveform families severely over-estimate the distance, with the true value barely included in the posterior, Fig.~\ref{Fig.dist_inc_BHNSq2s0_inc70}, ~\ref{Fig.dist_inc_BHNSq3s0_inc70}, and~\ref{Fig.dist_inc_BHNSq6s0_inc70}. 
This results in an orbital orientation measurement closer to face-on/off, and in turn affects the estimation of the source-frame chirp mass, as seen above.
This behavior is not unexpected, and can be explained with the strong Bayesian prior in the distance (proportional to $D_L^2$, and roughly uniform in comoving volume at the relatively small distances in our simulations), as well as the fact that the waveform approximants without HOMs do not strongly depend on the inclination angle. 
In a Bayesian framework it is thus often more advantageous to overestimate the distance (which comes with a prior boost) and compensate by measuring an orientation closer to face-on/off. This was explicitly shown for models without HOMs in~\cite{Chen:2018omi} (see also~\cite{Usman:2018imj}). 
It is also consistent with the fact that the only model with HOMs in our set, \XHM, usually recovers a posterior closer to the true value, and more and more so as the mass ratio increases, which as discussed above enhances the effect of HOMs. 
For the $q=3$ and $q=6$ runs, the \XHM posterior is clearly separated from all approximants, at both SNRs.
On the other hand, the statistical uncertainties on the distance for the high inclination runs are not significantly smaller than when the systems are closer to face-on. 
In fact, they can be larger. 
This is partially an artefact of quoting the 90\% credible intervals relative to the true value: as the inclination increases, the true distance of the source must be decreased to keep the same SNR. 
Since the absolute uncertainty can increase faster with the true distance than with inclination, the relative uncertainties on inclination can get larger (see Fig. 1 of ~\cite{Chen:2018omi}).
Overall, the medians for the high inclination runs are offset from the true value by significant fractions of the statistical uncertainty. 
The smallest offset we observe is $\si40\%$ of the statistical uncertainty for \XHM when $q=6$. 
Typical values are $50\%$ or larger.

It is interesting to compare non-HOM models based on the EOB vs the IMRPhenom formalisms. 
We see that the EOB-based models usually yield posteriors for the luminosity distance with a more pronounced tail toward small distances than IMRPhenom-based modes, while generally peaking at similar values. 
While we do not have a full explanation, we note the EOB-based modes we are using do not allow for spin-precession, unlike \IMRp and \IMRpT. 
Because spin precession causes amplitude (and phase) modulation that also breaks the distance-inclination degeneracy~\cite{Apostolatos:1994mx}, it is possible that the precessing models yield different posteriors due to the fact that some distance-spin configurations would be excluded when precession is not observed. 
This interpretation seems to be supported by the behavior of \IMRNSBH which is IMRPhenom-based but does not allow for precession. 
We see, for example in Fig.~\ref{Fig.q6_dist_inc_SNR30_inc30} how its posteriors follow closely those of the EOB models, rather than those of the other IMRPhenom's.

Overall, our results show that all models broadly agree for the runs with inclinations of $30$\dg. 
It is only with the high-inclination sources that we start seeing large intra-waveform differences for extrinsic parameters. 
We see a few instances where two posteriors are nearly disjoint: for $q=3$ and $q=6$ at SNR 70, the \XHM posterior is in strong tension with \IMRp and even more with \IMRpT (the tension with the EOB models is milder, since those have longer tails, e.g. Fig.~\ref{Fig.dist_inc_BHNSq6s0_inc70}).
The three NSBH-tuned models do not perform better than the other tidal-models when it comes to the measurement of distance and inclination.

\section{Conclusions}
\label{Sec.Conclusions}

Observations of neutron star black hole coalescences can lead to significant insights into the nature of neutron stars, for example yielding a precise measurement of their mass and radius, or providing information on their formation channels. 
However, GWs from NSBHs are very challenging to simulate with current numerical relativity tools. 
The presence of matter, of higher order modes enhanced by the high mass ratio, of the potential of BH spin precession, and the fact that the late inspiral and merger phases will be in a more sensitive frequency band of the detectors than that for BNSs, make it imperative to verify the role of waveform systematics.

In this work, we have created NSBH hybrid waveforms with recent NSBH NR simulations at three mass ratios, $q=2,3,6$.
We projected the signal into a three-interferometer network, and ran a full parameter estimation campaign, using most of the relevant waveform families available in the LIGO Algorithm Library~\cite{lalsuite}, including three that were especially built for NSBH systems.
For each mass ratio, we have considered 4 configurations, where the orbital orientation and the network SNR had all of the pairwise combinations of $\inc=30\dg,70\dg $ and $\mathrm{SNR}=30,70$.
This gave a total of 88 parameter estimation runs, making this study one of the most extensive analyses of statistical and systematic uncertainties in the analysis of NSBH systems to date.

\begin{figure}
\centering
\begin{subfigure}[b]{0.4\textwidth}
   \includegraphics[width=1\linewidth]{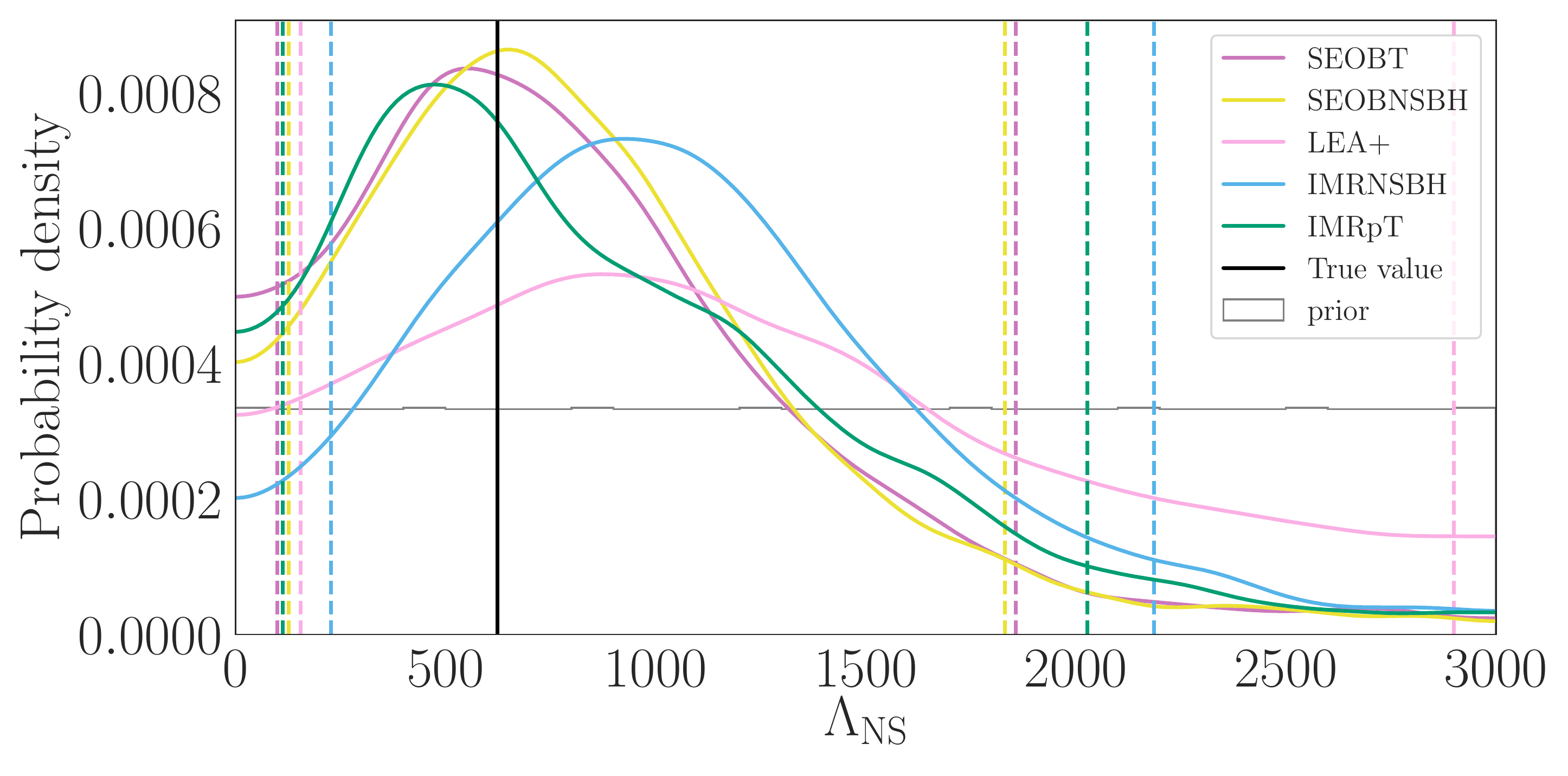}
   \vspace{-1.5\baselineskip}
   \caption{\LNS, SNR 30.}
   \label{Fig.q3_lambda2_SNR30_inc30} 
\end{subfigure}
\begin{subfigure}[b]{0.4\textwidth}
   \includegraphics[width=1\linewidth]{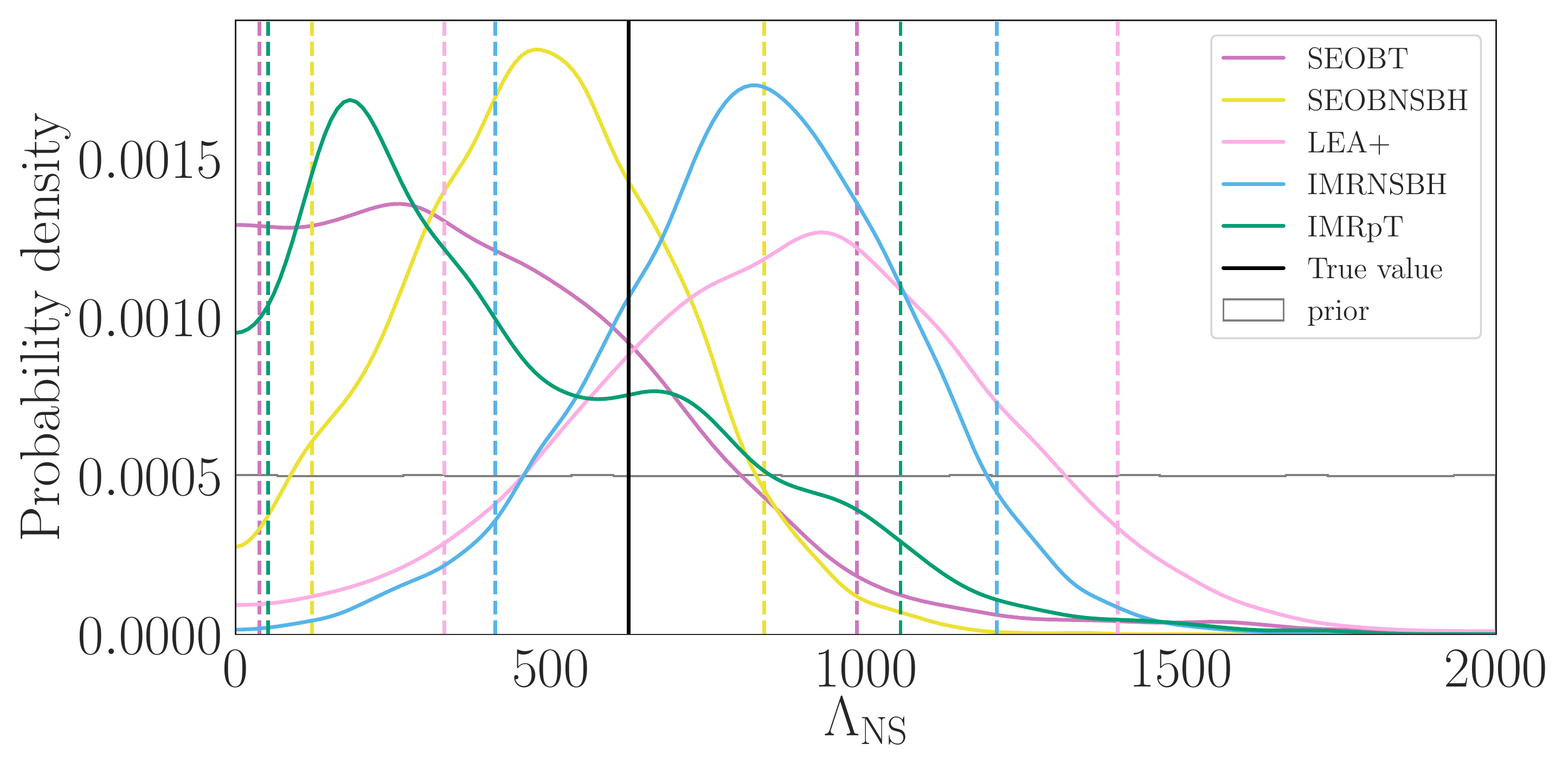}
   \vspace{-1.5\baselineskip}
   \caption{\LNS, SNR 70.}
   \label{Fig.q3_lambda2_SNR70_inc30}
\end{subfigure}
\begin{subfigure}[b]{0.4\textwidth}
   \includegraphics[width=1\linewidth]{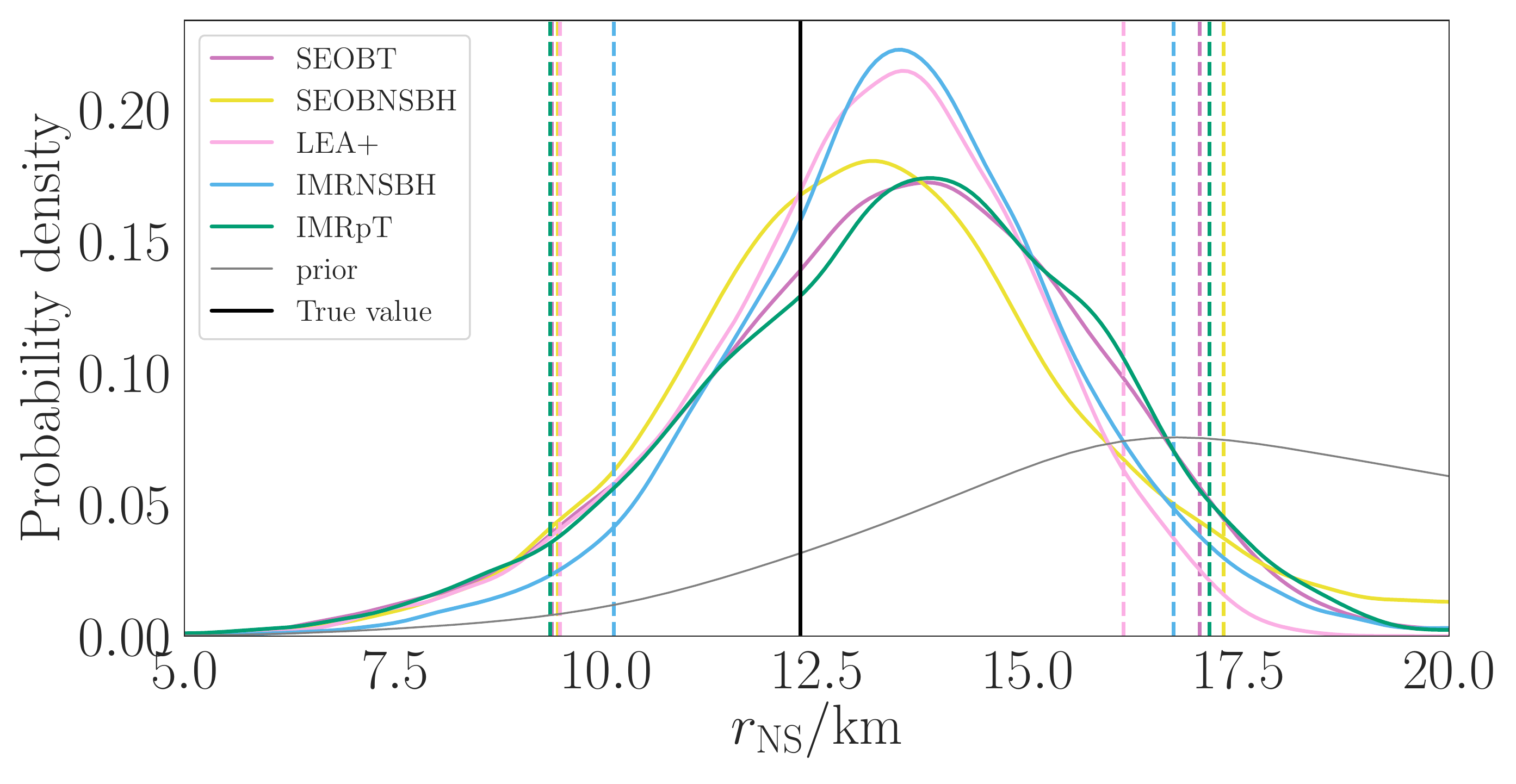}
   \vspace{-1.5\baselineskip}
   \caption{\rns, SNR 30.}
   \label{Fig.q3_rNS_SNR30_inc30} 
\end{subfigure}
\begin{subfigure}[b]{0.4\textwidth}
   \includegraphics[width=1\linewidth]{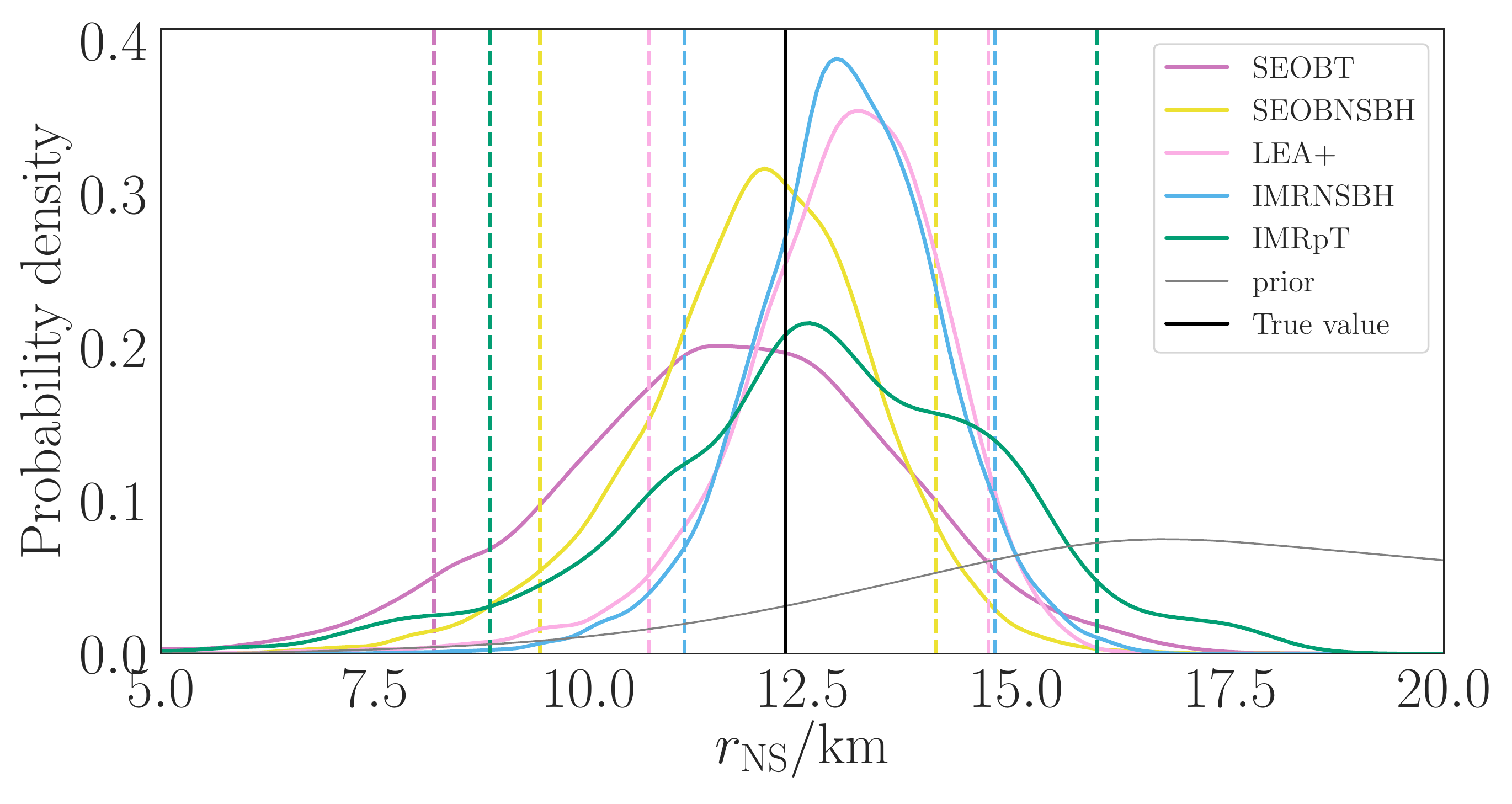}
   \vspace{-1.5\baselineskip}
   \caption{\rns, SNR 70.}
   \label{Fig.q3_rNS_SNR70_inc30}
\end{subfigure}
\caption{Posterior distributions for \LNS and \rns recovered by different approximants for $q=3$, inclination $30^{\circ}$. }
\label{Fig.q3_lambda2rNS_inc30}
\end{figure}

\begin{figure}
\centering
\begin{subfigure}[b]{0.4\textwidth}
   \includegraphics[width=1\linewidth]{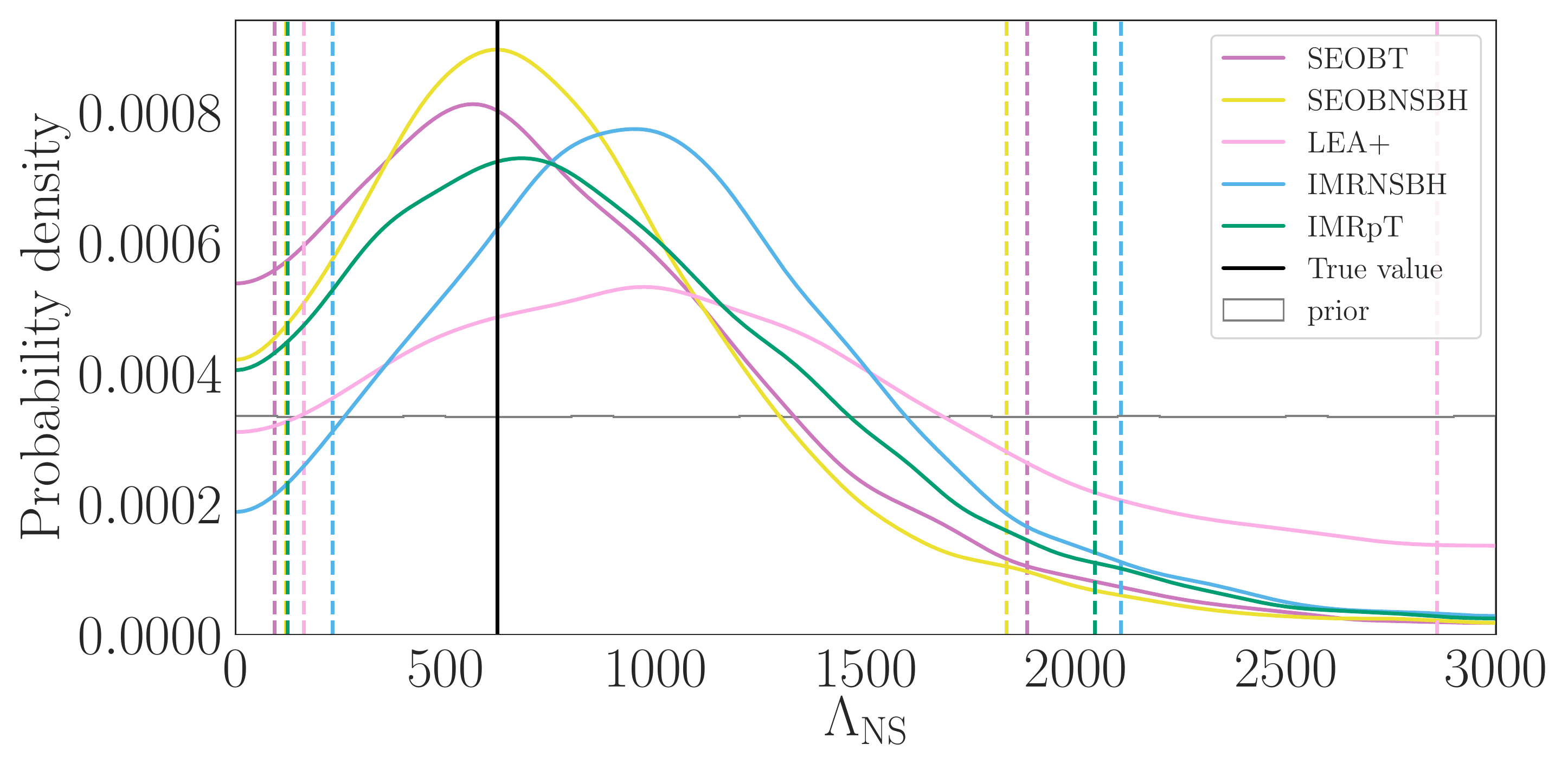}
   \vspace{-1.5\baselineskip}
   \caption{\LNS, SNR 30.}
   \label{Fig.q3_lambda2_SNR30_inc70} 
\end{subfigure}
\begin{subfigure}[b]{0.4\textwidth}
   \includegraphics[width=1\linewidth]{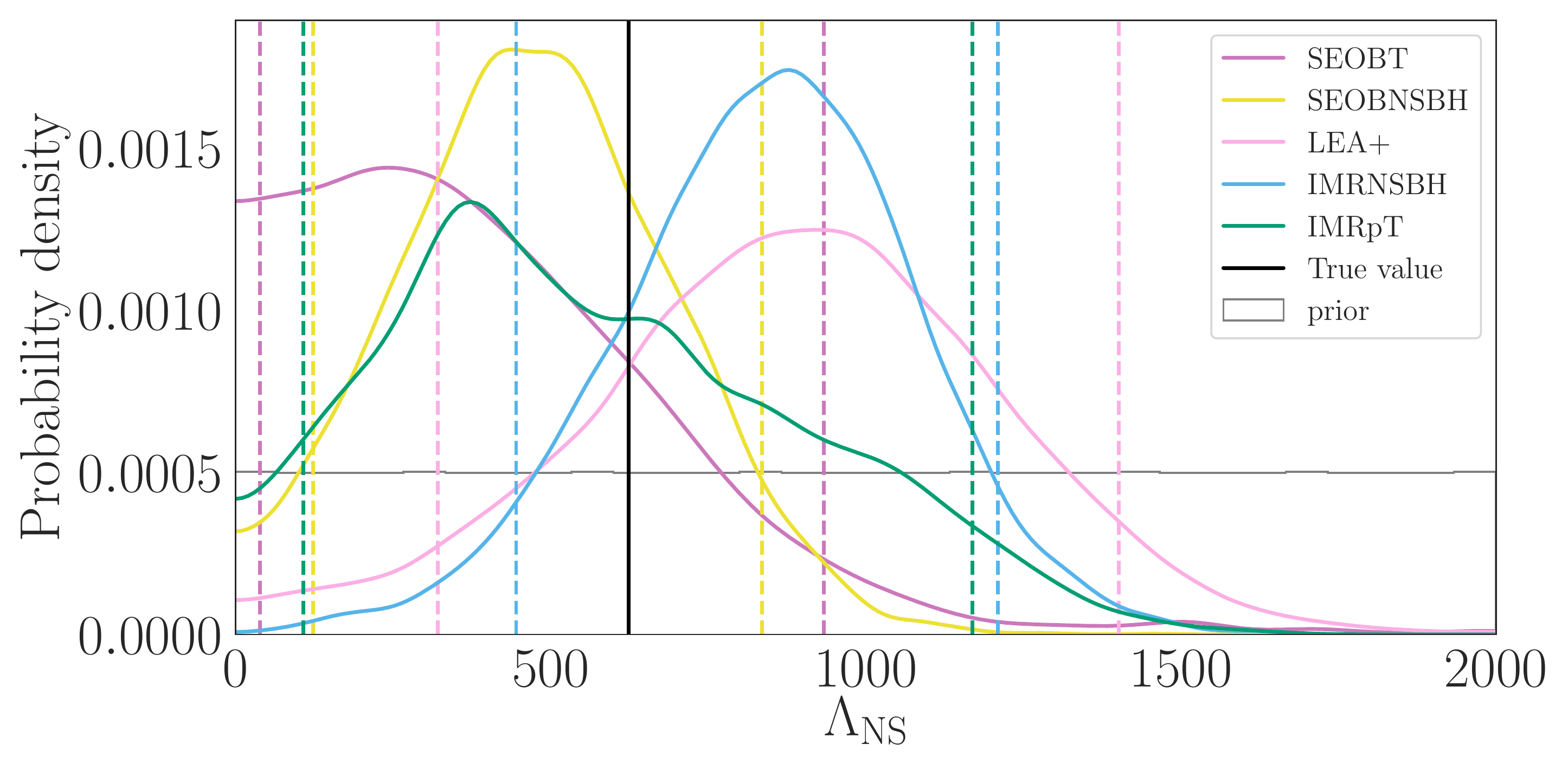}
   \vspace{-1.5\baselineskip}
   \caption{\LNS, SNR 70.}
   \label{Fig.q3_lambda2_SNR70_inc70}
\end{subfigure}
\begin{subfigure}[b]{0.4\textwidth}
   \includegraphics[width=1\linewidth]{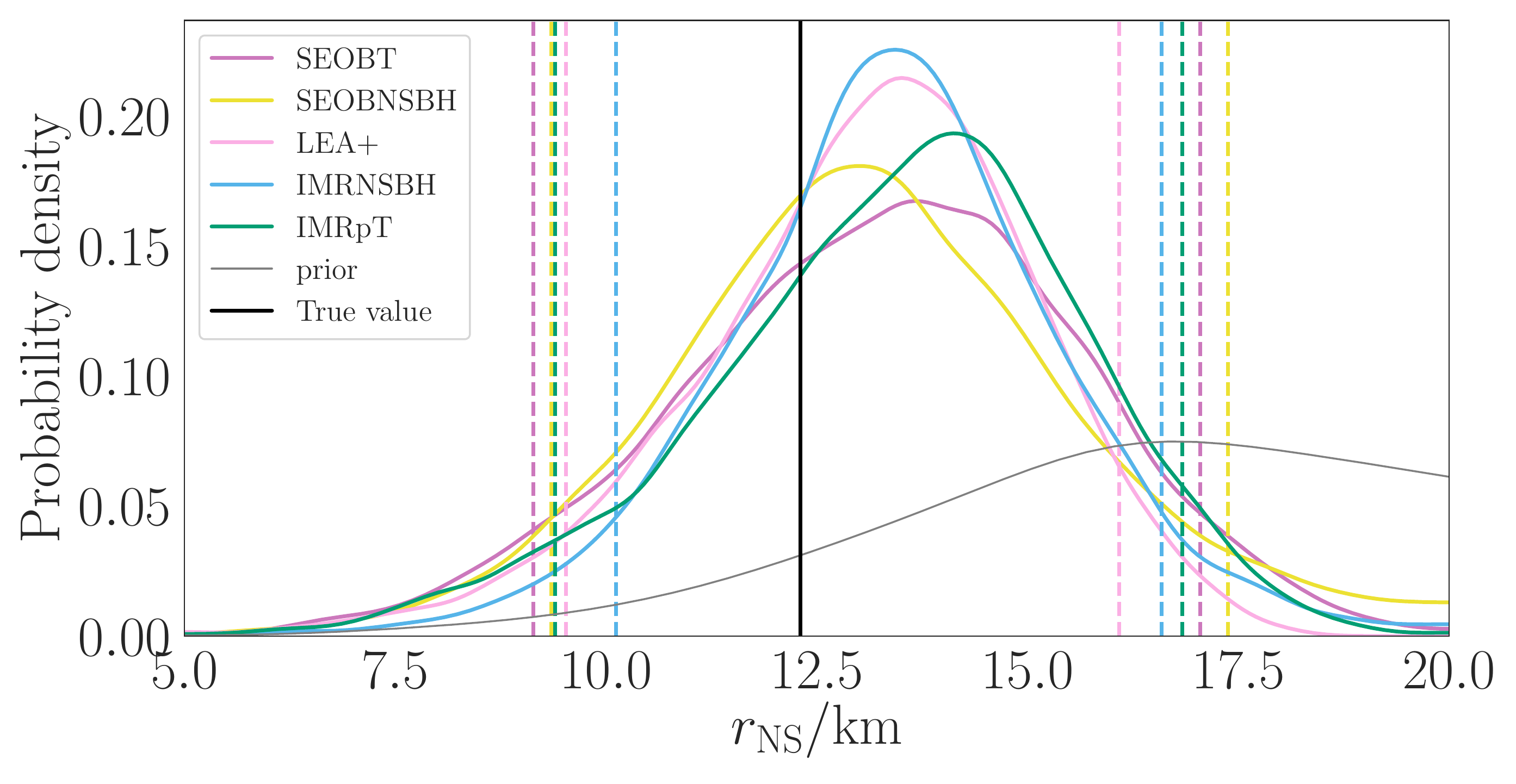}
   \vspace{-1.5\baselineskip}
   \caption{\rns, SNR 30.}
   \label{Fig.q3_rNS_SNR30_inc70} 
\end{subfigure}
\begin{subfigure}[b]{0.4\textwidth}
   \includegraphics[width=1\linewidth]{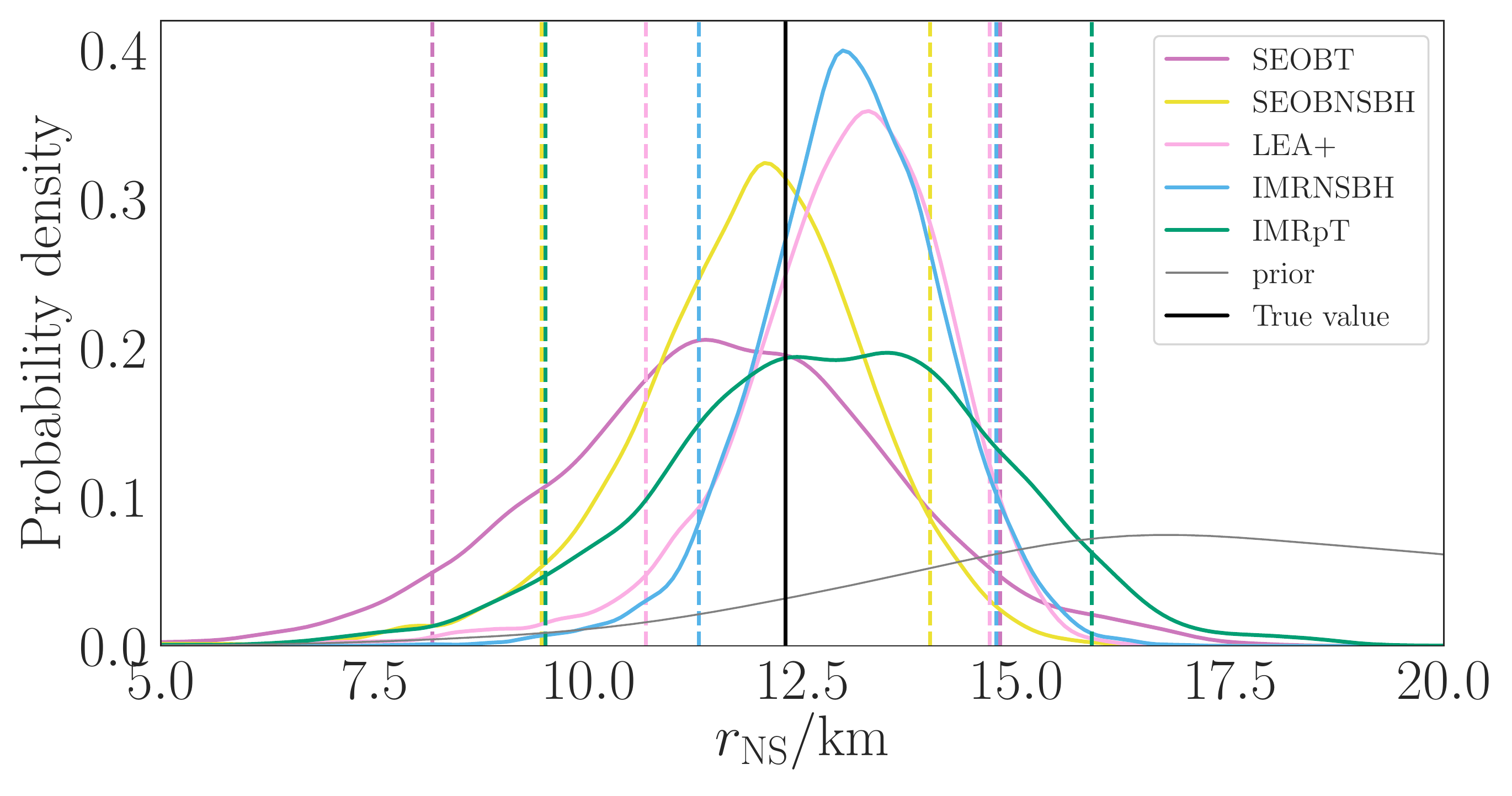}
   \vspace{-1.5\baselineskip}
   \caption{\rns, SNR 70.}
   \label{Fig.q3_rNS_SNR70_inc70}
\end{subfigure}
\caption{Posterior distributions for \LNS and \rns recovered by different approximants for $q=3$, inclination $70^{\circ}$. }
\label{Fig.q3_lambda2rNS_inc70}
\end{figure}

\begin{figure}
\centering
\begin{subfigure}[b]{0.4\textwidth}
   \includegraphics[width=1\linewidth]{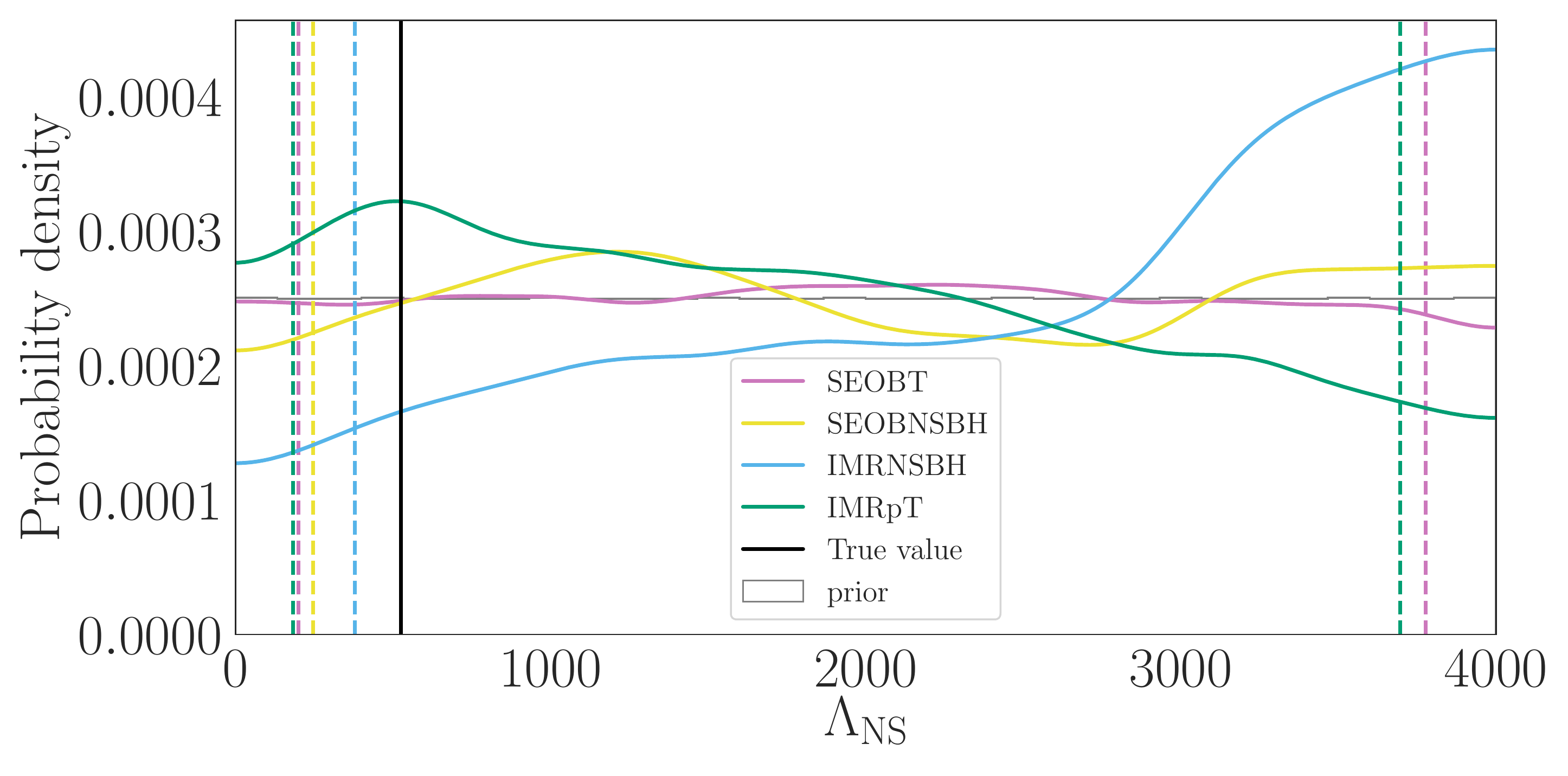}
   \vspace{-1.5\baselineskip}
   \caption{\LNS, SNR 30.}
   \label{Fig.q6_lambda2_SNR30_inc30} 
\end{subfigure}
\begin{subfigure}[b]{0.4\textwidth}
   \includegraphics[width=1\linewidth]{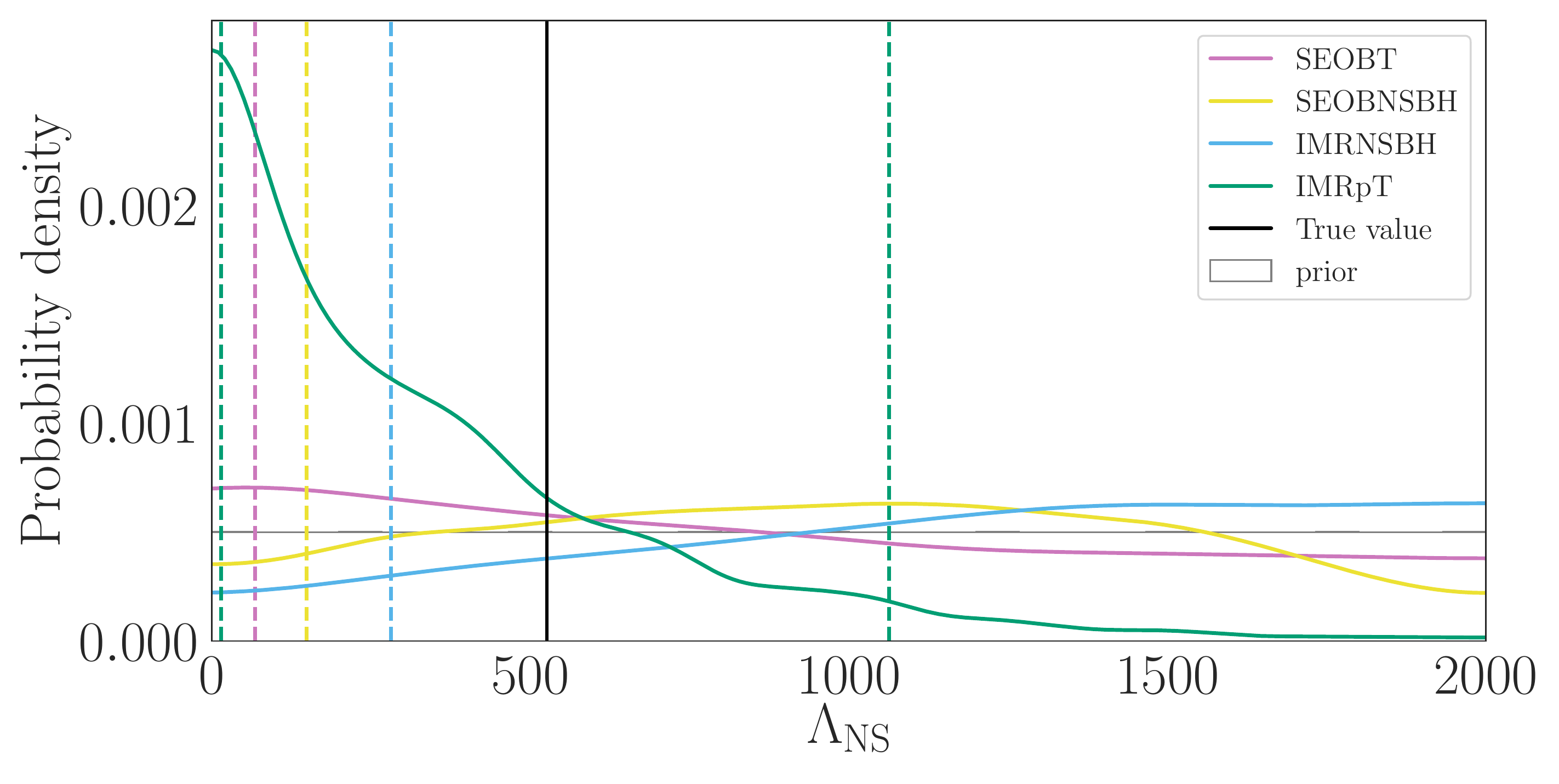}
   \vspace{-1.5\baselineskip}
   \caption{\LNS, SNR 70.}
   \label{Fig.q6_lambda2_SNR70_inc30}
\end{subfigure}
\begin{subfigure}[b]{0.4\textwidth}
   \includegraphics[width=1\linewidth]{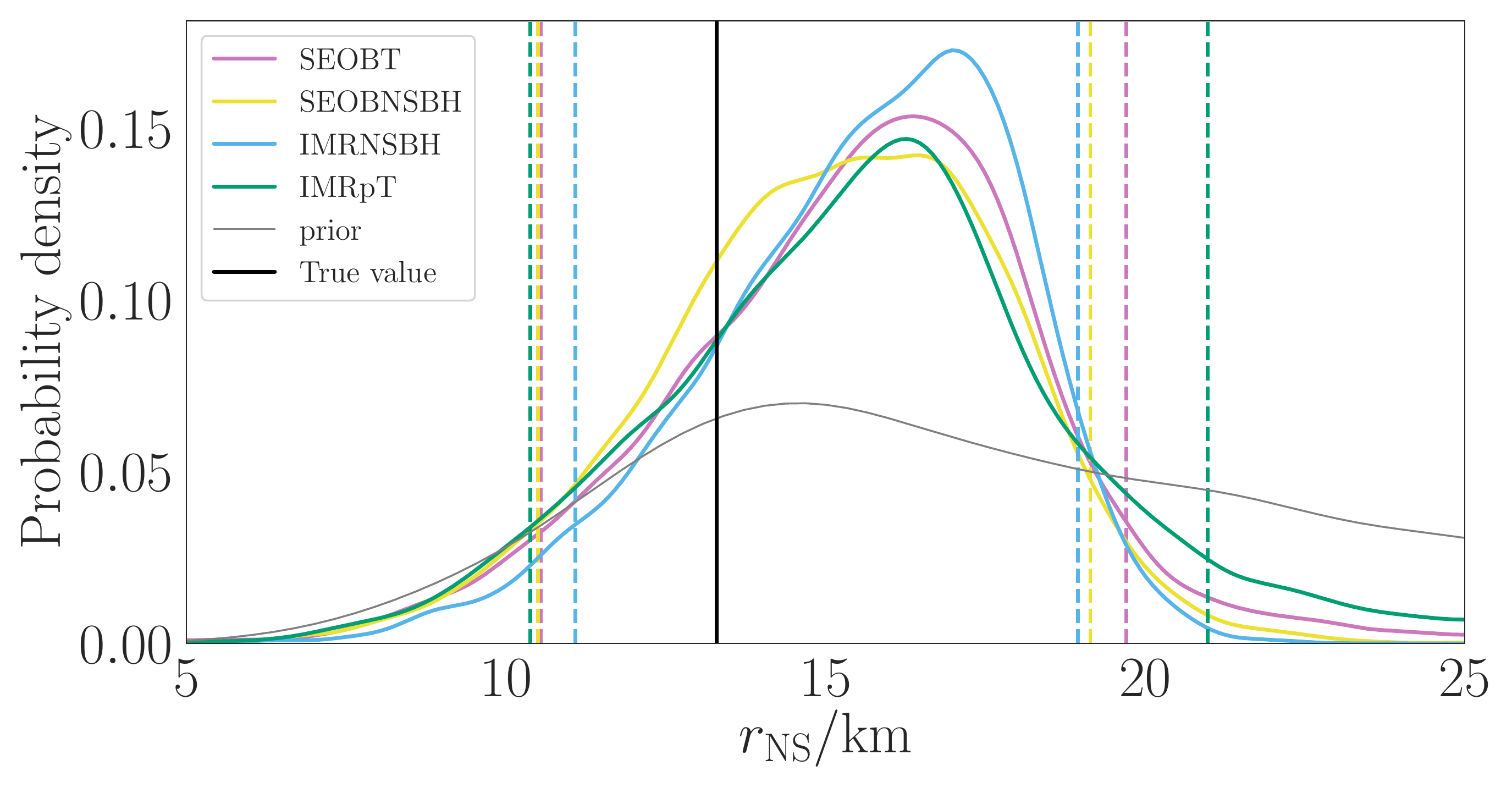}
   \vspace{-1.5\baselineskip}
   \caption{\rns, SNR 30.}
   \label{Fig.q6_rNS_SNR30_inc30} 
\end{subfigure}
\begin{subfigure}[b]{0.4\textwidth}
   \includegraphics[width=1\linewidth]{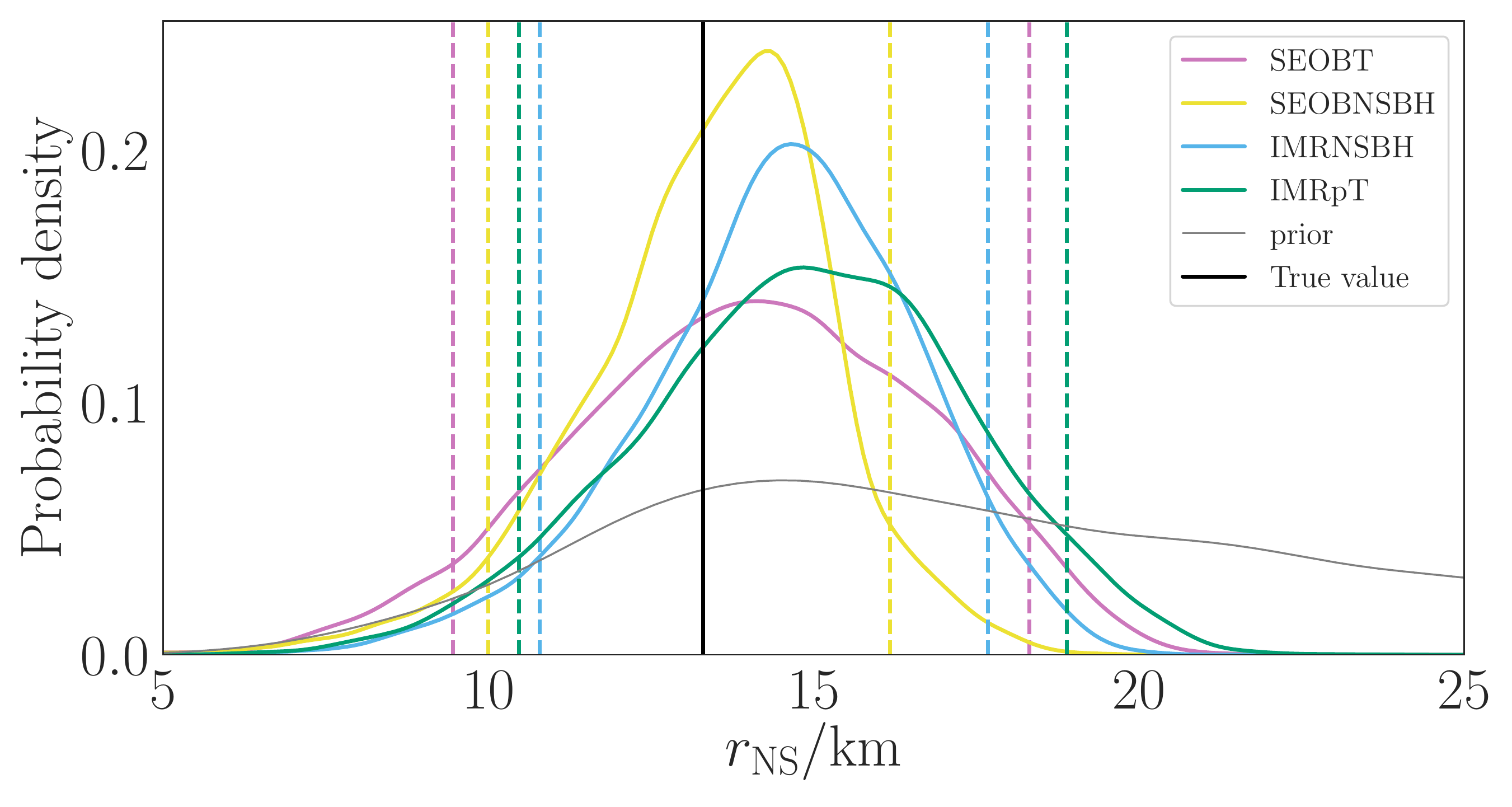}
   \vspace{-1.5\baselineskip}
   \caption{\rns, SNR 70.}
   \label{Fig.q6_rNS_SNR70_inc30}
\end{subfigure}
\caption{Posterior distributions for \LNS and \rns recovered by different approximants for $q=6$, inclination $30^{\circ}$. }
\label{Fig.q6_lambda2rNS_inc30}
\end{figure}

\begin{figure}
\centering
\begin{subfigure}[b]{0.4\textwidth}
   \includegraphics[width=1\linewidth]{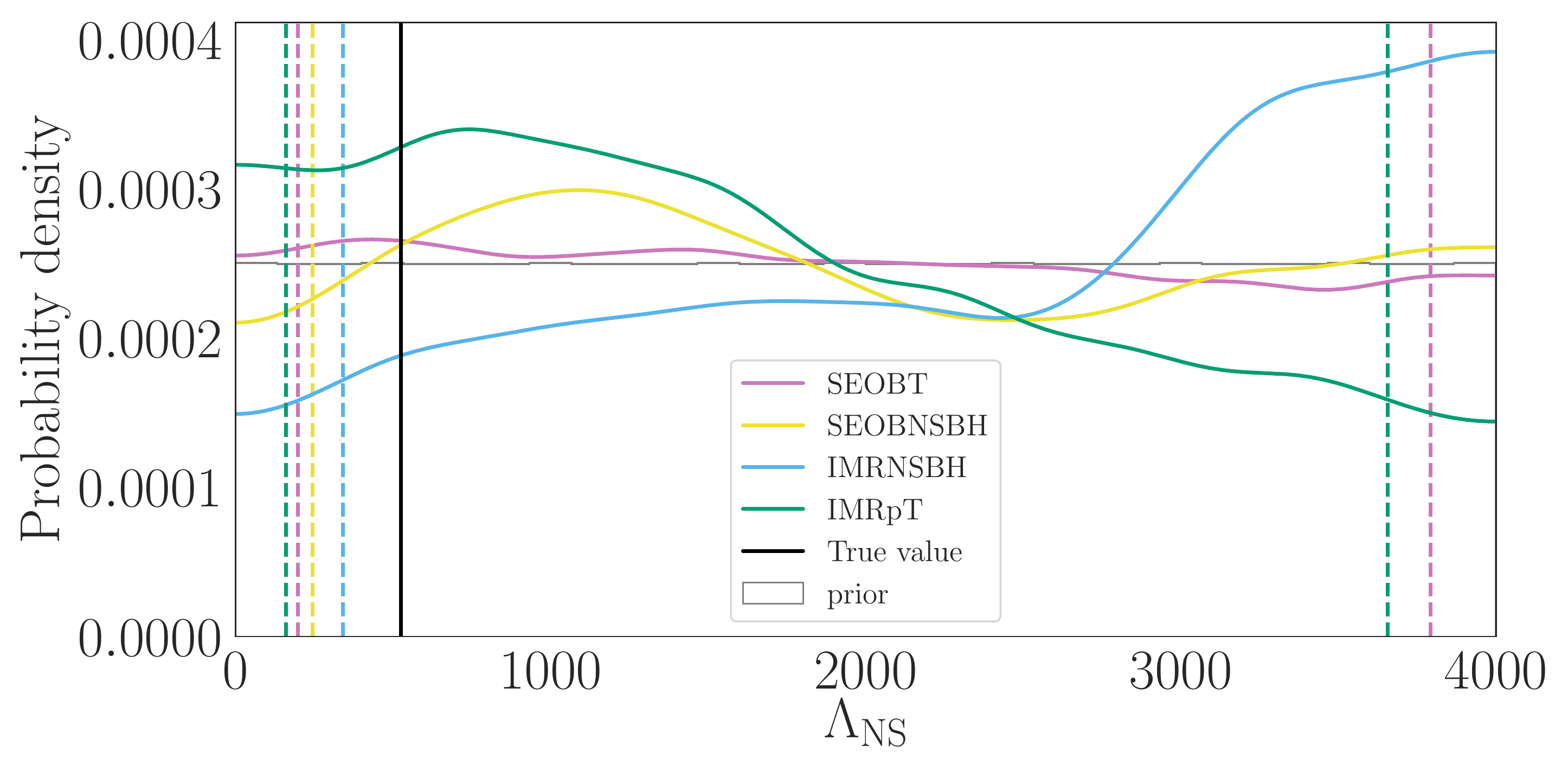}
   \vspace{-1.5\baselineskip}
   \caption{\LNS, SNR 30.}
   \label{Fig.q6_lambda2_SNR30_inc70} 
\end{subfigure}
\begin{subfigure}[b]{0.4\textwidth}
   \includegraphics[width=1\linewidth]{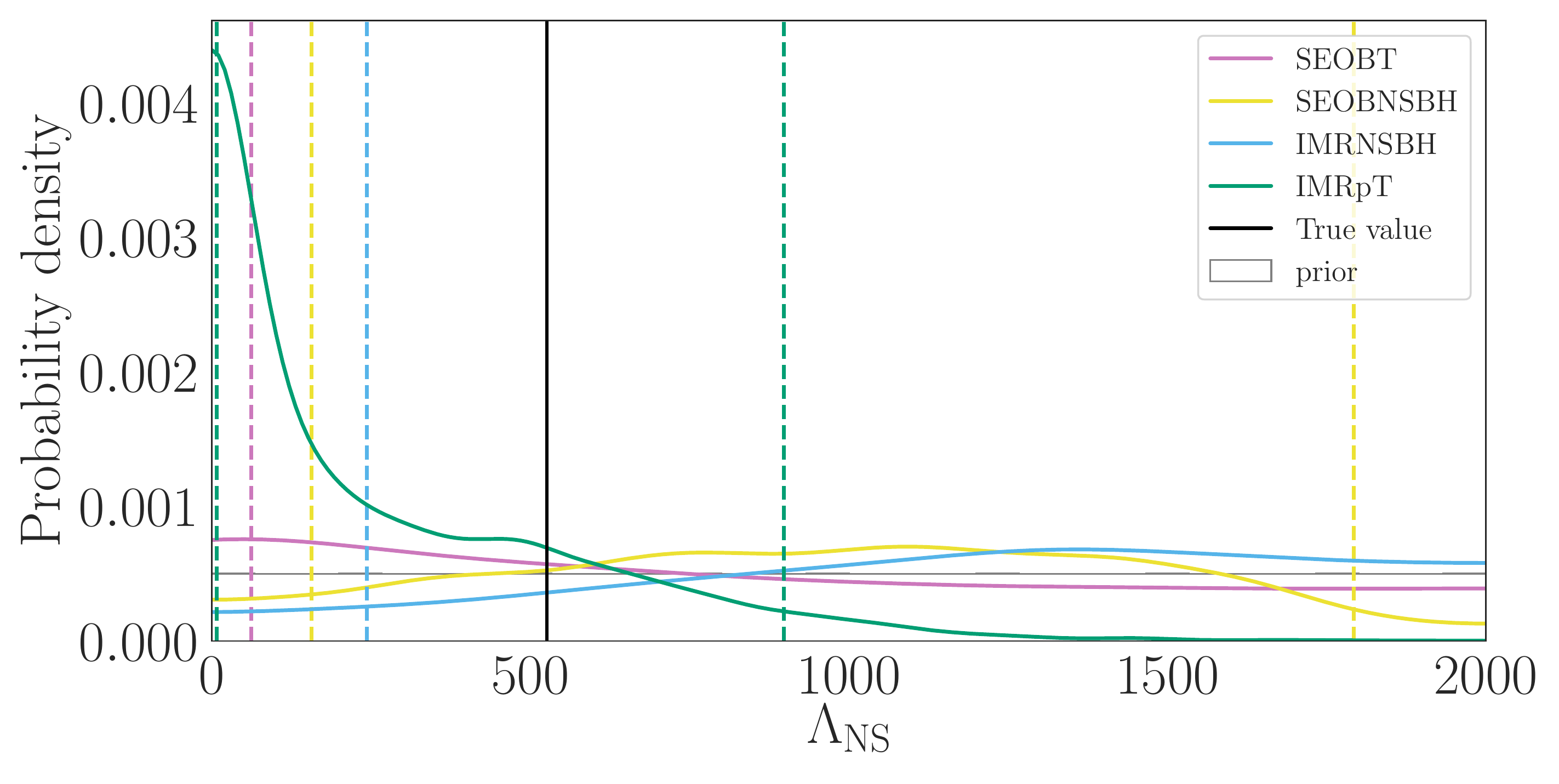}
   \vspace{-1.5\baselineskip}
   \caption{\LNS, SNR 70.}
   \label{Fig.q6_lambda2_SNR70_inc70}
\end{subfigure}
\begin{subfigure}[b]{0.4\textwidth}
   \includegraphics[width=1\linewidth]{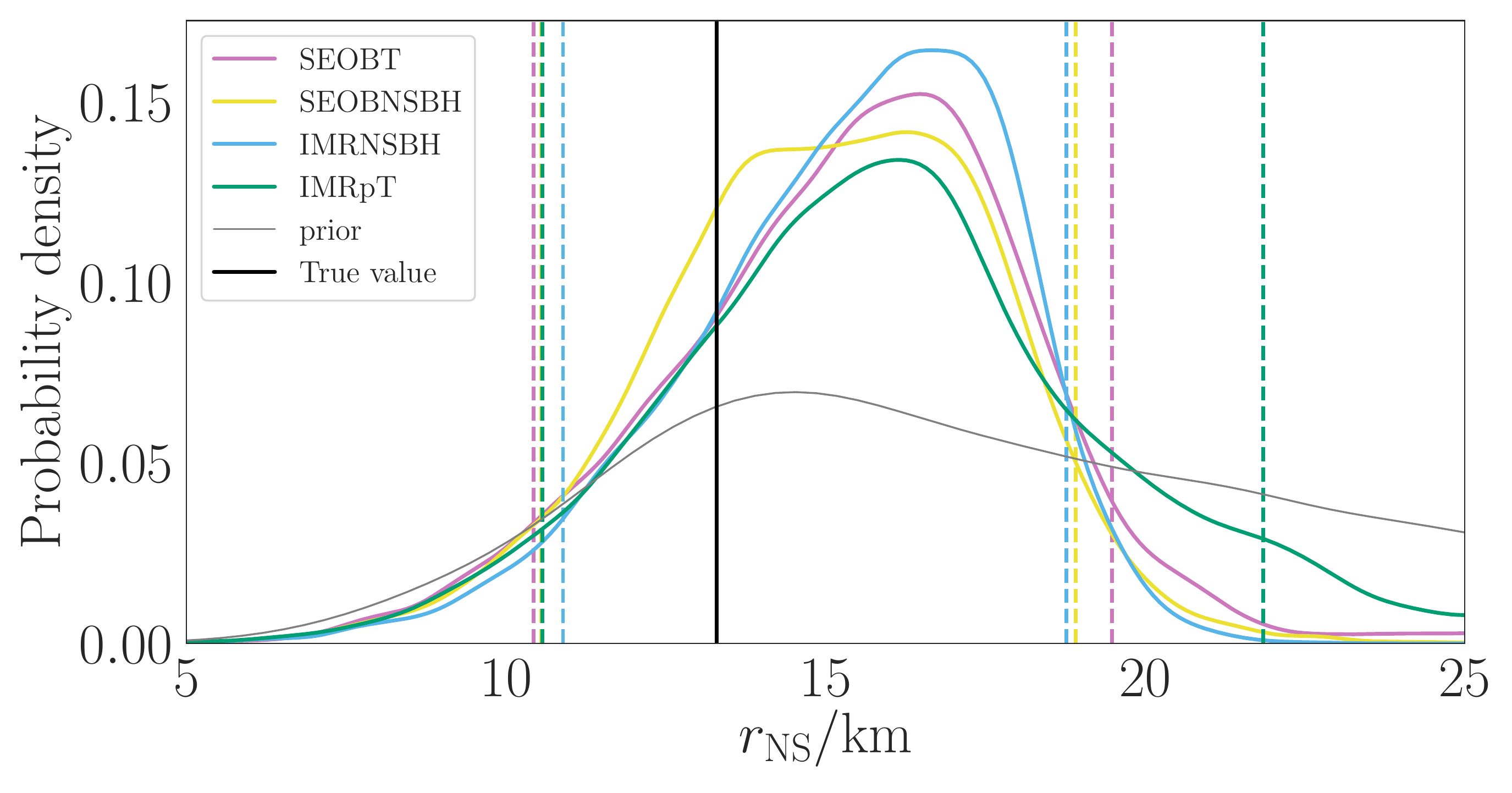}
   \vspace{-1.5\baselineskip}
   \caption{\rns, SNR 30.}
   \label{Fig.q6_rNS_SNR30_inc70} 
\end{subfigure}
\begin{subfigure}[b]{0.4\textwidth}
   \includegraphics[width=1\linewidth]{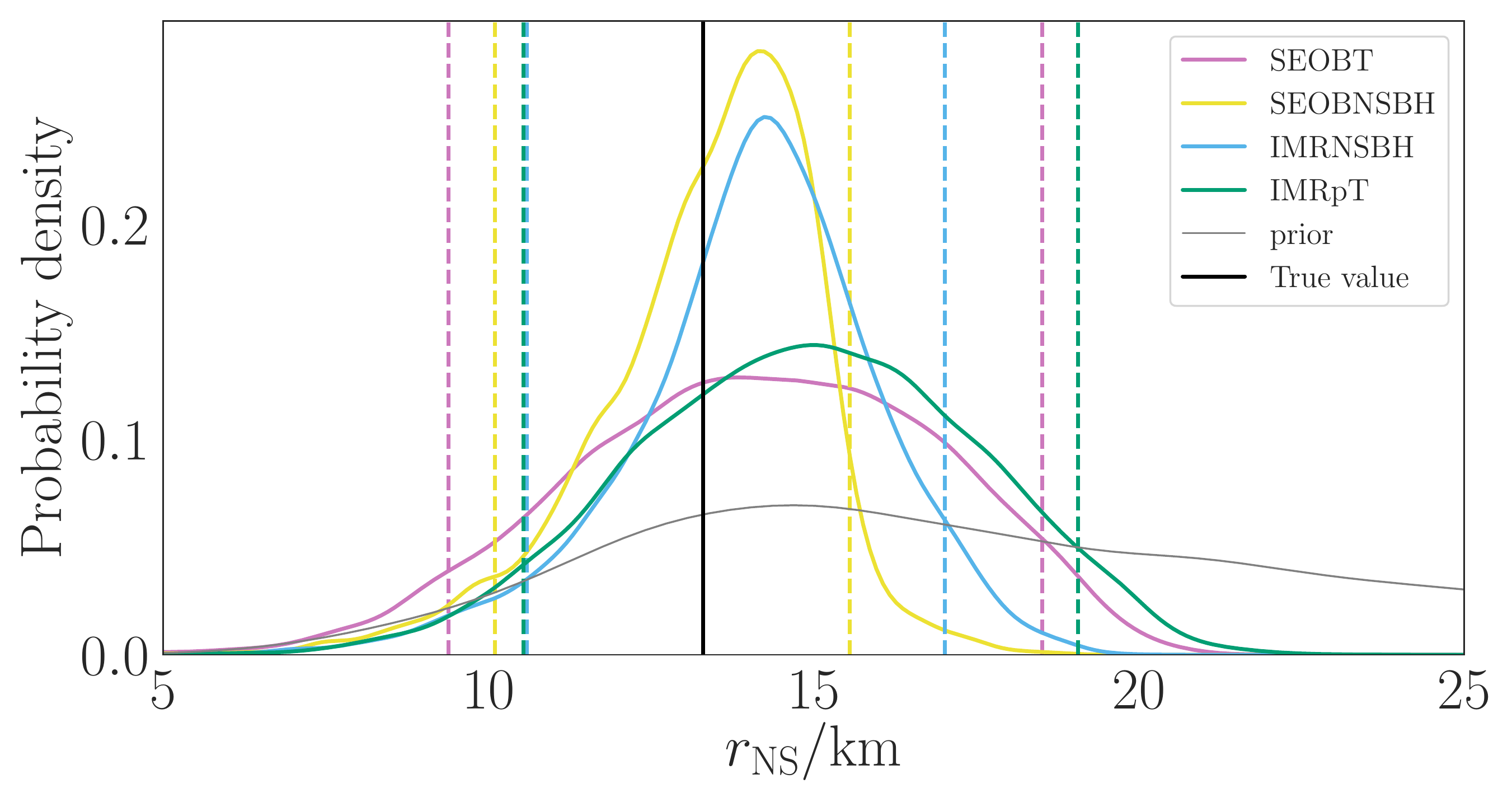}
   \vspace{-1.5\baselineskip}
   \caption{\rns, SNR 70.}
   \label{Fig.q6_rNS_SNR70_inc70}
\end{subfigure}
\caption{Posterior distributions for \LNS and \rns recovered by different approximants for $q=6$, inclination $70^{\circ}$. }
\label{Fig.q6_lambda2rNS_inc70}
\end{figure}

\begin{figure}
\centering
\begin{subfigure}[b]{0.4\textwidth}
   \includegraphics[width=1\linewidth]{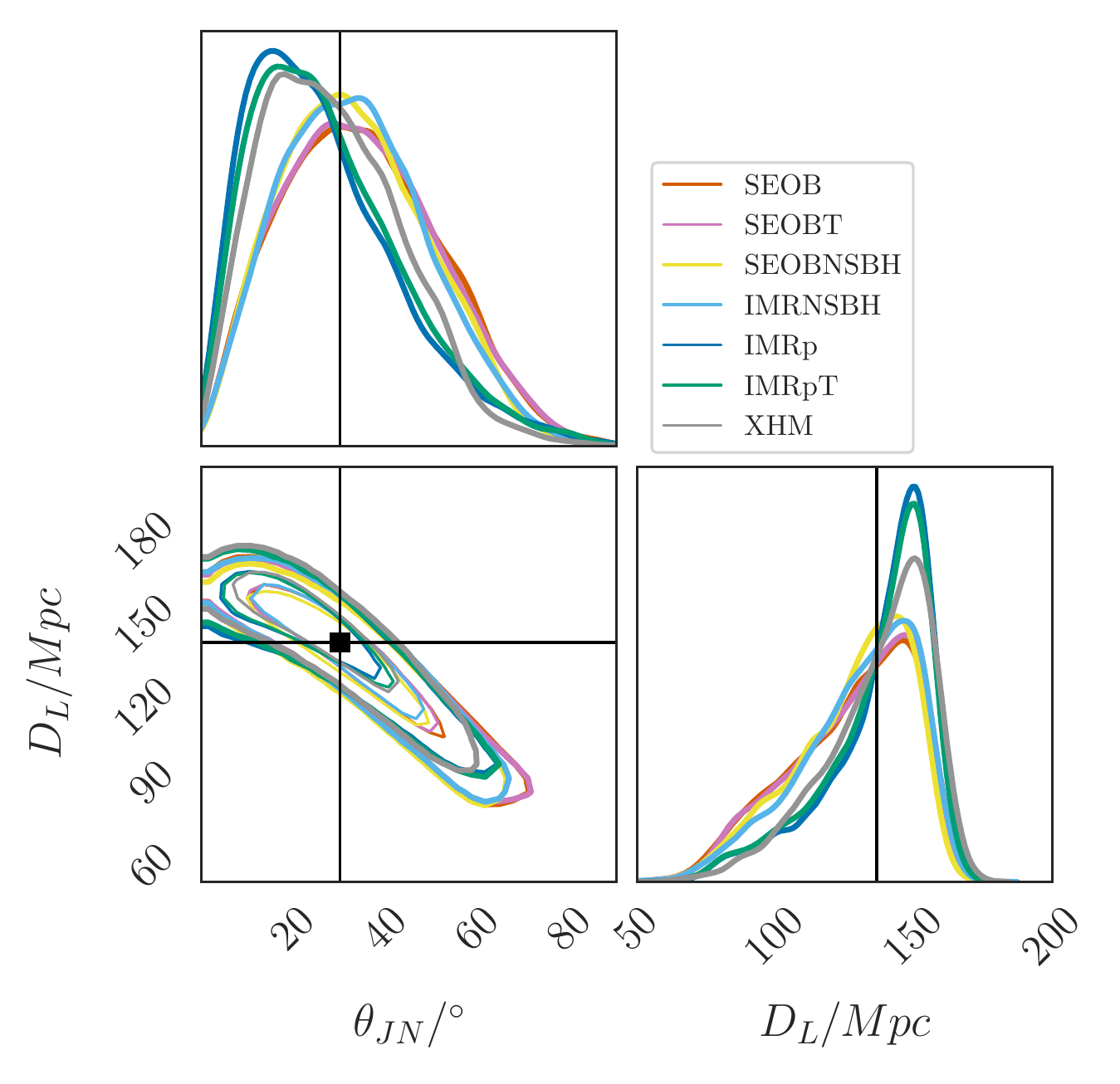}
    \vspace{-1.5\baselineskip}
   \caption{SNR 30.}
   \label{Fig.q2_dist_inc_SNR30_inc30} 
\end{subfigure}
\begin{subfigure}[b]{0.4\textwidth}
   \includegraphics[width=1\linewidth]{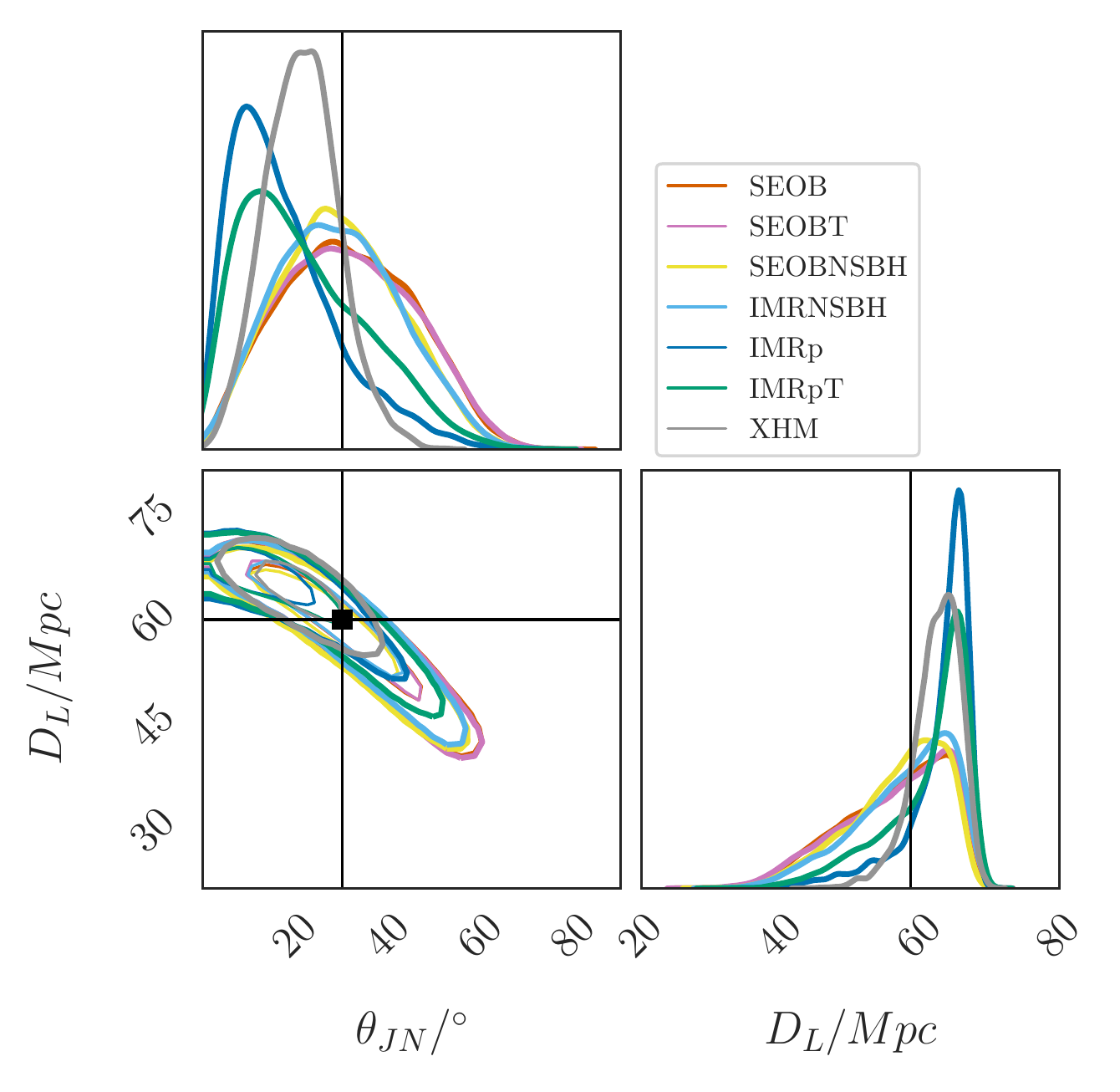}
   \vspace{-1.5\baselineskip}
   \caption{SNR 70.}
   \label{Fig.q2_dist_inc_SNR70_inc30}  
\end{subfigure}
\caption{2D contour plot of posterior distributions for luminosity distance \dl and the inclination angle \inc, recovered by different approximants for $q=2$, inclination $30^{\circ}$.}
\label{Fig.dist_inc_BHNSq2s0_inc30}
\end{figure}

\begin{figure}
\centering
\begin{subfigure}[b]{0.4\textwidth}
   \includegraphics[width=1\linewidth]{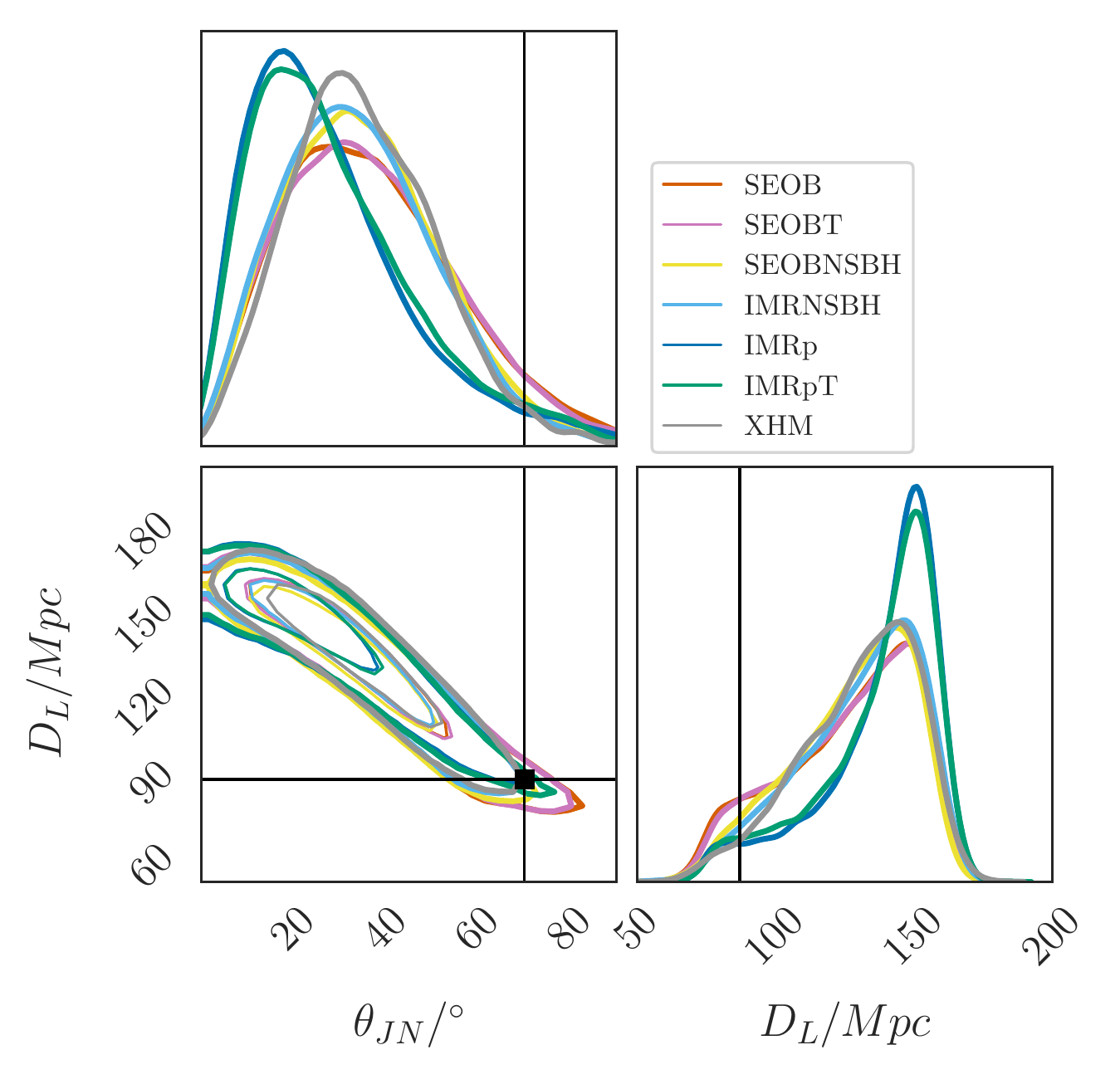}
   \vspace{-1.5\baselineskip}
   \caption{SNR 30.}
   \label{Fig.q2_dist_inc_SNR30_inc70} 
\end{subfigure}
\begin{subfigure}[b]{0.4\textwidth}
   \includegraphics[width=1\linewidth]{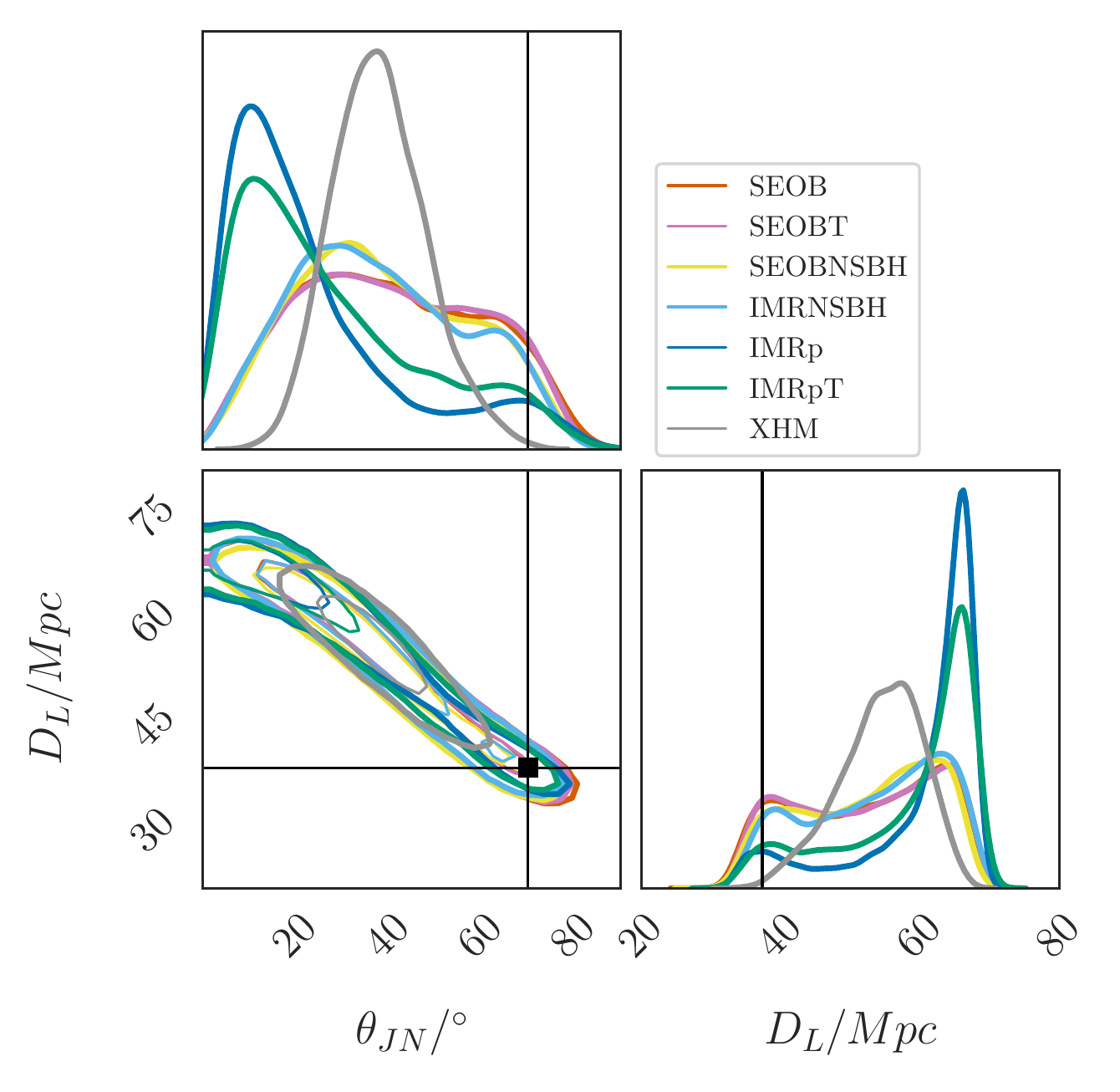}
   \vspace{-1.5\baselineskip}
   \caption{SNR 70.}
   \label{Fig.q2_dist_inc_SNR70_inc70}  
\end{subfigure}
\caption{2D contour plot of posterior distributions for luminosity distance \dl and the inclination angle \inc, recovered by different approximants for $q=2$, inclination $70^{\circ}$.}
\label{Fig.dist_inc_BHNSq2s0_inc70}
\end{figure}

\begin{figure}
\centering
\begin{subfigure}[b]{0.4\textwidth}
   \includegraphics[width=1\linewidth]{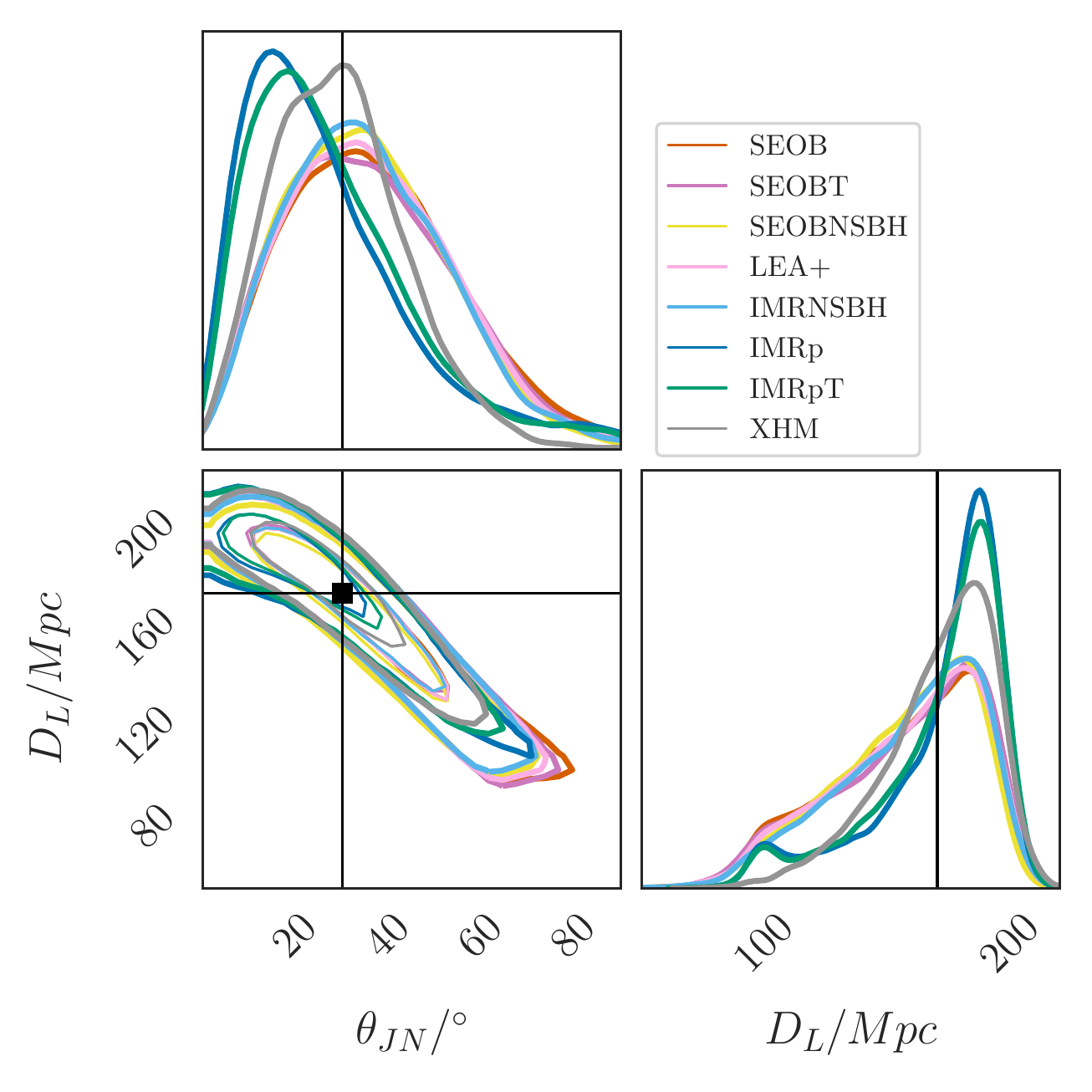}
   \vspace{-1.5\baselineskip}
   \caption{SNR 30.}
   \label{Fig.q3_dist_inc_SNR30_inc30} 
\end{subfigure}
\begin{subfigure}[b]{0.4\textwidth}
   \includegraphics[width=1\linewidth]{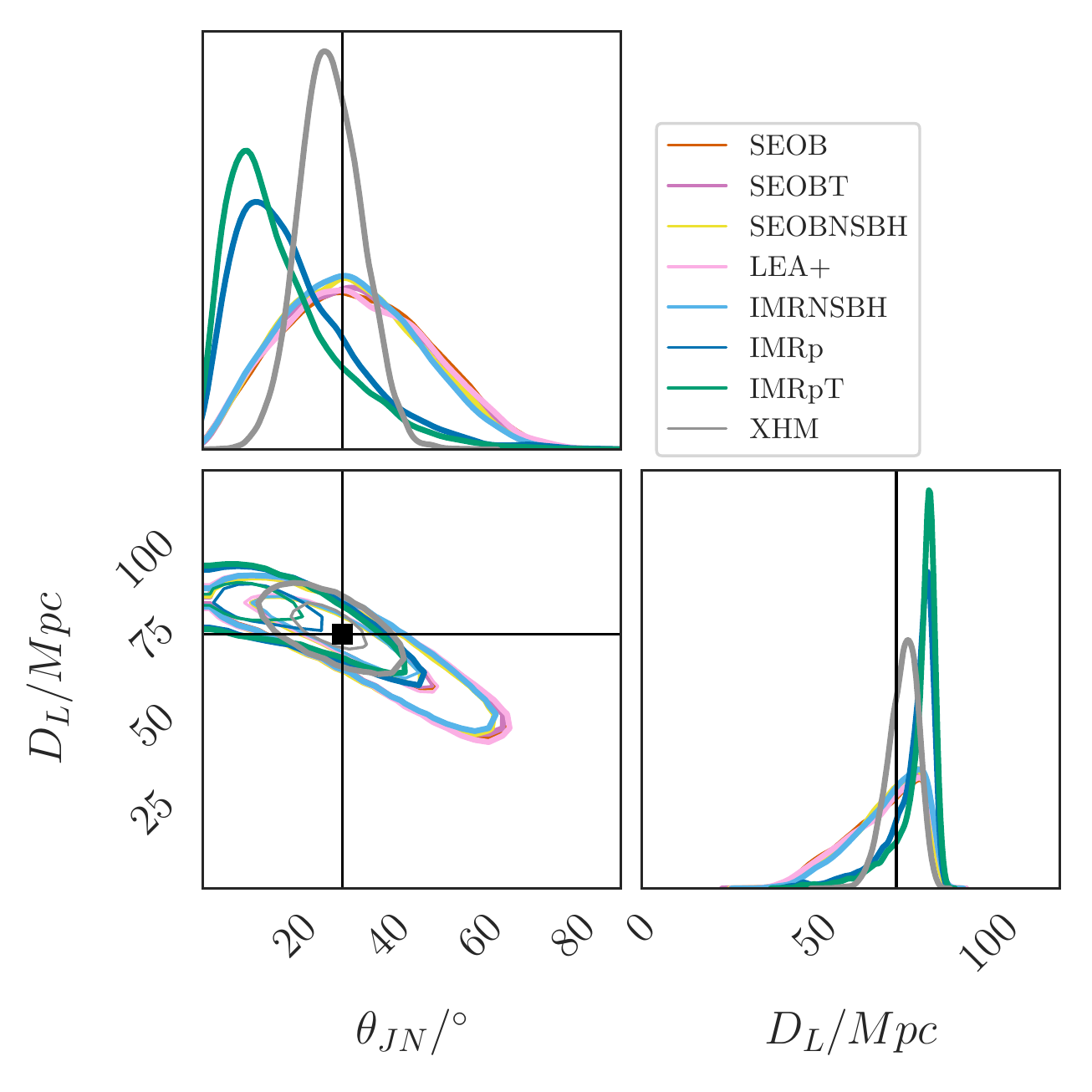}
   \vspace{-1.5\baselineskip}
   \caption{SNR 70.}
   \label{Fig.q3_dist_inc_SNR70_inc30}  
\end{subfigure}
\caption{2D contour plot of posterior distributions for luminosity distance \dl and the inclination angle \inc, recovered by different approximants for $q=3$, inclination $30^{\circ}$.}
\label{Fig.dist_inc_BHNSq3s0_inc30}
\end{figure}

\begin{figure}
\centering
\begin{subfigure}[b]{0.4\textwidth}
   \includegraphics[width=1\linewidth]{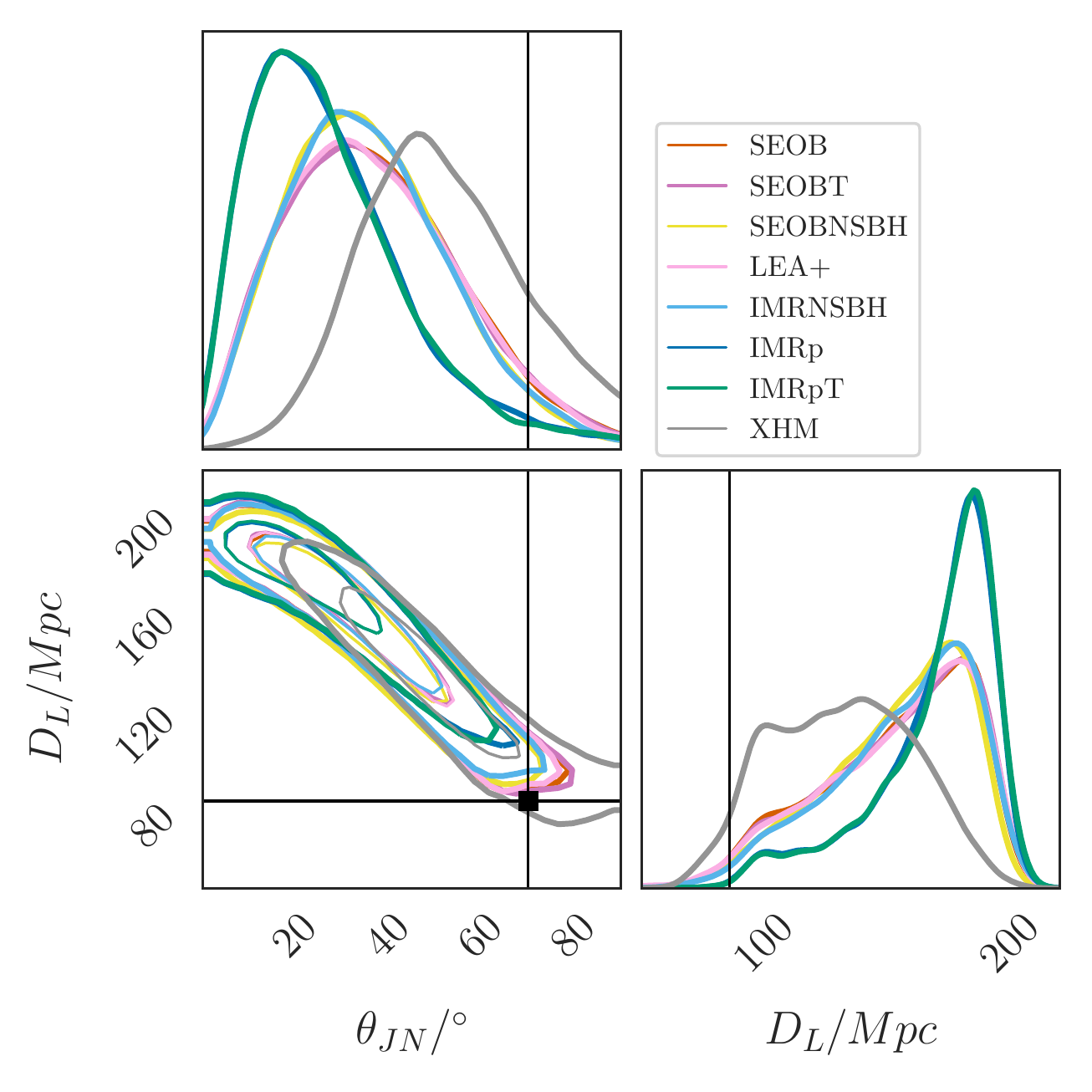}
   \vspace{-1.5\baselineskip}
   \caption{SNR 30.}
   \label{Fig.q3_dist_inc_SNR30_inc70} 
\end{subfigure}
\begin{subfigure}[b]{0.4\textwidth}
   \includegraphics[width=1\linewidth]{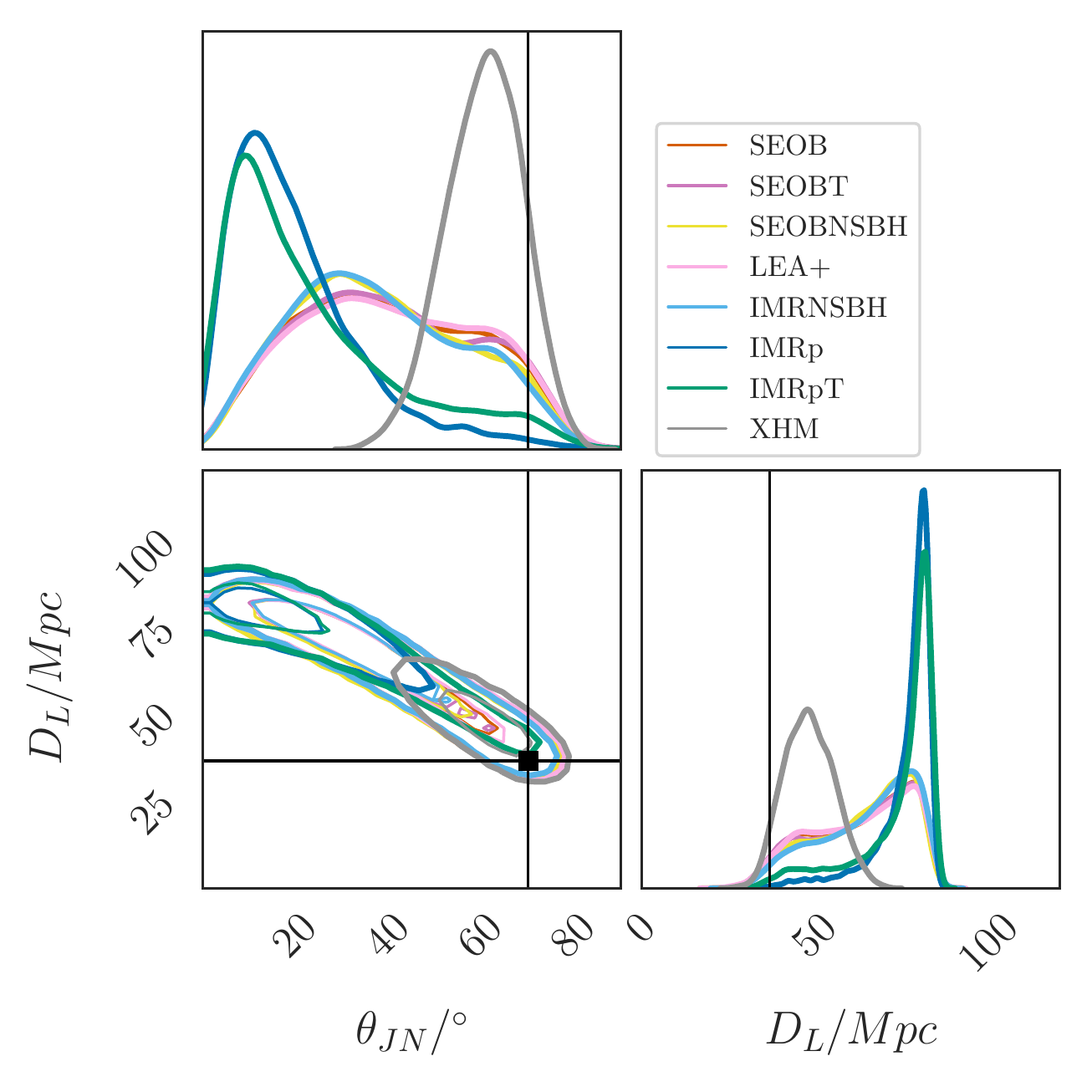}
   \vspace{-1.5\baselineskip}
   \caption{SNR 70.}
   \label{Fig.q3_dist_inc_SNR70_inc70}  
\end{subfigure}
\caption{2D contour plot of posterior distributions for luminosity distance \dl and the inclination angle \inc, recovered by different approximants for $q=3$, inclination $70^{\circ}$.}
\label{Fig.dist_inc_BHNSq3s0_inc70}
\end{figure}

\begin{figure}
\centering
\begin{subfigure}[b]{0.4\textwidth}
   \includegraphics[width=1\linewidth]{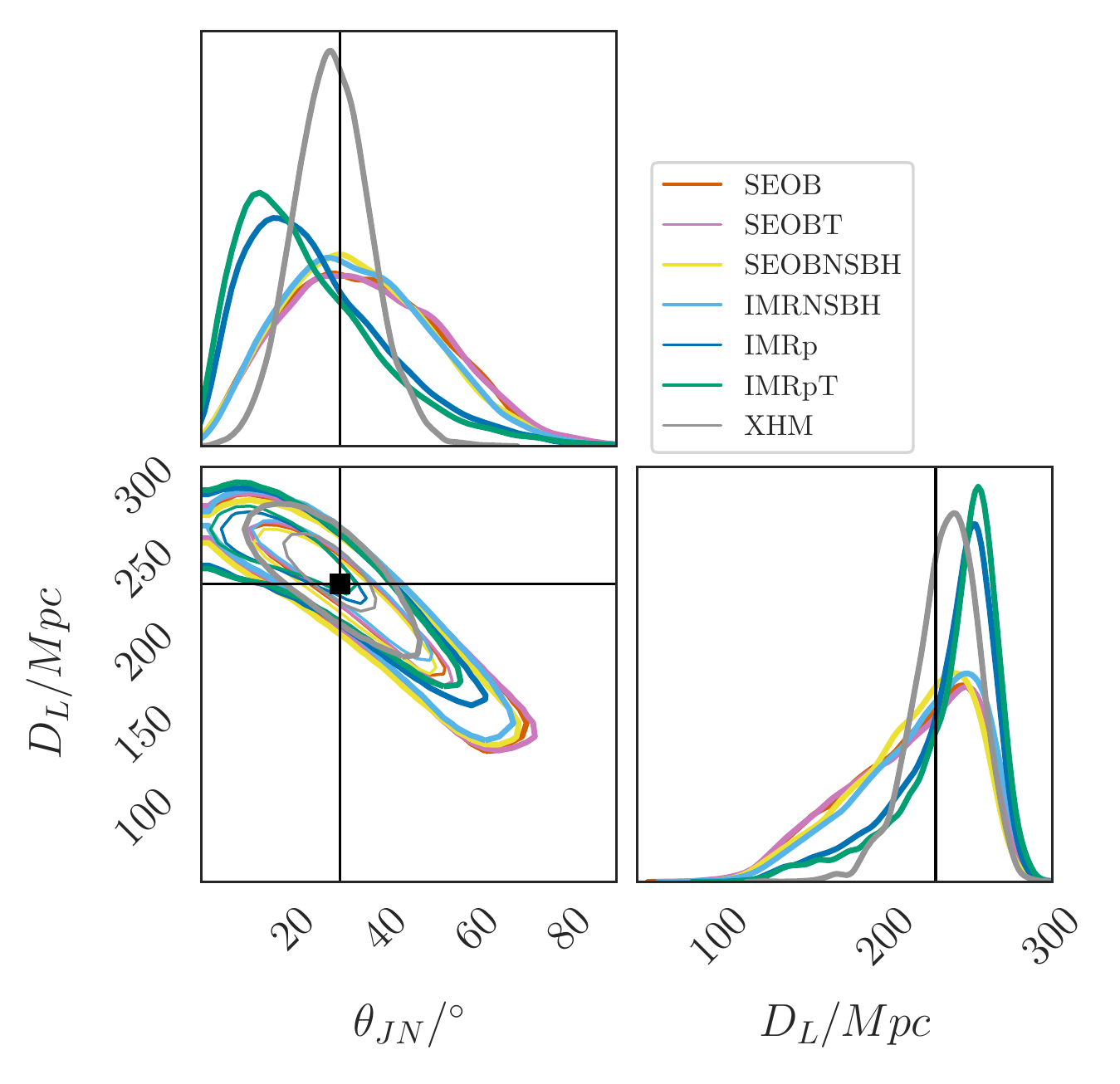}
   \vspace{-1.5\baselineskip}
   \caption{SNR 30.}
   \label{Fig.q6_dist_inc_SNR30_inc30} 
\end{subfigure}
\begin{subfigure}[b]{0.4\textwidth}
   \includegraphics[width=1\linewidth]{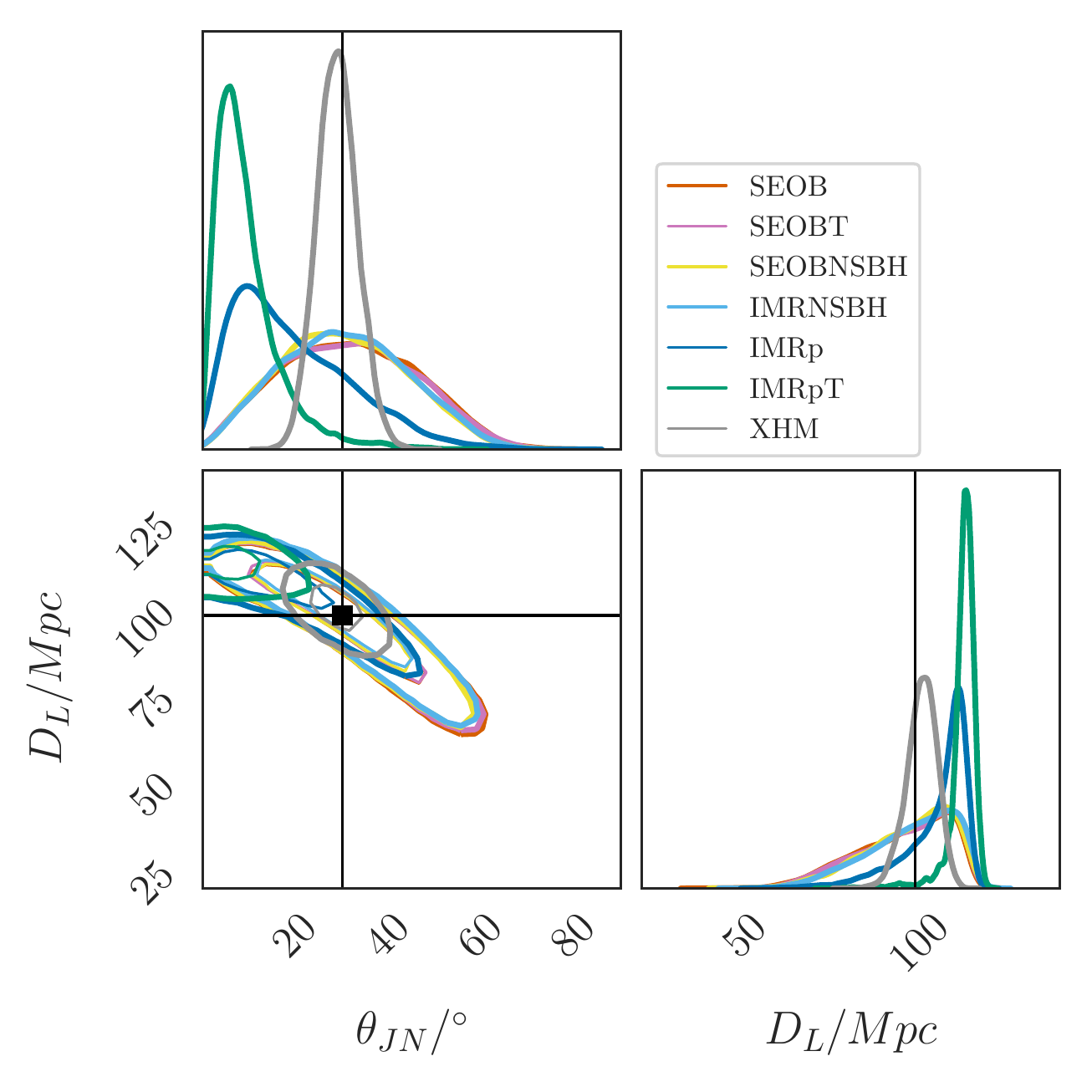}
   \vspace{-1.5\baselineskip}
   \caption{SNR 70.}
   \label{Fig.q6_dist_inc_SNR70_inc30}  
\end{subfigure}
\caption{2D contour plot of posterior distributions for luminosity distance \dl and the inclination angle \inc, recovered by different approximants for $q=6$, inclination $30^{\circ}$.}
\label{Fig.dist_inc_BHNSq6s0_inc30}
\end{figure}

\begin{figure}
\centering
\begin{subfigure}[b]{0.4\textwidth}
   \includegraphics[width=1\linewidth]{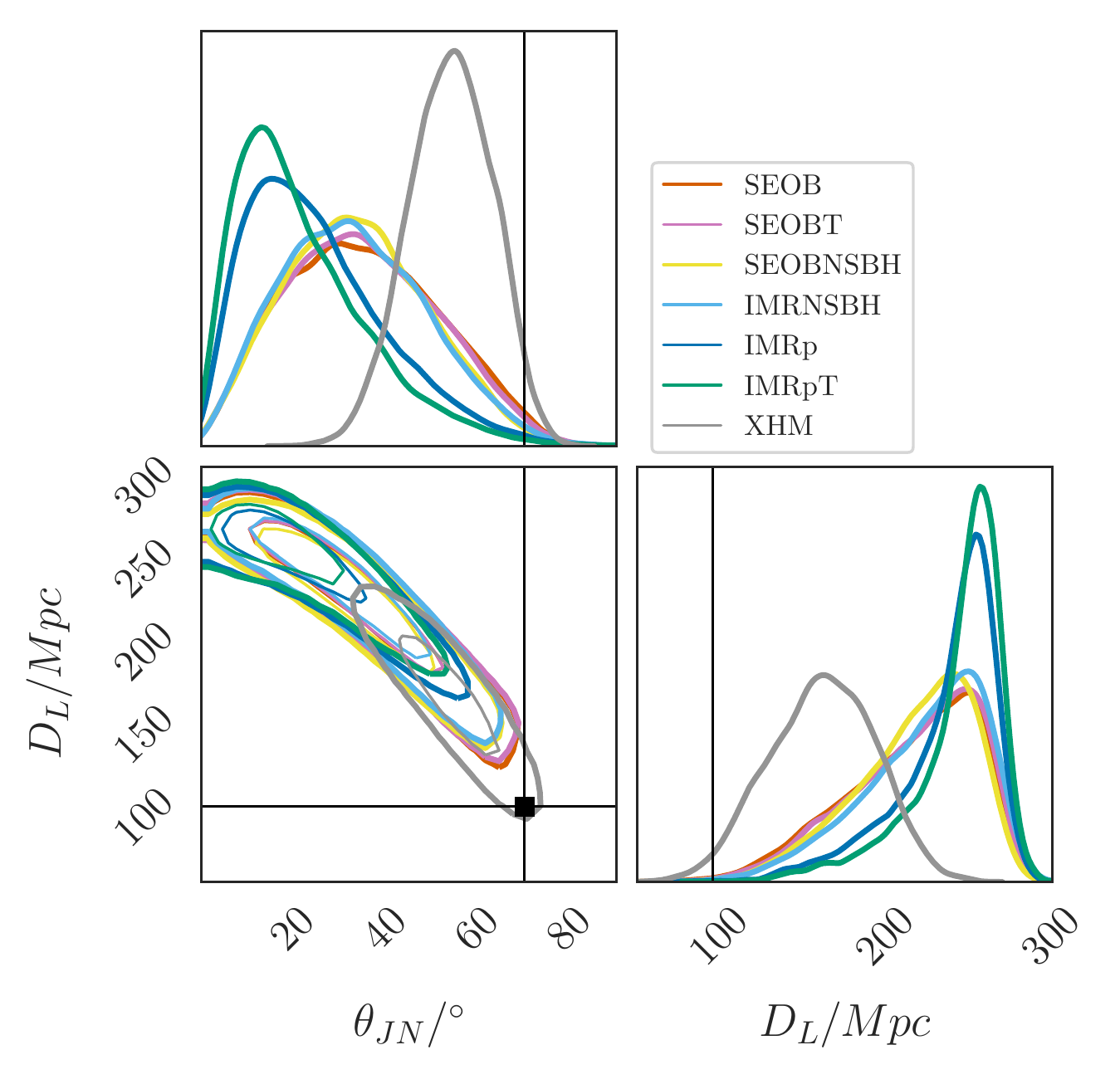}
   \vspace{-1.5\baselineskip}
   \caption{SNR 30.}
   \label{Fig.q6_dist_inc_SNR30_inc70} 
\end{subfigure}
\begin{subfigure}[b]{0.4\textwidth}
   \includegraphics[width=1\linewidth]{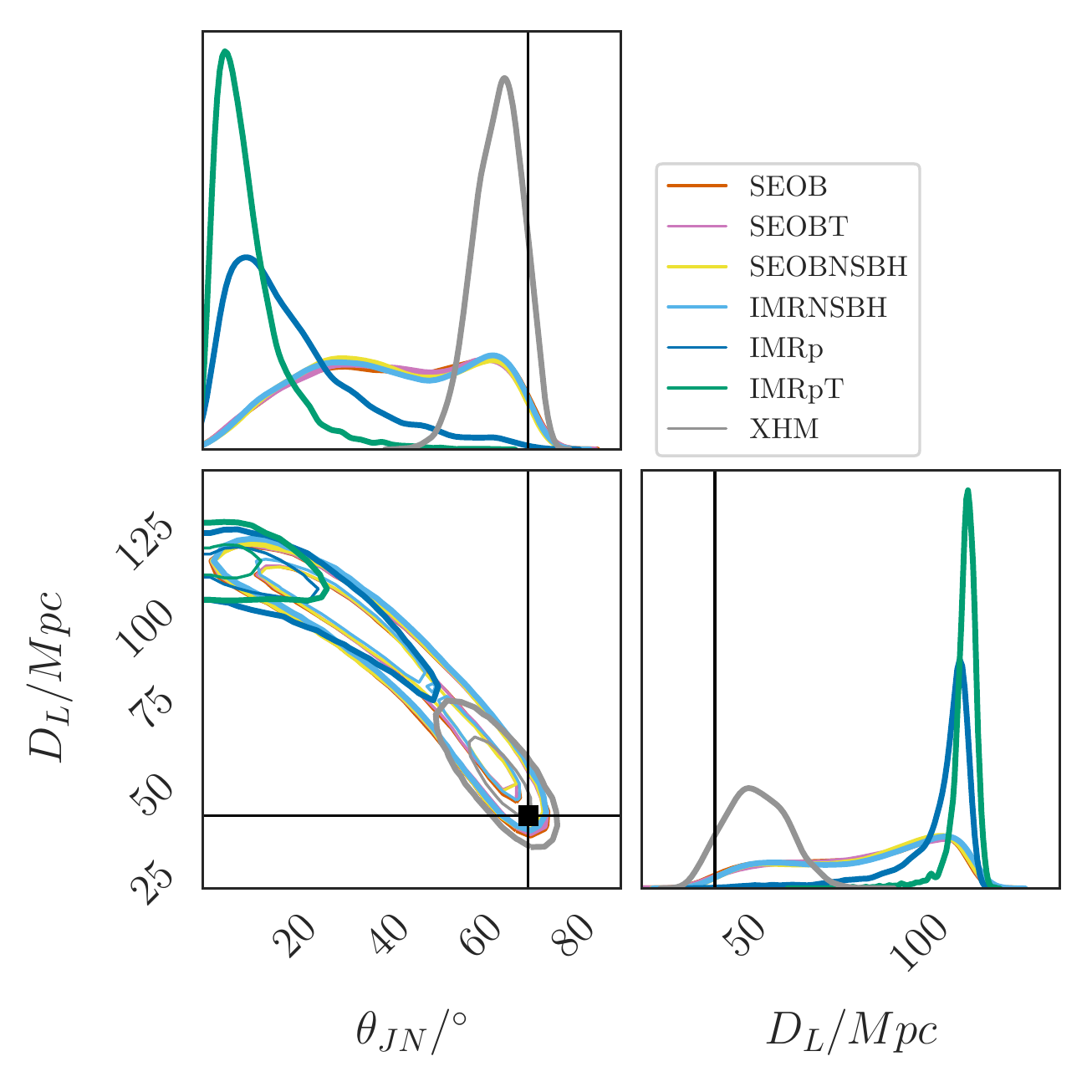}
   \vspace{-1.5\baselineskip}
   \caption{SNR 70.}
   \label{Fig.q6_dist_inc_SNR70_inc70}  
\end{subfigure}
\caption{2D contour plot of posterior distributions for luminosity distance \dl and the inclination angle \inc, recovered by different approximants for $q=6$, inclination $70^{\circ}$.}
\label{Fig.dist_inc_BHNSq6s0_inc70}
\end{figure}

We found that for signals with a SNR of 30, comparable to the loudest CBC signals detected to date, systematic uncertainties due to waveform modeling are smaller than statistical ones.
Some differences are visible, for example in the NS tidal deformability, \LNS, for which even at SNR 30, in some cases, the posterior distributions can cluster in two different groups. 
This is more visible for mass ratios of 3, Fig.~\ref{Fig.q3_lambda2_SNR30_inc30} than 2, Fig.~\ref{Fig.q2_lambda2_SNR30_inc30}.
Significant offsets are also found for the source-frame chirp mass, although they are not due to waveform modeling as much as to a failure to properly measure the source luminosity distance, which is required to convert the detector-frame masses (which are the quantities actually measured from GW data) to the source-frame ones. 
This is particularly visible for highly inclined sources, ~\Cref{Fig.q2_mc_SNR30_inc70,Fig.q3_mc_SNR30_inc70,Fig.q6_mc_SNR30_inc70}. 
The underlying reason, as discussed in Sec.~\ref{Sec:ExtrinsicParams}, is that the likelihood penalty for measuring an orientation closer to face-on, and hence a larger distance, can be more than compensated for by the fact that the Bayesian prior increases with distance, unless the true inclination angle is within $\sim15\dg$ from 90\dg~\cite{Chen:2018omi}.
This effect will not be seen for a typical detection, as most sources are expected to be detected at small inclination angles (i.e. close to 0\dg or 180\dg)~\cite{Schutz:2011tw}. 
It is also worth stressing that this offset is smaller for the \XHM waveform at $q=6$ since the detectable higher order mode contribution in the true signal help the \XHM model to break the distance-inclination degeneracy.
For the WF models tuned against NSBH systems, \IMRNSBH, \LEAplus and \SEOBNSBH, at SNR 30 we obtain 90\% statistical credible intervals on the NS source-frame mass of $\sim 0.2-0.5$~\msun. 
These uncertainties are comparable to those reported for the BNS GW170817~\cite{Abbott:2018wiz}. 
This comparison is not entirely fair, as both the dimensionality of the models and the priors used are different. 
The settings of our \IMRpT analyses are more directly comparable to Ref.~\cite{Abbott:2018wiz}: for the $q=2$ analysis and SNR 30 we find a 90\% CI uncertainty in the NS mass of 0.5~\msun. 
While this is nominally less constraining than GW170817's, it must be remembered that the mass posteriors for GW170817 have a hard prior bound (enforcing $m_1\leq m_2$, Fig. 5 of Ref.~\cite{Abbott:2018wiz}), which helps to explain why those posteriors appear narrower. 

The situation is starkly different at SNR 70, with biases comparable to or larger than the statistical uncertainties. 
At $q=2$, waveform models that do not account for tidal effects yield posterior measurements that do not include the true value in their 90\% CIs for the mass ratio, the component masses, and the effective inspiral spin. 
The overall trend is the same with $q=3$, but the biases are smaller due to the reduced impact of tides on the GW signal.
In this case, whether the true value is excluded depends on the exact approximant used.
At $q=6$, a configuration for which tidal effects, though formally included in the simulated source, do not play a significant role, the situation is somewhat reversed. 
Waveform approximants that do not include tides actually perform well, whilst waveforms with post-inspiral evolution tuned against nearly equal mass BNS NR simulations, \SEOBT and \IMRpT, yield the most severe biases.
For those, the recovered masses and spins are entirely different than those from all other waveforms, and systematically offset from the true value, with \IMRpT yielding a larger bias than \SEOBT.
We should stress that we are using these two waveform families in a region of mass ratios quite far from their calibration region, and hence these biases should not be surprising. 
However, we report them since they clearly show the importance of well-calibrated and faithful waveform models for the systems of interest.
While we have not done this test in our study, it would be interesting to show if the opposite is true, and NSBH-tuned waveforms would suffer from similar biases if used to characterize BNS sources.

Bayes factors between pairs of models can be used to reveal whether, and to what extent, some waveform models are inadequate at matching the data.
We reported them for a subset of the approximants we used, Sec.~\ref{Sec:TidalModelSelection} and show that the BNS-tuned models are clearly disfavored at $q=6$ even when compared to models that do not include matter effects.
These kinds of tests could be used to decide which waveform families should be used for specific analysis, or to combine samples from different waveforms as a way to marginalize over inaccuracies and differences between waveforms~\cite{Ashton:2019leq}. 

The effective inspiral spin is usually measured accurately and precisely by NSBH-tuned approximants, when applicable, for all configurations. 
At SNR 70, systematic biases are visible for $q=3$ from all non-tidal approximants, but are more important for $q=2$, where they are larger than statistical uncertainties. 
It is worth stressing that in none of the configurations and for none of the approximants can we constrain the NS spin to be smaller than $\sim 0.4$ (the true value was 0). 
This suggests that even with loud NSBH it might be challenging to set constraints on the NS spin to values comparable to those found in galactic pulsars. 
This conclusion might need to be checked against NSBH sources in which the BH has a large spin misaligned with the orbital angular momentum, though existing work suggests it might still hold true.

Finally, we found biases in the measurement of the NS tidal deformability, \LNS. 
For the $q=2$ and $q=3$ sources, the differences in the posteriors are visible even at SNR 30, though much smaller than the statistical uncertainty (which in itself is very large, in excess of $100\%$ of the true value). 
Perhaps the most interesting of the SNR 30 comparisons is the one shown in~\Cref{Fig.q3_lambda2_SNR30_inc30}, since it shows tension between two approximants that are tuned against NSBH NR simulations, \SEOBNSBH and \IMRNSBH. 
While at SNR 30, the offsets are still much smaller than the statistical uncertainties, they are worth stressing as one would have expected \IMRNSBH and \SEOBNSBH to perform similarly.
It is also worth stressing that \LEAplus, which belongs to the EOB-baseline family, agrees with \IMRNSBH, suggesting the differences we see are not merely due to the underlying difference between EOB or IMRPhenom models, but to the specific technical details such as the way each approximant implements tidal terms and the reference point-particle models.
This tension becomes much more visible at SNR 70, \Cref{Fig.q2_lambda2_SNR70_inc30,Fig.q2_lambda2_SNR70_inc70,Fig.q3_lambda2_SNR70_inc30,Fig.q3_lambda2_SNR70_inc70}, especially for $q=3$. 
Here again \LEAplus and \IMRNSBH roughly agree with each other (and are found to overestimate \LNS), while \IMRpT, \SEOBT and \SEOBNSBH recover a different, and smaller, value of \LNS.
Since measuring \LNS away from 0 is perhaps the best way of showing that the secondary object is not a BH when there is no EM counterpart detected, these differences are particularly interesting.
Whereas based on \IMRNSBH or \LEAplus, when available, one would exclude for nearly all of the q2 and q3 simulations that \LNS$=0$, the \SEOBNSBH, \IMRpT and \SEOBT have a larger support for $\LNS=0$
This said, none of the models exclude the true value of \LNS: for $q=2$ we find a general tendency to underestimate the tidal deformability, while for $q=3$ some approximants overestimate and other underestimate it, with the true value found roughly in the middle, e.g.~\Cref{Fig.q3_lambda2_SNR70_inc70}.
The most stringent constraints are found for the $q=2$ sources at SNR 70, with a 90\% CI of $500-600$. 
For the SNR 30 sources, only an upper bound can be placed. 
For $q=6$, the simulated signals do not carry information about tides. 
We indeed find that nearly all families return a posterior on \LNS that is very similar to the prior at SNR 30, and only exclude extremely large values at SNR 70, \Cref{Fig.q6_lambda2_SNR70_inc30,Fig.q6_lambda2_SNR70_inc70}. 
\IMRpT differs significantly from the other approximants and recovers a \LNS posterior that peaks at small values. 
As discussed above, the reason is that waveform approximants tuned for BNS systems are being used far from their calibration range.
The mass and \LNS posteriors can be converted, using phenomenological fits, to a measurement for the NS radius, \rns,~\cref{App.CalcRNS}. 
We find that, at SNR 30, all approximants yield comparable constraints on the radius, with statistical uncertainties of $5$~km or larger (which is larger than what was inferred for GW170817~\cite{Abbott:2018wiz}). 
Interestingly, even for sources where some discrepancy in \LNS was visible, the posteriors on the radius show a smaller spread. 
This shows that most of the information comes from the measurement of the NS mass, with \LNS contributing less to the inference of \rns.

Overall, we find that at least the three approximants that have especially been tuned against NSBH waveforms agree well with each other for most of the parameters, though, critically, they show differences in their measurement of the NS tidal parameters already at SNR 30. 
While this \emph{might} be enough for most of the sources detected in the next few years, it clearly is insufficient in the next-generation detectors era~\cite{Purrer:2019jcp}, where typical SNRs will be 10 times higher. 
This also shows the need for a larger set of numerical relativity simulations, covering a much larger fraction of the relevant parameter space than what is currently available, in order to further calibrate and verify future NSBH waveform models.
This will likely require further development of NR simulation codes~\cite{SpECTRE, Kidder:2016hev}, in order to balance the computational cost and the resolving power necessary to include all significant physical effects~\cite{Purrer:2019jcp, Samajdar:2018dcx,Dudi:2018jzn}.
Even before then, residual differences between approximants that are nominally on equal footing might be problematic when performing tests of general relativity with GWs from NSBHs.

It is worth remembering that all of the simulated signals used in this paper do not have spin. 
This certainly represents a best-case scenario, though not an unrealistic one since most of the black holes detected to date are consistent with not having spins: it seems likely that even the two NSBH-tuned models we are using would start showing biases if the true signal came from a NSBH source with a large precessing BH spin.
Work is ongoing, and will be presented in a forthcoming publication, to consider NSBH NR simulations with spinning BHs~\cite{Foucart:2014nda}, though even for those the spin is not precessing.

\section{Acknowledgments}

\textit{Software}: We acknowledge the use of the LIGO Algorithm Library~\cite{lalsuite}, and specifically of the \texttt{LALInference} inference package~\cite{Veitch:2014wba}, as released through \texttt{conda}~\cite{conda}. Plots were produced with \texttt{matplotlib}~\cite{Hunter:2007}, and \texttt{corner}~\cite{corner}. 
We acknowledge use of \texttt{iPython}~\cite{ipython}, \texttt{NumPy}~\cite{numpy2} and \texttt{SciPy}~\cite{scipy}. 
\vskip 0.2cm
The authors would like to thank T.~Dent, T.~Dietrich and R. Sturani for useful discussions. 
Y.~H., C.-~J.~H., S.~V., and S.~B.~acknowledge support of the National Science Foundation, and the LIGO Laboratory. 
S.B. is also supported by the Paul and Daisy Soros Fellowship for New Americans and the NSF Graduate Research Fellowship under Grant No. DGE-1122374. 
V.V.\ is generously supported by the Sherman Fairchild Foundation, and NSF
grants PHY–170212 and PHY–1708213 at Caltech, and by a Klarman Fellowship at
Cornell.
F. Foucart gratefully acknowledges support from NASA through grant 80NSSC18K0565, from the NSF through grant PHY-1806278, and from the DOE through CAREER grant DE-SC0020435.
The authors acknowledge usage of the LIGO Data Grid clusters and the MIT Engaging cluster.
LIGO was constructed by the California Institute of Technology and Massachusetts Institute of Technology with funding from the National Science Foundation and operates under cooperative agreement PHY-0757058.
Virgo is funded by the French Centre National de Recherche Scientifique (CNRS), the Italian Istituto Nazionale della Fisica Nucleare (INFN) and the Dutch Nikhef, with contributions by Polish and Hungarian institutes.
This is LIGO Document Number DCC-P2000176.

\clearpage

\clearpage
\null

\begin{table}[!tbp]
\centering
\begin{tabularx}{0.99\textwidth}{c|*{6}{b}}
\toprule\toprule
  Sim. & \SEOBNSBH/\SEOB & \SEOBNSBH/\SEOBT  & \SEOBNSBH/\LEAplus & \SEOBNSBH/\IMRNSBH & \SEOBNSBH/\IMRp & \SEOBNSBH/\IMRpT 
\begin{filecontents*}{CSV/BSN_SNR30_inc30_comp.csv}
waveform,SEOBNSBHvsSEOB,SEOBNSBHvsSEOBT,SEOBNSBHvsLEAplus,SEOBNSBHvsIMRNSBH,SEOBNSBHvsIMRp,SEOBNSBHvsIMRpT,SEOBNSBHvsXHM
BHNSq2s0,51.96,41.15,-,-0.35,42.54,42.64,-
BHNSq3s0,2.30,2.99,-0.06,-0.27,3.95,3.46,-
BHNSq6s0,2.62,4.81,-,0.24,4.29,6.35,-2.92
\end{filecontents*}
\csvreader[head to column names]{CSV/BSN_SNR30_inc30_comp.csv}{}%
{\\ [0.5ex]\midrule \waveform&\SEOBNSBHvsSEOB&\SEOBNSBHvsSEOBT&\SEOBNSBHvsLEAplus&\SEOBNSBHvsIMRNSBH&\SEOBNSBHvsIMRp&\SEOBNSBHvsIMRpT}%
\\
\bottomrule\bottomrule
\end{tabularx}
\caption{\BSN for different approximants, SNR 30, inclination $30^{\circ}$, reported as the odds ratio to the \BSN of \SEOBNSBH.}
\label{Table.BSN_SNR30_inc30_comp}
\end{table}

\begin{table}[!tbp]
\centering
\begin{tabularx}{0.99\textwidth}{c|*{6}{b}}
\toprule\toprule
  Sim. & \SEOBNSBH/\SEOB & \SEOBNSBH/\SEOBT  & \SEOBNSBH/\LEAplus & \SEOBNSBH/\IMRNSBH & \SEOBNSBH/\IMRp & \SEOBNSBH/\IMRpT  
\begin{filecontents*}{CSV/BSN_SNR30_inc70_comp.csv}
waveform,SEOBNSBHvsSEOB,SEOBNSBHvsSEOBT,SEOBNSBHvsLEAplus,SEOBNSBHvsIMRNSBH,SEOBNSBHvsIMRp,SEOBNSBHvsIMRpT,SEOBNSBHvsXHM
BHNSq2s0,51.62,40.71,-,-0.29,42.25,42.11,-
BHNSq3s0,2.43,2.96,0.06,-0.10,3.58,3.70,-
BHNSq6s0,2.42,5.11,-,0.01,3.90,6.13,-15.67
\end{filecontents*}
\csvreader[head to column names]{CSV/BSN_SNR30_inc70_comp.csv}{}%
{\\ [0.5ex]\midrule \waveform&\SEOBNSBHvsSEOB&\SEOBNSBHvsSEOBT&\SEOBNSBHvsLEAplus&\SEOBNSBHvsIMRNSBH&\SEOBNSBHvsIMRp&\SEOBNSBHvsIMRpT}%
\\
\bottomrule\bottomrule
\end{tabularx}
\caption{\BSN for different approximants, SNR 30, inclination $70^{\circ}$, reported as the odds ratio to the \BSN of \SEOBNSBH.}
\label{Table.BSN_SNR30_inc70_comp}
\end{table}

\begin{table}[!tbp]
\centering
\begin{tabularx}{0.99\textwidth}{c|*{6}{b}}
\toprule\toprule
  Sim. & \SEOBNSBH/\SEOB & \SEOBNSBH/\SEOBT  & \SEOBNSBH/\LEAplus & \SEOBNSBH/\IMRNSBH & \SEOBNSBH/\IMRp & \SEOBNSBH/\IMRpT 
  \begin{filecontents*}{CSV/BSN_SNR70_inc30_comp.csv}
waveform,SEOBNSBHvsSEOB,SEOBNSBHvsSEOBT,SEOBNSBHvsLEAplus,SEOBNSBHvsIMRNSBH,SEOBNSBHvsIMRp,SEOBNSBHvsIMRpT,SEOBNSBHvsXHM
BHNSq2s0,279.38,214.03,-,-0.27,220.98,218.13,-
BHNSq3s0,3.42,4.97,0.29,-0.27,6.15,6.61,-
BHNSq6s0,2.64,21.20,-,1.55,6.45,17.39,-
\end{filecontents*}
\csvreader[head to column names]{CSV/BSN_SNR70_inc30_comp.csv}{}%
{\\ [0.5ex]\midrule \waveform&\SEOBNSBHvsSEOB&\SEOBNSBHvsSEOBT&\SEOBNSBHvsLEAplus&\SEOBNSBHvsIMRNSBH&\SEOBNSBHvsIMRp&\SEOBNSBHvsIMRpT}%
\\
\bottomrule\bottomrule
\end{tabularx}
\caption{\BSN for different approximants, SNR 70, inclination $30^{\circ}$, reported as the odds ratio to the \BSN of \SEOBNSBH.}
\label{Table.BSN_SNR70_inc30_comp}
\end{table}

\begin{table}[!tbp]
\centering
\begin{tabularx}{0.99\textwidth}{c|*{6}{b}}
\toprule\toprule
  Sim. & \SEOBNSBH/\SEOB & \SEOBNSBH/\SEOBT  & \SEOBNSBH/\LEAplus & \SEOBNSBH/\IMRNSBH & \SEOBNSBH/\IMRp & \SEOBNSBH/\IMRpT 
\begin{filecontents*}{CSV/BSN_SNR70_inc70_comp.csv}
waveform,SEOBNSBHvsSEOB,SEOBNSBHvsSEOBT,SEOBNSBHvsLEAplus,SEOBNSBHvsIMRNSBH,SEOBNSBHvsIMRp,SEOBNSBHvsIMRpT,SEOBNSBHvsXHM
BHNSq2s0,277.24,212.81,-,-0.18,219.73,216.79,-
BHNSq3s0,3.46,5.41,0.40,-0.30,6.89,7.42,-
BHNSq6s0,2.67,23.47,-,1.53,6.21,17.87,-
\end{filecontents*}
\csvreader[head to column names]{CSV/BSN_SNR70_inc70_comp.csv}{}%
{\\ [0.5ex]\midrule \waveform&\SEOBNSBHvsSEOB&\SEOBNSBHvsSEOBT&\SEOBNSBHvsLEAplus&\SEOBNSBHvsIMRNSBH&\SEOBNSBHvsIMRp&\SEOBNSBHvsIMRpT}%
\\
\bottomrule\bottomrule
\end{tabularx}
\caption{\BSN for different approximants, SNR 70, inclination $70^{\circ}$, reported as the odds ratio to the \BSN of \SEOBNSBH.}
\label{Table.BSN_SNR70_inc70_comp}
\end{table} 
\clearpage

\appendix
\clearpage
\null
\section{Full parameter estimation results}
\label{App.FullPe}

We report the parameter estimation results for an extended set of parameters relative to Sec.~\ref{Sec.Results}.
The results are presented as the median of the marginalised 1D posterior distributions for different parameters, and the corresponding symmetric $90\%$ credible intervals.

The $q=2$ analyses are reported in Table~\ref{Table.Param_BHNSq2s0_SNR30} for SNR 30 and Table~\ref{Table.Param_BHNSq2s0_SNR70} for SNR 70.
For $q=3$, the SNR 30 results are shown in Table~\ref{Table.Param_BHNSq3s0_SNR30} and the SNR 70 results in Table~\ref{Table.Param_BHNSq3s0_SNR70}.
Finally, the $q=6$ results are reported in Table~\ref{Table.Param_BHNSq6s0_SNR30} for SNR 30 and Table~\ref{Table.Param_BHNSq6s0_SNR70} for SNR 70.

\section{Calculating neutron star radii}
\label{App.CalcRNS}

The compactness of the neutron star is estimated using a fit from~\cite{Yagi:2016bkt} (Eq. 78, in Section 4.4.1), 
\begin{equation}
\label{Eq.compactFit}
C_\mathrm{NS} = \sum^{2}_{k=0} a_k (\ln{\LNS})^k,
\end{equation}
with fitting parameters $a_0 = 0.371$, $a_1 = -0.0391$, and $a_2 = 0.001056$ from~\cite{Maselli:2013mva}.
As reported in~\cite{Yagi:2016bkt}, this fit, when compared to a large set of NS EoS models, has the largest deviation of $6.5\%$ that is significantly smaller than the statistical uncertainties reported in Sec.~\ref{Sec:rNS}.

The neutron star radius \rns is in turn related to the compactness through
\begin{equation}\label{Eq.compactToRadius}
\rns = \frac{m_\mathrm{NS}}{C_\mathrm{NS}}\;,
\end{equation}
with $m_\mathrm{NS}$ being the neutron star mass reported in the rest frame of the NSBH binary.
Again, we are here assuming $G = c = 1$.
\section{Prior}
\label{App.Priors}
We use priors which are routinely used in LVC publications~\cite{Abbott:2016blz, TheLIGOScientific:2016wfe, Abbott:2016nmj, Abbott:2017vtc, Abbott:2017gyy, Abbott:2017oio,TheLIGOScientific:2017qsa, Abbott:2018wiz}, in Table~\ref{Table.Prior_BHNSq2s0},~\ref{Table.Prior_BHNSq3s0} and~\ref{Table.Prior_BHNSq6s0}, for q = 2, 3, and 6 respectively.

We use uniform priors for \emph{detector-frame} component masses. 
When using a ROQ, additional prior constraints are imposed on the detector-frame chirp mass and mass ratio, which limit their range. 
For one system, the mass prior bounds are the same for all the aligned-spin waveforms, while slightly different from those for the precessing-spin waveforms due to different choices by the ROQ basis.
Note that the \LEAplus ROQ basis is constructed with prior constraints only on the component BH and NS masses.
The black hole spin prior is uniform in the dimensionless spin magnitude in the range $[0,0.99]$, and isotropic for the orientation for precessing-spin approximants. 
For non-precessing waveforms, the prior on the (aligned) spin magnitude is equal to the projection of an isotropic spin vector along the orbital angular momentum.
For waveform models that support tidal deformation of the neutron star, we use a prior uniform over \LNS within the range of validity.
We choose a prior for sky localization and the orientation of the orbital angular momentum with respect to the line of sight that is uniform over the sphere; a prior for the distance that is proportional to luminosity distance squared; a uniform prior over the arrival time and phase.

\clearpage
\begin{turnpage}
\begin{table}[!htbp]
\scriptsize
\centering

\begin{subtable}[]{1.1\textwidth}
		\centering
\begin{tabularx}{1.1\textwidth}{c|*{14}{b}}
\toprule\toprule
Approximant&$\mcs/\msun$&q&$\mBHs/\msun$&$\mNSs/\msun$&$\mts/\msun$&\chieff&\sBH&\sBHz&\sNS&\sNSz&\tidal&\rns/km&\dl/Mpc&\inc/\dg
\begin{filecontents*}{CSV/Param_BHNSq2s0_SNR30_inc30.csv}
Approximant,mc,q,mBH,mNS,mtotal,aeff,aBH,aBHz,aNS,aNSz,lambdaNS,rNS,dist,thetajn
True value,{ 1.65},{ 2.0},{ 2.72},{ 1.36},{ 4.08},{ 0.00},{ 0.0},{ 0.0},{ 0.0},{ 0.0},{ 791},{ 14.4},{ 136.5},{ 30}
SEOB,${ 1.66}_{-  0.01}^{+  0.02}$,${ 2.3}_{-  0.9}^{+  0.9}$,${ 2.9}_{-  0.7}^{+  0.6}$,${ 1.3}_{-  0.2}^{+  0.4}$,${ 4.2}_{-  0.3}^{+  0.4}$,${ 0.03}_{-  0.11}^{+  0.09}$,${ 0.07}_{-  0.07}^{+  0.17}$,${ 0.0}_{-  0.2}^{+  0.2}$,${ 0.13}_{-  0.12}^{+  0.41}$,${ 0.0}_{-  0.4}^{+  0.5}$,-,-,${ 131.4}_{-  46.0}^{+  25.9}$,${ 34}_{-  25}^{+  30}$
SEOBT,${ 1.66}_{-  0.01}^{+  0.02}$,${ 2.0}_{-  0.7}^{+  0.8}$,${ 2.7}_{-  0.6}^{+  0.5}$,${ 1.4}_{-  0.2}^{+  0.3}$,${ 4.1}_{-  0.2}^{+  0.3}$,${ 0.00}_{-  0.07}^{+  0.09}$,${ 0.07}_{-  0.06}^{+  0.19}$,${ 0.0}_{-  0.2}^{+  0.2}$,${ 0.12}_{-  0.12}^{+  0.41}$,${ 0.0}_{-  0.4}^{+  0.4}$,${ 503}_{-  426}^{+  957}$,${ 12.5}_{-  3.8}^{+  3.0}$,${ 131.8}_{-  45.9}^{+  25.1}$,${ 34}_{-  24}^{+  30}$
SEOBNSBH,${ 1.66}_{-  0.01}^{+  0.02}$,${ 1.9}_{-  0.8}^{+  0.7}$,${ 2.6}_{-  0.6}^{+  0.5}$,${ 1.4}_{-  0.2}^{+  0.4}$,${ 4.0}_{-  0.2}^{+  0.3}$,${ 0.00}_{-  0.07}^{+  0.07}$,${ 0.05}_{-  0.05}^{+  0.11}$,${-0.0}_{-  0.1}^{+  0.1}$,${ 0.00}_{-  0.00}^{+  0.00}$,${ 0.0}_{-  0.0}^{+  0.0}$,${ 547}_{-  415}^{+  773}$,${ 12.9}_{-  3.4}^{+  3.2}$,${ 131.9}_{-  44.1}^{+  22.8}$,${ 32}_{-  23}^{+  28}$
IMRNSBH,${ 1.65}_{-  0.01}^{+  0.02}$,${ 1.9}_{-  0.8}^{+  0.6}$,${ 2.7}_{-  0.6}^{+  0.5}$,${ 1.4}_{-  0.2}^{+  0.4}$,${ 4.1}_{-  0.2}^{+  0.3}$,${ 0.00}_{-  0.08}^{+  0.07}$,${ 0.04}_{-  0.04}^{+  0.11}$,${ 0.0}_{-  0.1}^{+  0.1}$,${ 0.00}_{-  0.00}^{+  0.00}$,${ 0.0}_{-  0.0}^{+  0.0}$,${ 694}_{-  492}^{+  744}$,${ 13.2}_{-  3.3}^{+  2.9}$,${ 133.4}_{-  44.5}^{+  23.5}$,${ 32}_{-  23}^{+  28}$
IMRp,${ 1.65}_{-  0.01}^{+  0.02}$,${ 2.2}_{-  0.8}^{+  0.8}$,${ 2.9}_{-  0.6}^{+  0.5}$,${ 1.3}_{-  0.2}^{+  0.3}$,${ 4.2}_{-  0.3}^{+  0.3}$,${ 0.02}_{-  0.09}^{+  0.08}$,${ 0.14}_{-  0.13}^{+  0.36}$,${ 0.0}_{-  0.2}^{+  0.2}$,${ 0.31}_{-  0.28}^{+  0.54}$,${ 0.0}_{-  0.3}^{+  0.4}$,-,-,${ 143.1}_{-  47.3}^{+  16.5}$,${ 24}_{-  18}^{+  34}$
IMRpT,${ 1.65}_{-  0.01}^{+  0.02}$,${ 1.9}_{-  0.7}^{+  0.7}$,${ 2.6}_{-  0.6}^{+  0.5}$,${ 1.4}_{-  0.2}^{+  0.4}$,${ 4.0}_{-  0.2}^{+  0.3}$,${-0.01}_{-  0.07}^{+  0.08}$,${ 0.13}_{-  0.12}^{+  0.40}$,${ 0.0}_{-  0.1}^{+  0.1}$,${ 0.25}_{-  0.23}^{+  0.53}$,${-0.0}_{-  0.2}^{+  0.3}$,${ 569}_{-  459}^{+  782}$,${ 13.0}_{-  3.5}^{+  2.7}$,${ 142.1}_{-  46.9}^{+  17.1}$,${ 25}_{-  19}^{+  33}$
XHM,${ 1.65}_{-  0.01}^{+  0.01}$,${ 2.2}_{-  0.9}^{+  0.9}$,${ 2.9}_{-  0.7}^{+  0.6}$,${ 1.3}_{-  0.2}^{+  0.4}$,${ 4.2}_{-  0.3}^{+  0.4}$,${ 0.03}_{-  0.10}^{+  0.09}$,${ 0.07}_{-  0.06}^{+  0.15}$,${ 0.0}_{-  0.2}^{+  0.2}$,${ 0.13}_{-  0.12}^{+  0.30}$,${ 0.0}_{-  0.3}^{+  0.4}$,-,-,${ 141.1}_{-  41.6}^{+  20.0}$,${ 27}_{-  20}^{+  28}$
\end{filecontents*}
\csvreader[head to column names]{CSV/Param_BHNSq2s0_SNR30_inc30.csv}{}%
{\\ [0.5ex]\midrule \texttt{\Approximant}&\mc&\q&\mBH&\mNS&\mtotal&\aeff&\aBH&\aBHz&\aNS&\aNSz&\lambdaNS&\rNS&\dist&\thetajn}
\\
\bottomrule\bottomrule
\end{tabularx}
 \caption{Inclination 30\dg.}
 \end{subtable}
 
\begin{subtable}[!htbp]{1.1\textwidth}
		\centering
\begin{tabularx}{1.1\textwidth}{c|*{14}{b}}
\toprule\toprule
Approximant&$\mcs/\msun$&q&$\mBHs/\msun$&$\mNSs/\msun$&$\mts/\msun$&\chieff&\sBH&\sBHz&\sNS&\sNSz&\tidal&\rns/km&\dl/Mpc&\inc/\dg
\begin{filecontents*}{CSV/Param_BHNSq2s0_SNR30_inc70.csv}
Approximant,mc,q,mBH,mNS,mtotal,aeff,aBH,aBHz,aNS,aNSz,lambdaNS,rNS,dist,thetajn
True value,{ 1.67},{ 2.0},{ 2.75},{ 1.37},{ 4.12},{ 0.00},{ 0.0},{ 0.0},{ 0.0},{ 0.0},{ 791},{ 14.4},{ 87.0},{ 70}
SEOB,${ 1.66}_{-  0.01}^{+  0.02}$,${ 2.3}_{-  0.9}^{+  0.9}$,${ 2.9}_{-  0.7}^{+  0.6}$,${ 1.3}_{-  0.2}^{+  0.4}$,${ 4.2}_{-  0.3}^{+  0.4}$,${ 0.03}_{-  0.11}^{+  0.09}$,${ 0.07}_{-  0.07}^{+  0.18}$,${ 0.0}_{-  0.2}^{+  0.2}$,${ 0.13}_{-  0.13}^{+  0.44}$,${ 0.0}_{-  0.4}^{+  0.5}$,-,-,${ 131.3}_{-  50.2}^{+  27.2}$,${ 35}_{-  25}^{+  37}$
SEOBT,${ 1.66}_{-  0.01}^{+  0.02}$,${ 2.0}_{-  0.8}^{+  0.8}$,${ 2.7}_{-  0.6}^{+  0.5}$,${ 1.4}_{-  0.2}^{+  0.4}$,${ 4.1}_{-  0.2}^{+  0.3}$,${ 0.00}_{-  0.07}^{+  0.09}$,${ 0.07}_{-  0.06}^{+  0.21}$,${ 0.0}_{-  0.2}^{+  0.2}$,${ 0.12}_{-  0.11}^{+  0.43}$,${ 0.0}_{-  0.4}^{+  0.4}$,${ 502}_{-  423}^{+  953}$,${ 12.5}_{-  3.8}^{+  3.1}$,${ 131.3}_{-  49.1}^{+  27.1}$,${ 35}_{-  25}^{+  36}$
SEOBNSBH,${ 1.66}_{-  0.01}^{+  0.02}$,${ 1.9}_{-  0.8}^{+  0.7}$,${ 2.6}_{-  0.6}^{+  0.5}$,${ 1.4}_{-  0.2}^{+  0.4}$,${ 4.0}_{-  0.2}^{+  0.3}$,${-0.01}_{-  0.07}^{+  0.07}$,${ 0.05}_{-  0.05}^{+  0.11}$,${-0.0}_{-  0.1}^{+  0.1}$,${ 0.00}_{-  0.00}^{+  0.00}$,${ 0.0}_{-  0.0}^{+  0.0}$,${ 550}_{-  406}^{+  782}$,${ 13.0}_{-  3.3}^{+  3.2}$,${ 131.0}_{-  44.9}^{+  25.2}$,${ 34}_{-  24}^{+  31}$
IMRNSBH,${ 1.65}_{-  0.01}^{+  0.02}$,${ 2.0}_{-  0.8}^{+  0.6}$,${ 2.7}_{-  0.6}^{+  0.4}$,${ 1.4}_{-  0.2}^{+  0.4}$,${ 4.1}_{-  0.2}^{+  0.3}$,${ 0.00}_{-  0.08}^{+  0.06}$,${ 0.04}_{-  0.04}^{+  0.11}$,${ 0.0}_{-  0.1}^{+  0.1}$,${ 0.00}_{-  0.00}^{+  0.00}$,${ 0.0}_{-  0.0}^{+  0.0}$,${ 707}_{-  486}^{+  782}$,${ 13.3}_{-  3.1}^{+  2.9}$,${ 133.8}_{-  45.6}^{+  24.9}$,${ 33}_{-  23}^{+  30}$
IMRp,${ 1.65}_{-  0.01}^{+  0.02}$,${ 2.3}_{-  0.8}^{+  0.8}$,${ 2.9}_{-  0.6}^{+  0.5}$,${ 1.3}_{-  0.2}^{+  0.3}$,${ 4.2}_{-  0.3}^{+  0.4}$,${ 0.03}_{-  0.10}^{+  0.08}$,${ 0.12}_{-  0.11}^{+  0.33}$,${ 0.0}_{-  0.1}^{+  0.1}$,${ 0.32}_{-  0.28}^{+  0.53}$,${ 0.0}_{-  0.3}^{+  0.4}$,-,-,${ 144.2}_{-  54.0}^{+  17.0}$,${ 24}_{-  18}^{+  41}$
IMRpT,${ 1.65}_{-  0.01}^{+  0.02}$,${ 1.8}_{-  0.7}^{+  0.7}$,${ 2.6}_{-  0.6}^{+  0.5}$,${ 1.4}_{-  0.2}^{+  0.4}$,${ 4.0}_{-  0.2}^{+  0.3}$,${-0.02}_{-  0.06}^{+  0.08}$,${ 0.14}_{-  0.12}^{+  0.43}$,${ 0.0}_{-  0.2}^{+  0.1}$,${ 0.24}_{-  0.21}^{+  0.49}$,${-0.0}_{-  0.2}^{+  0.2}$,${ 538}_{-  435}^{+  732}$,${ 12.9}_{-  3.6}^{+  2.8}$,${ 143.1}_{-  53.4}^{+  18.1}$,${ 25}_{-  19}^{+  40}$
XHM,${ 1.65}_{-  0.01}^{+  0.02}$,${ 2.2}_{-  0.7}^{+  1.0}$,${ 2.9}_{-  0.5}^{+  0.7}$,${ 1.3}_{-  0.2}^{+  0.2}$,${ 4.2}_{-  0.3}^{+  0.5}$,${ 0.03}_{-  0.08}^{+  0.10}$,${ 0.09}_{-  0.09}^{+  0.22}$,${ 0.0}_{-  0.3}^{+  0.2}$,${ 0.15}_{-  0.14}^{+  0.41}$,${ 0.1}_{-  0.3}^{+  0.5}$,-,-,${ 133.4}_{-  43.3}^{+  25.2}$,${ 34}_{-  22}^{+  30}$
\end{filecontents*}
\csvreader[head to column names]{CSV/Param_BHNSq2s0_SNR30_inc70.csv}{}%
{\\ [0.5ex]\midrule \texttt{\Approximant}&\mc&\q&\mBH&\mNS&\mtotal&\aeff&\aBH&\aBHz&\aNS&\aNSz&\lambdaNS&\rNS&\dist&\thetajn}
\\
\bottomrule\bottomrule
\end{tabularx}
 \caption{Inclination 70\dg.}
 \end{subtable}

\caption{Properties of the q=2, SNR 30 simulation, as recovered by the listed approximants. We report the 1D median, together with the symmetric the $90\%$ credible interval.}
\label{Table.Param_BHNSq2s0_SNR30}
\end{table}

\clearpage

\begin{table}[!htbp]
\scriptsize
\centering
\begin{subtable}[!htbp]{1.1\textwidth}
		\centering
\begin{tabularx}{1.1\textwidth}{c|*{14}{b}}
\toprule\toprule
Approximant&$\mcs/\msun$&q&$\mBHs/\msun$&$\mNSs/\msun$&$\mts/\msun$&\chieff&\sBH&\sBHz&\sNS&\sNSz&\tidal&\rns/km&\dl/Mpc&\inc/\dg
\begin{filecontents*}{CSV/Param_BHNSq2s0_SNR70_inc30.csv}
Approximant,mc,q,mBH,mNS,mtotal,aeff,aBH,aBHz,aNS,aNSz,lambdaNS,rNS,dist,thetajn
True value,{ 1.68},{ 2.0},{ 2.76},{ 1.38},{ 4.15},{ 0.00},{ 0.0},{ 0.0},{ 0.0},{ 0.0},{ 791},{ 14.4},{ 58.6},{ 30}
SEOB,${ 1.68}_{-  0.00}^{+  0.01}$,${ 2.5}_{-  0.4}^{+  0.4}$,${ 3.1}_{-  0.3}^{+  0.3}$,${ 1.2}_{-  0.1}^{+  0.1}$,${ 4.4}_{-  0.2}^{+  0.2}$,${ 0.06}_{-  0.05}^{+  0.04}$,${ 0.07}_{-  0.07}^{+  0.15}$,${ 0.1}_{-  0.2}^{+  0.2}$,${ 0.14}_{-  0.14}^{+  0.40}$,${ 0.1}_{-  0.4}^{+  0.4}$,-,-,${ 57.8}_{-  15.4}^{+  8.5}$,${ 31}_{-  22}^{+  24}$
SEOBT,${ 1.68}_{-  0.00}^{+  0.01}$,${ 2.0}_{-  0.6}^{+  0.5}$,${ 2.8}_{-  0.5}^{+  0.4}$,${ 1.4}_{-  0.1}^{+  0.3}$,${ 4.1}_{-  0.2}^{+  0.2}$,${ 0.00}_{-  0.07}^{+  0.06}$,${ 0.06}_{-  0.06}^{+  0.19}$,${ 0.0}_{-  0.2}^{+  0.2}$,${ 0.11}_{-  0.11}^{+  0.41}$,${-0.0}_{-  0.4}^{+  0.4}$,${ 415}_{-  291}^{+  264}$,${ 12.1}_{-  2.9}^{+  2.4}$,${ 57.9}_{-  16.0}^{+  8.4}$,${ 31}_{-  22}^{+  24}$
SEOBNSBH,${ 1.68}_{-  0.00}^{+  0.01}$,${ 1.9}_{-  0.6}^{+  0.4}$,${ 2.7}_{-  0.5}^{+  0.3}$,${ 1.4}_{-  0.1}^{+  0.3}$,${ 4.1}_{-  0.2}^{+  0.2}$,${ 0.00}_{-  0.06}^{+  0.04}$,${ 0.03}_{-  0.02}^{+  0.09}$,${ 0.0}_{-  0.1}^{+  0.1}$,${ 0.00}_{-  0.00}^{+  0.00}$,${ 0.0}_{-  0.0}^{+  0.0}$,${ 510}_{-  205}^{+  206}$,${ 12.7}_{-  1.7}^{+  2.1}$,${ 58.1}_{-  14.4}^{+  7.4}$,${ 29}_{-  20}^{+  23}$
IMRNSBH,${ 1.68}_{-  0.00}^{+  0.01}$,${ 2.0}_{-  0.5}^{+  0.4}$,${ 2.7}_{-  0.4}^{+  0.3}$,${ 1.4}_{-  0.1}^{+  0.2}$,${ 4.1}_{-  0.2}^{+  0.2}$,${ 0.00}_{-  0.05}^{+  0.04}$,${ 0.02}_{-  0.02}^{+  0.08}$,${ 0.0}_{-  0.1}^{+  0.1}$,${ 0.00}_{-  0.00}^{+  0.00}$,${ 0.0}_{-  0.0}^{+  0.0}$,${ 695}_{-  225}^{+  212}$,${ 13.2}_{-  1.4}^{+  1.6}$,${ 59.2}_{-  15.1}^{+  7.2}$,${ 29}_{-  20}^{+  23}$
IMRp,${ 1.68}_{-  0.00}^{+  0.00}$,${ 2.6}_{-  0.4}^{+  0.4}$,${ 3.2}_{-  0.3}^{+  0.3}$,${ 1.2}_{-  0.1}^{+  0.1}$,${ 4.4}_{-  0.2}^{+  0.2}$,${ 0.06}_{-  0.04}^{+  0.03}$,${ 0.12}_{-  0.11}^{+  0.21}$,${ 0.1}_{-  0.1}^{+  0.1}$,${ 0.28}_{-  0.24}^{+  0.50}$,${ 0.1}_{-  0.3}^{+  0.3}$,-,-,${ 64.4}_{-  12.6}^{+  3.0}$,${ 15}_{-  11}^{+  26}$
IMRpT,${ 1.68}_{-  0.00}^{+  0.01}$,${ 1.9}_{-  0.7}^{+  0.5}$,${ 2.7}_{-  0.6}^{+  0.3}$,${ 1.4}_{-  0.1}^{+  0.3}$,${ 4.1}_{-  0.2}^{+  0.2}$,${ 0.00}_{-  0.06}^{+  0.05}$,${ 0.09}_{-  0.08}^{+  0.25}$,${ 0.0}_{-  0.1}^{+  0.1}$,${ 0.18}_{-  0.16}^{+  0.46}$,${ 0.0}_{-  0.2}^{+  0.2}$,${ 502}_{-  269}^{+  284}$,${ 12.8}_{-  2.3}^{+  2.2}$,${ 63.2}_{-  15.2}^{+  4.0}$,${ 19}_{-  15}^{+  27}$
XHM,${ 1.68}_{-  0.00}^{+  0.00}$,${ 2.5}_{-  0.3}^{+  0.4}$,${ 3.1}_{-  0.2}^{+  0.2}$,${ 1.2}_{-  0.1}^{+  0.1}$,${ 4.4}_{-  0.1}^{+  0.2}$,${ 0.07}_{-  0.04}^{+  0.03}$,${ 0.06}_{-  0.05}^{+  0.12}$,${-0.0}_{-  0.2}^{+  0.1}$,${ 0.28}_{-  0.26}^{+  0.43}$,${ 0.3}_{-  0.3}^{+  0.4}$,-,-,${ 62.8}_{-  6.8}^{+  4.2}$,${ 21}_{-  12}^{+  13}$
\end{filecontents*}
\csvreader[head to column names]{CSV/Param_BHNSq2s0_SNR70_inc30.csv}{}%
{\\ [0.5ex]\midrule \texttt{\Approximant}&\mc&\q&\mBH&\mNS&\mtotal&\aeff&\aBH&\aBHz&\aNS&\aNSz&\lambdaNS&\rNS&\dist&\thetajn}
\\
\bottomrule\bottomrule
\end{tabularx}
 \caption{Inclination 30\dg.}
 \end{subtable}

\begin{subtable}[!htbp]{1.1\textwidth}
		\centering
\begin{tabularx}{1.1\textwidth}{c|*{14}{b}}
\toprule\toprule
Approximant&$\mcs/\msun$&q&$\mBHs/\msun$&$\mNSs/\msun$&$\mts/\msun$&\chieff&\sBH&\sBHz&\sNS&\sNSz&\tidal&\rns/km&\dl/Mpc&\inc/\dg
\begin{filecontents*}{CSV/Param_BHNSq2s0_SNR70_inc70.csv}
Approximant,mc,q,mBH,mNS,mtotal,aeff,aBH,aBHz,aNS,aNSz,lambdaNS,rNS,dist,thetajn
True value,{ 1.69},{ 2.0},{ 2.78},{ 1.39},{ 4.17},{ 0.00},{ 0.0},{ 0.0},{ 0.0},{ 0.0},{ 791},{ 14.4},{ 37.3},{ 70}
SEOB,${ 1.68}_{-  0.00}^{+  0.01}$,${ 2.5}_{-  0.4}^{+  0.4}$,${ 3.1}_{-  0.3}^{+  0.3}$,${ 1.2}_{-  0.1}^{+  0.1}$,${ 4.4}_{-  0.2}^{+  0.2}$,${ 0.06}_{-  0.05}^{+  0.04}$,${ 0.07}_{-  0.07}^{+  0.14}$,${ 0.1}_{-  0.2}^{+  0.1}$,${ 0.14}_{-  0.13}^{+  0.40}$,${ 0.0}_{-  0.4}^{+  0.4}$,-,-,${ 53.3}_{-  17.5}^{+  13.1}$,${ 39}_{-  29}^{+  33}$
SEOBT,${ 1.68}_{-  0.00}^{+  0.01}$,${ 2.0}_{-  0.6}^{+  0.5}$,${ 2.7}_{-  0.5}^{+  0.4}$,${ 1.4}_{-  0.1}^{+  0.3}$,${ 4.1}_{-  0.2}^{+  0.2}$,${ 0.00}_{-  0.06}^{+  0.06}$,${ 0.06}_{-  0.05}^{+  0.20}$,${ 0.0}_{-  0.2}^{+  0.2}$,${ 0.11}_{-  0.10}^{+  0.41}$,${ 0.0}_{-  0.4}^{+  0.4}$,${ 417}_{-  296}^{+  271}$,${ 12.2}_{-  3.0}^{+  2.3}$,${ 53.4}_{-  17.3}^{+  13.1}$,${ 39}_{-  29}^{+  32}$
SEOBNSBH,${ 1.68}_{-  0.00}^{+  0.01}$,${ 1.9}_{-  0.5}^{+  0.3}$,${ 2.7}_{-  0.4}^{+  0.2}$,${ 1.4}_{-  0.1}^{+  0.2}$,${ 4.1}_{-  0.2}^{+  0.1}$,${ 0.00}_{-  0.05}^{+  0.04}$,${ 0.02}_{-  0.02}^{+  0.08}$,${ 0.0}_{-  0.1}^{+  0.1}$,${ 0.00}_{-  0.00}^{+  0.00}$,${ 0.0}_{-  0.0}^{+  0.0}$,${ 507}_{-  200}^{+  203}$,${ 12.6}_{-  1.6}^{+  1.9}$,${ 54.1}_{-  17.7}^{+  11.4}$,${ 37}_{-  26}^{+  33}$
IMRNSBH,${ 1.68}_{-  0.00}^{+  0.01}$,${ 2.0}_{-  0.5}^{+  0.3}$,${ 2.7}_{-  0.4}^{+  0.2}$,${ 1.4}_{-  0.1}^{+  0.2}$,${ 4.1}_{-  0.2}^{+  0.1}$,${ 0.00}_{-  0.05}^{+  0.04}$,${ 0.02}_{-  0.02}^{+  0.07}$,${ 0.0}_{-  0.1}^{+  0.1}$,${ 0.00}_{-  0.00}^{+  0.00}$,${ 0.0}_{-  0.0}^{+  0.0}$,${ 693}_{-  216}^{+  215}$,${ 13.2}_{-  1.4}^{+  1.5}$,${ 55.2}_{-  18.2}^{+  11.3}$,${ 36}_{-  26}^{+  33}$
IMRp,${ 1.68}_{-  0.00}^{+  0.01}$,${ 2.5}_{-  0.3}^{+  0.4}$,${ 3.1}_{-  0.2}^{+  0.2}$,${ 1.2}_{-  0.1}^{+  0.1}$,${ 4.4}_{-  0.2}^{+  0.2}$,${ 0.06}_{-  0.04}^{+  0.03}$,${ 0.12}_{-  0.11}^{+  0.15}$,${ 0.0}_{-  0.1}^{+  0.1}$,${ 0.30}_{-  0.26}^{+  0.44}$,${ 0.1}_{-  0.2}^{+  0.3}$,-,-,${ 64.1}_{-  26.4}^{+  3.8}$,${ 18}_{-  14}^{+  51}$
IMRpT,${ 1.68}_{-  0.00}^{+  0.01}$,${ 1.9}_{-  0.7}^{+  0.4}$,${ 2.7}_{-  0.6}^{+  0.3}$,${ 1.4}_{-  0.1}^{+  0.3}$,${ 4.1}_{-  0.2}^{+  0.2}$,${ 0.00}_{-  0.06}^{+  0.05}$,${ 0.09}_{-  0.08}^{+  0.24}$,${ 0.0}_{-  0.1}^{+  0.1}$,${ 0.16}_{-  0.15}^{+  0.52}$,${ 0.0}_{-  0.2}^{+  0.2}$,${ 522}_{-  264}^{+  287}$,${ 12.9}_{-  2.3}^{+  2.1}$,${ 62.5}_{-  24.8}^{+  5.2}$,${ 22}_{-  18}^{+  46}$
XHM,${ 1.68}_{-  0.00}^{+  0.00}$,${ 2.6}_{-  0.3}^{+  0.4}$,${ 3.2}_{-  0.2}^{+  0.3}$,${ 1.2}_{-  0.1}^{+  0.1}$,${ 4.4}_{-  0.1}^{+  0.2}$,${ 0.07}_{-  0.04}^{+  0.03}$,${ 0.06}_{-  0.06}^{+  0.11}$,${ 0.1}_{-  0.1}^{+  0.1}$,${ 0.12}_{-  0.11}^{+  0.31}$,${ 0.1}_{-  0.3}^{+  0.3}$,-,-,${ 54.8}_{-  11.6}^{+  8.2}$,${ 38}_{-  15}^{+  18}$
\end{filecontents*}
\csvreader[head to column names]{CSV/Param_BHNSq2s0_SNR70_inc70.csv}{}%
{\\ [0.5ex]\midrule \texttt{\Approximant}&\mc&\q&\mBH&\mNS&\mtotal&\aeff&\aBH&\aBHz&\aNS&\aNSz&\lambdaNS&\rNS&\dist&\thetajn}
\\
\bottomrule\bottomrule
\end{tabularx}
 \caption{Inclination 70\dg.}
 \end{subtable}
 
\caption{Properties of the q=2, SNR 70 simulation, as recovered by the listed approximants. We report the 1D median, together with the symmetric the $90\%$ credible interval.}
\label{Table.Param_BHNSq2s0_SNR70}
\end{table} 

\clearpage
\vfill

\begin{table}[!htbp]
\scriptsize
\centering

\begin{subtable}[!htbp]{1.1\textwidth}
		\centering
\begin{tabularx}{1.1\textwidth}{c|*{14}{b}}
\toprule\toprule
Approximant&$\mcs/\msun$&q&$\mBHs/\msun$&$\mNSs/\msun$&$\mts/\msun$&\chieff&\sBH&\sBHz&\sNS&\sNSz&\tidal&\rns/km&\dl/Mpc&\inc/\dg
\begin{filecontents*}{CSV/Param_BHNSq3s0_SNR30_inc30.csv}
Approximant,mc,q,mBH,mNS,mtotal,aeff,aBH,aBHz,aNS,aNSz,lambdaNS,rNS,dist,thetajn
True value,{ 1.91},{ 3.0},{ 3.90},{ 1.30},{ 5.21},{ 0.00},{ 0.0},{ 0.0},{ 0.0},{ 0.0},{ 624},{ 12.3},{ 170.0},{ 30}
SEOB,${ 1.91}_{-  0.01}^{+  0.03}$,${ 3.0}_{-  0.6}^{+  0.4}$,${ 3.9}_{-  0.5}^{+  0.3}$,${ 1.3}_{-  0.1}^{+  0.1}$,${ 5.2}_{-  0.3}^{+  0.2}$,${ 0.00}_{-  0.08}^{+  0.05}$,${ 0.04}_{-  0.04}^{+  0.13}$,${ 0.0}_{-  0.2}^{+  0.1}$,${ 0.12}_{-  0.12}^{+  0.43}$,${ 0.0}_{-  0.4}^{+  0.4}$,-,-,${ 161.8}_{-  61.9}^{+  34.3}$,${ 36}_{-  26}^{+  41}$
SEOBT,${ 1.91}_{-  0.01}^{+  0.03}$,${ 2.6}_{-  1.1}^{+  0.7}$,${ 3.6}_{-  0.9}^{+  0.5}$,${ 1.4}_{-  0.1}^{+  0.4}$,${ 5.0}_{-  0.5}^{+  0.4}$,${-0.05}_{-  0.13}^{+  0.08}$,${ 0.07}_{-  0.07}^{+  0.24}$,${-0.0}_{-  0.3}^{+  0.2}$,${ 0.15}_{-  0.15}^{+  0.45}$,${-0.0}_{-  0.5}^{+  0.4}$,${ 726}_{-  626}^{+  1131}$,${ 13.5}_{-  4.2}^{+  3.5}$,${ 163.8}_{-  64.0}^{+  33.0}$,${ 35}_{-  26}^{+  42}$
SEOBNSBH,${ 1.91}_{-  0.01}^{+  0.02}$,${ 2.7}_{-  1.1}^{+  0.5}$,${ 3.7}_{-  0.9}^{+  0.4}$,${ 1.4}_{-  0.1}^{+  0.4}$,${ 5.1}_{-  0.5}^{+  0.3}$,${-0.03}_{-  0.13}^{+  0.06}$,${ 0.04}_{-  0.04}^{+  0.22}$,${-0.0}_{-  0.2}^{+  0.1}$,${ 0.00}_{-  0.00}^{+  0.00}$,${ 0.0}_{-  0.0}^{+  0.0}$,${ 745}_{-  618}^{+  1086}$,${ 13.2}_{-  3.7}^{+  4.2}$,${ 161.6}_{-  59.0}^{+  32.3}$,${ 35}_{-  25}^{+  35}$
LEA+,${ 1.91}_{-  0.01}^{+  0.02}$,${ 3.0}_{-  0.4}^{+  0.5}$,${ 3.9}_{-  0.3}^{+  0.4}$,${ 1.3}_{-  0.1}^{+  0.1}$,${ 5.2}_{-  0.2}^{+  0.3}$,${ 0.00}_{-  0.04}^{+  0.05}$,${ 0.02}_{-  0.02}^{+  0.05}$,${ 0.0}_{-  0.1}^{+  0.1}$,-,-,${ 1105}_{-  949}^{+  1795}$,${ 13.2}_{-  3.8}^{+  2.9}$,${ 162.7}_{-  61.3}^{+  33.6}$,${ 35}_{-  26}^{+  38}$
IMRNSBH,${ 1.91}_{-  0.01}^{+  0.02}$,${ 2.9}_{-  0.9}^{+  0.5}$,${ 3.8}_{-  0.7}^{+  0.3}$,${ 1.3}_{-  0.1}^{+  0.3}$,${ 5.1}_{-  0.4}^{+  0.3}$,${-0.01}_{-  0.11}^{+  0.05}$,${ 0.03}_{-  0.03}^{+  0.15}$,${-0.0}_{-  0.2}^{+  0.1}$,${ 0.00}_{-  0.00}^{+  0.00}$,${ 0.0}_{-  0.0}^{+  0.0}$,${ 1005}_{-  778}^{+  1181}$,${ 13.5}_{-  3.4}^{+  3.3}$,${ 164.3}_{-  60.7}^{+  31.5}$,${ 34}_{-  24}^{+  37}$
IMRp,${ 1.90}_{-  0.01}^{+  0.03}$,${ 3.2}_{-  0.7}^{+  0.7}$,${ 4.1}_{-  0.5}^{+  0.5}$,${ 1.3}_{-  0.1}^{+  0.1}$,${ 5.3}_{-  0.4}^{+  0.4}$,${ 0.02}_{-  0.07}^{+  0.07}$,${ 0.11}_{-  0.10}^{+  0.26}$,${ 0.0}_{-  0.1}^{+  0.1}$,${ 0.36}_{-  0.33}^{+  0.50}$,${ 0.0}_{-  0.4}^{+  0.4}$,-,-,${ 180.1}_{-  71.2}^{+  19.7}$,${ 23}_{-  18}^{+  59}$
IMRpT,${ 1.90}_{-  0.01}^{+  0.03}$,${ 2.7}_{-  1.4}^{+  0.7}$,${ 3.7}_{-  1.2}^{+  0.5}$,${ 1.4}_{-  0.1}^{+  0.5}$,${ 5.0}_{-  0.6}^{+  0.4}$,${-0.04}_{-  0.14}^{+  0.09}$,${ 0.11}_{-  0.10}^{+  0.33}$,${-0.0}_{-  0.2}^{+  0.1}$,${ 0.30}_{-  0.27}^{+  0.53}$,${-0.0}_{-  0.4}^{+  0.3}$,${ 739}_{-  627}^{+  1289}$,${ 13.6}_{-  4.3}^{+  3.5}$,${ 178.7}_{-  67.9}^{+  20.9}$,${ 25}_{-  19}^{+  59}$
XHM,${ 1.90}_{-  0.01}^{+  0.02}$,${ 3.1}_{-  0.7}^{+  0.8}$,${ 4.0}_{-  0.5}^{+  0.5}$,${ 1.3}_{-  0.1}^{+  0.2}$,${ 5.3}_{-  0.4}^{+  0.4}$,${ 0.01}_{-  0.09}^{+  0.08}$,${ 0.05}_{-  0.05}^{+  0.12}$,${ 0.0}_{-  0.1}^{+  0.1}$,${ 0.11}_{-  0.10}^{+  0.35}$,${ 0.0}_{-  0.3}^{+  0.4}$,-,-,${ 174.8}_{-  48.1}^{+  25.0}$,${ 30}_{-  21}^{+  119}$
\end{filecontents*}
\csvreader[head to column names]{CSV/Param_BHNSq3s0_SNR30_inc30.csv}{}%
{\\ [0.5ex]\midrule \texttt{\Approximant}&\mc&\q&\mBH&\mNS&\mtotal&\aeff&\aBH&\aBHz&\aNS&\aNSz&\lambdaNS&\rNS&\dist&\thetajn}
\\
\bottomrule\bottomrule
\end{tabularx}
 \caption{Inclination 30\dg.}
 \end{subtable}

\begin{subtable}[!htbp]{1.1\textwidth}
		\centering
\begin{tabularx}{1.1\textwidth}{c|*{14}{b}}
\toprule\toprule
Approximant&$\mcs/\msun$&q&$\mBHs/\msun$&$\mNSs/\msun$&$\mts/\msun$&\chieff&\sBH&\sBHz&\sNS&\sNSz&\tidal&\rns/km&\dl/Mpc&\inc/\dg
\begin{filecontents*}{CSV/Param_BHNSq3s0_SNR30_inc70.csv}
Approximant,mc,q,mBH,mNS,mtotal,aeff,aBH,aBHz,aNS,aNSz,lambdaNS,rNS,dist,thetajn
True value,{ 1.94},{ 3.0},{ 3.97},{ 1.32},{ 5.30},{ 0.00},{ 0.0},{ 0.0},{ 0.0},{ 0.0},{ 624},{ 12.3},{ 85.5},{ 70}
SEOB,${ 1.91}_{-  0.01}^{+  0.03}$,${ 3.0}_{-  0.6}^{+  0.4}$,${ 4.0}_{-  0.5}^{+  0.3}$,${ 1.3}_{-  0.1}^{+  0.1}$,${ 5.2}_{-  0.3}^{+  0.2}$,${ 0.00}_{-  0.08}^{+  0.05}$,${ 0.04}_{-  0.04}^{+  0.13}$,${ 0.0}_{-  0.2}^{+  0.1}$,${ 0.13}_{-  0.12}^{+  0.43}$,${ 0.0}_{-  0.4}^{+  0.5}$,-,-,${ 159.2}_{-  63.1}^{+  34.0}$,${ 36}_{-  26}^{+  40}$
SEOBT,${ 1.91}_{-  0.01}^{+  0.03}$,${ 2.6}_{-  1.0}^{+  0.7}$,${ 3.6}_{-  0.8}^{+  0.5}$,${ 1.4}_{-  0.1}^{+  0.3}$,${ 5.0}_{-  0.5}^{+  0.4}$,${-0.05}_{-  0.13}^{+  0.08}$,${ 0.07}_{-  0.06}^{+  0.21}$,${-0.0}_{-  0.2}^{+  0.1}$,${ 0.15}_{-  0.15}^{+  0.44}$,${-0.0}_{-  0.5}^{+  0.4}$,${ 712}_{-  618}^{+  1173}$,${ 13.4}_{-  4.3}^{+  3.7}$,${ 159.7}_{-  63.3}^{+  33.7}$,${ 36}_{-  26}^{+  41}$
SEOBNSBH,${ 1.91}_{-  0.01}^{+  0.02}$,${ 2.7}_{-  1.1}^{+  0.6}$,${ 3.7}_{-  0.9}^{+  0.4}$,${ 1.4}_{-  0.1}^{+  0.4}$,${ 5.1}_{-  0.5}^{+  0.3}$,${-0.03}_{-  0.13}^{+  0.06}$,${ 0.05}_{-  0.04}^{+  0.21}$,${-0.0}_{-  0.2}^{+  0.1}$,${ 0.00}_{-  0.00}^{+  0.00}$,${ 0.0}_{-  0.0}^{+  0.0}$,${ 718}_{-  597}^{+  1117}$,${ 13.1}_{-  3.7}^{+  4.3}$,${ 158.8}_{-  59.6}^{+  31.5}$,${ 35}_{-  24}^{+  35}$
LEA+,${ 1.91}_{-  0.01}^{+  0.03}$,${ 3.0}_{-  0.4}^{+  0.5}$,${ 3.9}_{-  0.3}^{+  0.4}$,${ 1.3}_{-  0.1}^{+  0.1}$,${ 5.2}_{-  0.2}^{+  0.3}$,${ 0.00}_{-  0.04}^{+  0.06}$,${ 0.02}_{-  0.02}^{+  0.05}$,${ 0.0}_{-  0.1}^{+  0.1}$,-,-,${ 1113}_{-  949}^{+  1747}$,${ 13.3}_{-  3.8}^{+  2.8}$,${ 159.9}_{-  64.3}^{+  33.7}$,${ 35}_{-  26}^{+  40}$
IMRNSBH,${ 1.91}_{-  0.01}^{+  0.02}$,${ 2.9}_{-  0.9}^{+  0.5}$,${ 3.8}_{-  0.7}^{+  0.3}$,${ 1.3}_{-  0.1}^{+  0.2}$,${ 5.2}_{-  0.4}^{+  0.2}$,${-0.01}_{-  0.10}^{+  0.05}$,${ 0.03}_{-  0.03}^{+  0.14}$,${-0.0}_{-  0.2}^{+  0.1}$,${ 0.00}_{-  0.00}^{+  0.00}$,${ 0.0}_{-  0.0}^{+  0.0}$,${ 985}_{-  753}^{+  1123}$,${ 13.4}_{-  3.3}^{+  3.2}$,${ 161.5}_{-  61.5}^{+  31.6}$,${ 35}_{-  24}^{+  37}$
IMRp,${ 1.90}_{-  0.01}^{+  0.03}$,${ 3.2}_{-  0.5}^{+  0.6}$,${ 4.1}_{-  0.4}^{+  0.4}$,${ 1.3}_{-  0.1}^{+  0.1}$,${ 5.3}_{-  0.3}^{+  0.3}$,${ 0.03}_{-  0.07}^{+  0.06}$,${ 0.08}_{-  0.07}^{+  0.21}$,${ 0.0}_{-  0.1}^{+  0.1}$,${ 0.33}_{-  0.30}^{+  0.48}$,${ 0.0}_{-  0.2}^{+  0.4}$,-,-,${ 176.2}_{-  64.5}^{+  20.3}$,${ 25}_{-  19}^{+  44}$
IMRpT,${ 1.90}_{-  0.01}^{+  0.03}$,${ 2.7}_{-  1.1}^{+  0.7}$,${ 3.7}_{-  0.9}^{+  0.5}$,${ 1.4}_{-  0.1}^{+  0.4}$,${ 5.0}_{-  0.5}^{+  0.4}$,${-0.04}_{-  0.12}^{+  0.08}$,${ 0.11}_{-  0.10}^{+  0.29}$,${-0.0}_{-  0.2}^{+  0.1}$,${ 0.30}_{-  0.28}^{+  0.51}$,${-0.0}_{-  0.4}^{+  0.3}$,${ 817}_{-  693}^{+  1228}$,${ 13.6}_{-  4.3}^{+  3.2}$,${ 176.7}_{-  64.4}^{+  20.3}$,${ 25}_{-  19}^{+  48}$
XHM,${ 1.92}_{-  0.02}^{+  0.02}$,${ 3.2}_{-  0.6}^{+  0.7}$,${ 4.0}_{-  0.5}^{+  0.5}$,${ 1.3}_{-  0.1}^{+  0.1}$,${ 5.3}_{-  0.3}^{+  0.4}$,${ 0.02}_{-  0.08}^{+  0.07}$,${ 0.05}_{-  0.05}^{+  0.12}$,${ 0.0}_{-  0.2}^{+  0.1}$,${ 0.15}_{-  0.15}^{+  0.41}$,${ 0.0}_{-  0.4}^{+  0.5}$,-,-,${ 131.9}_{-  44.8}^{+  45.6}$,${ 53}_{-  27}^{+  76}$
\end{filecontents*}
\csvreader[head to column names]{CSV/Param_BHNSq3s0_SNR30_inc70.csv}{}%
{\\ [0.5ex]\midrule \texttt{\Approximant}&\mc&\q&\mBH&\mNS&\mtotal&\aeff&\aBH&\aBHz&\aNS&\aNSz&\lambdaNS&\rNS&\dist&\thetajn}
\\
\bottomrule\bottomrule
\end{tabularx}
 \caption{Inclination 70\dg.}
 \end{subtable}
 \caption{Properties of the q=3, SNR 30 simulation, as recovered by the listed approximants. We report the 1D median, together with the symmetric the $90\%$ credible interval.}
\label{Table.Param_BHNSq3s0_SNR30}
\end{table} 

\clearpage
\vfill

\begin{table}[!htbp]
\scriptsize
\centering
\begin{subtable}[!htbp]{1.1\textwidth}
		\centering
\begin{tabularx}{1.1\textwidth}{c|*{14}{b}}
\toprule\toprule
Approximant&$\mcs/\msun$&q&$\mBHs/\msun$&$\mNSs/\msun$&$\mts/\msun$&\chieff&\sBH&\sBHz&\sNS&\sNSz&\tidal&\rns/km&\dl/Mpc&\inc/\dg
\begin{filecontents*}{CSV/Param_BHNSq3s0_SNR70_inc30.csv}
Approximant,mc,q,mBH,mNS,mtotal,aeff,aBH,aBHz,aNS,aNSz,lambdaNS,rNS,dist,thetajn
True value,{ 1.95},{ 3.0},{ 3.99},{ 1.33},{ 5.31},{ 0.00},{ 0.0},{ 0.0},{ 0.0},{ 0.0},{ 624},{ 12.3},{ 73.0},{ 30}
SEOB,${ 1.95}_{-  0.01}^{+  0.01}$,${ 3.1}_{-  0.3}^{+  0.3}$,${ 4.1}_{-  0.2}^{+  0.2}$,${ 1.3}_{-  0.1}^{+  0.1}$,${ 5.4}_{-  0.1}^{+  0.2}$,${ 0.02}_{-  0.03}^{+  0.03}$,${ 0.04}_{-  0.03}^{+  0.12}$,${ 0.0}_{-  0.1}^{+  0.1}$,${ 0.11}_{-  0.10}^{+  0.40}$,${ 0.0}_{-  0.4}^{+  0.4}$,-,-,${ 71.1}_{-  21.9}^{+  11.8}$,${ 33}_{-  24}^{+  26}$
SEOBT,${ 1.95}_{-  0.00}^{+  0.01}$,${ 2.8}_{-  0.6}^{+  0.4}$,${ 3.8}_{-  0.4}^{+  0.3}$,${ 1.4}_{-  0.1}^{+  0.1}$,${ 5.2}_{-  0.3}^{+  0.2}$,${-0.02}_{-  0.07}^{+  0.04}$,${ 0.05}_{-  0.04}^{+  0.15}$,${-0.0}_{-  0.2}^{+  0.1}$,${ 0.11}_{-  0.11}^{+  0.41}$,${ 0.0}_{-  0.4}^{+  0.4}$,${ 380}_{-  342}^{+  606}$,${ 11.7}_{-  3.5}^{+  3.1}$,${ 71.7}_{-  21.6}^{+  11.3}$,${ 32}_{-  23}^{+  27}$
SEOBNSBH,${ 1.95}_{-  0.00}^{+  0.01}$,${ 2.8}_{-  0.5}^{+  0.3}$,${ 3.9}_{-  0.4}^{+  0.2}$,${ 1.4}_{-  0.1}^{+  0.1}$,${ 5.2}_{-  0.2}^{+  0.2}$,${-0.02}_{-  0.05}^{+  0.04}$,${ 0.03}_{-  0.02}^{+  0.08}$,${-0.0}_{-  0.1}^{+  0.0}$,${ 0.00}_{-  0.00}^{+  0.00}$,${ 0.0}_{-  0.0}^{+  0.0}$,${ 481}_{-  359}^{+  358}$,${ 12.0}_{-  2.6}^{+  2.0}$,${ 71.7}_{-  21.2}^{+  10.7}$,${ 31}_{-  22}^{+  26}$
LEA+,${ 1.95}_{-  0.00}^{+  0.01}$,${ 3.0}_{-  0.3}^{+  0.3}$,${ 4.0}_{-  0.3}^{+  0.2}$,${ 1.3}_{-  0.1}^{+  0.1}$,${ 5.3}_{-  0.2}^{+  0.2}$,${ 0.00}_{-  0.04}^{+  0.04}$,${ 0.02}_{-  0.02}^{+  0.04}$,${ 0.0}_{-  0.1}^{+  0.0}$,-,-,${ 888}_{-  556}^{+  512}$,${ 13.1}_{-  2.4}^{+  1.6}$,${ 71.5}_{-  22.7}^{+  11.5}$,${ 32}_{-  23}^{+  28}$
IMRNSBH,${ 1.95}_{-  0.00}^{+  0.01}$,${ 2.9}_{-  0.5}^{+  0.3}$,${ 3.9}_{-  0.4}^{+  0.2}$,${ 1.3}_{-  0.1}^{+  0.1}$,${ 5.3}_{-  0.3}^{+  0.2}$,${ 0.00}_{-  0.05}^{+  0.03}$,${ 0.02}_{-  0.02}^{+  0.07}$,${-0.0}_{-  0.1}^{+  0.0}$,${ 0.00}_{-  0.00}^{+  0.00}$,${ 0.0}_{-  0.0}^{+  0.0}$,${ 831}_{-  419}^{+  377}$,${ 13.0}_{-  1.9}^{+  1.7}$,${ 72.5}_{-  21.3}^{+  10.6}$,${ 31}_{-  22}^{+  26}$
IMRp,${ 1.94}_{-  0.00}^{+  0.01}$,${ 3.3}_{-  0.2}^{+  0.3}$,${ 4.2}_{-  0.2}^{+  0.2}$,${ 1.3}_{-  0.1}^{+  0.0}$,${ 5.5}_{-  0.1}^{+  0.2}$,${ 0.04}_{-  0.03}^{+  0.03}$,${ 0.07}_{-  0.06}^{+  0.14}$,${ 0.0}_{-  0.1}^{+  0.1}$,${ 0.28}_{-  0.22}^{+  0.43}$,${ 0.1}_{-  0.2}^{+  0.2}$,-,-,${ 79.9}_{-  18.8}^{+  4.2}$,${ 17}_{-  13}^{+  28}$
IMRpT,${ 1.94}_{-  0.00}^{+  0.01}$,${ 2.5}_{-  1.3}^{+  0.7}$,${ 3.6}_{-  1.1}^{+  0.5}$,${ 1.4}_{-  0.1}^{+  0.6}$,${ 5.1}_{-  0.6}^{+  0.4}$,${-0.04}_{-  0.11}^{+  0.07}$,${ 0.17}_{-  0.15}^{+  0.34}$,${-0.1}_{-  0.3}^{+  0.1}$,${ 0.34}_{-  0.31}^{+  0.52}$,${ 0.0}_{-  0.4}^{+  0.3}$,${ 371}_{-  319}^{+  685}$,${ 12.8}_{-  3.9}^{+  3.2}$,${ 81.2}_{-  16.8}^{+  3.4}$,${ 14}_{-  11}^{+  28}$
XHM,${ 1.95}_{-  0.00}^{+  0.00}$,${ 3.3}_{-  0.3}^{+  0.3}$,${ 4.2}_{-  0.2}^{+  0.2}$,${ 1.3}_{-  0.1}^{+  0.1}$,${ 5.5}_{-  0.2}^{+  0.2}$,${ 0.04}_{-  0.03}^{+  0.03}$,${ 0.04}_{-  0.04}^{+  0.10}$,${ 0.0}_{-  0.1}^{+  0.1}$,${ 0.13}_{-  0.13}^{+  0.39}$,${ 0.1}_{-  0.3}^{+  0.4}$,-,-,${ 75.3}_{-  8.3}^{+  6.0}$,${ 27}_{-  10}^{+  12}$
\end{filecontents*}
\csvreader[head to column names]{CSV/Param_BHNSq3s0_SNR70_inc30.csv}{}%
{\\ [0.5ex]\midrule \texttt{\Approximant}&\mc&\q&\mBH&\mNS&\mtotal&\aeff&\aBH&\aBHz&\aNS&\aNSz&\lambdaNS&\rNS&\dist&\thetajn}
\\
\bottomrule\bottomrule
\end{tabularx}
 \caption{Inclination 30\dg.}
 \end{subtable}

\begin{subtable}[!htbp]{1.1\textwidth}
		\centering
\begin{tabularx}{1.1\textwidth}{c|*{14}{b}}
\toprule\toprule
Approximant&$\mcs/\msun$&q&$\mBHs/\msun$&$\mNSs/\msun$&$\mts/\msun$&\chieff&\sBH&\sBHz&\sNS&\sNSz&\tidal&\rns/km&\dl/Mpc&\inc/\dg
\begin{filecontents*}{CSV/Param_BHNSq3s0_SNR70_inc70.csv}
Approximant,mc,q,mBH,mNS,mtotal,aeff,aBH,aBHz,aNS,aNSz,lambdaNS,rNS,dist,thetajn
True value,{ 1.96},{ 3.0},{ 4.02},{ 1.34},{ 5.36},{ 0.00},{ 0.0},{ 0.0},{ 0.0},{ 0.0},{ 624},{ 12.3},{ 36.6},{ 70}
SEOB,${ 1.95}_{-  0.01}^{+  0.01}$,${ 3.1}_{-  0.3}^{+  0.3}$,${ 4.1}_{-  0.2}^{+  0.2}$,${ 1.3}_{-  0.1}^{+  0.1}$,${ 5.4}_{-  0.1}^{+  0.2}$,${ 0.02}_{-  0.03}^{+  0.03}$,${ 0.04}_{-  0.03}^{+  0.11}$,${ 0.0}_{-  0.1}^{+  0.1}$,${ 0.10}_{-  0.10}^{+  0.39}$,${ 0.0}_{-  0.4}^{+  0.4}$,-,-,${ 65.5}_{-  27.1}^{+  15.8}$,${ 39}_{-  28}^{+  32}$
SEOBT,${ 1.95}_{-  0.01}^{+  0.01}$,${ 2.8}_{-  0.6}^{+  0.4}$,${ 3.8}_{-  0.5}^{+  0.3}$,${ 1.4}_{-  0.1}^{+  0.2}$,${ 5.2}_{-  0.3}^{+  0.2}$,${-0.02}_{-  0.07}^{+  0.04}$,${ 0.05}_{-  0.05}^{+  0.18}$,${-0.0}_{-  0.2}^{+  0.1}$,${ 0.13}_{-  0.12}^{+  0.45}$,${ 0.0}_{-  0.4}^{+  0.5}$,${ 357}_{-  318}^{+  577}$,${ 11.6}_{-  3.4}^{+  3.2}$,${ 65.8}_{-  27.8}^{+  15.6}$,${ 39}_{-  28}^{+  33}$
SEOBNSBH,${ 1.95}_{-  0.01}^{+  0.01}$,${ 2.9}_{-  0.5}^{+  0.3}$,${ 3.9}_{-  0.4}^{+  0.2}$,${ 1.4}_{-  0.1}^{+  0.1}$,${ 5.3}_{-  0.2}^{+  0.2}$,${-0.01}_{-  0.05}^{+  0.03}$,${ 0.02}_{-  0.02}^{+  0.07}$,${-0.0}_{-  0.1}^{+  0.0}$,${ 0.00}_{-  0.00}^{+  0.00}$,${ 0.0}_{-  0.0}^{+  0.0}$,${ 474}_{-  350}^{+  362}$,${ 12.0}_{-  2.5}^{+  2.0}$,${ 66.7}_{-  27.5}^{+  14.0}$,${ 37}_{-  26}^{+  33}$
LEA+,${ 1.95}_{-  0.01}^{+  0.01}$,${ 3.0}_{-  0.3}^{+  0.4}$,${ 4.0}_{-  0.2}^{+  0.3}$,${ 1.3}_{-  0.1}^{+  0.1}$,${ 5.3}_{-  0.2}^{+  0.2}$,${ 0.00}_{-  0.04}^{+  0.04}$,${ 0.02}_{-  0.02}^{+  0.04}$,${ 0.0}_{-  0.1}^{+  0.0}$,-,-,${ 891}_{-  570}^{+  510}$,${ 13.1}_{-  2.4}^{+  1.6}$,${ 64.7}_{-  26.9}^{+  16.6}$,${ 40}_{-  29}^{+  32}$
IMRNSBH,${ 1.95}_{-  0.01}^{+  0.01}$,${ 2.9}_{-  0.5}^{+  0.3}$,${ 3.9}_{-  0.4}^{+  0.2}$,${ 1.3}_{-  0.1}^{+  0.1}$,${ 5.3}_{-  0.3}^{+  0.2}$,${ 0.00}_{-  0.05}^{+  0.03}$,${ 0.02}_{-  0.02}^{+  0.06}$,${ 0.0}_{-  0.1}^{+  0.0}$,${ 0.00}_{-  0.00}^{+  0.00}$,${ 0.0}_{-  0.0}^{+  0.0}$,${ 852}_{-  407}^{+  357}$,${ 13.1}_{-  1.8}^{+  1.7}$,${ 67.7}_{-  28.3}^{+  13.7}$,${ 36}_{-  26}^{+  33}$
IMRp,${ 1.94}_{-  0.00}^{+  0.01}$,${ 3.4}_{-  0.2}^{+  0.3}$,${ 4.3}_{-  0.2}^{+  0.2}$,${ 1.3}_{-  0.0}^{+  0.0}$,${ 5.5}_{-  0.1}^{+  0.1}$,${ 0.04}_{-  0.03}^{+  0.02}$,${ 0.08}_{-  0.06}^{+  0.16}$,${ 0.0}_{-  0.1}^{+  0.1}$,${ 0.27}_{-  0.24}^{+  0.51}$,${ 0.1}_{-  0.3}^{+  0.3}$,-,-,${ 78.7}_{-  21.5}^{+  4.1}$,${ 17}_{-  13}^{+  32}$
IMRpT,${ 1.94}_{-  0.00}^{+  0.01}$,${ 2.8}_{-  1.4}^{+  0.4}$,${ 3.8}_{-  1.2}^{+  0.3}$,${ 1.4}_{-  0.1}^{+  0.5}$,${ 5.2}_{-  0.7}^{+  0.2}$,${-0.02}_{-  0.13}^{+  0.05}$,${ 0.07}_{-  0.07}^{+  0.30}$,${-0.0}_{-  0.3}^{+  0.1}$,${ 0.23}_{-  0.21}^{+  0.55}$,${-0.0}_{-  0.4}^{+  0.3}$,${ 530}_{-  422}^{+  639}$,${ 12.9}_{-  3.4}^{+  3.0}$,${ 78.7}_{-  33.3}^{+  4.5}$,${ 18}_{-  14}^{+  47}$
XHM,${ 1.96}_{-  0.01}^{+  0.00}$,${ 3.3}_{-  0.3}^{+  0.3}$,${ 4.3}_{-  0.2}^{+  0.2}$,${ 1.3}_{-  0.0}^{+  0.1}$,${ 5.5}_{-  0.2}^{+  0.2}$,${ 0.04}_{-  0.03}^{+  0.03}$,${ 0.04}_{-  0.04}^{+  0.09}$,${ 0.0}_{-  0.1}^{+  0.1}$,${ 0.15}_{-  0.14}^{+  0.38}$,${ 0.1}_{-  0.3}^{+  0.4}$,-,-,${ 47.8}_{-  10.8}^{+  12.3}$,${ 61}_{-  15}^{+  13}$
\end{filecontents*}
\csvreader[head to column names]{CSV/Param_BHNSq3s0_SNR70_inc70.csv}{}%
{\\ [0.5ex]\midrule \texttt{\Approximant}&\mc&\q&\mBH&\mNS&\mtotal&\aeff&\aBH&\aBHz&\aNS&\aNSz&\lambdaNS&\rNS&\dist&\thetajn}
\\
\bottomrule\bottomrule
\end{tabularx}
 \caption{Inclination 70\dg.}
 \end{subtable}
 
\caption{Properties of the q=3, SNR 70 simulation, as recovered by the listed approximants. We report the 1D median, together with the symmetric the $90\%$ credible interval.}
\label{Table.Param_BHNSq3s0_SNR70}
\end{table}

\clearpage
\vfill

\begin{table}[!htbp]
\scriptsize
\centering

\begin{subtable}[!htbp]{1.1\textwidth}
		\centering
\begin{tabularx}{1.1\textwidth}{c|*{14}{b}}
\toprule\toprule
Approximant&$\mcs/\msun$&q&$\mBHs/\msun$&$\mNSs/\msun$&$\mts/\msun$&\chieff&\sBH&\sBHz&\sNS&\sNSz&\tidal&\rns/km&\dl/Mpc&\inc/\dg
\begin{filecontents*}{CSV/Param_BHNSq6s0_SNR30_inc30.csv}
Approximant,mc,q,mBH,mNS,mtotal,aeff,aBH,aBHz,aNS,aNSz,lambdaNS,rNS,dist,thetajn
True value,{ 2.65},{ 6.0},{ 8.00},{ 1.33},{ 9.33},{ 0.00},{ 0.0},{ 0.0},{ 0.0},{ 0.0},{ 526},{ 13.3},{ 229.5},{ 30}
SEOB,${ 2.652}_{-  0.023}^{+  0.043}$,${ 6.1}_{-  0.8}^{+  0.9}$,${ 8.1}_{-  0.6}^{+  0.7}$,${ 1.3}_{-  0.1}^{+  0.1}$,${ 9.4}_{-  0.5}^{+  0.6}$,${ 0.01}_{-  0.07}^{+  0.06}$,${ 0.03}_{-  0.03}^{+  0.08}$,${ 0.0}_{-  0.1}^{+  0.1}$,${ 0.15}_{-  0.14}^{+  0.45}$,${ 0.0}_{-  0.5}^{+  0.5}$,-,-,${ 221.4}_{-  77.4}^{+  44.0}$,${ 34}_{-  25}^{+  30}$
SEOBT,${ 2.650}_{-  0.024}^{+  0.043}$,${ 5.4}_{-  1.4}^{+  1.1}$,${ 7.5}_{-  1.2}^{+  0.8}$,${ 1.4}_{-  0.1}^{+  0.2}$,${ 8.9}_{-  1.0}^{+  0.7}$,${-0.05}_{-  0.14}^{+  0.09}$,${ 0.06}_{-  0.06}^{+  0.20}$,${-0.1}_{-  0.2}^{+  0.1}$,${ 0.18}_{-  0.17}^{+  0.49}$,${ 0.0}_{-  0.6}^{+  0.5}$,${ 1990}_{-  1790}^{+  1788}$,${ 15.7}_{-  5.2}^{+  4.0}$,${ 221.5}_{-  78.6}^{+  44.5}$,${ 34}_{-  25}^{+  31}$
SEOBNSBH,${ 2.651}_{-  0.022}^{+  0.041}$,${ 5.8}_{-  1.2}^{+  0.7}$,${ 7.9}_{-  0.9}^{+  0.6}$,${ 1.4}_{-  0.1}^{+  0.1}$,${ 9.2}_{-  0.8}^{+  0.5}$,${-0.01}_{-  0.10}^{+  0.05}$,${ 0.03}_{-  0.03}^{+  0.11}$,${-0.0}_{-  0.1}^{+  0.1}$,${ 0.00}_{-  0.00}^{+  0.00}$,${ 0.0}_{-  0.0}^{+  0.0}$,${ 1938}_{-  1691}^{+  2524}$,${ 15.3}_{-  4.8}^{+  3.8}$,${ 221.9}_{-  75.4}^{+  40.9}$,${ 33}_{-  23}^{+  29}$
IMRNSBH,${ 2.648}_{-  0.022}^{+  0.042}$,${ 6.0}_{-  1.0}^{+  0.7}$,${ 8.0}_{-  0.8}^{+  0.5}$,${ 1.3}_{-  0.1}^{+  0.1}$,${ 9.3}_{-  0.7}^{+  0.5}$,${ 0.00}_{-  0.08}^{+  0.05}$,${ 0.02}_{-  0.02}^{+  0.08}$,${ 0.0}_{-  0.1}^{+  0.1}$,${ 0.00}_{-  0.00}^{+  0.00}$,${ 0.0}_{-  0.0}^{+  0.0}$,${ 2591}_{-  2212}^{+  2120}$,${ 15.9}_{-  4.8}^{+  3.0}$,${ 227.3}_{-  76.9}^{+  40.8}$,${ 33}_{-  23}^{+  29}$
IMRp,${ 2.640}_{-  0.014}^{+  0.040}$,${ 6.2}_{-  0.5}^{+  0.6}$,${ 8.1}_{-  0.4}^{+  0.5}$,${ 1.3}_{-  0.1}^{+  0.0}$,${ 9.5}_{-  0.3}^{+  0.4}$,${ 0.01}_{-  0.04}^{+  0.05}$,${ 0.05}_{-  0.04}^{+  0.12}$,${ 0.0}_{-  0.0}^{+  0.1}$,${ 0.32}_{-  0.28}^{+  0.54}$,${ 0.0}_{-  0.3}^{+  0.4}$,-,-,${ 243.9}_{-  75.2}^{+  25.8}$,${ 22}_{-  17}^{+  33}$
IMRpT,${ 2.636}_{-  0.014}^{+  0.040}$,${ 5.6}_{-  3.2}^{+  0.8}$,${ 7.6}_{-  2.9}^{+  0.6}$,${ 1.4}_{-  0.1}^{+  0.6}$,${ 9.0}_{-  2.3}^{+  0.6}$,${-0.04}_{-  0.26}^{+  0.07}$,${ 0.09}_{-  0.08}^{+  0.43}$,${-0.0}_{-  0.4}^{+  0.1}$,${ 0.37}_{-  0.33}^{+  0.53}$,${ 0.0}_{-  0.6}^{+  0.5}$,${ 1710}_{-  1527}^{+  1986}$,${ 15.7}_{-  5.3}^{+  5.3}$,${ 248.3}_{-  73.6}^{+  23.9}$,${ 20}_{-  15}^{+  33}$
XHM,${ 2.644}_{-  0.014}^{+  0.020}$,${ 6.3}_{-  0.7}^{+  0.8}$,${ 8.2}_{-  0.5}^{+  0.6}$,${ 1.3}_{-  0.1}^{+  0.1}$,${ 9.5}_{-  0.5}^{+  0.5}$,${ 0.02}_{-  0.05}^{+  0.05}$,${ 0.03}_{-  0.03}^{+  0.07}$,${ 0.0}_{-  0.1}^{+  0.1}$,${ 0.13}_{-  0.13}^{+  0.34}$,${ 0.0}_{-  0.4}^{+  0.4}$,-,-,${ 237.1}_{-  37.0}^{+  27.3}$,${ 28}_{-  13}^{+  14}$
\end{filecontents*}
\csvreader[head to column names]{CSV/Param_BHNSq6s0_SNR30_inc30.csv}{}%
{\\ [0.5ex]\midrule \texttt{\Approximant}&\mc&\q&\mBH&\mNS&\mtotal&\aeff&\aBH&\aBHz&\aNS&\aNSz&\lambdaNS&\rNS&\dist&\thetajn}
\\
\bottomrule\bottomrule
\end{tabularx}
 \caption{Inclination 30\dg.}
 \end{subtable}

\begin{subtable}[!htbp]{1.1\textwidth}
		\centering
\begin{tabularx}{1.1\textwidth}{c|*{14}{b}}
\toprule\toprule
Approximant&$\mcs/\msun$&q&$\mBHs/\msun$&$\mNSs/\msun$&$\mts/\msun$&\chieff&\sBH&\sBHz&\sNS&\sNSz&\tidal&\rns/km&\dl/Mpc&\inc/\dg
\begin{filecontents*}{CSV/Param_BHNSq6s0_SNR30_inc70.csv}
Approximant,mc,q,mBH,mNS,mtotal,aeff,aBH,aBHz,aNS,aNSz,lambdaNS,rNS,dist,thetajn
True value,{ 2.72},{ 6.0},{ 8.23},{ 1.37},{ 9.60},{ 0.00},{ 0.0},{ 0.0},{ 0.0},{ 0.0},{ 526},{ 13.3},{ 95.4},{ 70}
SEOB,${ 2.651}_{-  0.023}^{+  0.045}$,${ 6.1}_{-  0.8}^{+  0.9}$,${ 8.1}_{-  0.6}^{+  0.7}$,${ 1.3}_{-  0.1}^{+  0.1}$,${ 9.4}_{-  0.5}^{+  0.6}$,${ 0.01}_{-  0.07}^{+  0.06}$,${ 0.03}_{-  0.03}^{+  0.08}$,${ 0.0}_{-  0.1}^{+  0.1}$,${ 0.15}_{-  0.15}^{+  0.47}$,${ 0.0}_{-  0.5}^{+  0.5}$,-,-,${ 222.5}_{-  80.5}^{+  44.3}$,${ 34}_{-  25}^{+  29}$
SEOBT,${ 2.648}_{-  0.023}^{+  0.043}$,${ 5.4}_{-  1.4}^{+  1.0}$,${ 7.6}_{-  1.2}^{+  0.8}$,${ 1.4}_{-  0.1}^{+  0.2}$,${ 9.0}_{-  1.0}^{+  0.7}$,${-0.05}_{-  0.14}^{+  0.09}$,${ 0.06}_{-  0.05}^{+  0.19}$,${-0.0}_{-  0.2}^{+  0.1}$,${ 0.16}_{-  0.16}^{+  0.48}$,${ 0.0}_{-  0.5}^{+  0.5}$,${ 1936}_{-  1738}^{+  1857}$,${ 15.6}_{-  5.2}^{+  3.9}$,${ 224.3}_{-  79.0}^{+  43.3}$,${ 34}_{-  24}^{+  29}$
SEOBNSBH,${ 2.650}_{-  0.021}^{+  0.040}$,${ 5.8}_{-  1.1}^{+  0.7}$,${ 7.9}_{-  0.9}^{+  0.6}$,${ 1.4}_{-  0.1}^{+  0.1}$,${ 9.2}_{-  0.8}^{+  0.5}$,${-0.01}_{-  0.10}^{+  0.05}$,${ 0.03}_{-  0.03}^{+  0.11}$,${-0.0}_{-  0.1}^{+  0.1}$,${ 0.00}_{-  0.00}^{+  0.00}$,${ 0.0}_{-  0.0}^{+  0.0}$,${ 1864}_{-  1619}^{+  2526}$,${ 15.1}_{-  4.6}^{+  3.8}$,${ 223.0}_{-  73.8}^{+  40.8}$,${ 33}_{-  23}^{+  27}$
IMRNSBH,${ 2.647}_{-  0.021}^{+  0.042}$,${ 6.0}_{-  0.9}^{+  0.7}$,${ 8.0}_{-  0.7}^{+  0.5}$,${ 1.3}_{-  0.1}^{+  0.1}$,${ 9.3}_{-  0.6}^{+  0.5}$,${ 0.00}_{-  0.08}^{+  0.05}$,${ 0.02}_{-  0.02}^{+  0.07}$,${ 0.0}_{-  0.1}^{+  0.1}$,${ 0.00}_{-  0.00}^{+  0.00}$,${ 0.0}_{-  0.0}^{+  0.0}$,${ 2449}_{-  2108}^{+  2227}$,${ 15.7}_{-  4.8}^{+  3.1}$,${ 229.5}_{-  77.5}^{+  40.2}$,${ 32}_{-  23}^{+  28}$
IMRp,${ 2.639}_{-  0.014}^{+  0.039}$,${ 6.2}_{-  0.5}^{+  0.7}$,${ 8.2}_{-  0.4}^{+  0.5}$,${ 1.3}_{-  0.1}^{+  0.0}$,${ 9.5}_{-  0.3}^{+  0.5}$,${ 0.01}_{-  0.04}^{+  0.05}$,${ 0.05}_{-  0.04}^{+  0.12}$,${ 0.0}_{-  0.0}^{+  0.1}$,${ 0.30}_{-  0.28}^{+  0.56}$,${ 0.0}_{-  0.3}^{+  0.4}$,-,-,${ 245.0}_{-  72.9}^{+  25.9}$,${ 23}_{-  17}^{+  31}$
IMRpT,${ 2.634}_{-  0.013}^{+  0.037}$,${ 5.4}_{-  3.1}^{+  1.0}$,${ 7.5}_{-  2.9}^{+  0.8}$,${ 1.4}_{-  0.1}^{+  0.6}$,${ 8.9}_{-  2.2}^{+  0.7}$,${-0.05}_{-  0.28}^{+  0.09}$,${ 0.11}_{-  0.10}^{+  0.45}$,${-0.0}_{-  0.4}^{+  0.1}$,${ 0.41}_{-  0.36}^{+  0.50}$,${ 0.0}_{-  0.5}^{+  0.5}$,${ 1549}_{-  1389}^{+  2108}$,${ 15.8}_{-  5.3}^{+  5.9}$,${ 250.8}_{-  67.5}^{+  22.8}$,${ 19}_{-  14}^{+  31}$
XHM,${ 2.685}_{-  0.027}^{+  0.031}$,${ 6.4}_{-  0.6}^{+  0.8}$,${ 8.4}_{-  0.5}^{+  0.6}$,${ 1.3}_{-  0.1}^{+  0.1}$,${ 9.8}_{-  0.4}^{+  0.5}$,${ 0.03}_{-  0.05}^{+  0.05}$,${ 0.04}_{-  0.04}^{+  0.08}$,${ 0.0}_{-  0.1}^{+  0.1}$,${ 0.14}_{-  0.13}^{+  0.37}$,${ 0.0}_{-  0.4}^{+  0.4}$,-,-,${ 162.3}_{-  54.6}^{+  48.7}$,${ 54}_{-  15}^{+  14}$
\end{filecontents*}
\csvreader[head to column names]{CSV/Param_BHNSq6s0_SNR30_inc70.csv}{}%
{\\ [0.5ex]\midrule \texttt{\Approximant}&\mc&\q&\mBH&\mNS&\mtotal&\aeff&\aBH&\aBHz&\aNS&\aNSz&\lambdaNS&\rNS&\dist&\thetajn}
\\
\bottomrule\bottomrule
\end{tabularx}
 \caption{Inclination 70\dg.}
 \end{subtable}
 
 \caption{Properties of the q=6, SNR 30 simulation, as recovered by the listed approximants. We report the 1D median, together with the symmetric the $90\%$ credible interval.}
\label{Table.Param_BHNSq6s0_SNR30}
\end{table} 

\clearpage
\vfill

\begin{table}[!htbp]
\scriptsize
\centering
\begin{subtable}[!htbp]{1.1\textwidth}
		\centering
\begin{tabularx}{1.1\textwidth}{c|*{14}{b}}
\toprule\toprule
Approximant&$\mcs/\msun$&q&$\mBHs/\msun$&$\mNSs/\msun$&$\mts/\msun$&\chieff&\sBH&\sBHz&\sNS&\sNSz&\tidal&\rns/km&\dl/Mpc&\inc/\dg
\begin{filecontents*}{CSV/Param_BHNSq6s0_SNR70_inc30.csv}
Approximant,mc,q,mBH,mNS,mtotal,aeff,aBH,aBHz,aNS,aNSz,lambdaNS,rNS,dist,thetajn
True value,{ 2.72},{ 6.0},{ 8.22},{ 1.37},{ 9.59},{ 0.00},{ 0.0},{ 0.0},{ 0.0},{ 0.0},{ 526},{ 13.3},{ 98.4},{ 30}
SEOB,${ 2.721}_{-  0.009}^{+  0.016}$,${ 6.1}_{-  0.4}^{+  0.4}$,${ 8.3}_{-  0.3}^{+  0.3}$,${ 1.4}_{-  0.0}^{+  0.0}$,${ 9.7}_{-  0.3}^{+  0.3}$,${ 0.01}_{-  0.03}^{+  0.03}$,${ 0.02}_{-  0.02}^{+  0.06}$,${ 0.0}_{-  0.1}^{+  0.1}$,${ 0.13}_{-  0.13}^{+  0.42}$,${ 0.0}_{-  0.4}^{+  0.5}$,-,-,${ 97.2}_{-  26.5}^{+  14.9}$,${ 31}_{-  23}^{+  24}$
SEOBT,${ 2.719}_{-  0.009}^{+  0.016}$,${ 5.4}_{-  0.6}^{+  0.6}$,${ 7.7}_{-  0.5}^{+  0.5}$,${ 1.4}_{-  0.1}^{+  0.1}$,${ 9.1}_{-  0.4}^{+  0.4}$,${-0.04}_{-  0.05}^{+  0.05}$,${ 0.08}_{-  0.07}^{+  0.13}$,${-0.1}_{-  0.1}^{+  0.1}$,${ 0.21}_{-  0.20}^{+  0.55}$,${ 0.1}_{-  0.4}^{+  0.6}$,${ 921}_{-  852}^{+  2273}$,${ 14.1}_{-  4.6}^{+  4.2}$,${ 97.8}_{-  26.6}^{+  14.6}$,${ 31}_{-  22}^{+  24}$
SEOBNSBH,${ 2.720}_{-  0.008}^{+  0.015}$,${ 6.0}_{-  0.4}^{+  0.4}$,${ 8.2}_{-  0.4}^{+  0.3}$,${ 1.4}_{-  0.0}^{+  0.0}$,${ 9.6}_{-  0.3}^{+  0.3}$,${ 0.00}_{-  0.04}^{+  0.03}$,${ 0.01}_{-  0.01}^{+  0.03}$,${ 0.0}_{-  0.0}^{+  0.0}$,${ 0.00}_{-  0.00}^{+  0.00}$,${ 0.0}_{-  0.0}^{+  0.0}$,${ 979}_{-  831}^{+  1225}$,${ 13.6}_{-  3.6}^{+  2.6}$,${ 99.0}_{-  25.7}^{+  13.6}$,${ 30}_{-  21}^{+  24}$
IMRNSBH,${ 2.720}_{-  0.008}^{+  0.015}$,${ 6.1}_{-  0.5}^{+  0.4}$,${ 8.3}_{-  0.4}^{+  0.3}$,${ 1.4}_{-  0.0}^{+  0.0}$,${ 9.7}_{-  0.3}^{+  0.3}$,${ 0.00}_{-  0.03}^{+  0.03}$,${ 0.02}_{-  0.01}^{+  0.03}$,${ 0.0}_{-  0.0}^{+  0.0}$,${ 0.00}_{-  0.00}^{+  0.00}$,${ 0.0}_{-  0.0}^{+  0.0}$,${ 1562}_{-  1281}^{+  2058}$,${ 14.6}_{-  3.9}^{+  3.1}$,${ 99.5}_{-  25.8}^{+  13.9}$,${ 30}_{-  21}^{+  23}$
IMRp,${ 2.715}_{-  0.003}^{+  0.013}$,${ 6.3}_{-  0.2}^{+  0.4}$,${ 8.4}_{-  0.2}^{+  0.3}$,${ 1.3}_{-  0.0}^{+  0.0}$,${ 9.8}_{-  0.2}^{+  0.3}$,${ 0.02}_{-  0.02}^{+  0.03}$,${ 0.04}_{-  0.03}^{+  0.09}$,${ 0.0}_{-  0.0}^{+  0.1}$,${ 0.24}_{-  0.21}^{+  0.51}$,${ 0.1}_{-  0.3}^{+  0.4}$,-,-,${ 108.1}_{-  23.2}^{+  5.6}$,${ 17}_{-  13}^{+  27}$
IMRpT,${ 2.708}_{-  0.003}^{+  0.004}$,${ 2.9}_{-  0.8}^{+  0.9}$,${ 5.5}_{-  0.9}^{+  0.9}$,${ 1.9}_{-  0.2}^{+  0.3}$,${ 7.3}_{-  0.6}^{+  0.7}$,${-0.18}_{-  0.11}^{+  0.09}$,${ 0.42}_{-  0.14}^{+  0.16}$,${-0.4}_{-  0.1}^{+  0.1}$,${ 0.66}_{-  0.50}^{+  0.29}$,${ 0.4}_{-  0.3}^{+  0.3}$,${ 262}_{-  247}^{+  801}$,${ 14.9}_{-  4.5}^{+  3.9}$,${ 112.8}_{-  7.2}^{+  3.5}$,${ 8}_{-  6}^{+  16}$
XHM,${ 2.720}_{-  0.004}^{+  0.005}$,${ 6.4}_{-  0.4}^{+  0.5}$,${ 8.5}_{-  0.3}^{+  0.4}$,${ 1.3}_{-  0.0}^{+  0.0}$,${ 9.9}_{-  0.3}^{+  0.3}$,${ 0.03}_{-  0.03}^{+  0.03}$,${ 0.03}_{-  0.03}^{+  0.06}$,${ 0.0}_{-  0.1}^{+  0.1}$,${ 0.14}_{-  0.14}^{+  0.41}$,${ 0.1}_{-  0.4}^{+  0.5}$,-,-,${ 100.5}_{-  8.0}^{+  6.4}$,${ 29}_{-  7}^{+  7}$
\end{filecontents*}
\csvreader[head to column names]{CSV/Param_BHNSq6s0_SNR70_inc30.csv}{}%
{\\ [0.5ex]\midrule \texttt{\Approximant}&\mc&\q&\mBH&\mNS&\mtotal&\aeff&\aBH&\aBHz&\aNS&\aNSz&\lambdaNS&\rNS&\dist&\thetajn}
\\
\bottomrule\bottomrule
\end{tabularx}
 \caption{Inclination 30\dg.}
 \end{subtable}

\begin{subtable}[!htbp]{1.1\textwidth}
		\centering
\begin{tabularx}{1.1\textwidth}{c|*{14}{b}}
\toprule\toprule
Approximant&$\mcs/\msun$&q&$\mBHs/\msun$&$\mNSs/\msun$&$\mts/\msun$&\chieff&\sBH&\sBHz&\sNS&\sNSz&\tidal&\rns/km&\dl/Mpc&\inc/\dg
\begin{filecontents*}{CSV/Param_BHNSq6s0_SNR70_inc70.csv}
Approximant,mc,q,mBH,mNS,mtotal,aeff,aBH,aBHz,aNS,aNSz,lambdaNS,rNS,dist,thetajn
True value,{ 2.75},{ 6.0},{ 8.32},{ 1.39},{ 9.71},{ 0.00},{ 0.0},{ 0.0},{ 0.0},{ 0.0},{ 526},{ 13.3},{ 40.9},{ 70}
SEOB,${ 2.727}_{-  0.014}^{+  0.025}$,${ 6.2}_{-  0.4}^{+  0.4}$,${ 8.4}_{-  0.3}^{+  0.3}$,${ 1.4}_{-  0.0}^{+  0.0}$,${ 9.7}_{-  0.3}^{+  0.3}$,${ 0.01}_{-  0.03}^{+  0.03}$,${ 0.02}_{-  0.02}^{+  0.07}$,${ 0.0}_{-  0.1}^{+  0.1}$,${ 0.14}_{-  0.13}^{+  0.43}$,${ 0.0}_{-  0.4}^{+  0.5}$,-,-,${ 87.1}_{-  42.1}^{+  24.6}$,${ 41}_{-  30}^{+  27}$
SEOBT,${ 2.725}_{-  0.014}^{+  0.025}$,${ 5.3}_{-  0.6}^{+  0.7}$,${ 7.7}_{-  0.5}^{+  0.6}$,${ 1.4}_{-  0.1}^{+  0.1}$,${ 9.2}_{-  0.5}^{+  0.5}$,${-0.04}_{-  0.05}^{+  0.05}$,${ 0.08}_{-  0.08}^{+  0.14}$,${-0.1}_{-  0.1}^{+  0.1}$,${ 0.24}_{-  0.23}^{+  0.56}$,${ 0.2}_{-  0.5}^{+  0.6}$,${ 973}_{-  911}^{+  2407}$,${ 14.2}_{-  4.8}^{+  4.3}$,${ 87.6}_{-  42.0}^{+  24.5}$,${ 41}_{-  30}^{+  27}$
SEOBNSBH,${ 2.725}_{-  0.013}^{+  0.026}$,${ 6.0}_{-  0.4}^{+  0.3}$,${ 8.2}_{-  0.3}^{+  0.3}$,${ 1.4}_{-  0.0}^{+  0.0}$,${ 9.6}_{-  0.3}^{+  0.2}$,${ 0.00}_{-  0.03}^{+  0.03}$,${ 0.01}_{-  0.01}^{+  0.03}$,${ 0.0}_{-  0.0}^{+  0.0}$,${ 0.00}_{-  0.00}^{+  0.00}$,${ 0.0}_{-  0.0}^{+  0.0}$,${ 986}_{-  829}^{+  807}$,${ 13.7}_{-  3.5}^{+  1.9}$,${ 89.5}_{-  43.0}^{+  22.5}$,${ 39}_{-  28}^{+  28}$
IMRNSBH,${ 2.726}_{-  0.014}^{+  0.026}$,${ 6.2}_{-  0.4}^{+  0.4}$,${ 8.4}_{-  0.3}^{+  0.3}$,${ 1.4}_{-  0.0}^{+  0.0}$,${ 9.7}_{-  0.3}^{+  0.3}$,${ 0.01}_{-  0.03}^{+  0.03}$,${ 0.02}_{-  0.02}^{+  0.03}$,${ 0.0}_{-  0.0}^{+  0.0}$,${ 0.00}_{-  0.00}^{+  0.00}$,${ 0.0}_{-  0.0}^{+  0.0}$,${ 1350}_{-  1106}^{+  1672}$,${ 14.2}_{-  3.6}^{+  2.8}$,${ 89.7}_{-  43.2}^{+  23.5}$,${ 40}_{-  29}^{+  28}$
IMRp,${ 2.715}_{-  0.003}^{+  0.018}$,${ 6.4}_{-  0.3}^{+  0.4}$,${ 8.5}_{-  0.2}^{+  0.3}$,${ 1.3}_{-  0.0}^{+  0.0}$,${ 9.8}_{-  0.2}^{+  0.2}$,${ 0.03}_{-  0.02}^{+  0.03}$,${ 0.04}_{-  0.04}^{+  0.08}$,${ 0.0}_{-  0.0}^{+  0.1}$,${ 0.29}_{-  0.26}^{+  0.47}$,${ 0.1}_{-  0.3}^{+  0.3}$,-,-,${ 108.9}_{-  32.0}^{+  5.2}$,${ 16}_{-  12}^{+  34}$
IMRpT,${ 2.708}_{-  0.003}^{+  0.005}$,${ 2.6}_{-  0.8}^{+  1.0}$,${ 5.1}_{-  0.9}^{+  1.0}$,${ 2.0}_{-  0.3}^{+  0.4}$,${ 7.1}_{-  0.6}^{+  0.8}$,${-0.22}_{-  0.11}^{+  0.12}$,${ 0.47}_{-  0.16}^{+  0.24}$,${-0.4}_{-  0.2}^{+  0.1}$,${ 0.48}_{-  0.36}^{+  0.41}$,${ 0.3}_{-  0.2}^{+  0.3}$,${ 186}_{-  178}^{+  713}$,${ 14.9}_{-  4.4}^{+  4.2}$,${ 113.2}_{-  7.5}^{+  3.6}$,${ 8}_{-  6}^{+  17}$
XHM,${ 2.748}_{-  0.009}^{+  0.008}$,${ 6.4}_{-  0.4}^{+  0.5}$,${ 8.6}_{-  0.3}^{+  0.4}$,${ 1.3}_{-  0.0}^{+  0.0}$,${ 10.0}_{-  0.3}^{+  0.3}$,${ 0.03}_{-  0.03}^{+  0.02}$,${ 0.03}_{-  0.03}^{+  0.07}$,${ 0.0}_{-  0.1}^{+  0.1}$,${ 0.16}_{-  0.15}^{+  0.41}$,${ 0.1}_{-  0.4}^{+  0.5}$,-,-,${ 52.5}_{-  13.8}^{+  15.4}$,${ 64}_{-  9}^{+  8}$
\end{filecontents*}
\csvreader[head to column names]{CSV/Param_BHNSq6s0_SNR70_inc70.csv}{}%
{\\ [0.5ex]\midrule \texttt{\Approximant}&\mc&\q&\mBH&\mNS&\mtotal&\aeff&\aBH&\aBHz&\aNS&\aNSz&\lambdaNS&\rNS&\dist&\thetajn}
\\
\bottomrule\bottomrule
\end{tabularx}
 \caption{Inclination 70\dg.}
 \end{subtable}
 
\caption{Properties of the q=6, SNR 70 simulation, as recovered by the listed approximants. We report the 1D median, together with the symmetric the $90\%$ credible interval.}
\label{Table.Param_BHNSq6s0_SNR70}
\end{table} 

\end{turnpage}

\clearpage
\null
\begin{table}[!tbp]
\tiny
\centering
\begin{tabularx}{0.99\textwidth}{c|*{8}{b}}
\toprule\toprule
Approximant&$\mcd/\msun$&q&$\mBHd/\msun$&$\mNSd/\msun$&$s_1$&$s_2$&\tidal&\dl/Mpc
\begin{filecontents*}{CSV/Prior_BHNSq2s0.csv}
Approximant;mc;q;mBH;mNS;aBH;aNS;lambdaNS;dist
SEOB;[1.480,2.711];[1.0,8.0];[1.043, 9.664];[1.043, 9.664];[-0.99,0.99];[-0.99,0.99];-;[0,500]
SEOBT;[1.480,2.711];[1.0,8.0];[1.043, 9.664];[1.043, 9.664];[-0.99,0.99];[-0.99,0.99];[0,4000];[0,500]
SEOBNSBH;[1.480,2.711];[1.0,8.0];[1.043, 9.664];[1.043, 3.000];[-0.5,0.8];-;[0,5000];[0,500]
IMRNSBH;[1.480,2.711];[1.0,8.0];[1.043, 9.664];[1.043, 3.000];[-0.5,0.5];-;[0,5000];[0,500]
IMRp;[1.421,2.602];[1.0,8.0];[1.001, 9.277];[1.001, 9.277];[-0.99,0.99];[-0.99,0.99];-;[0,500]
IMRpT;[1.421,2.602];[1.0,8.0];[1.001, 9.277];[1.001, 9.277];[-0.99,0.99];[-0.99,0.99];[0,4000];[0,500]
XHM;[1.480,2.711];[1.0,8.0];[1.043, 9.664];[1.043, 9.664];[-0.99,0.99];[-0.99,0.99];-;[0,500]
\end{filecontents*}
\csvreader[head to column names,
  separator=semicolon]{CSV/Prior_BHNSq2s0.csv}{}%
{\\ [0.5ex]\midrule \texttt{\Approximant}&\mc&\q&\mBH&\mNS&\aBH&\aNS&\lambdaNS&\dist}
\\
\bottomrule\bottomrule
\end{tabularx}
 \caption{Prior bounds for the q=2 simulation.}
\label{Table.Prior_BHNSq2s0}
\end{table} 

\begin{table}[!tbp]
\tiny
\centering
\begin{tabularx}{0.99\textwidth}{c|*{8}{b}}
\toprule\toprule
Approximant&$\mcd/\msun$&q&$\mBHd/\msun$&$\mNSd/\msun$&$s_1$&$s_2$&\tidal&\dl/Mpc
\begin{filecontents*}{CSV/Prior_BHNSq3s0.csv}
Approximant;mc;q;mBH;mNS;aBH;aNS;lambdaNS;dist
SEOB;[1.776,3.253];[1.0,8.0];[1.252,11.597];[1.252,11.597];[-0.99,0.99];[-0.99,0.99];-;[0,500]
SEOBT;[1.776,3.253];[1.0,8.0];[1.252,11.597];[1.252,11.597];[-0.99,0.99];[-0.99,0.99];[0,4000];[0,500]
SEOBNSBH;[1.776,3.253];[1.0,8.0];[1.252,11.597];[1.252, 3.000];[-0.5,0.8];-;[0,5000];[0,500]
LEA+;-;[2.0,5.0];[3.0,7.2];[1.2,1.45];[-0.5,0.5];-;[0,4000];[0,500]
IMRNSBH;[1.776,3.253];[1.0,8.0];[1.252,11.597];[1.252, 3.000];[-0.5,0.5];-;[0,5000];[0,500]
IMRp;[1.421,2.602];[1.0,8.0];[1.001, 9.277];[1.001, 9.277];[-0.99,0.99];[-0.99,0.99];-;[0,500]
IMRpT;[1.421,2.602];[1.0,8.0];[1.001, 9.277];[1.001, 9.277];[-0.99,0.99];[-0.99,0.99];[0,4000];[0,500]
XHM;[1.776,3.253];[1.0,8.0];[1.252,11.597];[1.252,11.597];[-0.99,0.99];[-0.99,0.99];-;[0,500]
\end{filecontents*}
\csvreader[head to column names,
  separator=semicolon]{CSV/Prior_BHNSq3s0.csv}{}%
{\\ [0.5ex]\midrule \texttt{\Approximant}&\mc&\q&\mBH&\mNS&\aBH&\aNS&\lambdaNS&\dist}
\\
\bottomrule\bottomrule
\end{tabularx}
 \caption{Prior bounds for the q=3 simulation.}
\label{Table.Prior_BHNSq3s0}
\end{table}

\begin{table}[!tbp]
\tiny
\centering
\begin{tabularx}{0.99\textwidth}{c|*{8}{b}}
\toprule\toprule
Approximant&$\mcd/\msun$&q&$\mBHd/\msun$&$\mNSd/\msun$&$s_1$&$s_2$&\tidal&\dl/Mpc
\begin{filecontents*}{CSV/Prior_BHNSq6s0.csv}
Approximant;mc;q;mBH;mNS;aBH;aNS;lambdaNS;dist
SEOB;[2.184,4.016];[1.000,17.944];[1.000,22.953];[1.000,22.953];[-0.99,0.99];[-0.99,0.99];-;[0,500]
SEOBT;[2.184,4.016];[1.000,17.944];[1.000,22.953];[1.000,22.953];[-0.99,0.99];[-0.99,0.99];[0,4000];[0,500]
SEOBNSBH;[2.184,4.016];[1.000,17.944];[1.000,22.953];[1.000,3.000];[-0.5,0.8];-;[0,5000];[0,500]
IMRNSBH;[2.184,4.016];[1.000,17.944];[1.000,22.953];[1.000,3.000];[-0.5,0.5];-;[0,5000];[0,500]
IMRp;[2.184,4.016];[1.000,8.000];[1.001,14.317];[1.001,14.317];[-0.99,0.99];[-0.99,0.99];-;[0,500]
IMRpT;[2.184,4.016];[1.000,8.000];[1.001,14.317];[1.001,14.317];[-0.99,0.99];[-0.99,0.99];[0,4000];[0,500]
XHM;[2.184,4.016];[1.000,17.944];[1.000,22.953];[1.000,22.953];[-0.99,0.99];[-0.99,0.99];-;[0,500]
\end{filecontents*}
\csvreader[head to column names,
  separator=semicolon]{CSV/Prior_BHNSq6s0.csv}{}%
{\\ [0.5ex]\midrule \texttt{\Approximant}&\mc&\q&\mBH&\mNS&\aBH&\aNS&\lambdaNS&\dist}
\\
\bottomrule\bottomrule
\end{tabularx}
 \caption{Prior bounds for the q=6 simulation.}
\label{Table.Prior_BHNSq6s0}
\end{table}

\clearpage

\bibliographystyle{apsrev4-1}
\bibliography{ThisPaperBib}
\end{document}